\documentclass[12pt,a4paper,oneside,bibliography=totoc]{scrbook}
\usepackage{amsmath,amsthm,amssymb}   
\usepackage{hyperref}
\usepackage{float}
\usepackage{xcolor} 
\usepackage{tikz}
\usepackage{pgfplots}
\usepackage{amsmath,amssymb,amstext}
\usepackage{mathtools}
\usepackage{graphicx}
\usepackage{caption}
\usepackage{capt-of}
\usepackage{textcomp}
\usepackage{paralist}
\usepackage[vcentermath]{youngtab}
\usepackage{dsfont}
\usepackage{float}
\usepackage{latexsym}
\usepackage[english]{babel}
\usepackage[ansinew]{inputenc}
\usepackage{setspace}
\usepackage{lmodern}
\usepackage[T1]{fontenc}
\usepackage{amsmath}
\usepackage{tikz}

\newcommand{\Ms}{M_{\mathrm{s}}}

\begin{document}
\pagestyle{plain}
\frontmatter
\thispagestyle{empty}
\begin{flushleft}
{\Huge\textbf{Lectures on ultrathin film ferromagnetism.}}\\
\rule{\textwidth}{2mm}
\end{flushleft}
\begin{center}
\vspace{2cm}
by\\
\vspace{2cm} 
\Large{D. Pescia\\
Laboratory for Solid State Physics, ETH Zurich, 8093 Zurich, Switzerland\\
e-mail: pescia@solid.phys.ethz.ch\\
\url{https://orcid.org/0000-0001-7436-418X}\\
\vspace{1cm}
August, 2026}
\end{center}

\chapter{Abstract}
\pagestyle{plain}
These Lecture Notes report the content of a set of elective lectures held
repeatedly for the students of the Departments of Physics and of Materials during
my tenure at ETH Zurich.\\
In these Lecture Notes, we review some of the fundamental principles that have emerged from research on the ferromagnetism of ``ultrathin'' films consisting of
3d transition-metal overlayers. The adjective ``ultrathin'' means that the
thickness of the films amounts to only a few atomic layers. Their growth, as
determined e.g. by Scanning Tunneling Microscopy and Electron Diffraction,
is often layer-by-layer. This growth mode produces quantum wells along the vertical direction that profoundly impact any physical property of the materials. In addition, the vertical confinement establishes spin ensembles that extend to macroscopic distances along the in-plane directions and are finite along the vertical (perpendicular) direction, i.e. they are two-dimensional. Accordingly, they display ground state properties
that originate from the two-dimensionality, such as ``dead'' magnetic layers
or ``enhanced magnetic moments'', an oscillatory interlayer magnetic coupling
and an anomalous perpendicular versus in-plane magnetic anisotropy that
produces, in some specific situations, a perpendicular collective orientation
of the spins. At finite temperatures, ferromagnetic order is observed to
persist and an analysis of the magnetic order of ultrathin films in terms of
the renormalization group provides a suitable framework for explaining this
observation. The ferromagnetic order is lost at a phase transition which
follows closely the two-dimensional Ising universality class, as shown by an
accurate analysis of data in the vicinity of the critical point. The
perpendicular spin orientation is often observed to turn in-plane by a
reorientation phase transition which is also properly described by a
renormalization group argument. Finally, the perpendicular spin orientation
introduces topological excitations of the ferromagnetic order, consisting of
``stripes'' of reversed perpendicular spin direction. The stripe order
undergoes a phase transition to the paramagnetic state that is not yet
completely understood.

\tableofcontents
\mainmatter

\chapter{Introduction}

\begin{figure}[H]
\begin{center}
\includegraphics[width=0.8\textwidth]{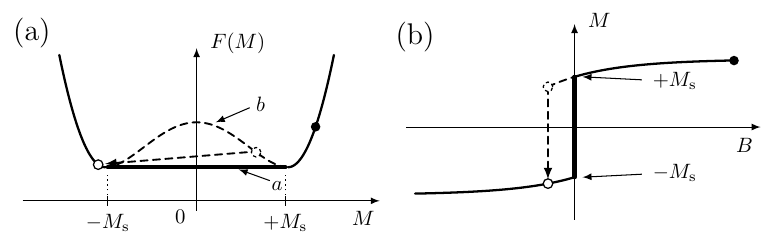}
\caption{(a) Sketch of the Helmholtz free energy as a function of $M$. The thick
continuous curve is the equilibrium free energy: it has a flat portion joining
the two values $\pm \Ms$ of the spontaneous magnetization (path a), see
Ref.~\cite{LL_spont}. Within Landau theory the interval $[-\Ms,+\Ms]$ admits, in
addition, the higher lying dashed curve (path b), along which a barrier separates
the two minima. (b) Sketch of the graph of $M$ as a function of the applied field
$B$. Along the equilibrium path a, $M$ jumps by $2\Ms$ at $B=0$ (thick vertical
line). Along path b the magnetization persists in a metastable state (dashed
curve) until it decays, at a finite negative field -- the coercive field -- into
the equilibrium state with reversed magnetization (dashed arrow). The full circle
marks the starting point of the ``experiment'' described in the text, the dashed
open circle the metastable state, the open circle the final equilibrium state.
The same three states are marked in both panels.}
\label{Fig:spont}
\end{center}
\end{figure}

\begin{figure}[htb]
  \centering
  \includegraphics[width=0.6\textwidth]{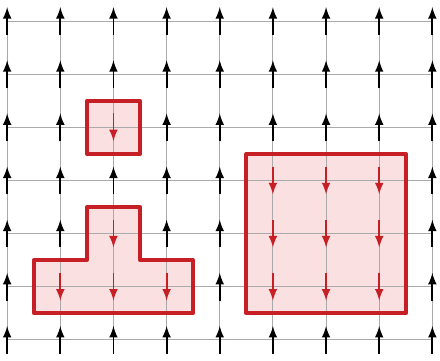}
  \caption{A state of the two-dimensional Ising model, of the type used by
  Peierls in his proof of the existence of a spontaneous magnetization\cite{Pei}:
  within a sea of aligned spins (black), droplets of overturned spins (red) are
  enclosed by closed boundaries. Along each unit of boundary length two
  antiparallel spins meet, so that a droplet costs an exchange energy
  proportional to the length of its boundary.}
  \label{fig:droplets}
\end{figure}

\paragraph{Materials.} The ultrathin films and multilayers dealt with in these
Lectures are mainly built by overlayers, a few atomic layers thick, of the
ferromagnetic 3d transition metals Fe, Co and Ni deposited onto non-magnetic
substrates such as the noble metals Cu, Ag and Au\cite{Brune,Thamm}. The essential knowledge about the epitaxial
growth of transition metal ultrathin films is summarized in Chapter 2. That Chapter informs the reader about the overall morphology of the metallic deposits, and appropriate references will guide the reader to the specialized literature on the subject. It is important because it informs about the state of the art of epitaxial transition metal growth, to be compared with the entirely new class of two-dimensional systems, introduced at the beginning of the 2020's and produced by the method of exfoliation of bulk samples\cite{Gong,Huang,Burch,Li,Novo,Dai}. These novel two-dimensional materials have a morphology that fundamentally contrasts with the morphology of epitaxially grown transition metal ultrathin films: they are almost perfectly flat over lateral sizes of many hundreds of $\mu$m and therefore achieve a two-dimensionality that is unthinkable with epitaxial transition metal films. Despite the contrasting morphology, the physical properties that have emerged from the study of the exfoliated materials resemble very much the leitmotifs encountered during the many years of study of epitaxial films.
Because of the similarity of the fundamental physical aspects, these Notes are  bound to provide a research insight that can be used immediately for
dealing with the exfoliated type of materials as well.
\paragraph{The spontaneous magnetization in ultrathin films.}
We introduce, right at the beginning of these Lectures, the key experimental
finding that has allowed research on transition metal ultrathin films to
provide new insights into condensed matter physics: the existence of a
spontaneous magnetization\cite{LL_spont,Argentina} in transition metal
ultrathin films. This concept is not an obvious one and its treatment has occupied statistical physics for many decades, indeed centuries, starting from a famous paper by Maxwell in which he described the celebrated Maxwell construction\cite{Maxwell_cont}. To define the spontaneous magnetization we refer to the Landau Theory of Phase Transitions\cite{LL_spont}. This theory allows computing, phenomenologically, the thermodynamic potential called Helmholtz free
energy of a finite volume ferromagnet. We designate this potential as $F(M,T)$, with $M$ being the magnetization. Within Landau theory, and below the critical temperature separating the paramagnetic (spin disordered) phase from the ferromagnetic phase (with the collective order of the spins), the ferromagnetic order is characterized by an $F(M)$ that follows the thick black, respectively the dashed line in Fig.~\ref{Fig:spont}(a).
\noindent How do we read the graphs (a) and (b)? Have in mind that the derivative of the $F(M)$ with respect to $M$ is the applied field $B$ and start a measurement of $M$ from some point on the right hand side of
$F(M)$ (full circle). At that point, the value of $M$ corresponds uniquely to a
positive value of the applied field. The point is reproduced in the part (b) of
Fig.~\ref{Fig:spont}, which plots $M$ versus $B$. Continue the experiment by
moving from the full circle toward the left hand side: $F(M)$ and $M(B)$
follow the black continuous curve. At $B=0$ the magnetization assumes a value
$+\Ms$. This value defines a possible value for the spontaneous magnetization:
$\Ms(B=+0)$. A second possible value for the spontaneous magnetization is observed when one starts from the left
hand side of the continuous graph in (a) and moves toward $B=-0$, $\Ms(B=-0)$. But let us continue our ``experiment''. When one arrives at $\Ms(B=+0)$, Fig.~\ref{Fig:spont}(a) and (b) indicate two possible paths our experiment could follow in the interval $[-\Ms,+\Ms]$, both leading to magnetization reversal.\\
To discuss path b (dashed line) and path a (thick continuous line) we refer to the states used by Peierls in his original proof of the existence of a spontaneous magnetization in the two-dimensional Ising model\cite{Pei}, as a way of rendering the actual states of a spin system, see Fig.~\ref{fig:droplets}.
Reversing the magnetization along b proceeds by nucleating droplets of overturned spins (red), each of which pays a free energy proportional to the length of its boundary. At the boundary, in fact, two opposite spins meet and this meeting costs exchange energy. The droplet (or \textbf{domain}) state (dashed circle in (a) and (b)) is, however, metastable\footnote{The states close to the maximum of the dashed curve are, strictly speaking, forbidden in Nature as they are thermodynamically unstable.}: its free energy is higher than the free energy of an equilibrium solution at the same negative applied field (full circle in (a) and (b)). Accordingly, after some relaxation time, the domain state decays (``switches'') into the equilibrium state with reversed magnetization.
This decay is indicated by a dashed arrow in (a) and (b) and occurs at a finite negative magnetic field. This field is called the \textbf{coercive field}. One therefore expects, because of the existence of a spontaneous magnetization, that the $M(B)$-curve follows a so called hysteresis loop, where $M$ is not a unique function of $B$. This is actually the hallmark of ferromagnetic order. One can find a clear example for this behaviour in a reference system for ferromagnetic transition metal films, i.e. overlayers of Fe on W(110)\cite{Vaterlaus}.\footnote{The nature of the metastable states and the processes that lead to their decay are the subject of a wide field of research and cannot be discussed in these Lectures.} The hysteresis curve of overlayers of Fe on W(110) has a sharp-edged rectangular shape, typical, actually, of the hysteresis curve in transition metal overlayers. Magnetic imaging\cite{Vaterlaus} reveals that both states on either side of the vertical line are single domain within the resolution of the experiment. The exact nature of the almost vertical switch at a well defined coercive field has so far not been investigated in detail. One suspects that small defects with reversed spins, appearing e.g. at the border of practical samples, act as nucleation centres for the opposite magnetization state. One thinks that these defects are pinned and only invade the entire sample at a well defined negative magnetic field, at which they suddenly depin.\\
Path a in (a) (the thick horizontal line), along which $\Ms(B=+0)$ switches to
$\Ms(B=-0)$ exactly at $B=0$ (thick vertical line in (b)) has similar droplet
states as path b but with an essential difference: building domains of reversed
magnetization along the horizontal line does not cost free energy. Why is this
the case? Because path a is the free energy in the thermodynamic limit of
infinite volume, where the length of the boundary has a vanishing ``weight'' with
respect to the area of the system. Path a is therefore the equilibrium free
energy. Mathematically, it is the ``convex'' envelope of the Landau free energy.
The equilibrium curve was anticipated by Maxwell\cite{Maxwell_cont} but the
definitive proof of it was given in a famous paper by Lebowitz and
Penrose\cite{Penrose}.

\paragraph{Fundamental aspects.} Epitaxial transition metal overlayers are  complex materials and the result of epitaxial growth is a bouquet of epitaxial systems that can have minute and sometimes even unresolved differences. Accordingly, research on this topic has produced a myriad of results which underline various aspects of ferromagnetism, some of them being so peculiar that they are still unexplained. It is not the aim of these Lectures to provide an exhaustive overview of all possible
results and of the subtle aspects that have emerged in the past years. There is a
number of Review papers that, taken together, cover the variety of acquired results, specific to each material investigated. These Reviews will be quoted at the appropriate place. These Lectures attempt something different: they synthesize, from the wealth of disparate results, those main fundamental aspects that have a universal character and can be, accordingly, captured by simple models. In these Lectures, we would like to dispense with the
unnecessary details, to let the main principles of ultrathin film ferromagnetism,
which will defeat oblivion, emerge. The reader will learn that apparently complex aspects of the ferromagnetism of ultrathin films can be understood by ``back of the envelope'' arguments that find their roots in e.g. elementary quantum mechanics.\\
One can roughly identify seven main leitmotifs that have emerged from the research on the ferromagnetism of epitaxial 3d transition metal ultrathin films.\\
\textit {I. ``Dead'' layers and enhanced magnetic moments.} Thin films of
ferromagnetic materials were actually the subject of a pioneering paper by L.
N\'eel in the early 1950's\cite{Neel}. In this paper, the property of an enhanced
magnetic anisotropy produced by the breaking of the translational symmetry along
the direction perpendicular to the surface was predicted. A significant paper by
U. Gradmann\cite{Grad} reported early experimental results on thin magnetic
films. Simultaneously to Gradmann's work, the technology of Molecular Beam
Epitaxy was introduced during the 1970's decade, with the aim of producing
novel semiconductor nanostructures\cite{MBE,Dots}. In the
1980's, it was realized that this technology could be also applied to grow metal
overlayers of highest precision. Together with the invention of the Scanning
Tunneling Microscope by G. Binnig and H. Rohrer\cite{STM}, metal MBE developed
into a main-stream surface science method to achieve the nanoscale accuracy that was envisaged by N\'eel\cite{Brune}. This precision allowed controlling, at the atomic scale, the thickness of the deposit, from the submonolayer range up to the few
monolayer range\footnote{1 ML corresponds to a layer with the thickness of the
size of one single atom. We therefore use also the terminology ``1'' atomic
layer (AL)}.\\
At the few ML thickness range, overlayers of the ferromagnetic 3d transition
metals, extending over macroscopic in-plane lengths, develop physical
properties that are dictated by their electron system  being quantum confined between vacuum on one side and a non-magnetic surface on the other\cite{Roy}. One of such properties are the so
called ``dead magnetic layers'', in which the atoms completely lose their
atomic magnetic moment. This was the content of two  famous papers by
Liebermann et al., in 1969, reporting on magnetic dead layers of Fe and Ni films
with atomic scale thickness deposited on top of a Cu single crystal surface\cite
{Lieber}. It was suggested that the contact with an almost free electron gas
represented by the non-magnetic Cu substrate produces a quenching of the local magnetic moments of Fe and Ni (known to exist in Fe and Ni atoms and in their three-dimensional assembly). It took almost 16 years to provide accurate
computations of the ground state energy of magnetic overlayers of 3d transition
metals on noble metal surfaces. These computations\cite{Freeman}\footnote{The precise band structure calculations were based on the local spin density approximation, a computational technology developed by W. Kohn\cite{Kohn} et al.} did explain
the existence of dead layers. However, they also found a second and most surprising aspect of ultrathin magnetic films, i.e. situations\footnote{Most of the 3d bcc transition metal overlayers and/or sandwiches and superlattices with Au and Ag show this phenomenon.} in which the magnetic moments in single monolayers were enhanced with respect to their bulk
counterparts. Although research on ultrathin films of transition metals was
actually initiated by the reports that, in the limit of only few atomic layers, films of Fe and Ni deposited on noble metals were actually magnetically ``dead'', it was the perspective of finding enhanced magnetic moments that inspired many researchers to enter the field. Chapter 3 deals specifically with the subject of local magnetic moments in transition metal overlayers.\\
\textit{II. N\'eel magnetic anisotropy.} Chapter 4 introduces a fundamental interaction that played a key role in the research on transition metal ultrathin films: the single ion N\'eel\cite{Neel} magnetic anisotropy. It originates from the highly asymmetric environment provided by the breaking of the translational symmetry perpendicular to the film plane. A landmark paper by Gay and Richter\cite{Gay,GR} produced, many years after N\'eel predicted it, a precise computation of this interaction for the monolayer of Fe. This paper motivated a string of theoretical and experimental work dealing with the magnetic anisotropy in transition metal thin films.\\
\textit{III. Interlayer coupling.} The research on ultrathin film magnetism
received a new twist in the 1980's: the discovery that two atomically thin
films separated by a non-magnetic interlayer can couple ferromagnetically or
antiferromagnetically and that the coupling could be controlled by suitably
tuning various parameters\cite{Grun}. The immediate discovery that followed --
the Giant Magnetoresistance originating from the coupling -- was finally awarded the Nobel prize in 2007\cite{Fert}. Chapter 4 explores as well, by simple modeling, the interlayer magnetic coupling as originating from the interatomic exchange interaction. Interlayer coupling introduced a novel, previously unobserved phenomenon in  magnetism and paved the way to important technological applications known as ``Spintronics'' and ``Magnonics''\cite{Spintronics}\footnote{These topics go beyond the scope of these Lectures.}. One can safely state that interlayer coupling actually turned the tide in the research on the magnetism of transition metal overlayers and rendered this topic a central one in condensed matter physics.\\
\textit{IV. The spontaneous magnetization in ultrathin films.}
A most extraordinary theorem of mathematical physics\cite{Mermin} rules out
long ranged ferromagnetism or antiferromagnetism in one- and two-dimensional systems with continuous rotational symmetry. In contrast, the
famous exact solution of the  two-dimensional Ising\cite{Ising} model by Onsager\cite{Ons} predicts a spontaneous magnetization at finite temperatures and a second order phase transition to a disordered state. The observations regarding the spontaneous magnetization in ultrathin films point to very small symmetry breaking components being able to install a conventional long range order in ultrathin films. In Chapters 6 and 7  we expand on the Mermin-Wagner theorem and use the Polyakov renormalization group algorithm to deal with symmetry breaking interactions and the finite temperature ferromagnetism of two-dimensional transition metal ultrathin films.\\
\textit{V. The reorientation transition.}
In some particular materials, like Fe on Cu(100)\cite{Pescia_Fe_Cu,FMR}, the perpendicular single ion magnetic anisotropy produces a perpendicular orientation of the spins in the state of spontaneous magnetization. Inherent to the perpendicular spin orientation is the existence of a
new type of phase transition, associated with the turning of the preferential
spin direction from perpendicular to in-plane -- the ``reorientation phase
transition''\cite{PP,Pappas,Rolf}. Chapters 7 and 8 discuss the ``reorientation'' transition in ultrathin films with perpendicular spin orientation. The reorientation transition originates from the competition between the N\'eel magnetic anisotropy and the dipolar interaction, through a mechanism which is best explained within Polyakov's renormalization group algorithm. These Chapters contain also new ideas on how the transition evolves as a function of the thickness.\\
\textit{VI. Phase transitions in two-dimensional systems.}
The specific geometry
of ultrathin ferromagnetic films -- extremely thin in one spatial
direction, but extended over macroscopic sizes along the remaining two
directions -- allows a further fundamental aspect to emerge.  As
ferromagnetism is the result of a collective behaviour of
matter, this geometry opens the possibility of exploring those aspects of
collective order that are specific to two-dimensional systems, such as the
behaviour at phase transitions\cite{Ons,Back_Nature}. Chapter 9 reports on the phase transition in ultrathin films. Phase
transitions are, technically speaking, the subject of both the Renormalization
Group method\cite{Wilson,Poki} and the phenomenological approach known as
``scaling hypothesis''\cite{Scale,Poki}. Chapter 10 explains how these tools
apply to the situation of ultrathin films.\\
\textit{VII. The stripe domain structure.}
Ferromagnetism is typically discussed in terms of the
exchange interaction and magnetic single ion anisotropies. Yet real samples are inevitably affected by the magnetostatic dipole-dipole interaction. Because of this interaction, a theorem  by Griffiths\cite{Griff_dipole} forbids a spontaneous
magnetization in, nota bene, three-dimensional bodies.
This means that, strictly speaking, ferromagnetism, in  three-dimensional
bodies, is actually only local and globally the magnetization must vanish by some topological excitations. In these Lectures (Chapter 11) we discuss some topological aspects of ultrathin magnetic films of transition metal overlayers, related to the presence of alternating stripe domains with opposite perpendicular spins. These stripes undergo a
complex phase transition, either to an in-plane orientation or to the
paramagnetic phase\cite{NiculinPRB}. The properties of the stripes and their
disordering processes are not yet fully understood, so that these Lectures
should provide the proper key for exploring this active field of research.\\
A special remark: Chapter 5 deals with establishing those energy functionals that are most useful to describe magnetic properties of ultrathin ferromagnetic films. These functionals are used e.g. in the Landau-Lifshitz equation of motion of the local spin. We discuss the application of the Landau-Lifshitz equation in determining the various magnetic coupling parameters and list typical values for parameters characteristic of ultrathin transition metal thin films.\\
A final chapter is still being written: the community is learning how to use the spin of the electrons for performing various operations which are usually performed so far exploiting their
charge -- this new chapter of physics, called ``Spintronics'', is just emerging as
a strong new direction of research. Paradoxically, all emerged from the finding
of ``dead magnetic layers'', showing how results, also beneficial to technology
and society at large, can emerge even from apparently sterile results, provided
one is curious enough to take some risks.

\chapter{Morphological aspects of transition metal overlayers.}
\label{Growth}
\begin{figure}[H]
\begin{center}
\setlength{\fboxsep}{1pt}
	\setlength{\fboxrule}{1pt}
\includegraphics*[width=0.7\textwidth]{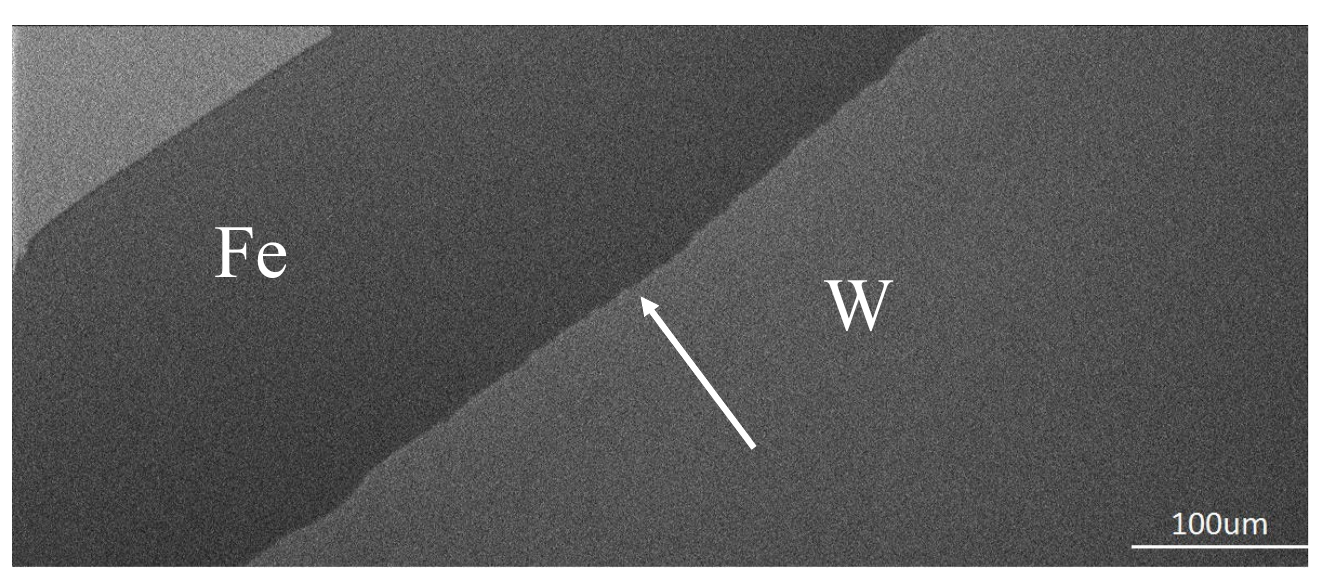}
\caption{View from above of a section of a 2-3 ML thick Fe stripe (dark) on W(110) (brighter), imaged by Scanning Electron Microscopy. In this specific experiment, the primary SEM-beam has an energy of about 10 keV. When the electron beam hits the surface, it produces a set of scattered electrons. The scattering cross section is material dependent, producing a contrast between the intensity of the electrons emitted from Fe and the intensity of the electrons emitted from W. This contrast is seen in the Figure. The white arrow indicates the direction along which the STM scans of Fig.~\ref{Fig:STM_Fe} are performed. The scale bar corresponds to 100 $\mu$m. This figure and the next one were obtained during the final years of my tenure at ETH Zurich by members of my group.  I acknowledge the collective effort of the latest members of the team: A.-K. Thamm, J. Wei, J. Zhou, C. G. H. Walker, H. Cabrera, M. Demydenko, and, most prominently, U. Ramsperger. External help came from A. Suri, A. Pratt, S. P. Tear and M. M. El-Gomati. 
}
\label{Fig:SEM_Fe}
\end{center}
\end{figure}

\begin{figure}[H]
\begin{center}
\includegraphics*[width=0.8\textwidth]{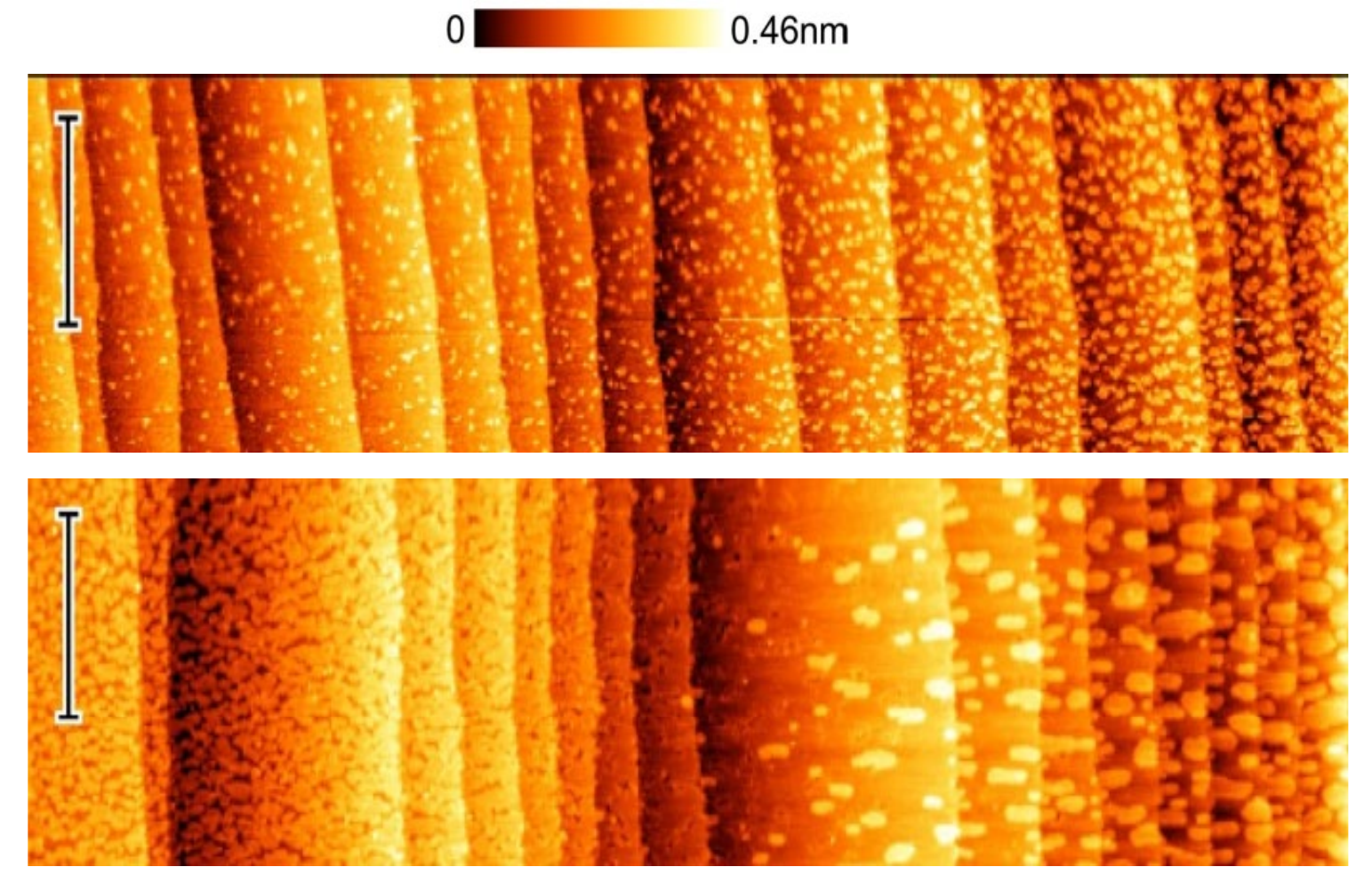}
\caption{Images of a section of the boundary between Fe and W. The nominal thickness of the Fe film increases from left to right and from the top image to the bottom image, from the submonolayer range on the left-hand side of the top image to $\approx$ 2 ML on the right-hand side of the bottom image. The length of the STM images is 500 nm; the white vertical bar in each image corresponds to 100 nm. Experiments are performed at room temperature. STM is performed in constant current mode, i.e., during the scan, the tip displaces vertically to keep the tunneling current constant. The vertical displacement is, thus, a direct measure of the surface
topography. The vertical displacement is encoded with the colour scale shown above the images. The colour scale spans 0.46 nm, i.e. about two atomic steps -- the interlayer distance of W(110) is 0.224 nm -- so that the number of completed layers can be read off the colours directly.
}
\label{Fig:STM_Fe}
\end{center}
\end{figure}

\begin{figure}[H]
\begin{center}
\def\primaryE{14}
\begin{tikzpicture}
  \node[anchor=south west,inner sep=0] (qw) at (0,0)
       {\includegraphics[width=0.7\textwidth]{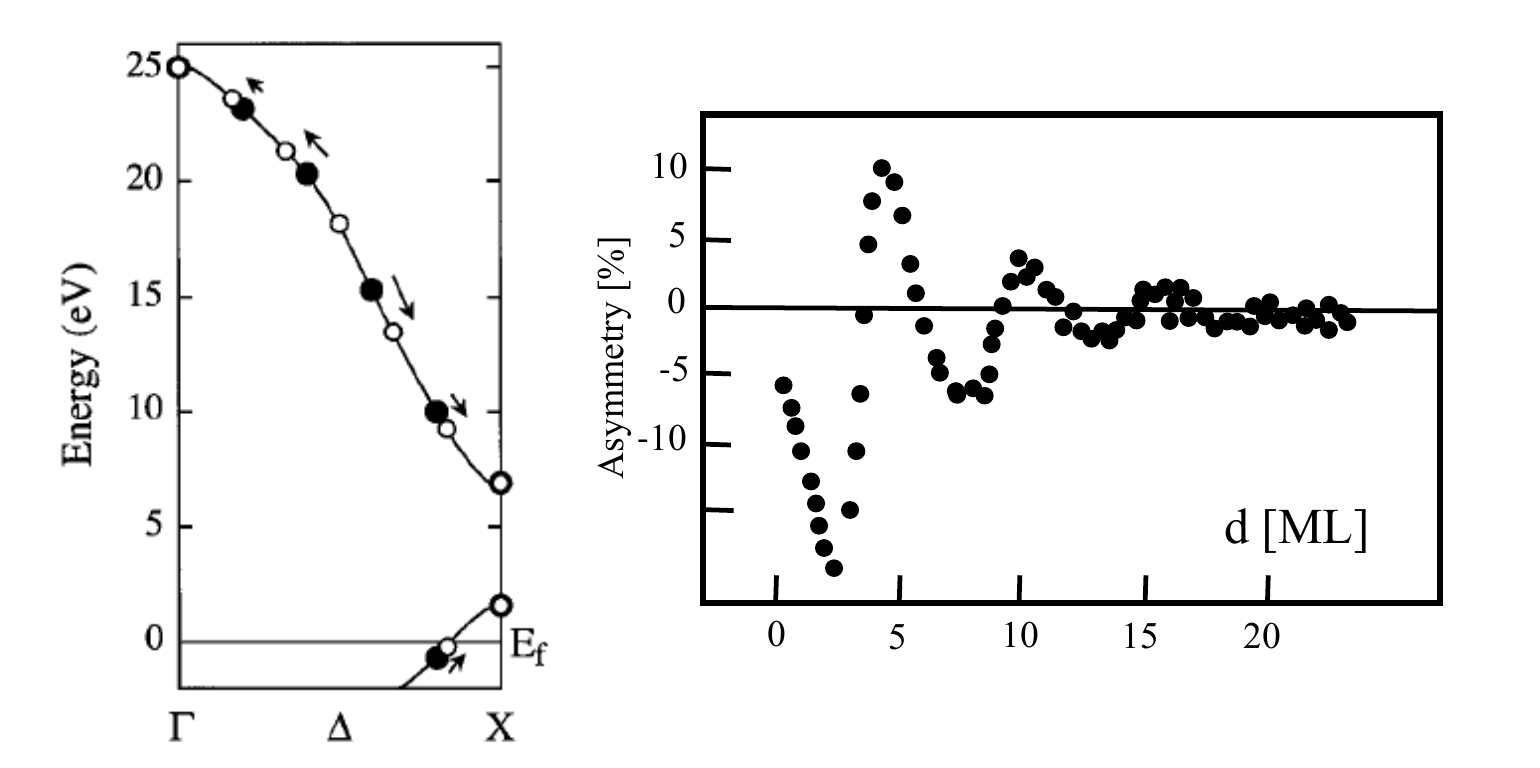}};
  \begin{scope}[x={(qw.south east)},y={(qw.north west)}]
    \draw[dashed,line width=0.6pt]
      (0.147,{0.135+0.0308*\primaryE}) -- (0.325,{0.135+0.0308*\primaryE});
    \node[anchor=west,inner sep=1pt,font=\footnotesize]
      at (0.330,{0.135+0.0308*\primaryE}) {$E_{\mathrm{p}}$};
  \end{scope}
\end{tikzpicture}
\caption{Left: The free electron band along the $\Delta$ direction in Cu. The confinement produces quantum well states (black dots). Their position changes (open dots) when the thickness of the Cu-overlayer is varied; the arrows indicate the direction in which the states move as the thickness increases. The dashed line marks the energy $E_{\mathrm{p}}$ of the primary electrons, which is kept fixed while the thickness is changed: the oscillations of the right panel arise as the quantum well states sweep through it. Right: Oscillatory component of the reflected intensity as a function of the thickness $d$ (in ML) of the Cu-top layer. The use of spin polarized electrons (plotted on the vertical scale is the asymmetry between the reflected intensity of incoming spin-up and spin-down electrons) serves the purpose of enhancing the oscillatory component over the monotonous reflected background.  Adapted from Ref.\cite{Egger}.}
\label{Fig:QW}
\end{center}
\end{figure}

\paragraph{Molecular Beam Epitaxy.} Molecular Beam Epitaxy (MBE)\cite{MBE} is a material science process in which
atoms and molecules are evaporated from pure sources and are transported as a
beam in an Ultra High Vacuum\cite{UHV} environment toward a target. MBE was
invented, originally, for semiconductor technology applications, where it has been
used routinely since the beginning of the 1970's\cite{MBE,Dots}. However, it turned out to be a valuable method for growing metal overlayers on a metallic
substrate (see e.g. Refs.~\cite{Venable,Brune,Thin} and the numerous references
therein). Once the atoms or molecules arrive at the surface of the target, they
build a deposit. Depending on the growth mode\cite{Venable,Brune}, the
interfaces between the deposit and vacuum on one side, deposit and substrate on
the other, are more or less well defined. Deposits with optimal flatness are
typically obtained when the growth mode is ``layer-by-layer'': the next atomic
layer starts to grow only after the one being deposited is complete.
By its very nature, MBE is used not only to grow standard crystalline
structures but is also intended to achieve completely novel materials
whose geometry, morphology and composition are defined at will, on the
nanoscale\cite{Dots}. One example of such nanoscale-defined crystals are ultrathin films of Fe on W(110). Because of its magnetic properties, the system Fe/W(110) represents a reference system for ultrathin film magnetism\cite{Vaterlaus} and will be used often in these Lectures.
\paragraph{Scanning Tunneling Microscopy. } The
growth of Fe is detected in this specific case by means of Scanning Tunneling
Microscopy (STM)\cite{STM}, a method of surface science that has become a
standard for the determination of the surface structure with atomic scale
precision. In STM, a metallic tip is brought to within subnanometer distances of
a surface. At these distances, electrons can tunnel from or to the tip.
Scanning the tip parallel to the surface and recording the tunnel current allow
one to map the surface geometry, often with atomic precision. In the experiments leading to the observation of the growth of Fe/W(110), 
the Fe film was deposited through a slit positioned very close to the surface. The resulting film was, actually, an Fe stripe, a few hundred $\mu$m wide and about two to three ML thick\cite{Thamm}, see Fig.~\ref{Fig:SEM_Fe}.
The mask inserted to block the Fe atoms around the slit has a finite thickness:
the resulting penumbra produces a non-abrupt ending of the Fe stripe. Along the penumbra, the Fe film develops as a wedge, with thickness changing slowly along a lateral scale of a few micrometers. STM images of the Fe along this wedge can be used to detect the growth of the film as its nominal thickness varies. For instance, the lower boundary of the stripe is imaged by STM in Fig.~\ref{Fig:STM_Fe} along the direction specified by the white arrow in Fig.~\ref{Fig:SEM_Fe}. The STM imaging starts from the Fe-uncovered side, observed on the left-hand side of the top image, and moves toward the 2-ML thick side of the stripe, observed on the right-hand side of the bottom image.
In Fig.~\ref{Fig:STM_Fe} one recognizes, most prominently, the flat terraces
(width: 20-50 nm) separated by monoatomic steps that typically occur at the
surface of a metallic single crystal. In the images, the steps run along the
vertical direction. On top of the flat terraces, the epitaxial deposit
develops. On the left-hand side of the top image in Fig.~\ref{Fig:STM_Fe}, where
the actual Fe film starts, small, one ML-thick Fe elements are visible,
rendered in the image as brighter spots as they are closer to the tip. The
region of the image consisting of small, one monolayer thick elements that are far away from each other is referred to as the ``submonolayer'' range. The
density of the Fe elements increases when moving toward the right-hand side of
the image, signifying that the nominal film thickness increases. However, the
actual thickness of the individual Fe elements remains the same, namely
one monolayer. Toward the end of the top image and within the first quarter of the bottom image, the Fe
elements are so dense that the one monolayer is almost completed. Only a few
``voids'' are still visible, and a small number of Fe elements cover the underlying Fe-deposit. They are said to ``occupy the second
layer''. These defects represent a deviation from the so called
``layer-by-layer'' growth. Proceeding toward the right-hand side of the bottom image, the second layer builds on top of the first one. Again, Fe-elements (see the centre of the bottom image) develop to form the second layer and their density grows toward the right-hand end of the bottom image, where the second layer is almost completed (albeit with a larger
number of ``voids'' and ``third layer'' elements, see also \cite{Back_Nature}).
\paragraph{Low Energy Electron Diffraction.}
Throughout this range of growth -- both when a layer is almost complete and when ``voids'' and ``antivoids'' (in the terminology used above: holes left in the layer being completed, and elements already occupying the next one) coexist -- one observes the persistence of the diffraction pattern recorded in a so called Low Energy Electron Diffraction (LEED) experiment. LEED is a further method of
surface science \cite{Venable} used to
determine the crystalline structure of the top surface layers. The lattice provides centres at which the incoming electrons are diffracted. One observes bright back-diffracted spots at angles that depend on the periodicity of the lattice parallel to the surface of the target. The intensity of the bright spots oscillates with energy and is used to determine the periodicity perpendicular to the crystal surface \cite{Amiri}. In the specific case of ultrathin Fe films on W(110), the bright spots
observed in a diffraction experiment from the Fe-uncovered W(110)-surface
continue to exist with approximately the same brightness and position when the
Fe film is deposited\cite{Back_Nature} up to about 2 monolayers. Accordingly,
the Fe film grows on W(110) in registry with the underlying surface
structure, i.e. the Fe deposit assumes the same bcc crystal structure as the
underlying W-crystal, with the same lattice parameters. A similar in-registry
growth is observed in a variety of epitaxial overlayers\cite{Roy}, e.g., during
the growth of fcc Fe and Co ultrathin films on Cu(100)\cite{Amiri}.\\
In-registry, layer-by-layer growth is often the sought-for growth mode, aimed
at producing well-defined metal overlayers, precise on an atomic scale. It is,
however, the exception in the field of epitaxial metal and semiconductor
growth, where inter-diffusion processes between
overlayers and substrate and clustering of the deposit are often
observed\cite{Venable,Brune} and produce, typically, a ``blurring'' of the
interfaces. Yet, interdiffusion or clustering might also have significant
physical impact and produce important applications\cite{Thin,Dots}\footnote{This subject is beyond the scope of these Lectures.}.
\paragraph{Quantum well states.}
As we will discuss in depth later, in transition metal overlayers one has two
types of electrons. The $d$-states  are localized around the core and their
band structure has a small bandwidth. The electrons originating from the $sp$-atomic levels assume, in the solid, a free electron character and build so
called free-electron like bands. In a situation where the transition metal is
deposited on top of a different metal, as it happens in epitaxial overlayers and
multilayers, the Fermi levels of the various components are aligned along the
entire vertical stack structure. This causes a misalignment of the origin
of the $sp$ free-electron-like band: this so called optical potential is a
characteristic of the material and therefore changes when proceeding along the
vertical direction. Because of the changing optical potential, the $sp$-electrons
sense a potential jump at the interface between two different materials --
provided the interface is defined
on a subnanoscale. For the free electrons within the stack, the potential
steps act to build a one-dimensional sequence of so called ``quantum wells''.
The one-dimensional quantum well problem was solved at the dawn of quantum
mechanics. It entails a set of quantized energy levels which correspond to true
bound states when the energy is below the material-vacuum potential barrier and
to resonances when the energy is above the material-vacuum potential barrier.
For the practical purpose of a stack of metal overlayers, we expect the free
electron like bands to partially condense into a set of quantized states called
``quantum  well states'' \cite{Roy,Egger} (for a review of the physics and
applications of quantum wells in semiconductor technology and numerous
references about the subject see e.g. Ref.~\cite{QW}). In the ideal case of ``perfect
confinement'', i.e. infinite potential walls, the quantum states resulting from
the condensation would be infinitely sharp. In practice, one observes broad
quantum well states, but their very observation is indicative that the potential steps
appearing at the interface of the stack of metal overlayers are sharp enough for
quantum interference of electron waves to take place. The capability of quantum
interference of electron waves at the interfaces is the basis for
the observation of magnetic properties such as the interlayer coupling, that was
used in concrete applications of metal overlayers\cite{Fert} and will be
discussed later in Chapter 4. For the purpose of illustrating
the quantum interference phenomenon in epitaxial overlayers we discuss the
observation of quantum well states in the reflection of low energy electrons during the growth of Cu on an ultrathin Co film, deposited previously on Cu(100)\cite{Egger}. The principle of the experiment is sketched on the left-hand side of Fig.~\ref{Fig:QW}.
The Cu-film is confined between vacuum on one side and the Co-film on the
other. This confinement produces quantum well states (black dots) which,
ideally, represent the only available energy levels in the Cu-film. The dashed line
indicates the energy of the primary electrons, which is kept fixed while the
Cu-thickness is changed.
When the thickness of the Cu is increased, the quantum states move along the
free electron band to assume a different energy (open dots). If the primary
energy of the electrons is such that it does not coincide with the energy of a
quantum  state, the incoming electrons do not find any state that channels them
into the underlying material and are, accordingly, totally reflected (in the
ideal case of infinitely sharp quantum well states). When a quantum well state
intercepts the energy level of the incoming electrons, there is a finite
transmission probability for the electrons, so that
the reflected intensity decreases. In such an experiment one indeed observes an
oscillation of the reflected intensity as a function of the Cu thickness. The
use of spin polarized primary electrons in Ref.~\cite{Egger} serves the purpose of enhancing the oscillatory component. The interested reader is referred to the original literature.

\chapter{Ground state magnetic moment of transition metal overlayers.}
\label{Chap:ground:moment}

\renewcommand{\thesection}{\thechapter.\arabic{section}}
\renewcommand{\theequation}{\thechapter.\arabic{equation}}

\noindent Almost every atom of the $3d$, $4d$ and $5d$ series carries a magnetic
moment, and almost no metal built out of those same atoms does: iron, cobalt and
nickel are the only elemental ferromagnets among them. The magnetic moment is
therefore not a property an atom simply brings along into the solid -- it is the
outcome of a competition that the solid state usually wins. Ultrathin films sit
exactly at the tipping point of that competition. Reducing the number of
neighbours pushes toward the atomic behaviour, contact with the substrate pushes
back, and both ``dead'' layers, which have lost the moment altogether, and
layers with moments larger than the bulk ones have been found in the same
materials.

\noindent This Chapter builds the argument needed to understand that outcome, in
four steps. We first establish what a magnetic moment is, classically and then
quantum mechanically (Secs.~3.1 and 3.2). We then assemble the magnetic moment
of an atom, which requires a hierarchy of interactions of decreasing strength
(Sec.~3.3). We ask next what survives when atoms are condensed into a solid:
the orbital part of the moment is quenched (Sec.~3.4) and the spin part is put
at risk by the broadening of the atomic levels into bands, which is the content
of the Stoner model (Sec.~3.5). Only then are we in a position to discuss dead
and enhanced moments in impurities and overlayers (Sec.~3.6). Six Appendices
collect the derivations that would otherwise interrupt the line of reasoning.

\section{Some results on the magnetism of matter from classical physics.}
\paragraph{The magnetic moment in classical physics.}
Classical mechanics, applied to the Rutherford model of an atom, consisting of
electrons  circulating around a positively charged point charge, provides us with an important relation for the magnetic moment generated by the circulating electrons. Let an electron  with charge $-e$ move  with an orbital
angular momentum vector $\vec L$. Its magnetic moment vector, defined as the
circulating current $\times$ the area defined by the loop $\times$ the normal
unit vector to the loop (the definition is derived from Amp\`ere's hypothesis in
Appendix~\ref{App:Ampere}) amounts to
\begin{equation}
\label{Eq:mu_class}
\vec \mu = -\frac{e}{2m}\cdot \vec L
\end{equation}
Because of the negative sign of the electron charge, $\vec \mu$ and $\vec L$ are antiparallel to each other. The ratio between $\vert \vec \mu\vert$ and $\vert \vec L\vert$ ($\frac{\vert e\vert}{2m}$ in the case of a single
electron) is known as the gyromagnetic ratio.
Classical physics provides us with an alternative, more precise definition of the magnetic moment of an electron. Again, we have the Rutherford model of an atom in mind. The Hamilton function of an electron of charge $-e$ embedded in a scalar field $\Phi(\vec r)$ and in a magnetic field $\vec B$ writes:
\begin{eqnarray}
\label{Eq:Hamilton_class}
H(\vec p, \vec r, \vec B) =
\frac {\vec p^2}{2m} -e\Phi(r)+\frac {e}{2m} \cdot \vec L\cdot \vec B+\frac {e^2}{8m} (\vec B\times \vec r)^2
\end{eqnarray}
In the context of Hamiltonian mechanics, the  magnetic moment vector is defined as
\begin{equation}
\vec \mu \doteq -\frac{\partial H(\vec B)}{\partial \vec B}=-\frac {e}{2m}
\cdot \vec L-\frac {e^2}{8m} \frac{\partial(\vec B\times \vec
r)^2}{\partial \vec B}
\end{equation}
The magnetic moment is, accordingly, the sum of two terms. The first one can be
either positive or negative, the second one is strictly negative. In classical
physics  a most peculiar phenomenon appears: the average of the first term with
the suitable Gibbs canonical probability at any finite temperature over all
possible classical orbits cancels exactly the canonical average of the
second term. This is made possible by the fact that, in classical physics, the
angular momentum is not quantized and the average of the first component of the
magnetic moment is dominated by those values of $\vec L$ which are close to
zero.\footnote{The same conclusion is reached by the standard argument: the
magnetic field enters the Hamilton function only through the combination $\vec
p+e\vec A(\vec r)$, and the momentum of each electron is integrated over the
whole of $\mathbb{R}^3$, so that the substitution $\vec p\rightarrow\vec p-e\vec
A(\vec r)$ removes $\vec B$ from the canonical partition function altogether.
The free energy is then independent of $\vec B$ and the magnetization
$-\partial F/\partial \vec B$ vanishes identically.} This exact compensation is
the content of a remarkable theorem, the so-called Bohr-van Leeuwen
theorem\cite{Bohr}, and is the reason why classical  physics does not explain
magnetism in matter. Notice what has to fail for magnetism to exist: the
angular momentum must be quantized.

\section{The magnetic moment in quantum mechanics.}
By the correspondence principle, one can define, starting from the classical relations, quantum mechanical operators. The Hamilton operator of a charge $-e$ embedded in a scalar field $\Phi(\vec r)$ and in a magnetic field $\vec B$ can be translated to quantum mechanics by the usual translation keys
\begin{equation}
\vec p\rightarrow -i\cdot \hbar \vec \nabla \quad ;\quad \vec L\rightarrow -i\cdot \hbar\cdot \vec r\times \vec \nabla
\end{equation}
From here on $\vec L$ and $\vec S$ denote \emph{dimensionless} angular momentum
operators, i.e. angular momenta in units of $\hbar$; the factor $\hbar$ is
carried by the Bohr magneton introduced below. The magnetic moment operator is
defined, using the correspondence principle, by means of the operator
\begin{equation}
\label{Eq:muop}
\vec \mu = -\underbrace{\frac{\vert e\vert \cdot \hbar}{2m}}_{\mu_B}\cdot \vec L -\frac {e^2}{8m} \frac{\partial(\vec B\times \vec
r)^2}{\partial \vec B}
\end{equation}
In Eq.~\eqref{Eq:muop} we have introduced the Bohr magneton
\begin{equation}
\mu_B\doteq \frac{e\cdot \hbar}{2m}\approx  5.8\cdot 10^{-5} \frac{\mathrm{eV}}{\mathrm{T}}
\end{equation}
i.e. the unit for the magnetic moment in quantum mechanics.\\
We claim that the magnetic moment produced by the second term in
Eq.~\eqref{Eq:muop} can be neglected in solving most problems when the electron
has a finite angular momentum. To prove this claim, we estimate the quantum mechanical expectation values for the various terms occurring in the Hamilton operator for an electron in  the Coulomb field of a proton and embedded in a magnetic field that, for convenience and without loss of generality, we align along the $z$-direction.
\begin{eqnarray}
\label{Eq:Hamilton_qm}
H(\vec p, \vec r, B) &=&
\underbrace{\frac {\vec p^2}{2m} -e\Phi(r)}_{\text{(a)}}+\underbrace{\mu_B\cdot L_z\cdot B}_{\text{(b)}} \underbrace{+\frac {e^2}{8m} (x^2+y^2)\cdot B_z^2}_{\text{(c)}}
\end{eqnarray}
1. The order of magnitude of $(a)$ can be estimated using hydrogen like wave functions to compute that the energy difference between the 2s- and 1s- energy levels of the hydrogen atom amounts to $\approx 10.2$ eV.\\
2. By assuming an orbital angular momentum $\hbar$ we find
$(b)\approx 10^{-4}\cdot \mathrm{eV}\cdot \frac{B}{\mathrm{T}}$, i.e. $(b)<<(a)$, even when magnetic field strengths of a few Tesla (typical for a laboratory using superconducting coils) are used.\\
3. For a  $1s$-state of Hydrogen we find $(c)\approx 10^{-11}\,\mathrm{eV}\cdot(\frac{B}{\mathrm{T}})^2$, i.e. $(c)<<(b)$.\\
We conclude that, in quantum mechanics, where $L$ is quantized, unless
the level under scrutiny has exactly $L=0$ (it can happen, e.g. for noble gas
elements), the term (b) always dominates in strength with respect to (c) and is
therefore the only one needed to be taken into account for explaining e.g.
spectroscopic observations of atoms in a magnetic field. The term $(c)$ produces the so called \textbf{diamagnetism} of atoms. All the core electrons in atoms, e.g., contribute to their diamagnetism. The free electrons of metals are also diamagnetic (Landau diamagnetism, \cite{Landau_dia}). For the understanding of observations relating to the ferromagnetism in epitaxial layers, the term $(c)$, however, is negligible, so that we will continue with the definition of the magnetic moment vector operator as $-\mu_B\cdot \vec L$.\\
The Hamilton operator obtained by the correspondence principle is incomplete. The low energy Hamiltonian obtained, in a spherically symmetric electrostatic field and in a magnetostatic situation of a uniform magnetic field, from the Dirac equation of an electron with charge $-e$ by Foldy and Wouthuysen\cite{Foldy} writes
\begin{eqnarray}
{\cal H} &=& m_e\cdot c^2 +\frac{\vec p^2}{2m_e} -e\cdot \Phi(\vec r)\nonumber\\
&+& \frac {e}{2m} \cdot \vec L\cdot \vec B+\frac {e^2}{8m} (\vec B\times \vec r)^2\nonumber\\
&+&2\cdot \frac {e}{2m} \cdot \vec S\cdot \vec B\nonumber\\
&-& \frac{e}{2m_e\cdot c^2}\cdot \frac{1}{r}\frac{d\Phi(r)}{dr}\vec S\cdot \vec L\nonumber\\
&\pm& .....
\end{eqnarray}
(in this equation, and in this equation only, $\vec S$ and $\vec L$ are in physical units).
The third line entails the correction to the interaction with a magnetic field brought about by a degree of freedom that was unknown to Schr\"odinger quantum mechanics: the intrinsic \textbf{spin} of the electron. Remarkably, this contribution has a factor ``2'' with respect to the contribution from the orbital angular momentum\footnote{
It is not even exactly ``2'': Quantum electrodynamics provides the value
$2.00231930436182$ -- and this is one of the physical quantities which have been measured with the highest precision. We will continue using ``2''.}. The operator for the quantum mechanical magnetic moment of an electron writes, accordingly,
\begin{equation}
\vec \mu = -\mu_B\cdot (\vec L +2\cdot\vec S)
\end{equation}
Because of the new degree of freedom, the Hilbert space must be, formally,  extended by the operation known as the tensor or Kronecker product\cite{DP_QM,DP_Group}, so that the composite wave functions are linear combinations of a new set of basis functions (here are three different writings)
\begin{equation}
\psi_{l,m_l}(\vec r)\cdot Y(m_s) \quad ;\quad \psi_{l,m_l}\otimes Y^{\pm} \quad ; \quad \vert l,m_l>\vert \pm \frac{1}{2}>
\end{equation}
On the left, the new basis functions are obtained by standard multiplication of functions of some variables ($\vec r$ for the orbital component, $m_s=\pm \frac{1}{2}$ for the spin component). The middle and right writings are formal ``multiplications'' of states, the one on the right hand side originating from the famous book on Quantum Mechanics by P.A.M. Dirac\cite{Dirac}. The operators act in the Kronecker product space with their Kronecker product. In simple words, each operator ``operates'' onto that component of the wave function it is supposed to operate on. For example, the operator for the magnetic moment of one electron writes, formally,
\begin{equation}
-\mu_B\cdot (\vec L\otimes \mathds 1 + 2\cdot \mathds 1 \otimes \vec S)
\end{equation}
$\mathds 1$ is the identity operator and
\begin{equation}
\vec J\dot=(\vec L\otimes \mathds 1 + \mathds 1 \otimes \vec S)
\end{equation}
is the operator for the \textbf{total angular momentum} of the single electron.
Having in mind this formal aspect, we will proceed using a simplified writing that avoids the ``$\otimes$'' sign but should be univocal enough.\\
The fourth line of the Foldy and Wouthuysen operator is the so called spin-orbit operator. It will play an essential role in determining the magnetic anisotropy in epitaxial layers and will be discussed in Chapter 4.

\section{The magnetic moment of atoms.}
Atoms are the essential components of solids and the understanding of the
magnetic moment in solids and epitaxial overlayers starts with the
understanding of the magnetic moment in atoms. Atoms are made of many
electrons and the computation of the magnetic moment of a many-electron system
is a non-trivial extension of the Schr\"odinger (or Foldy and Wouthuysen)
one-electron equation.\\
The extension is organized as a \textbf{hierarchy of interactions}, and it pays
to state the principle before the details. One starts from the strongest
interaction and treats each weaker one as a perturbation of the level scheme
produced by the previous step. Each step lifts part of the degeneracy left by
the step before, on an energy scale smaller than the one before: the
configuration Hamiltonian produces shells; the Coulomb repulsion between the
electrons splits each configuration into multiplets $^{2S+1}L$; the spin-orbit
coupling splits each multiplet into fine structure levels $^{2S+1}L_J$; and the
magnetic field finally splits each fine structure level into its $2J+1$ Zeeman
sublevels. Figure~\ref{Fig:Atom} at the end of this Section is the map of this
ladder, and the four paragraphs that follow are its four rungs. Only at the last
rung does a number for the magnetic moment appear.

\begin{figure}[H]
\begin{center}
\includegraphics[width=6cm]{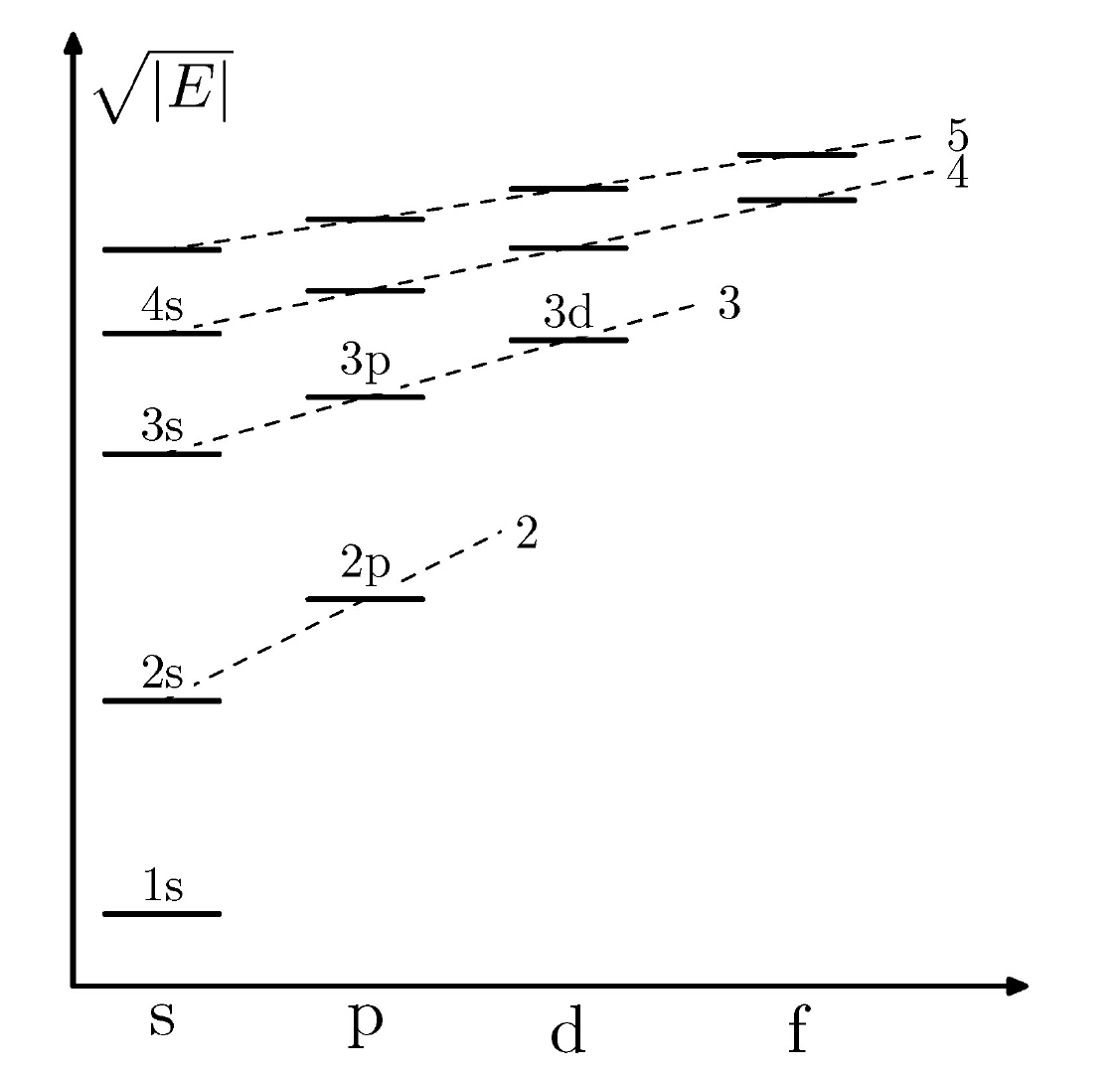}
\caption{Schematic energy diagram of an atom, resulting from single electron numerical computations.}
\label{Fig:atom_level}
\end{center}
\end{figure}

\begin{figure}[H]
\begin{center}
\includegraphics[width=0.55\textwidth]{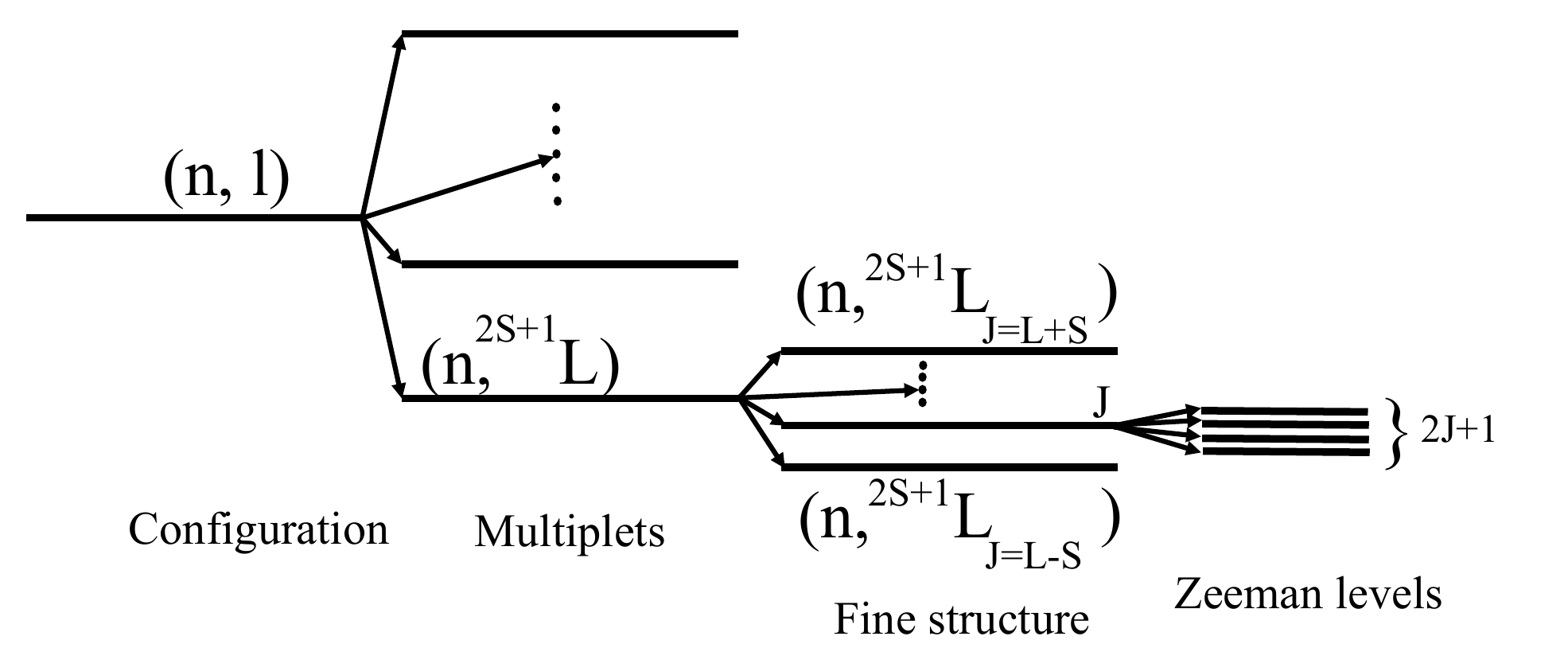}
\caption{Schematic energy diagram summarizing the atomic levels: the hierarchy
of interactions, from the configuration Hamiltonian on the left to the Zeeman
sublevels on the right, each interaction lifting part of the degeneracy left by
the previous one and on a smaller energy scale.} \label{Fig:Atom}
\end{center}
\end{figure}

\paragraph{The configuration Hamiltonian.} For determining the electronic structure of atoms and its magnetic moment, one solves first numerically the eigenvalue problem of the so-called single-electron
``configuration Hamiltonian''. This Hamiltonian entails a mean
field single electron potential summarizing the Coulomb potential exerted by the
nucleus and the ``average'' potential exerted by the remaining $N-1$ electrons.
Being a single electron approximation, the levels computed with the
configuration Hamiltonian can be labeled by a main quantum number $n$,
reminding us of the solution of Schr\"odinger equation of the hydrogen atom. The
next configuration quantum number $l$  designates the orbital momentum quantum
number, originally introduced to label the energy levels in the hydrogen atom.
It is also a good quantum number in atoms, as the mean potential entering the
configuration Hamiltonian is, approximately, spherically symmetric.
The typical set of atomic single-electron energy levels of the configuration Hamiltonian is sketched in Fig.~\ref{Fig:atom_level}.
The $(n,l)$-shell, resulting from the solution of the eigenvalue problem of the
single-electron configuration Hamiltonian, is $(2l+1)\cdot 2$-times essentially
degenerate. The $(2l+1)$-degeneracy originates from  the magnetic quantum
number $m_l$, designating the $z$-component of the orbital angular momentum. To
each orbital state one can attach one of the spin states ``up'' and ``down''.
This doubles the orbital degeneracy. The basis set of the $(n,l)$-shell thus
consists of the states $\{\vert n,l,m_l>\vert m_s>\}$, with $m_l=-l,...,l$
and $m_s=\pm \frac{1}{2}$. Knowing that there are such degenerate basis
states attached to each shell is an essential piece of information when it comes to
filling the energy levels with the available electrons. One starts filling the
basis states of the lowest energy level and one continues filling the basis
states of the next energy levels, taking into account the \textbf{Pauli
principle}: each single electron basis state can be filled by just one
electron. Thus, the shell $(n,l)$ can host a maximum number of electrons
corresponding to its degeneracy $(2l+1)\cdot 2$. Let us consider the example of
the so called $3d$-transition metals $V,Cr,Fe,Co, ....$. When one gets to the
main quantum number $n=3$, one notices an exception: first $3s$ is filled, then
$3p$ is filled further, but the $3d$ shell has a higher energy than the $4s$-shell, which is filled before the $3d$-shell ($K$: $(4s)^1$, then $Ca$ with
$(4s)^2$). Then the $3d$-shell starts being filled, whereby the configuration
of the outer shell is kept fixed to the one of $Ca$. Therefore, those
elements which are formed by the filling of the $3d$-shell and have the same
outer shell configuration belong to the same column of the periodic table (the
second one) and must be separately ordered along an exceptional line -- on
paper this is rendered by setting the $d$-transition metals together with the
suitable element in the second column. Thus, the $3d$ elements, for instance,
can be regarded as ``offshoots'' of $Ca$ ($Ca$ is in the column of the
so-called ``alkaline earth metals''). The $d$ transition metals all have
similar outer electron configurations $s^2$ (with a few exceptions), and can
therefore be considered chemically equivalent. Their configuration thus writes
(up to a few exceptions)
\begin{equation}
(1s)^2(2s)^2(2p)^6(3s)^2(3p)^6(3d)^x(4s)^2
\end{equation}

\paragraph{Multiplets.} Knowing the configuration is not yet enough to find whether the atom has a magnetic moment or not. The complexity is one of properly adding the spins and the orbital angular momenta of the many electrons occupying each shell to find the quantum numbers $S$, $L$ and $J$ that characterize the ground state of the atom. The solution of this problem is deep within the realm of the theory of angular momentum\cite{DP_Group}. In short: there are two ways of classifying the total states of the $x$ electrons within the $(n,l)$-shell. One consists in building all possible products between the $x$ single electron states, compatible with the Pauli principle. There are many such products, e.g. the configuration $(2p)^2$ contains 15 such two-particle product states. This way of labeling the states is straightforward but non-informative. There is a better way of organizing the many particle product states. This second method relies on the Schur-Weyl theorem\cite{DP_Group}. The method starts with finding first
all possible values for the total spin $S$ (using e.g. the Clebsch-Gordan
series). Schur and Weyl have demonstrated that the values allowed for the total
orbital angular momentum $L$ are fixed, once $S$ is given. In other words: the Hilbert space of the
configuration $(n,l)^x$ is built by product spaces, each belonging to a couple
$(S,L)$. Each couple $(S,L)$ is called a \textbf{multiplet} and builds one of
the eigenspaces of the configuration $(n,l)^x$. The symbol for the multiplets is $^{2S+1}L$.

\paragraph{Multiplet splitting.}
Within the configuration Hamiltonian, multiplets are degenerate.
When the Coulomb interaction between the electrons is switched on explicitly (typically using perturbation theory), one obtains the so called multiplet Hamiltonian, which lifts the degeneracy of the multiplets. By means of the
multiplet splitting, a hierarchy of energy levels is produced, each of which is
labeled by well defined values of $L$ and $S$. The multiplet splitting is of
the same order of magnitude as the splitting of configurational energy levels. The energy difference between multiplets is known as the \textbf{intra-atomic exchange interaction} (more about it in Appendix~\ref{App:exchange}). Multiplet splitting on the basis of the Coulomb interaction has all the
appearances of being a difficult problem. It is indeed, and can only be solved
numerically with a large amount of computing. Yet, its solution obeys simple
rules known as the (first and second) \textbf{Hund} rules\cite{Hund}. Hund's
first rule establishes the highest possible value of $S$, compatible with the
Pauli principle, as the total spin quantum number of the ground state. Hund's second rule associates to the highest value of $S$ the highest value for $L$ as the total orbital angular momentum quantum number.

\paragraph{The fine structure of the multiplets.} Knowing the ground state
multiplet is not yet enough, in atomic physics, for assigning a value of the
magnetic moment. Relevant to this purpose are the possible quantum numbers for
the total angular momentum operator. Each multiplet level has, in fact, a
degeneracy of $(2S+1)\cdot (2L+1)$ and hosts eigenspaces of the total angular
momentum operator which can be assigned one of the values $L+S,L+S-1,...,\vert L-S\vert$
(Clebsch-Gordan series). Within the multiplet Hamiltonian, these values are
degenerate and no final assignment for the value of the ground state $J$ would
be possible. However, when spin-orbit coupling is introduced, each multiplet
level splits into  fine structure levels, each  carrying  one of the quantum
numbers of the total angular momentum operator $\vec J$. The symbol for a fine
structure level is the spectroscopic symbol $^{2S+1}L_{J}$. The ordering of the
fine structure levels is empirically given by the third Hund rule. Notice that
the fine structure splitting is, for most atoms, several orders of magnitude
smaller than the multiplet splitting. This is due to the fact that the
spin-orbit coupling operator is a relativistic effect\cite{Foldy}, i.e. it is
much smaller than the Coulomb interaction.

\paragraph{The Zeeman operator.} The last operator in the hierarchical build-up is the so called Zeeman operator\footnote{Pieter Zeeman, Nobel prize 1902.}, generated by the coupling of the total magnetic moment operator $-\mu_B\cdot (\vec L + 2\cdot \vec S)$ to the magnetic field. Without loss of generality, we assume $\vec B = (0,0,B)$ so that the Zeeman operator writes
\begin{equation}
H_Z = \mu_B\cdot (L_z + 2\cdot S_z)\cdot B
\end{equation}
It should be considered as a small perturbation of a given fine structure level $(L,S,J)$ and has eigenvalues (see Ref.~\cite{DP_QM}, p.~113-116)
\begin{eqnarray}
E_{LSJM_J}&=& g_{LSJ}\cdot \mu_B\cdot M_J\cdot B \quad M_J= J,J-1,...,-J\nonumber\\
g_{LSJ}&=&1+\frac{J(J+1)+S(S+1)-L(L+1)}{2J(J+1)}
\end{eqnarray}
$g_{LSJ}$ being the Land\'e factor. The magnetic field therefore lifts the $2J+1$ degeneracy of the fine structure levels. The Zeeman sublevels can be observed spectroscopically in atomic vapors (see Ref.~\cite{DP_QM}, p.~115-116).
The eigenvalues of the $z$-component of the magnetic moment operator write
\begin{equation}
\label{Eq:mu_atom}
-\frac{\partial E_{LSJM_J}}{\partial B}= -g_{LSJ}\cdot \mu_B\cdot M_J \quad M_J= J,J-1,...,-J
\end{equation}
Fig.~\ref{Fig:Atom} summarizes, schematically, the typical sequence of energy levels that develop in an atom when, successively, all relevant interactions are switched on, starting from the strongest ones (see e.g. Ref.~\cite{DP_QM}, p.~93-108 for a detailed explanation of this diagram).
\noindent Equation~\eqref{Eq:mu_atom} is the result we carry into the solid state:
the magnetic moment of a free atom is $g_{LSJ}\mu_B\cdot J$, with
$L$, $S$ and $J$ fixed by the Hund rules. The next two Sections examine what
happens to the two ingredients of that result -- first to $L$, then to $S$ --
when the atoms are brought together to form a solid.

\section{From atoms to solids: quenching of the orbital moment.}
The origin of the magnetic moment in solids resides, primarily, within the quantum numbers  $L,S,J$ of the ground state of the atom. However, there is an important effect that takes place when the atoms are bound together to form a solid: the quantum number $L$ is reduced to zero, a phenomenon known as ``quenching of the orbital angular momentum''. We follow the classical route to this result, which starts from a measurement on paramagnetic salts.

\begin{figure}[H]
\begin{center}
\includegraphics[width=0.35\textwidth]{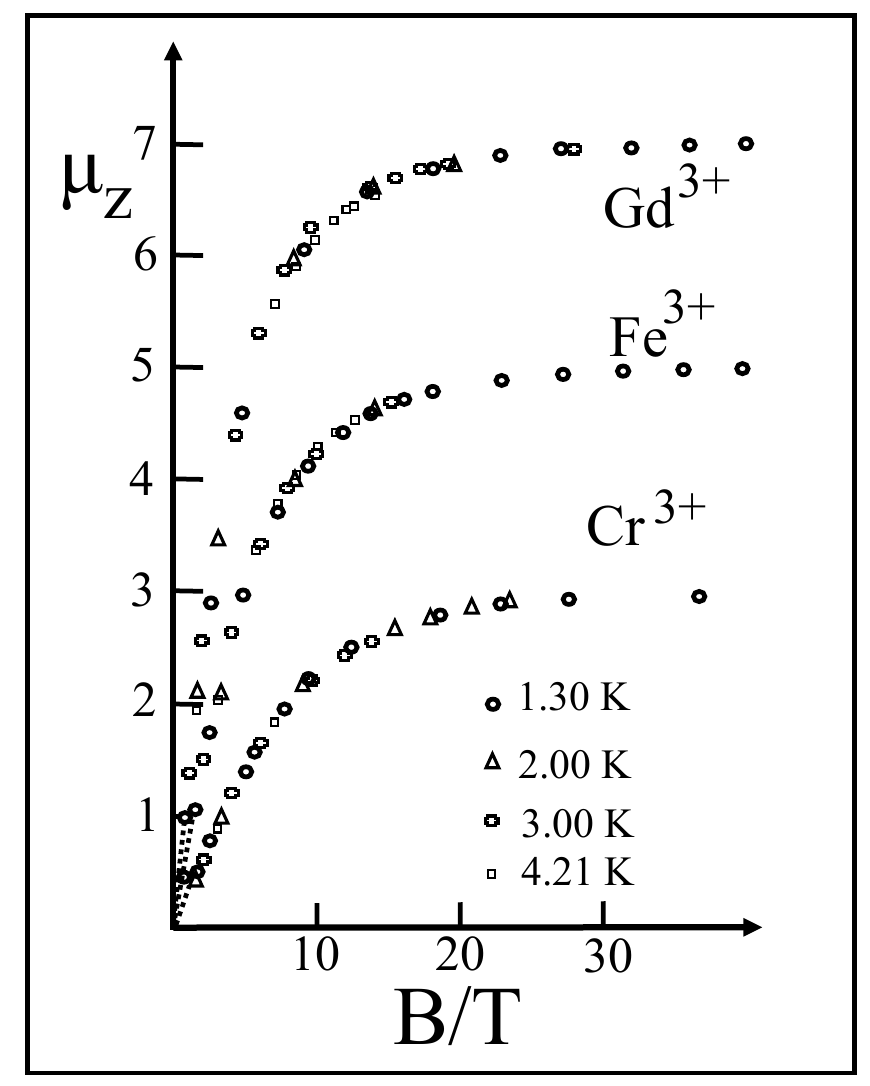}
\caption{Average magnetic moment $\mu_z$ (in units of $\mu_B$) per ion as a function of $B_z/T$ (in units of $10^{-1}\,\text{Tesla}/K$) for the three paramagnetic ions $Gd^{3+}$ (caged into a salt $GdSO_4\cdot 8H_2O$), $Fe^{3+}$ (caged into $NH_4Fe(SO_4)_2\cdot 12H_2O$) and $Cr^{3+}$ (caged into $KCr(SO_4)_2\cdot 12 H_2O$). Adapted from Ref.\cite{Kittel}.
\label{Fig:para}}
\end{center}
\end{figure}

\paragraph{Reading a magnetic moment off a paramagnetic salt.}
The relative population of the Zeeman sublevels changes with temperature and
magnetic field and it produces a temperature and magnetic field
dependence of the thermal average of the magnetic moment operator $-g_{LSJ}\cdot
\mu_B\cdot J_z$\cite{Kittel} -- typical for the so-called paramagnetism of atoms, as shown for three ions in Fig.~\ref{Fig:para}.
Plotted as a function of $\frac{B_z}{T}$, the experimental data taken at
different temperatures (indicated in the figure) collapse onto one single curve
(mathematically speaking: a Brillouin function). At sufficiently low temperatures
and high magnetic field only one Zeeman level is occupied, namely the one with
the highest magnetic moment, at which the curves saturate. Accordingly, one can
read out the magnetic moment corresponding to the ground state: $7\mu_B$ for
$Gd^{3+}$, $5\mu_B$ for $Fe^{3+}$ and $3\mu_B$ for $Cr^{3+}$. Two of these three
values are exactly what the Hund rules predict. The configuration of $Gd^{3+}$ is
$4f^7$. The first Hund rule fills the configuration with all spin up electrons,
i.e. $S=\frac{7}{2}$. The second Hund rule finds $L=0$ in the ground state, so
that we expect a saturation moment of $2\cdot \frac{7}{2}\cdot \mu_B$, as indeed
observed.  $Fe^{3+}$ has configuration $3d^5$. The first Hund rule fills the 5
orbitals with spin ``up'', i.e. $S=\frac{5}{2}$. The second Hund rule finds also
$L=0$, for a total magnetic moment of $5\mu_B$, as observed. Notice that in both
cases the shell is exactly half filled, so that $L=0$ anyway: these two ions
cannot tell us anything about the fate of the orbital moment.

\paragraph{The $Cr^{3+}$ anomaly and the quenching of $L$.}
The Cr salt  (Chromium(III) potassium sulfate dodecahydrate or Chrom-alaun\cite{Alun}) is the telling example. It is a crystal with cubic unit cell. The Cr
atoms are on an fcc lattice. The unit cell is filled by non-magnetic ligand atoms
that separate the Cr-ions, which are accordingly non-interacting. The electronic configuration of $Cr^{3+}$ is $3d^3$. If we evaluate this situation following
the Hund rules, we obtain $S=\frac{3}{2}$, $L=3$ and $J = \frac{3}{2}$. The
Land\'e factor amounts to $\frac{2}{5}$ and we expect a total magnetic moment of
$\frac{3}{5}\mu_B$. Observed is a magnetic moment of $3\mu_B$, i.e. the
electrons in the $Cr^{3+}$ ions act as a multiplet state with $L=0$ rather than
$L=3$: the observed moment is the spin-only value $2S\mu_B$. This is a
contradiction to Hund's second rule which needs understanding. The understanding follows the following path. We observe that the salt environment that renders the Cr-ions non-interacting provides also an octahedral potential that breaks the spherical symmetry of the atomic potential. In an octahedral ligand field (one speaks of
``crystal field'' when the atoms carrying the 3d configuration are embedded into
a crystalline environment) a $d$-level splits into a three fold degenerate
level $t_{2g}$ and a two fold
degenerate level $e_g$, see e.g. Ref.~\cite{DP_Group}, p.~471-472. The $t_{2g}$ level has, typically, lower energy than the $e_g$ level. Their single particle, symmetry adapted orbital basis functions write
\begin{eqnarray}
\label{Eq:t2g}
i\cdot\sqrt{\frac{1}{2}} \left( Y_2^{- 2} - Y_2^2\right)\doteq xy \quad\quad
i \sqrt{\frac{1}{2}} \left( Y_2^{- 1} + Y_2^1 \right)\doteq yz\quad\quad
\sqrt{\frac{1}{2}} \left( Y_2^{- 1} - Y_2^1 \right)\doteq xz
\end{eqnarray}
The $e_g$ level has symmetry adapted basis functions
\begin{eqnarray}
\label{Eq:Eg}
\sqrt{\frac{1}{2}} \left( Y_2^{- 2} + Y_2^2 \right)\doteq x^2-y^2\quad\quad
Y_2^0\doteq 2z^2-x^2-y^2
\end{eqnarray}
The decisive property of these combinations is that they are \emph{real}
functions of position, in contrast to the spherical harmonics they are built
from. The first Hund rule establishes the four spin states with total spin quantum
number $S=\frac{3}{2}$, symmetric with respect to particle exchange, as the
total spin wave function of the three electron system of $Cr^{3+}$. The
Slater determinant computed using the single particle states $xy,xz,yz$ provides
the total orbital wave function for the ground state. The matrix element of the
operator $L_z(1)+L_z(2)+L_z(3)$\footnote{More precisely: $L_z\otimes\mathds
1\otimes\mathds 1+ \mathds 1\otimes L_z\otimes\mathds 1+ \mathds
1\otimes\mathds 1\otimes L_z$.} computed over the total orbital wave function
gives exactly $0$, thus ``quenching'' the orbital component of the magnetic
moment in the ground state, in agreement with experiment.\\
The mechanism is worth isolating, because it is completely general: $\vec L$ is
a purely imaginary operator ($\vec L=-i\,\vec r\times\vec\nabla$), so that for
any \emph{real} wave function $\psi$ the expectation value obeys
$\overline{(\psi,L_{x,y,z}\psi)} = -(\psi,L_{x,y,z}\psi)$, and a quantity equal
to minus its own complex conjugate, being real, must vanish. Whenever the
environment of an ion is low-symmetric enough that its ground state can be
built from real orbitals, the orbital moment is quenched. This is the situation
in a crystal field, and it is the reason why the quenching of the orbital
angular momentum is a leitmotif for the magnetism in solids\cite{g}.\\
That a real basis can always be chosen holds just as well in a solid, and for
the same reason: the Hamiltonian of a non-magnetic crystal is a real operator,
so that with $\psi$ also $\overline\psi$ is an eigenstate with the same energy.
In a crystal $\overline{\psi_{n\vec k}}=\psi_{n,-\vec k}$, i.e. the complex
conjugate of a Bloch state is not that state itself but its partner at $-\vec
k$. A single Bloch state is a travelling wave and does carry a finite
$(\psi,\vec L\psi)$. Its partner, however, is degenerate with it, so that the
two span a two-dimensional eigenspace, and out of that pair one may form the
two \emph{real} standing waves
\begin{equation}
\label{Eq:realbasis}
\psi_{n\vec k}+\psi_{n,-\vec k}\quad\quad \frac{1}{i}\left(\psi_{n\vec k}-\psi_{n,-\vec k}\right)
\end{equation}
each of which has, by the argument just given, a vanishing expectation value of
$\vec L$. In the ground state $\vec k$ and $-\vec k$ are occupied together, so
the occupied subspace admits throughout a basis of real functions and the
orbital moment of the band vanishes. One may equally well say that the
contributions of $\vec k$ and $-\vec k$ cancel pairwise: the sum of
$(\psi,\vec L\psi)$ over a degenerate subspace is a trace, and a trace does not
depend on the basis in which it is evaluated, so the two readings are the same
computation. Only the spin-orbit
coupling, which correlates the orbital motion with the spin direction and makes
the Hamiltonian complex, can
restore a small orbital moment -- typically a few hundredths of $\mu_B$ in the
$3d$ metals -- and this is the same interaction that will produce the magnetic
anisotropy of Chapter 4.\\
What survives of the atomic moment is therefore the spin, $2S\mu_B$. Whether
even that survives is the subject of the next Section.

\section{Band ferromagnetism: the Stoner-Slater model.}
\begin{figure}[H]
\begin{center}
\includegraphics[width=0.5\textwidth]{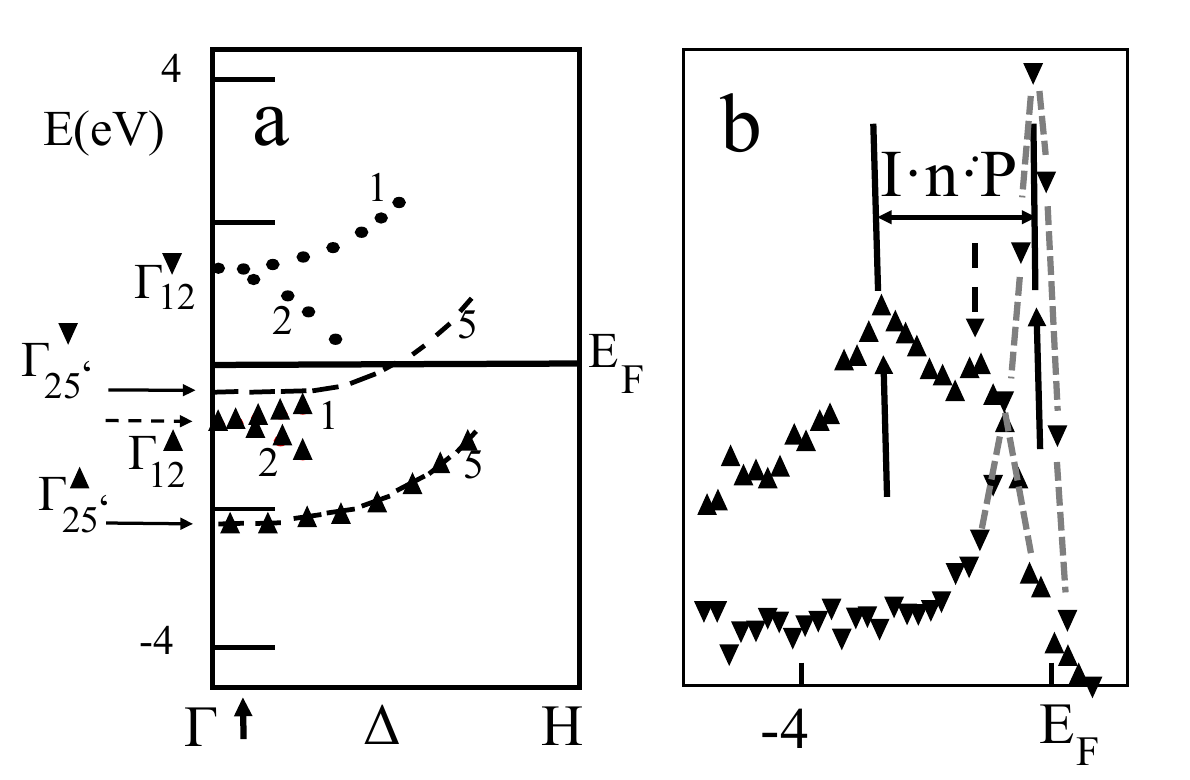}
\caption{a: Selected bands of ferromagnetic Fe at the $\Gamma$ point and along the $\Delta$ direction, drawn from \cite{Wang}. b:
Intensity of energy-analyzed photoemitted electrons, excited along the normal to
the $(100)$-surface at a photon energy of $60$ eV, as a function of their energy
with respect to $E_F$ and analyzed according to their spin.
\label{Fig:Kisker}
}
\end{center}
\end{figure}

Even if the orbital angular momentum is quenched, the 1st Hund rule takes care
that almost all 3d, 4d and 5d transition elements have a non-vanishing $S$
quantum number that, ultimately, leads to a spin magnetic moment in the ground
state of the atoms: the Hund rule of maximizing the spin state makes almost all
such atoms ``magnetic''. In striking contrast to this outcome, almost all 3d, 4d
and 5d metals are non-magnetic in the solid state, i.e. they lose completely
their atomic spin moment when the atoms are brought together to form a solid.
Magnetism is therefore an absolute exception in the solid state, as much as it
is the rule in the atomic state: only Fe, Co, Ni and their alloys keep actually
a non-vanishing magnetic moment\cite{Moruzzi}. This is a quite mysterious
outcome that we need to explore more thoroughly. The solution of this
``mystery'' is the preliminary step toward understanding the formation of
magnetic moments in epitaxial overlayers.\\
A further observation is that magnetic moments per atom are expected to be an
integer (the so called ``magneton number'') times $\mu_B$, the integer being
$2S$ and the factor $2$ the spin $g$ factor. This would lead, for example, to
atomic magnetic moments of
2, 3, respectively 4 $\mu_B$ for Ni, Co and Fe. The ascertained values in bulk Ni,
Co and Fe are $0.616$, $1.715$ and $2.216$ $\mu_B$\cite{Moruzzi}, i.e.
considerably smaller than the values expected from the full atomic spin moment
and corresponding to non-integer magneton numbers\footnote{Overlayers of these metals display a further singular behavior, ``dead'' versus ``enhanced'' magnetic moments, see Sec.~3.6.}.
\\
This variety of behaviours finds a comprehensive explanation within the framework of the ``Stoner model of ferromagnetism''\cite{Stoner,Gunnarson}.
The key observation of the Stoner model is that the sharp atomic energy levels
carrying the magnetic moments (the $d$-levels for transition metals) acquire a
certain finite energy width (see Appendix~\ref{App:vbs}) when they are embedded into a ``sea''
of free electrons, provided by the $sp$-electrons. In the language of solid
state physics, the atomic levels broaden into energy bands. In addition, the $d$-shell carrying the magnetic moment is partially filled and one can envisage that, in the solid state, it will be pinned around the Fermi level to allow
partial occupation. We therefore have a situation of energy levels which are
potentially magnetically active, have a finite energy width and are in the
vicinity of the Fermi level. The Stoner model indicates exactly how to deal
with this situation, with the aim of establishing whether the ground state of
the metal has a net magnetic moment per atom and how strong the magnetic moment
is.\\
In order to expose the essential results, we need to introduce a few
parameters that characterize the Stoner model. One parameter is the number $n$
of $d$-electrons per atom that are responsible, in the atom, for the magnetic
moment. For Fe, for instance, $n=4$. A further parameter is the spin
polarization $P$ of these electrons. One distinguishes between the paramagnetic
state, with $P=0$, and the fully polarized state, with $P=1$. $P$ is actually a
quantity which is sought for by the self-consistent calculation entailed by the
Stoner model (Appendix~\ref{App:Stoner}). The magnetic
moment is, accordingly, $\mu_B\cdot n\cdot P$. The exchange splitting between
spin ``up'' and spin ``down'' bands amounts to $I\cdot n\cdot P$. $I$ is the
intra-atomic exchange energy parameter used in the literature\cite{Gunnarson,Moruzzi};
it is the same quantity denoted by $J$ in Appendices~\ref{App:exchange}
and~\ref{App:HDvV}, where it is computed. The intra-atomic exchange interaction is an atomic
property and is a measure of the multiplet splitting in atoms. It remains
approximately the same even when the atoms build a solid\cite{Ded}. Its value
is approximately $1$ eV. A quantitative  measure of the narrowness of the
states, used within the Stoner model, is the density of states $n_0(E_F^0)$ at
the Fermi level of the paramagnetic state (the $^0$ in $E_F^0$ refers to
the Fermi level of the paramagnetic metal). On the basis of these parameters, the
Stoner criterium of ferromagnetism states that ferromagnetic order in a
band-like transition metal only exists if
\begin{equation}
\label{Eq:criterium}
I\cdot n_0(E_F^0)\geq 1
\end{equation}
(Stoner criterium for ferromagnetism of metals).
In other words: provided the broadening of the atomic energy levels due to band formation in solids reduces the density of states at the Fermi level below a certain threshold, the atomic magnetic moment vanishes (in the sense that the paramagnetic state has a lower energy than the ferromagnetic one). On the other side, if the density of states at the Fermi level of the orbitals
carrying the atomic spins remains large enough, the ferromagnetic state (with finite magnetic moment) has a lower energy than the paramagnetic state (with vanishing magnetic moment). The ferromagnetic state is then found to have a spin polarization $P$. Owing to the fact that $P$ is some real number between $0$ and $1$, the magnetic moment is, in general, less than the atomic value of $\mu_B\cdot n$ and is not necessarily an integer multiple of $\mu_B$: this is the answer to the second of the two puzzles raised at the beginning of this Section. Ref.~\cite{Gunnarson} (see Table 1) finds that Fe, Co and Ni are the only transition metals that fulfil the Stoner criterium (see also Ref.~\cite{Moruzzi} for extended band structure  calculations of transition metals) -- the answer to the first puzzle.\\
Band structure calculations\cite{Wang} and experimental results\cite{Kisker} support the key elements of the Stoner model. Fig.~\ref{Fig:Kisker} reports (a) the section of the band structure of Fe probed by a photoemission experiment\cite{Kisker} and (b) the actual experimental
observation. In (a) the majority spin bands are marked by upwards triangles. The bands with $\Delta_5$-symmetry are dashed, the bands with $\Delta_2$ and $\Delta_1$ symmetry are given by upwards triangles (majority) and circles (minority). The $\Gamma'_{25}$ and $\Gamma_{12}$ energy levels are also indicated. The initial states close to the $\Gamma$-point observed by the photoemission experiment (b) are pointed at by horizontal arrows in figure (a) and by vertical arrows in figure (b). The spin polarized photoemission experiment\cite{Kisker} shows the intensity of the majority (black up-triangles) and minority (black down-triangles) photoemitted electrons separately. The exchange splitting between $\Gamma'_{25}(+)$ and $\Gamma'_{25}(-)$ is indicated. Assuming $I\approx 1$ eV and reading out $I\cdot n\cdot P\approx 2$ eV from the figure on the right, one obtains $n\cdot P \approx 2.2$, in agreement with the computed\cite{Small} value of $S\approx 1.1$ and the measured magnetic moment of $2.2 \mu_B$. The
unusual value of $S$ is typical for Stoner ferromagnets, where $P$ (and
accordingly $S$) can have a non-integer value\cite{Small}. Notice that
computations based on first principles (see (a) in
Fig.~\ref{Fig:Kisker}) indicate a (roughly) rigid exchange splitting between majority and minority bands, in agreement with the Stoner model.

\section{Dead layers and enhanced magnetic moments.}
\begin{figure}[H]
\begin{center}
\includegraphics*[width=0.7\textwidth]{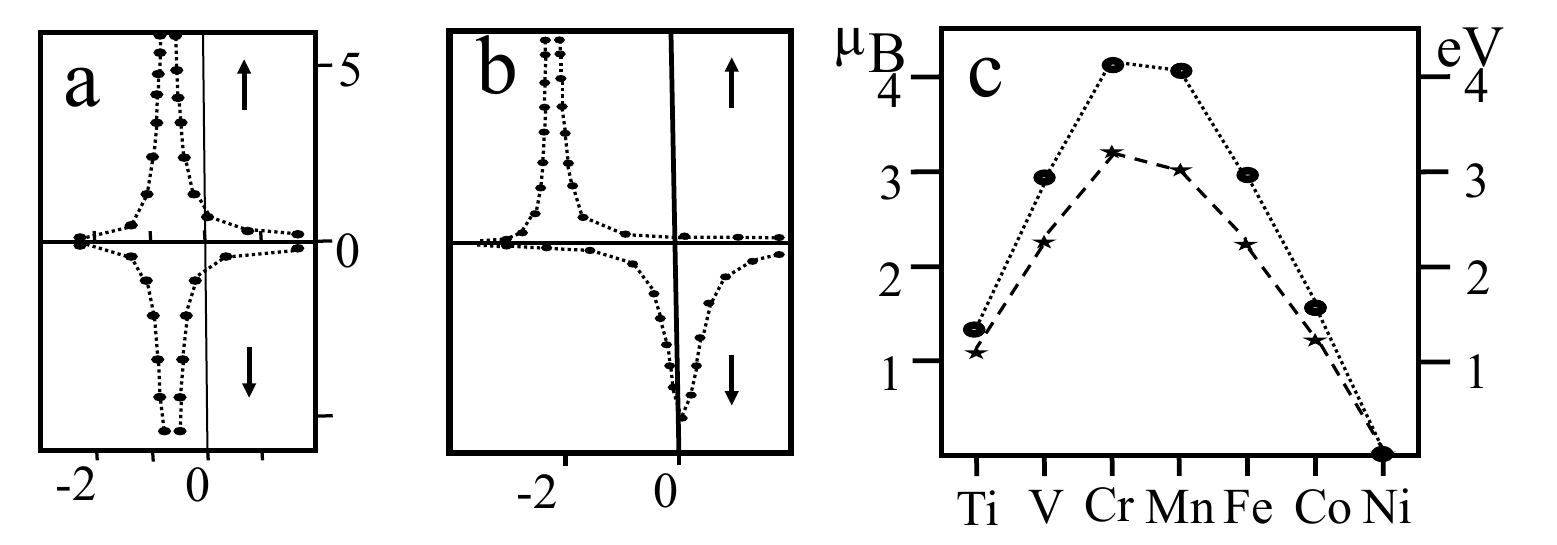}
\caption{a: Spin resolved density of states for a Ni impurity in Ag. (b):
Spin resolved density of states for a Fe impurity in Ag. (c): Computed magnetic moment per atom (circles)  and exchange splitting (asterisks) for impurities of the 3d transition metals (horizontal scale) in Ag. The exchange splitting is proportional to the magnetic moment, as implied by the Stoner model. Drawn from Ref.\cite{Ded}.
\label{Fig:dederichs}}
\end{center}
\end{figure}

The Stoner criterium establishes the two parameters that determine the occurrence
of ferromagnetic order and, if this occurs, the value of the magnetic moment:
the intra-atomic exchange energy $I$ and the density of states at the Fermi
level. $I$ is an atomic parameter and is known to remain close to the atomic
value even when atoms are embedded as an impurity\cite{Ded} or are replicated into
a crystal\cite{Gunnarson,Moruzzi}. The density of states at the Fermi level,
instead, depends very much on the broadening of the atomic level. That
broadening is precisely what a surface, an interface or a dilute environment
changes, and this is why the question of the magnetic moment becomes a question
about dimensionality.\\
There are at
least two situations in the literature, other than solids, where the formation
of a magnetic moment was studied: a single impurity of a $d$-transition metal in
a free electron like material\cite{Ded} like $Cu$ and $Ag$, and overlayers\cite{Freeman} on non-magnetic substrates. In both these situations, there are
two mechanisms that intervene to potentially modify the density of states at
the Fermi level, and they pull in opposite directions. On one side, the reduction of the number of nearest
neighbours, provided by the low-dimensionality of the overlayers, tends to
decrease the band width and thus increase the density of states at the Fermi
level. This mechanism can be understood e.g. within the framework of a
tight-binding band structure calculation, where the bandwidth simply scales
with the number of nearest neighbours. On the other side, the contact of the $d$-orbitals with a jellium-like free electron gas, provided by the non-magnetic
host or substrate, tends to increase the spreading of the $d$-orbital (we will
study this less trivial situation in Appendix~\ref{App:vbs}, on the concept of ``virtual bound states''). These two mechanisms compete in determining the density of states of the $d$-levels at the Fermi level. If this density decreases below the value required to sustain ferromagnetism, a magnetically ``dead'' layer or diluted atom is produced. On the other side, if this density is enhanced, the spin polarization $P$ of the $d$-electrons is increased (Appendix~\ref{App:Stoner}) and this produces an enhanced magnetic moment. The result of this competition is a challenge for the most advanced numerical computation schemes\cite{Ded,Freeman}.\\
We first comment on $d$-metal impurities\cite{Ded}, see Fig.~\ref{Fig:dederichs}.
One finds, for instance,
that an impurity of $Ni$ in a Ag-host completely loses its magnetic moment (Fig.~\ref{Fig:dederichs}a), while $Fe$ keeps it (Fig.~\ref{Fig:dederichs}b). The position of the peaks in (b) shows that the exchange splitting is actually enhanced with respect to bulk Fe (compare Fig.~\ref{Fig:dederichs}b with Fig.~\ref{Fig:Kisker}b). Fig.~\ref{Fig:dederichs}c summarizes the magnetic moment and the exchange splitting of the ground state of $d$-transition metal impurities in Ag. For the $Ni$-impurity, the paramagnetic solution has the lowest energy and, accordingly, magnetic moment and exchange splitting vanish. The situation
for an Fe impurity is different: there, the ground state magnetic moment (black
circle) is enhanced with respect to the situation of an Fe atom in bulk Fe. A
similar enhancement is observed in other 3d transition metals. Both outcomes of
the competition are thus realized within one and the same host, and which one
occurs is decided by the position and the width of the $d$ resonance -- exactly
the two quantities entering the Stoner criterium.\\
A similar trend was discovered in transition metal overlayers on non-magnetic
substrates. Research on ultrathin films of transition metals was initiated by
the reports that, in the limit of only a few atomic layers, films of Fe and Ni
deposited on noble metals were actually magnetically ``dead'', i.e. their
magnetic moment appeared to be suppressed in this limit\cite{Lieber}. It took
about 15 years and the development of extremely precise numerical total
energy band structure calculations based on the so called ``local spin density
approximation'' to find that the magnetic moment of most of the 3d bcc
transition metal overlayers and/or sandwiches and superlattices with Au and Ag
is actually enhanced\cite{Freeman}. The work by Freeman's group provided one of
the main impulses toward research on the ground state magnetic properties of
transition metal overlayers, by a wide range of experimental methods\cite{Bader_review,Wuttig}. The perspective of finding ``alive'' layers, instead of
the ``dead'' layers found by previous research\cite{Lieber}, opened up a new
subject of magnetism.

\setcounter{section}{0} 
\renewcommand{\thesection}{\thechapter.\Alph{section}} 
\setcounter{equation}{0}
\renewcommand{\theequation}{\thesection\arabic{equation}}
\section*{Appendices}

\section{The vector potential of a current loop and the definition of the magnetic moment.}
\label{App:Ampere}
One introduces the definition of the magnetic moment when the $\vec A$-field of a current circulating along a small loop is computed, based on Amp\`ere's hypothesis. We consider a thin wire forming a small, closed loop $C$ that bounds a surface $S$. For such a thin wire, the $\vec A$-field is computed from a line integral along the loop $C$ that is subsequently transformed, via Stokes' theorem, into an integral over the surface $S$:
\begin{equation}
\vec{A}(\vec{r}) = \frac{\mu_0 I}{4\pi} \oint_C \frac{1}{|\vec{r}-\vec{r}'|} \,\mathrm{d}\vec{s}' = \frac{\mu_0 I}{4\pi} \int_{S} \vec{n} \times \vec{\nabla}' \left( \frac{1}{|\vec{r}-\vec{r}'|} \right) \mathrm{d}S'.
\label{A_App}
\end{equation}
In equation \eqref{A_App}, the coordinates along the loop are denoted by $\vec{r}'$, $\mathrm{d}\vec{s}'$ is the vector line element along the path, $\vec{n}$ is the unit normal vector to the surface $S$, and $\vec{\nabla}'$ denotes the gradient operator with respect to the source coordinates $\vec{r}'$.
Evaluating the gradient with respect to the primed coordinates via the chain rule yields:
\begin{equation}
\vec{\nabla}' \left( \frac{1}{|\vec{r}-\vec{r}'|} \right) = \frac{\vec{r}-\vec{r}'}{|\vec{r}-\vec{r}'|^3}.
\label{Eq:Nabla_App}
\end{equation}
For an ideal point dipole located at the origin, the spatial dimensions of the loop shrink to zero ($|\vec{r}'| \to 0$), while the observation point remains far away ($|\vec{r}| \gg |\vec{r}'|$). We can approximate equation \eqref{Eq:Nabla_App} to first order by setting $\vec{r}' \approx \vec{0}$:
\begin{equation}
\vec{\nabla}' \left( \frac{1}{|\vec{r}-\vec{r}'|} \right) \approx \frac{\vec{r}}{|\vec{r}|^3} + \mathcal{O}\left(\frac{|\vec{r}'|}{|\vec{r}|^3}\right).
\end{equation}
Substituting this expansion back into the surface integral, the term $\vec{r}/|\vec{r}|^3$ depends solely on the unprimed observation coordinate and can be factored outside of the integration:
\begin{equation}
\vec{A}(\vec{r}) \approx \frac{\mu_0 I}{4\pi} \left( \int_{S} \vec{n} \,\mathrm{d}S' \right) \times \frac{\vec{r}}{|\vec{r}|^3}.
\end{equation}
We define the magnetic dipole moment $\vec{\mu}$ as the product of the current and the vector area of the loop:
\begin{equation}
\vec{\mu} = I \int_{S} \vec{n} \,\mathrm{d}S' = I S \vec{n}.
\end{equation}
Using this definition, the classic expression for the vector potential of a point magnetic dipole takes the familiar form:
\begin{equation}
\vec{A}(\vec{r}) = \frac{\mu_0}{4\pi} \frac{\vec{\mu} \times \vec{r}}{|\vec{r}|^3} = \frac{\mu_0}{4\pi} \vec{\nabla} \times \frac{\vec{\mu}}{|\vec{r}|}.
\end{equation}
The expression on the right-hand side is established directly using a standard identity of vector analysis. Further details about the magnetic field produced by a single magnetic dipole can be found in Ref.~\cite{Jack,Point_dipole}.

\section{The intra-atomic exchange interaction.}
\label{App:exchange}
Our aim is the quantitative computation of the strength of the multiplet splitting in a mathematically simple enough situation. For this purpose, we conduct the analysis of the energy eigenvalues of the two configurations of the $He$-atom: the $1s^2$ (the actual ground state) and the $(1s2s)$ excited state configuration.
\paragraph{The configuration energy eigenvalues.} 
The configuration Hamiltonian for two electrons in the central field of two point-like protons consists of the simplified Hartree equations (\cite{Hartree} \cite{Landau_239}, p.239) where the reciprocal interaction of the two electrons is completely neglected ($e^2$ standing for  $\frac{e^2}{4\pi \epsilon _0}$):
\begin{eqnarray}
-\frac{\hbar ^2}{2m}(\mathbf \nabla_1^2)\psi_1(r_1)+\big[- \frac{2\cdot e^2}{r_1}\big]\psi_1(r_1)&=& E_1\cdot \psi_1(r_1)\nonumber\\
-\frac{\hbar ^2}{2m}(\mathbf \nabla_2^2)\psi_2(r_2)+\big[- \frac{2\cdot e^2}{r_2}\big]\psi_2(r_2)&=& E_2\cdot \psi_2(r_2)
\end{eqnarray}
The separate Schr\"odinger-like single electron differential equations are solved by the hydrogen-like orbitals. For the $(1s^2)$ configuration we have
\begin{equation}
\psi_1(r_1)=\frac{R_{1,s}}{r_1}\cdot Y_{0,0}(\vartheta,\varphi)\quad \quad \psi_2(r_2)=\frac{R_{1,s}}{r_2}\cdot Y_{0,0}(\vartheta,\varphi)
\end{equation}
and
\begin{equation}
E_{(1s)^2}= 2\cdot E_{1s}
\end{equation}
For the $(1s)(2s)$ configuration we have
\begin{equation}
\psi_1(r_1)=\frac{R_{1,s}}{r_1}\cdot Y_{0,0}(\vartheta,\varphi)\quad \quad \psi_2(r_2)=\frac{R_{2,s}}{r_2}\cdot Y_{0,0}(\vartheta,\varphi)
\end{equation}
and
\begin{equation}
E_{(1s)(2s)}= E_{1s} + E_{2s}
\end{equation}
\paragraph{The multiplet splitting.} We construct the multiplets of the two configurations $(1s)^2$ and $(1s2s)$.\\
\noindent $\quad (1s)^2$: The configurational basis states
\begin{eqnarray}
\psi_{1s}Y^+\psi_{1s}Y^{+}\quad \psi_{1s}Y^+\psi_{1s}Y^{-}\quad \psi_{1s}Y^-\psi_{1s}Y^{+}\quad \psi_{1s}Y^-\psi_{1s}Y^{-}
\end{eqnarray}
can be organized into basis states of the multiplet $(L=0,S=0)$ (the spin state of the singlet is antisymmetric with respect to particle exchange, so that the orbital state must be symmetric with respect to particle exchange):
\begin{equation}
\psi_{1s}\psi_{1s}\otimes \frac{1}{\sqrt{2}}\big(Y^+Y^--Y^-Y^+\big)
\end{equation}
and $(L=0,S=1)$ (the spin states of the triplet are symmetric with respect to particle exchange, so that the orbital state must be antisymmetric with respect to particle exchange):
\begin{eqnarray}
\frac{1}{\sqrt{2}}\big(\psi_{1s}\psi_{1s}-\psi_{1s}\psi_{1s}\big)\otimes Y^+Y^+\nonumber\\
\frac{1}{2}\big(\psi_{1s}\psi_{1s}-\psi_{1s}\psi_{1s}\big)\otimes\big(Y^+Y^-+Y^-Y^+\big)\nonumber\\
\frac{1}{\sqrt{2}}\big(\psi_{1s}\psi_{1s}-\psi_{1s}\psi_{1s}\big)\otimes Y^-Y^-
\end{eqnarray}
The problem with the basis states of the multiplet $(0,1)$ is that their orbital component vanishes exactly, so that only the multiplet $(0,0)$
exists. The ground state of the two-electron system is a spin singlet, in agreement with the most simple version of the Pauli principle. It is represented in the literature in a simplified way as $(1sY^+,1sY^-)$.\\
\noindent $\quad (1s2s)$. The multiplet basis function for $(0,0)$ writes
\begin{equation}
\frac{1}{\sqrt{2}}\big(\psi_{1s}\psi_{2s}+\psi_{2s}\psi_{1s}\big)\otimes \frac{1}{\sqrt{2}}\big(Y^+Y^--Y^-Y^+\big)
\end{equation}
The multiplet basis functions for $(0,1)$ write
\begin{eqnarray}
\frac{1}{\sqrt{2}}\big(\psi_{1s}\psi_{2s}-\psi_{2s}\psi_{1s}\big)\otimes Y^+Y^+\nonumber\\
\frac{1}{2}\big(\psi_{1s}\psi_{2s}-\psi_{2s}\psi_{1s}\big)\otimes\big(Y^+Y^-+Y^-Y^+\big)\nonumber\\
\frac{1}{\sqrt{2}}\big(\psi_{1s}\psi_{2s}-\psi_{2s}\psi_{1s}\big)\otimes Y^-Y^-
\end{eqnarray}

\noindent The multiplet Hamiltonian considers perturbatively the Coulomb interaction
\begin{equation}
\frac{e^2}{\mid \mathbf r_1-\mathbf r_2\mid }\dot=\frac{e^2}{r_{12}}
\end{equation}
between the two electrons and finds its eigenvalues in the subspace spanned by the four multiplet basis states. The Ritz-matrix of the complete Hamiltonian, including the configuration Hamiltonian and the electron-electron interaction  Hamiltonian, writes
\begin{equation}
  \begin{psmallmatrix}
     E_{1s}+E_{2s}+Q_{1s2s}+J_{1s2s}&0&0&0\\
      0&E_{1s}+E_{2s}+Q_{1s2s}-J_{1s2s}&0&0\\
      0&0&E_{1s}+E_{2s}+Q_{1s2s}-J_{1s2s}&0\\
      0&0&0&E_{1s}+E_{2s}+Q_{1s2s}-J_{1s2s}\\
      \end{psmallmatrix}
\end{equation}
with
\begin{equation}
\label{Eq:Q}
Q_{(1s)(2s)}= \int dV \frac{\mid \psi_{1s}(\mathbf r_1)\mid^2\cdot e^2\cdot\mid \psi_{2s}(\mathbf r_2)\mid^2}{r_{12}}
\end{equation}
being the electrostatic interaction between the two charge densities $\mid \psi_{1s}\mid ^2,\mid \psi_{2s}\mid ^2$. $Q_{(1s)(2s)}$
is also expected from classical electrostatics and produces, as for the ground state energy, a shift of the energy of the
$(1s)(2s)$ configuration. The integral $J$
\begin{equation}
J_{(1s)(2s)}= \int dV \frac{\psi_{1s}(\mathbf r_1)^*\psi_{2s}^*(\mathbf r_2)\cdot e^2\cdot\psi_{1s}(\mathbf r_2)\psi_{2s}(\mathbf r_1)}{r_{12}}
\label{Eq:J}
\end{equation}
is referred to as the intra-atomic exchange integral. ``Intra-atomic'' because the wave functions appearing in it are centered at the same
atom. The exchange energy $J_{(1s)(2s)}$ produces a removal of the degeneracy
between the multiplets $(0,0)$ and $(0,1)$ and, accordingly, a split of the configuration energy level. The parameters $Q$ and $J$ have the same order of magnitude ($\approx$ eV) as they both originate from the Coulomb repulsion between the two electrons\cite{Dav}.
\begin{figure}[H]
\begin{center}
\includegraphics[width=0.4\textwidth]{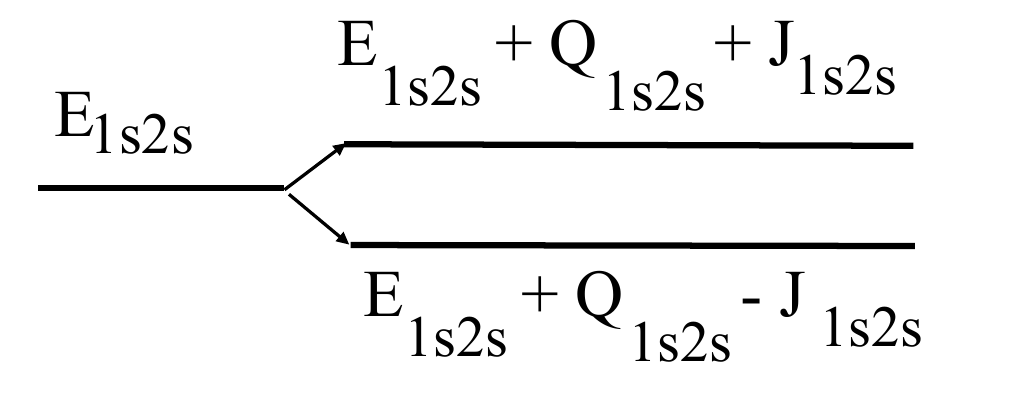}
\caption{Schematic energy diagram summarizing the triplet-singlet splitting.} \label{Fig:1s2s}
\end{center}
\end{figure}
\noindent The sign of the intra-atomic exchange integral is positive so that the triplet
has a lower energy than the singlet (the positivity of $J$ is the first Hund
rule). There is an intuitive explanation for this positivity. The Coulomb energy
is large when the two electrons are closer to each other. In a triplet spin
state the antisymmetric orbital wave function takes care that
the two electrons are as far as possible from each other, and this reduces the
Coulomb repulsion with respect to the symmetric orbital wave function, where the
two electrons are, on the average, allowed to be closer to each other. The
kernel of the exchange integral is the Coulomb interaction, but the ordering of
the wave functions in it originates from the antisymmetrizing requirement,
which is purely of quantum mechanical origin. One explains this hybrid content
by introducing a novel fundamental interaction of Nature, the so called (intra-atomic) \textbf{exchange interaction}, see Appendix \ref{App:HDvV}.

\section{Heisenberg-Dirac-Van Vleck representation of the exchange interaction.}
\label{App:HDvV}
We seek an operator that formally acts in the space of the two spin $\frac{1}{2}$ particles
\begin{equation}
Y^+\otimes Y^+\quad\quad Y^+\otimes Y^-\quad\quad Y^-\otimes Y^+\quad\quad Y^-\otimes Y^-
\end{equation}
and has the same eigenvalues as the multiplet operator. This operator is known as the Heisenberg-Dirac-Van Vleck operator.\\
Dirac defined an operator acting in spin space
\begin{equation}
H_{Spin} = (E_{1s} + E_{2s} + Q_{1s2s})\cdot \mathds{1} - J_{1s2s}\cdot P_{12}
\end{equation}
with the exchange operator $P_{12}$
\begin{equation}
P_{12}Y^{\pm }Y^{\pm }=Y^{\pm }Y^{\pm }\quad\quad P_{12}Y^{\pm }Y^{\mp }=Y^{\mp }Y^{\pm }
\end{equation}
The eigenvalues of the operator $H_{spin}$, when restricted to the $1s2s$ eigenspace, are identical with the eigenvalues of the multiplet operator. A useful way of writing $P_{12}$ is
\begin{equation}
P_{12} = \frac{\mathds{1}+\mathbf \sigma _1\otimes\mathbf \sigma _2}{2}
\end{equation}
or, setting $\mathbf S = \frac{1}{2}\mathbf \sigma$
\begin{equation}
P_{12} = \frac{1}{2}[\mathds{1} + 4\cdot \mathbf S_1\otimes\mathbf S_2]
\end{equation}
where $\mathbf \sigma$ are the two-by-two Pauli matrices and the product $\sigma_i\otimes \sigma_j$ must be taken as a Kronecker product of matrices.
The spin Hamiltonian simulating the original Hamiltonian acting in the orbital space reads
\begin{equation}
\label{Eq:Heis}
\left[E_{1s} + E_{2s}\right]\cdot \mathds 1\otimes \mathds 1 + \left[Q_{1s2s}-\frac{1}{2}J_{1s2s}\right]\cdot \mathds 1\otimes \mathds 1 - 2\cdot J_{1s2s}\cdot  \vec S _1\otimes\vec S_2
\end{equation}
Eq.~\eqref{Eq:Heis} entails the use of the (dimensionless) spin operators
$\vec S$, defined using the Pauli matrices: $\vec S \doteq \frac{1}{2}\vec \sigma$. The non-diagonal part $- 2\cdot J_{1s2s}\cdot \vec S _1\otimes\vec S_2$
is also called the Heisenberg-Dirac  exchange (vector) operator\cite{Heisenberg,Dirac}.
A simplified version of it
\[-2\cdot J\cdot S^z_1\otimes S^z_2\]
involving only the $z$-component of the spin operator is called the ``Ising model''\cite{Ising}.
Versions of the Heisenberg and Ising models where $\mathbf S$ is a classical vector are also used. In physical science, but also in any other branch of natural science and even in the humanities, the Heisenberg and Ising models, in particular in their generalization to many degrees of freedom
\[-2\cdot J\cdot \sum_{i\not=j} \mathbf S_i\otimes \mathbf S_j\]
respectively
\[-2\cdot J\cdot \sum_{i\not=j}S_{z_i}\otimes S_{z_j}\]
have a widespread use. They represent the simplest way of capturing and describing two-body interactions -- such as those appearing in a community of fishes or in the traffic of cars moving within restricted roads -- and lead to non-trivial complexity.\\
Dirac\cite{Dirac} has shown that, to first order perturbation theory, the energy level $E_0$ of a two-body system is modified, quite generally, by the interaction between the two bodies to
\begin{equation}
E_0\cdot\mathds 1\otimes \mathds 1 + U\cdot\mathds 1\otimes \mathds 1 - 2\cdot J\cdot \vec S _1\otimes\vec S_2
\end{equation}
$U$ and $J$ being parameters that suit the specific situation.

\section{Mean field approximation of the Heisenberg interaction.}
\label{App:MFA}
The Dirac-Heisenberg interaction is a two-body interaction. It can be
transformed into a single body interaction in order to accommodate the exchange
interaction within the single particle approximation of energy levels in atoms
and later in solids. The way of translating the two-body interaction into a
single body one is to replace, in the two-body product operator, one operator
$\vec S$ with its quantum mechanical expectation value $<S_z>$ (one
assumes $z$ as the quantization axis, established, e.g., by an applied magnetic
field, so that $<\vec S>=<S_z>$) -- the so called Mean Field Approximation,
``MFA'':
\begin{eqnarray}
\label{Eq:MFA}
H_{\text{Heis}}&=& E_0\cdot\mathds 1\otimes \mathds 1 + U\cdot\mathds 1\otimes \mathds 1 - 2\cdot J\cdot \vec S _1\otimes\vec S_2 \nonumber\\
&\longrightarrow &\nonumber\\
H_{MFA}&=&
(E_0+U)\cdot\mathds 1 - 2\cdot J\cdot <S_z>\cdot S_z
\end{eqnarray}
One often uses, instead of $<S_z>$, the identical quantity $n\cdot \frac{P}{2}$, with $P\doteq 2<S_{z}>/n$ the spin polarization of the electrons and $n$ the number of electrons that participate in the formation of the total spin, so that, writing $I$ for $J$ as is customary in the band-magnetism literature,
\begin{equation}
H_{MFA}=(E_0+U)\cdot\mathds 1 - I\cdot n\cdot P\cdot S_z
\end{equation}
In the MFA, one convenes that $<S_z>$ (the mean local spin state) is taken to be
positive for so-called ``majority'' electrons. In atomic Fe, for instance, $<S_z>= 2$, i.e. $n=4$ and $P=1$. The quantity
\begin{equation}
I\cdot n\cdot P
\end{equation}
provides a mean field (with the units of ``energy'') for the single particle operator $S_z$, and is the exchange splitting between the majority and the minority states. The parameter $I$ is adjusted to fit the triplet-singlet splitting. For determining $I$ in Fe, for instance, one uses the triplet-singlet splitting between the two Fe-multiplets $^5D$ (the ground state, with $S=2$) and $^3D$, which amounts to $3.6$ eV\cite{NIST}, and sets $P=1$, to obtain $I\approx 0.9$ eV for Fe. This empirical estimate compares favorably with the results of state-of-the-art electronic structure computations for 3d transition metals, see e.g. Ref.~\cite{Gunnarson}.

\section{Stoner criterium with free electron bands.}
\label{App:Stoner}
For the sake of mathematically illustrating
the Stoner model\cite{Stoner}, we consider an electron gas described by the energy levels $E=\hbar ^2\vec k^2/2m$ and the density of states
\begin{equation}
n_0(E) =
\frac{3}{4}\cdot n\cdot \frac{1}{E_{F}^{\frac{3}{2}}}\cdot \sqrt{E}
\label{Eq:free}
\end{equation}
$n$ being the total number of electrons per unit cell and $E_{F}$ being the Fermi level. Without any exchange interaction, both spin channels $\uparrow$ and $\downarrow $ have the same energy values and the same density of states
$n_0(E)$. We now introduce, by means of the mean field approximation, an energy split between majority and minority spins:
\begin{equation}
E^+=\hbar ^2\vec k^2/2m\quad \quad E^-=\hbar ^2\vec k^2/2m + I\cdot n\cdot P
\end{equation}
Extended band structure calculations show the value of the exchange
integral $I$ to be used in the solid state to be very much the atomic one. $n$ is
also known from the atomic configuration (in the case of Fe one would assume
$n=4$, i.e. the number of unpaired electrons in the $d$-band). $P$, instead, is
the sought for quantity, and $\mu_B\cdot n\cdot P$ is the magnetic moment per
atom. $P$ for an atom is just ``1'', but for a solid it might be smaller,
depending on the energy balance. In fact, one finds $P$ by minimizing the total
energy of the electron gas. The energy consists of two terms. The first term is its kinetic energy
\begin{eqnarray}
\label{Eq:Ekin}
E_{kin}\!=\!\int_0^{E_F(P)}dE\cdot E\cdot n_0(E)
\!+\!\int_{I\cdot n\cdot P}^{E_F(P)}dE\cdot (E-I\cdot n\cdot P)\cdot n_0(E-I\cdot n\cdot P)
\end{eqnarray}
The total kinetic energy contains the integration up to the Fermi level, which
is a function of $P$: $E_F(P)$. There is a need of changing the
Fermi level when a finite $P$ is introduced. By the process of emptying states
in the minority spin channel, a finite $P$ is produced. Simultaneously, the
total number of electrons (given the fact that each level in the minority spin
channel is occupied by exactly one electron, in virtue of the Pauli principle)
is reduced. This must be compensated by increasing the Fermi level by a certain amount, which must be determined.\\
The second term is the self-magnetic energy. The self-magnetic energy per unit cell is minus $\frac{1}{2}$ the product of the exchange field $\frac{I\cdot n\cdot P}{2\mu_B}$ and the magnetic moment per unit cell, $\mu_B\cdot n\cdot P$, giving
\begin{equation}
E_{\text{magn}}= -\frac{1}{4}\cdot I\cdot n^2\cdot P^2
\end{equation}
The equilibrium value of $P$ results from finding the minimum of the total energy $E_{kin}(P)+ E_{magn}(P)$. In the process of minimization, one must take two boundary conditions into account:
\begin{eqnarray}
\label{Eq:boundary1}
n&=&\int_0^{E_F(P)}dE\cdot n_0(E)+ \int_{I\cdot n\cdot P}^{E_F(P)}dE\cdot n_0(E-I\cdot n\cdot P)\nonumber\\
n\cdot P &=& \int_0^{E_F(P)}dE\cdot n_0(E)- \int_{I\cdot n\cdot P}^{E_F(P)}dE\cdot n_0(E-I\cdot n\cdot P)
\end{eqnarray}
These two boundary conditions can be manipulated, mathematically, to produce two useful relations for $E_F(P)$:
\begin{equation}
2(E_F^0)^\frac{3}{2}=(E_F^P)^\frac{3}{2}+ (E_F^P-InP)^\frac{3}{2}\quad 2\cdot P\cdot (E_F^0)^\frac{3}{2}= (E_F^P)^\frac{3}{2}-(E_F^P-InP)^\frac{3}{2}\end{equation}
i.e.
\begin{equation}
2= \left(\frac{E_F^P}{E_F^0}\right)^\frac{3}{2}+ \left(\frac{E_F^P}{E_F^0}-\frac{InP}{E_F^0}\right)^\frac{3}{2}\quad \quad 2P= \left(\frac{E_F^P}{E_F^0}\right)^\frac{3}{2}-\left(\frac{E_F^P}{E_F^0}-\frac{InP}{E_F^0}\right)^\frac{3}{2}
\end{equation}
i.e.
\begin{equation}
\left(\frac{E_F^P}{E_F^0}\right)^\frac{3}{2}= 1+P \quad \quad 1-P= \left(\frac{E_F^P}{E_F^0}-\frac{InP}{E_F^0}\right)^\frac{3}{2}
\end{equation}
i.e.
\begin{equation}
\left(\frac{E_F(P)}{E_F(0)}\right)^\frac{3}{2}= 1+P \quad \quad
\frac{E_F(P)-InP}{E_F(0)}=(1-P)^{\frac{2}{3}}
\end{equation}
We use these last two relations for the computation of the integrals appearing in the expression for the total kinetic energy:
\begin{equation}
\int_0^{E_F(P)}dE\cdot E\cdot n_0(E)=\frac{3}{10}\cdot n\cdot E_F(0)\cdot \left(\frac{E_F(P)}{E_F(0)}\right)^{\frac{5}{2}}= \frac{3}{10}\cdot n\cdot E_F(0)\cdot (1+P)^{\frac{5}{3}}
\end{equation}
\begin{equation}
\int_{0}^{E_F(P)-I\cdot n\cdot P}dE\cdot E\cdot n_0(E)=\frac{3}{10}\cdot n \cdot E_F(0)\cdot \left(\frac{E_F(P)-I\cdot n\cdot P}{E_F(0)}\right)^{\frac{5}{2}}= \frac{3}{10}\cdot n\cdot E_F(0)\cdot (1-P)^{\frac{5}{3}}
\end{equation}
The total kinetic energy writes, accordingly, (in the limit of small $P$)
\begin{equation}
\frac{3}{10}\cdot n\cdot E_F(0)\cdot \left(2 + \frac{10}{9}\cdot P^2 + \frac{10}{243}\cdot P^4 + O(P^6)\right)
\end{equation}
It is usual to write this expression by replacing $E_F(0)$ with
$\frac{3}{4}\cdot n\cdot \frac{1}{n_0(E_F(0))}$, so that the total energy, consisting of the kinetic energy and the magnetic energy, writes
\begin{equation}
\frac{9}{20}\cdot n^2\cdot \frac{1}{n_0(E_F(0))}+ \frac{1}{4}\cdot n^2\cdot \left(\frac{1}{n_0(E_F(0))}-I\right)\cdot P^2 + \frac{1}{4}\cdot\frac{1}{27}\cdot n^2\cdot \frac{1}{n_0(E_F(0))}\cdot P^4 +....
\end{equation}
The minimization of this functional provides three solutions for $P$: a trivial one, $P=0$, and
\begin{equation}
P = \pm\sqrt{\frac{27}{2}}\cdot \sqrt{n_0(E_F(0))\cdot I -1}
\end{equation}
The argument of the square root is positive and the non-trivial finite $P$ solutions represent possible physical states if
\begin{equation}
n_0(E_F(0))\cdot I -1>0
\end{equation}
(Stoner criterium for ferromagnetism in metals). Notice that the criterium
decides the \emph{existence} of the ferromagnetic state, while the size of $P$ --
and with it the size of the magnetic moment $\mu_B nP$ -- is fixed by how far
above threshold $n_0(E_F(0))\cdot I$ lies. This is the quantitative content of
the statement, made in Sec.~3.6, that a modest change of the density of states
at the Fermi level can turn a layer from ``dead'' into ``enhanced''.

\section{The physical principles of virtual bound states.}
\label{App:vbs}
\begin{figure}[H]
\begin{center}
\includegraphics[width=0.6\textwidth]{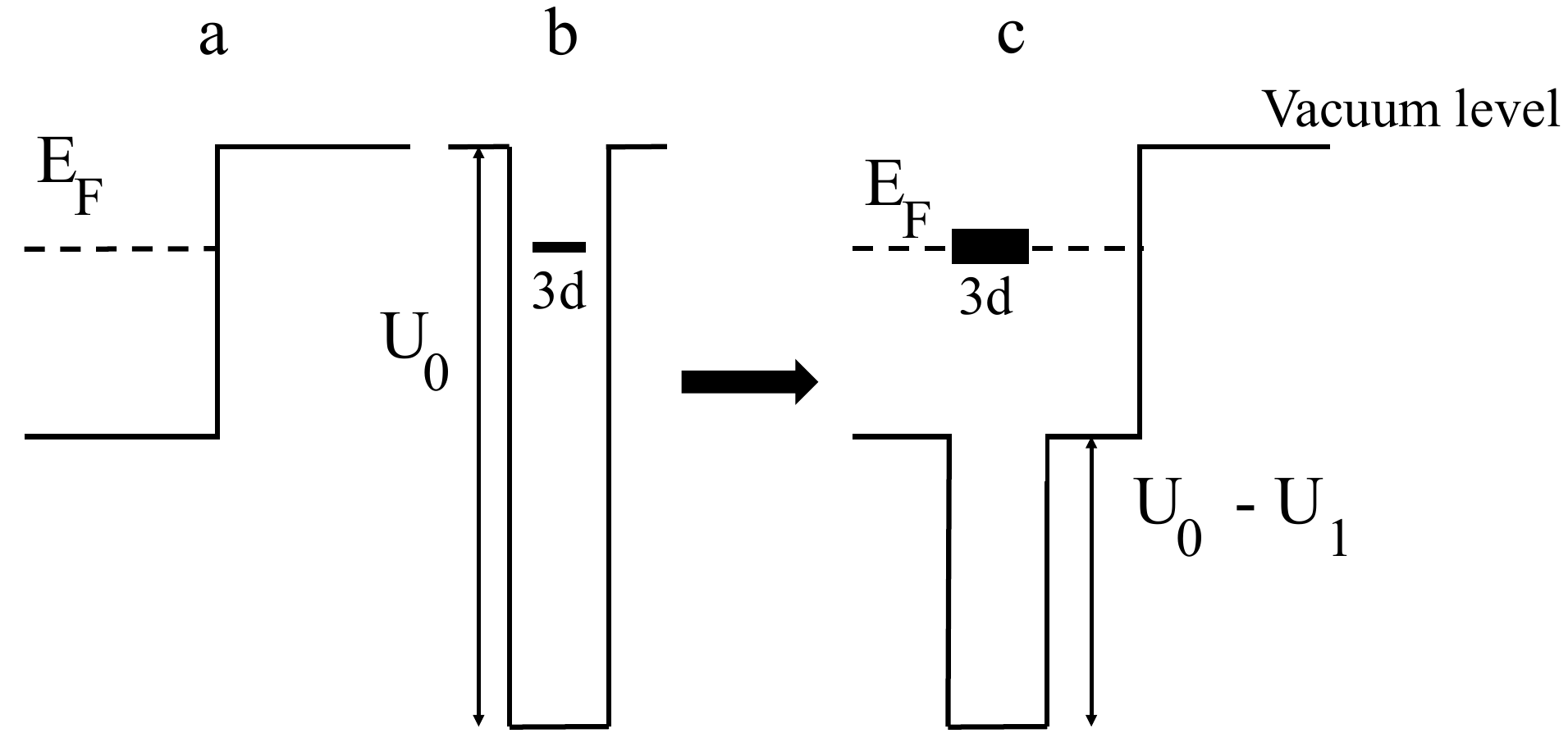}
\caption{Energy diagram (potential as a function of a coordinate) for a solid (a), an isolated atom (b) and an atom embedded into the solid (c).
\label{Fig:Friedel}}
\end{center}
\end{figure}

The situation of a $3d$-transition metal embedded as an impurity into a solid with $sp$-like electrons must take into account that the $sp$-like electrons are hosted by plane-wave like energy levels. The Fermi level controls the occupation of the plane-wave like energy levels. The $3d$-levels of the transition metal impurity are pinned to the Fermi level of the host $E_F$ by the requirement of partial occupation. For instance, the Fe atom has 6 $3d$-electrons and must develop, when embedded into an $sp$-like solid as an impurity, into a ``level'' that hosts 6 $3d$-electrons at or below $E_F$. The remaining four empty places in the Fe-$3d$-configuration must be accommodated above $E_F$, as unoccupied states. This means that the $3d$-atomic levels must be found close to the Fermi level and within the continuum spectrum of a suitable quantum mechanical problem. The most appropriate model to describe the quantum mechanical situation of a sharp atomic level degenerate with a continuum of energy levels was proposed by J. Friedel\cite{Friedel} and later developed by P.W. Anderson\cite{Anderson}: the atomic level becomes a ``virtual bound state''.\\
Fig.~\ref{Fig:Friedel} illustrates a back-of-the-envelope model of the physical principles spelled out by Friedel. In (a) the solid is illustrated by an energy diagram. The horizontal direction is some coordinate across the solid. The solid terminates at a boundary at which the potential, represented along the vertical scale, jumps to the vacuum level. The potential inside the solid is referred to as the ``optical'' potential, hosting free electrons up to the Fermi level $E_F$. In (b), an isolated atom is represented by a potential well. The $3d$-levels (indicated) are bound states of the atom and are, accordingly, sharp. In (c) the atom is embedded into the solid. The potential well joins the optical potential. The atomic $3d$-levels, originally below the potential well threshold, find themselves above the potential well threshold and become ``virtual bound states''.
As we will show below on the basis of a one-dimensional back-of-the-envelope
model, virtual bound states have a finite energy width, which tends to reduce
their density of states. The Stoner model then translates this reduced density
into a reduced magnetic moment. Friedel's original argument explains the finite
width as follows. By inserting the atom into a host crystal, a perturbing
potential is added (think of the optical potential) that renders the atomic
potential less attractive. It is convenient to think that the bound state still
exists but finds itself degenerate with some plane-wave like state with exactly
the same energy. One can imagine that the orbital corresponding to the bound
state has a finite matrix element with the plane wave state of the same
energy. One can also imagine that this matrix element is related to the
Fourier transform of the atomic orbital. The atomic orbital is very
localized in space, so that we expect its Fourier transform (and the matrix
element) at any $k$ to be small. But it is finite, so that by virtue of a
simple minded degenerate perturbation theory, it splits the degenerate original
level into two closely spaced adjacent sublevels. These sublevels are
themselves degenerate with further extended state levels with the same energy,
which, accordingly, split into two closely spaced sublevels, and so on. The
result of this process is that the original ``virtual bound state'' acquires
a finite width and, accordingly, distributes its density of states over a
finite energy interval. Formally, the states within this structure are
spatially extended, but the wave function has a strong amplitude at the site
of the impurity. Accordingly, provided some amount of magnetic moment is kept
alive, it is quite localized at the atomic site.\\
Here are now Friedel's arguments, rendered within a one dimensional potential well model $U_0(x)$, with depth $\vert U_0\vert$ and width $2a_0$. In this model, one describes the energy of the bound states (for the computation of the energy of a potential box see e.g. Ref.~\cite{DP_QM}, p.~24) as
\begin{equation}
E_n = \frac{\hbar^2}{2m}\cdot k_n^2\quad\;\quad k_n= \frac{\pi\cdot n}{2a_0}
\end{equation}
$n=1,2,3,...$ (for this formula to be precise, $n$ must be small enough for the infinite potential well approximation to hold true). Close to the potential well threshold the values for $E_n$ must be computed using the exact solution\cite{Dav}). We recall that the wave functions of the bound states for an infinite potential well vanish outside the interval $[-a_0,a_0]$ and are given by
\begin{equation}
\sqrt{\frac{1}{a_0}}\cos k_n \cdot x
\end{equation}
($n$ odd) respectively
\begin{equation}
\sqrt{\frac{1}{a_0}}\sin k_n \cdot x
\end{equation}
for $n$ even.
In the concrete case of a potential well with finite depth, the wave functions of the bound states have a non-vanishing, exponentially decreasing amplitude on both sides of the interval. We normalize the exponential tails of the total wave function to unity and obtain, in $\vert x\vert>a_0$,
\begin{equation}
\sqrt{\eta}\cdot e^{a_0\cdot \eta}\cdot e^{-\eta\cdot \vert x\vert}
\end{equation}
with $\eta= \sqrt{k_0^2-k_n^2}$, $k_0=\sqrt{\frac{2m U_0}{\hbar ^2}}$ (the exact normalization to unity would entail taking
into account the wave function inside the interval as well, so that the
coefficient of the exponential function would be slightly modified accordingly.
Again, we do not need this kind of precision here). We call the wave function
of the $n$-th bound state $\psi_n(x)$, $\psi_n(x)$ entailing both the oscillating
component within the interval and the exponentially decreasing components
outside it. We now lower the threshold of the square well potential by an
amount $-U_1$ (in Fig.~\ref{Fig:Friedel}, the potential well becomes $U_1(x)$ with a depth $U_0-U_1$, lowered to
meet the optical potential inside the solid). By this, the original square
well is rendered less attractive and thus becomes suitable for the
emergence of a virtual bound state. We consider, in fact, one of the bound
states $n=i$ which, when the potential outside the interval is lowered by some
amount, acquires an energy in the range of the extended states of the modified
potential. Its wave function, originally localized at the potential well site,
will decay into some extended solution according to an exponential law and with
a typical decay rate, given by the Fermi golden rule. The exponential decay produces a finite width of the virtual bound state which amounts to $\hbar$ times the decay rate and writes\cite{Ded}
\begin{equation}
\Delta_{b\rightarrow q}= 2\pi\cdot \vert \left(\psi_q(x),U_1(x)\psi_i(x)\right)\vert^2\cdot \rho(E_q)
\end{equation}
$E_q$ is the energy of the extended state, $E_q=E_i-(U_0-U_1)\geq 0$. $q$ is the corresponding wave number:
\begin{equation}
q= \sqrt{\frac{2m\cdot E_q}{\hbar^2}}
\end{equation}
$\rho(E_q)$ is the density of extended states with energy $E_q$.
For computing the matrix element, we approximate the extended function with an escaping plane wave, e.g. to the right: $\frac{1}{\sqrt{L}}\cdot e^{i\cdot q\cdot x}$. The plane wave is normalized within an interval $[-\frac{L}{2},\frac{L}{2}]$, with $L>>a_0$ and with periodic boundary conditions at $L$. We compute the matrix element according to
\begin{equation}
 -U_1\cdot \int_{a_0}^\infty \overline{\psi_q(x)}\psi_i(x)\,dx
 \end{equation}
 and find, solving the integral,
 \begin{equation}
\vert\left(\psi_q(x),U_1(x)\psi_i(x)\right)\vert^2= \frac{1}{L}\cdot U_1^2\cdot \eta\cdot \frac{1}{q^2+\eta^2}
\end{equation}
The probability to escape into the interval to the left is exactly the same, so that the escape probability enters the expression for the width with a factor of two. For finding the width of the virtual bound state, we need also the density of states at the energy $E_q$. We assume free electrons as extended waves and find that
$\rho(E_q) = \frac{L}{2\pi}\cdot \frac{m}{\hbar^2\cdot q}$. The parameter $L$, defining artificial walls at $-\frac{L}{2},\frac{L}{2}$, cancels out when computing the width, which finally amounts to
\begin{equation}
\Delta_{i\rightarrow q}= 4\cdot \frac{m}{\hbar^2\cdot q}\cdot U_1^2 \cdot \frac{\eta}{q^2+\eta^2}
\end{equation}
For estimating the significance of this one-dimensional model for a realistic three-dimensional virtual bound state, we consider the Fourier transform of an atomic orbital. This Fourier transform provides us with an estimate of the matrix element between a bound state decaying as $\propto e^{-\eta\cdot r}$ ($r$ being the radial variable) and a plane-wave like extended state. The qualitative form of these Fourier transforms is the same and is already contained in the simplest case of the $1s$-orbital\cite{Gem_1s}:
\begin{equation}
\Phi(\vec q) = \frac{8\cdot \sqrt{\pi}\cdot \eta^{\frac{5}{2}}}{\left(\eta^2+4\pi^2\cdot\vert\vec q\vert^2\right)^2}
\end{equation}
The Fourier transform (and with it the width of the virtual bound state) decays
like some power of the parameter $\eta$, which fixes the amount of localization
of the orbital. Orbitals with non-vanishing quantum number $L$ have a Fourier
transform that depends on the direction of $\vec q$, but it replicates the
rotational symmetry underlying the atomic orbitals. $\vert q\vert \geq 0$ corresponds (approximately) to an energy between the origin of the ``optical potential'' of a solid (or the ``muffin tin zero'') and the Fermi energy (in our 1D model, the zero of the optical potential is at $U_0-U_1$).\\
The exact extended-state solution of the potential well problem, which we solve now, provides a justification of the perturbational approach based on the Fermi golden rule.
We seek, for the sake of illustration, the solution of even parity of the Schr\"odinger equation with the Ansatz
\begin{eqnarray}
\psi _{I} &=& A\cdot\frac{1}{\sqrt{L}}e^{i q\cdot x} + B\cdot \frac{1}{\sqrt{L}} e^{-i q\cdot x}\nonumber\\
\psi _{II} &=& \alpha\cdot \frac{2}{\sqrt{L}}\cdot \cos (k\cdot x)\nonumber\\
\psi _{III} &=& B\cdot\frac{1}{\sqrt{L}}e^{i q\cdot x} + A\cdot \frac{1}{\sqrt{L}} e^{-i q\cdot x}
\end{eqnarray}
The relation between $q$ and $k$ is important and is given, in the present notation, by
\begin{equation}
q^2 = k^2-(k_0^2-k_1^2)
\end{equation}
We find the relations between the sought for amplitudes $A,B,\alpha$ by matching the wave functions and their derivatives e.g. at $-a_0$:
\begin{eqnarray}
A\cdot e^{-i q\cdot a_0}+ B\cdot e^{i q\cdot a_0}&=& 2\cdot \alpha\cdot \cos (k\cdot a_0)\nonumber\\
A\cdot i\cdot q\cdot e^{-i q\cdot a_0}-B\cdot i\cdot q\cdot e^{i q\cdot a_0}&=&k\cdot 2\cdot \alpha \cdot \sin k\cdot a_0
\end{eqnarray}
We eliminate from these coupled algebraic equations the parameter $B$, respectively $A$, to find the relation
\begin{equation}
\vert\frac{\alpha}{A}\vert^2 = \vert\frac{\alpha}{B}\vert^2= \frac{1}{1+\frac{(k_0^2-k_1^2)\cdot a_0^2}{(q\cdot a_0)^2}\cdot \sin^2(k\cdot a_0)}
\end{equation}
This relation is particularly relevant because
\begin{equation}
\vert\frac{\alpha}{A}\vert^2\cdot\frac{2}{L}\cos^2(k\cdot x)
\end{equation}
is proportional to the local density of states at the site of the potential well. The function $\vert\frac{\alpha}{A}\vert^2$ has a peculiar feature: it goes through maximum values of ``1'' periodically, when the quantization condition $k\cdot a_0 = n\cdot \pi$ is fulfilled. This ``resonant behaviour'' occurs, to a very high degree of precision, exactly in correspondence with the quantization relation that determines the bound states in the potential $U_0(x)$, so that one speaks of the states with energy $q^2_n= k^2_n-(k_0^2-k_1^2)$ as the ``virtual bound states'' of the potential $U_0(x)+U_1(x)$. This is the microscopic content of the level broadening that the Stoner criterium of Sec.~3.5 feels only through the density of states at the Fermi level.

\chapter{Magnetic interactions in transition metals ultrathin films.}
\label{Chap:interactions}

\renewcommand{\thesection}{\thechapter.\arabic{section}}
\renewcommand{\theequation}{\thechapter.\arabic{equation}}

\noindent Chapter 3 established what carries the magnetism of a transition
metal film: a magnetic moment of spin origin, of the order of $\mu_B$, sitting
on each atomic site. This Chapter asks what these moments do to each other, and
the answer comes in three interactions of very different strength.\\
The first is the \textbf{exchange interaction} between moments on neighbouring
sites. It decides whether the moments align at all, and it is by far the
strongest: for Fe it amounts to some tens of meV per pair of atoms. It is,
however, blind to direction -- it depends only on the angle between two
neighbouring spins, not on where they point with respect to the crystal.\\
The two remaining interactions are the ones that break that isotropy, and both
are smaller by one to two orders of magnitude. The \textbf{spin-orbit coupling},
combined with the local symmetry of the crystal, produces a single-ion magnetic
anisotropy of the order of $1$ meV per atom in a film -- the anisotropy
postulated by N\'eel. The magnetostatic \textbf{dipole-dipole interaction}
between the moments produces a second anisotropy of comparable size, but of
opposite preference: the first can favour a magnetization perpendicular to the
film, the second always favours an in-plane one.\\
That near-degeneracy is not an accident of bookkeeping: it is the reason why
ultrathin films, and only ultrathin films, show a reorientation transition
(Chapter 8) and stripe domain patterns (Chapter 11). The three Sections that
follow treat the three interactions in turn, and the last two end where those
later Chapters begin.

\section{Interatomic exchange coupling.}
The picture emerging from our discussion of the ferromagnetic ground state of transition metal ultrathin films is
one of possibly non-integer magnetic moments which can be enhanced with respect to their bulk counterparts. As in the bulk, the magnetic moments appear to be mainly localized at the atomic sites. The ground state energy considerations compare the total energy of a state where the magnetic moments are aligned along a certain direction (the ``ferromagnetic'' ground state) with the total energy of a state where the magnetic moments are assumed to be vanishing (in band structure calculations, the state with vanishing magnetic
moments is often defined as the ``paramagnetic'' state).\\
There are, accordingly, two distinct ways of exciting the ferromagnetic ground
state, and it is worth keeping them apart from the beginning: one can change
the \emph{length} of the local moment, or one can change its \emph{direction}.
The first costs the exchange energy of Chapter 3 and is the subject of the next
paragraph; the second costs the interatomic exchange energy, which is the
subject of this Section and the quantity we are after.\\
The first excitation foresees that at one atomic site the magnetic moment is rendered vanishing. In a simple triplet-singlet state language, the vanishing of the spin at the atomic site means that the triplet state has been excited into the singlet state. This excitation changes the total value of the spin by $1\cdot\hbar$. The excitation is related to the first Hund's rule, which has its validity mainly in atomic physics. For instance, the triplet-singlet excitation is used e.g. for producing metastable ortho-helium atoms and has a wide application in different fields of science\cite{Julich}.\\
In solids, one uses a band structure point of view. In such a one-electron picture, the triplet-singlet excitation is represented as one majority electron being excited into an unoccupied minority state and is a so-called spin-flip excitation. The excitation in the solid is often referred to, in the literature, as a ``Stoner'' excitation. There are differences between the excitation in atoms and the Stoner excitation in solids. One important difference is that the typical energy of a
Stoner excitation is $I\cdot n\cdot P$ and is smaller than the energy of the
corresponding  excitation in an isolated atom ($I\cdot n$). In addition, Stoner
excitations are distributed over the entire Brillouin zone and can occur with a
finite momentum transfer, so that their energy is not as sharp as in atomic physics but is spread out about the typical value of $I\cdot n\cdot P$. In electron scattering from surfaces, the Stoner excitation is revealed by a very weak loss  feature centered at $I\cdot n\cdot P$ in the vicinity of the elastic peak\cite{Modesti}.\\
In a ferromagnet, there are other types of fundamental excitations. A most
relevant one foresees that the spin at the atomic site is \textbf{rotated} away from the $z$-quantization axis. In a stringent quantum mechanical viewpoint, such a rotation is realized by the spin at a site being assigned values of $S_z$ between $+S$ and $-S$. The (possibly non-integer) spin quantum number itself is not changed, i.e. the spin is not rendered vanishing, just ``rotated''. In an atom, such a ``rotated'' state and
the state with $+S$ are degenerate in energy. In a solid, the presence of some
neighbours lifts this degeneracy by the mechanism of the exchange interaction
between electrons centered at different atomic sites. One can best appreciate this new interaction in an easily (but representatively) computable situation, consisting of a two-electron, two-site molecule (see e.g. Chapter 27 in Ref.~\cite{DP_QM}). In this problem the energies of four states are computed. In three of them, the two neighboring spins are ferromagnetically aligned: these three states are the two configurations $Y^+Y^+$ and $Y^-Y^-$ together with the symmetric combination given below. By rotating the spin on one site, one produces the two spin configurations
\begin{equation}
\sqrt{\frac{1}{2}}\left(Y^+Y^-+Y^-Y^+\right) \quad \quad
\sqrt{\frac{1}{2}}\left(Y^+Y^--Y^-Y^+\right)
\end{equation}
The one with the ``$+$'' sign is degenerate with the previous two (they build a
spin triplet with total spin ``1''). The coupling between the spins at the two
sites is said to be ``ferromagnetic''. The one with the ``$-$'' sign belongs to
the total spin $S=0$. This state is called an ``antiferromagnetic state''.
Notice that, in the antiferromagnetic state, the total spin of the couple is
vanishing while the individual spins at the sites remain non-vanishing. The aim
of the computation of the energy of the two spin configurations is to establish
the possibility of the lifting of the degeneracy between the antiferromagnetic and ferromagnetic spin configuration and, if applicable, the finding of the underlying interaction. One finds that the configurational energy level $E_{nL}+E_{nL}$ of the two-site system is changed by the various Coulomb interactions to a two-separate-level system with energy eigenvalues
\begin{equation}
2\cdot E(n,L) + Q + J
\end{equation}
corresponding to the singlet state
and
\begin{equation}
2\cdot E(n,L) + Q - J
\end{equation}
corresponding to the ferromagnetic triplet state.
$Q$ is some typical Coulomb integral containing the classical Coulomb interactions between electrons and nuclei. $J$ is the \textbf{interatomic} exchange energy, given by\cite{Marshall}
\begin{eqnarray*}
J = \int dV_1dV_2 \psi _A(r_1)\psi _B(r_2)\left[\frac{e^2}{R}+\frac{e^2}{\vert\vec r_2-\vec r_1\vert}
-\frac{e^2}{r_{B1}}-\frac{e^2}{r_{A2}}\right]\psi_A(r_2)\psi_B(r_1)
\end{eqnarray*}
$e^2$ stands for $\frac{e^2}{4\pi\epsilon_0}$. $r_1$ and $r_2$ are the spatial
coordinates of the two electrons. $\psi_A(r)$ and $\psi_B(r)$ are the  wave
functions (assumed real) of the orbitals carrying the spins. In contrast to the
situation underlying the intra-atomic exchange energy of Chapter 3, the orbitals in this
situation are centered at separate atomic sites (for practical purposes, they
do not need to be orthogonalized,
but, in principle, one can think of these orbitals as being suitably
orthogonalized by e.g. the Wannier orthogonalization protocol). The first two
terms  of this expression are positive, the last two are negative. All four are
of the same order of magnitude, so that the sign of this integral is difficult
to estimate. In the specific example of the hydrogen molecule, the wave
functions $\psi_A$ and $\psi_B$ are the $1s$ orbitals of hydrogen. For this
specific problem $J$ is known to be negative, and its strength is in the range
of a few eV, so that the singlet state is energetically favored. This result
produces a robust fundamental law of chemistry, which foresees that the
covalent bonding between two identical atoms is a singlet. For ferromagnetic
order, instead, $J$ must be positive, otherwise the rotation of a spin would
lead to a state with lower energy.\\
Notice that the singlet coupling underlying the chemical bond and the absence of
a net ferromagnetic order in the ground state is a ``robust'' result: there exists a very strong theorem by Lieb and Mattis\cite{Mattis}
stating that, in a linear arrangement of atoms, the non-magnetic state, i.e. the
state with lowest total spin, is indeed the ground state. One needs to go to
spatial dimensions higher than one in order to circumvent this theorem. The
need for higher dimensions is that only in dimensions higher than one can states
with different symmetry (i.e. atomic orbitals with different quantum numbers)
hybridize and be degenerate. Close to the Fermi level, one can therefore
envisage a situation where electrons of a certain symmetry are involved in
making the chemical bond (and couple antiferromagnetically) while electrons of
different symmetry but degenerate with the orbitals carrying the chemical bond
are involved in establishing the ferromagnetic coupling between the different
atomic sites\cite{Slater}. In other words: this degeneracy between orbital wave
functions with different symmetry provides a route to escape the strong Pauli
principle that favors antiparallel alignment between orbital states with the
same quantum number. Whether, ultimately, neighboring spins are actually
oriented parallel or antiparallel depends on which one of the terms in the
exchange energy actually prevails.\\
In 3d-transition metals, the orbitals hosting the magnetic moments are the $d$-orbitals and the question is, at first glance, one of determining the exchange
energy between these orbitals. It turns out that these orbitals are very
localized. Their direct exchange energy is positive, thus it energetically
favors the triplet state. However, at the distances corresponding to the nearest
neighbours in bcc Fe, it is of the order of $10^{-4}$ eV\cite{Marshall}, and
this value is several hundred times smaller than the value observed e.g.
in ferromagnetic Fe\cite{Marshall,Small}. The explanation for the proper sign
and strength of the interatomic exchange interaction in Fe was discussed in a review article by M.B. Stearns\cite{Stearn}. That paper establishes the two main
mechanisms for ferromagnetic order. The first mechanism foresees that the \textbf{intra-atomic exchange energy}  acts to build the localized magnetic moments
found in the localized $d$-orbitals of the transition metals. This mechanism is
taken care of, in principle, within the framework of the Stoner model of Chapter 3. The
second mechanism finds a so called  \textbf{indirect interatomic exchange
coupling} between the localized Fe spins. This indirect exchange coupling
involves several steps.\\
\textbf{First step.} The first step foresees that the actual spins are carried by the localized $d$-orbital, but the metal has also conduction $sp$-electrons. Their wave function is spread out over the entire unit cell and has, as such, a substantial overlap with the localized $d$-orbitals. Through this overlap, the (positive) exchange integral -- let us call this exchange $J_{sd}$ -- between the $d$-orbitals and the $sp$-orbitals has a large single-atomic like component and acquires a value that is comparable with the atomic Hund's rule exchange energy. As $J_{sd}$ is positive, the $sp$-electrons at the site of the
atoms have majority character.\\
\textbf{Second step.} In the second step, we need to figure out how this exchange interaction that occurs between a $d$ spin
at one site and the surrounding $sp$-electrons is propagated to a $d$-spin at
neighboring sites. The answer to this question is related to the
Ruderman-Kittel-Kasuya-Yosida (RKKY) interaction\cite{RKKY,Friedel}.
Within the scope of this lecture, we provide a simple, computable mean field
model that simulates the situation we are confronted with in real metals.
Within this model, the localized $d$-electrons provide a potential sink with a
width corresponding to their spatial extent (approximately the size of the
atom, i.e. the Bohr radius $a_0$) and a depth corresponding to $J_{sd}$. The
sink is applied to the majority spin electrons, while the minority ones move in a
spatially constant potential. For simplicity, we use a delta-function
representation of this potential sink and solve the corresponding Schr\"odinger equation in Appendix~\ref{App:RKKY}. The outcome are the charge densities $\rho^{\pm}\dot= \vert\psi^{\pm}\vert^2$ for majority (``$+$'') and minority (``$-$'') spins as a function of the distance $d$ from the potential sink, and the electron spin polarization produced by the potential sink, defined as
\begin{equation}
{\cal P} \doteq \frac{\rho ^+-\rho ^-}{\rho ^++\rho ^-}
\end{equation}
We find
\begin{eqnarray}
{\cal P}(d\approx 0)&\approx & \frac{\pi^2}{16}\cdot \frac{J_{sd}}{E_F}\cdot \frac{a_0}{a}\cdot n\nonumber\\
{\cal P}(d\rightarrow \infty)&\approx &\frac{\pi}{4}\cdot  \frac{J_{sd}}{E_F}\cdot \frac{a_0}{a}\cdot n\cdot \frac{\cos(2k_F\cdot d)}{(k_F\cdot d)}
\end{eqnarray}
Hereby $a$ is the lattice constant of the lattice underlying the free electron
gas, $E_F$ is the Fermi level of the one dimensional free electron gas and
$n$ the number of $sp$-electrons per unit cell. $d$ measures the distance from
the location of the impurity magnetic moment. $k_F$ is the Fermi wave number of
the electron gas. This result shows that a local perturbation in one spin
channel produces an oscillating density in the affected spin channel, while the
other spin remains uniformly distributed. This unbalance produces a local spin
polarization of the electron gas surrounding the impurity that propagates far
away from the perturbing magnetic moment. At large distances, the spin
polarization propagates periodically in space, with a spatial wavelength corresponding to $\frac{\pi}{k_F}$. The decay of the oscillatory polarization is as $\frac{1}{k_F\cdot d}$ for a point source in one dimension. For a point source in $3$ dimensions (a situation which describes more suitably the coupling in the electron gas of a 3D-crystal like Fe), the decay is $\propto \frac{1}{(k_F\cdot d)^3}$.\\
\textbf{Third step.} In the third step, one must define the role of this spin imbalance for determining the sign and strength of the indirect exchange coupling. Within a mean field approach, the spin polarization of the $sp$-electrons provides an effective magnetic field $-J_{sd}\cdot \frac{{\cal P}(d)}{2}$ for the spin component $S_z$ of the spin residing at a distance $d$ from the potential sink. Depending on
the sign of the spin polarization, the spin at a distance $d$ will align
parallel or antiparallel with respect to the spin at the site of the potential
sink. By this propagating spin polarization, the exchange interaction,
originally arising at the site ``0'', can propagate away from the site,
oscillating between positive and negative depending on the position $d$, and can
couple spins which are quite distant from each other.\\
In the specific case of Fe, the realistic computations by Stearns\cite{Stearn}
demonstrate that the spin polarization of the delocalized electrons provides just the right sign at the nearest neighbour distance to sustain the observed ferromagnetic order. More advanced computations by Small and Heine\cite{Small} have produced a quantitative  picture of ferromagnetism in Fe.\\
The origin of the magnetic moment are the four $3d$-electrons, which are localized enough to fulfil the Stoner criterium. Yet, the broadening of the $d$-band in the solid state reduces the magnetic moment to
about 2.2 $\mu_B$ per atom. The magnetic moment has purely spin origin so that the $g$-factor is about 2. Regarding the representation of the exchange coupling, Small et al.\cite{Small} find that the local spin is best described by classical vectors $\vec S_i$ residing at the site $i$, with length $S=1.1$ (in units of $\hbar$). The exchange interaction between two spins $\vec S_i$ and $\vec S_j$ at the sites $(i,j)$ is then given as the elastic energy of classical vectors\cite{Small}, i.e. as
\begin{equation}
\label{Eq:Heis_class}
-J_{ij}\cdot S^2\cdot \cos\theta_{ij}
\end{equation}
whereby $\theta_{ij}$ is the angle between the vectors $\vec S_i$ and $\vec S_j$. $J_{ij}$ up to the fifth nearest neighbours are computed and tabulated (Table 1 in \cite{Small}). For the comparison with experiment, a suitable effective value is computed from the values of $J_{ij}$ (see Eq. 4.1 in Ref.~\cite{Small}). This equation produces an effective exchange coupling between two Fe spins that amounts to $J\approx + 46$ meV. Accordingly, the interatomic exchange coupling is about twenty times smaller than the intra-atomic exchange coupling $I\approx 0.9$ eV found for Fe in Chapter 3 -- small on the atomic scale, but still two orders of magnitude larger than the anisotropies of Secs.~4.3 and 4.4.

\section{The interlayer exchange coupling.}
\begin{figure}[H]
\begin{center}
\includegraphics[width=0.8\textwidth]{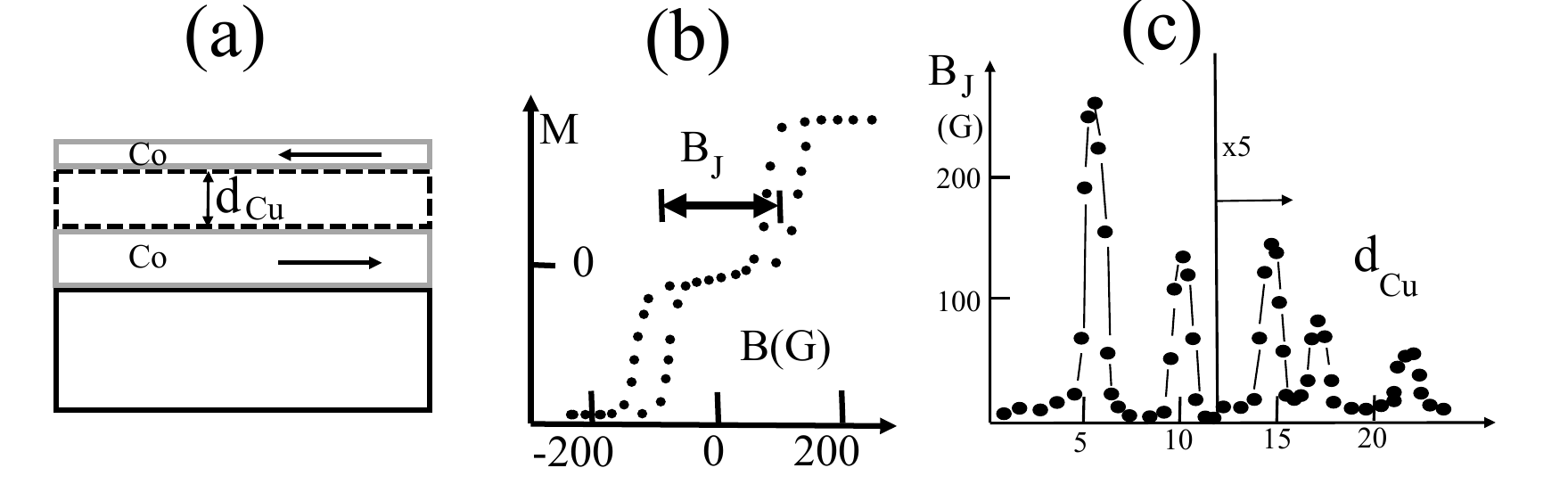}
\caption{{\footnotesize (a): Schematic stacked structure used for coupling experiments. The grey rectangles are the magnetically active films, in the specific case Co thin films\cite{Our}. The spacer is made of Cu(100). The black arrows indicate the orientation of the magnetization vector. (b): Typical hysteresis curve ($M$ versus magnetic field $B$) recorded for antiferromagnetically exchange-coupled Co films. At the shift field $B_J$ the magnetizations of the individual films are aligned to the direction
specified by the external magnetic field. (c): The shift field $B_J$ is measured as a function of the Cu spacer thickness $d_{Cu}$ by means of the optical Kerr effect\cite{Our}. A finite shift field means antiferromagnetic coupling. A vanishing shift field means ferromagnetic coupling. The  data points on the right of the thin vertical line are multiplied by 5 for better visibility. Drawn from Ref.\cite{Our}.} }
\label{Fig:Fert}
\end{center}
\end{figure}

The oscillatory interatomic exchange interaction became directly measurable by
the technology called \textbf{interlayer exchange coupling}, developed by
P. Gr\"unberg and A. Fert (see Ref.~\cite{BrunoJsd} and the references therein).
Fig.~\ref{Fig:Fert} provides a summary of the main aspects underlying this technology.
Typically, a ferromagnetic film (represented by the lower gray rectangle in (a)) is deposited onto a non-magnetic substrate (black rectangle in (a)). In the present case it is an fcc Co film deposited on top of a Cu(100)-single crystal surface by Molecular Beam Epitaxy. The ferromagnetic film, typically a few nm thick, is magnetized e.g. along an in-plane direction (black arrow pointing right in (a)). The magnetization is typically spatially uniform across the lower film. The lower ferromagnetic film provides a planar source of magnetic sites. On top of the ferromagnetic film a non-magnetic spacer (in this case made of Cu) is deposited. The non-magnetic spacer provides a reservoir of $sp$-like free electrons that is used to propagate the spin polarization originating from the planar source of ferromagnetic sites. The thickness $d_{Cu}$ of the non-magnetic spacer is then varied. One finds\cite{BrunoJsd} that the spin polarization at the coordinate $z$, measured from the surface of the lower thin film, undergoes oscillations from negative to positive and so on,  with a typical spatial period that is related to the Fermi wave vector of the non-magnetic spacer\footnote{In the present case Cu. In fact, because of the warped shape of the Fermi sphere of the non-magnetic spacer, the spin polarization can contain more than one characteristic period, which propagate with different amplitude and phase\cite{BrunoJsd,Our}.}
For probing these oscillations, a second magnetic film
(represented by the top gray rectangle in (a)) is deposited onto the non-magnetic spacer. The spins at the boundary to the non-magnetic spacer couple to the spin polarization of the $sp$-electrons at the boundary with the $J_{sd}$ coupling constant. This produces an oscillatory effective exchange coupling $J$ per boundary spin which
assumes theoretical values in the range 1-10 meV, Ref.~\cite{BrunoJsd}.
Experimental values of the coupling are obtained by measuring the magnetization
of the entire stacked structure (Fig.~\ref{Fig:Fert}b). The upper magnetic film is the
``sensor'' of the coupling and has, typically, a lower coercivity than the
lower magnetic film (this is obtained e.g. by tuning the thickness of the
Co-films). When a magnetic field is applied in-plane, the lower magnetic film
undergoes an almost square hysteresis curve. When the thickness of the Cu is
such that the coupling between the Co-films is ferromagnetic, the entire stacked
structure undergoes a square hysteresis curve. On the other side, when the
thickness of the Cu is such that the coupling is antiferromagnetic, the
application of an external in-plane magnetic field produces a so-called shifted
hysteresis curve (see Fig.~\ref{Fig:Fert}b). The shift field $B_J$ is related to the effective antiferromagnetic exchange coupling $J(z,d_{Cu})$ by the relation
\begin{equation}
\label{Eq:BJ}
J(z,d_{Cu})= B_J\cdot g\mu_B\cdot S
\end{equation}
with $g\mu_B\cdot S$ being the magnetic moment per atom of the ferromagnetic layer at the boundary with the non-magnetic spacer\cite{Bloemen}. At $B_J$, the applied in-plane magnetic field is able to rotate the magnetization of the top ferromagnetic film along the direction of the magnetization of the lower film, against the antiferromagnetic coupling. The recording of $B_J$ as a function of $d_{Cu}$ (Fig.~\ref{Fig:Fert}c) produces a set of sharp maxima at those $d_{Cu}$ at which the coupling is antiferromagnetic.\\
One important experimental finding refers to the imaging of the oscillatory nature of the interaction. Imaging relies on the non-magnetic spacer being deposited with a wedge-like shape. In this particular geometry, it provides a continuous range of $d_{Cu}$ that can be observed by imaging with spatial resolution the magnetic state of the probe layer deposited on top of the wedge. One observes that, because of the oscillatory behaviour of $J(d,z)$, the top Co film develops stripes of spins that are parallel, respectively antiparallel, oriented to the spins of the lower layer. The position of the stripes along the wedge depends on the coordinate $z$ at which the spins at the boundary with the non-magnetic spacer reside. The sequence of parallel/antiparallel coupled stripes imaged by a scanning focussed
electron beam along the wedge-like multilayered structure is shown e.g. in Ref.~\cite{Unguris}.

\section{Magnetic anisotropy: spin-orbit coupling.}
\begin{figure}[H]
\begin{center}
\includegraphics[width=0.5\textwidth]{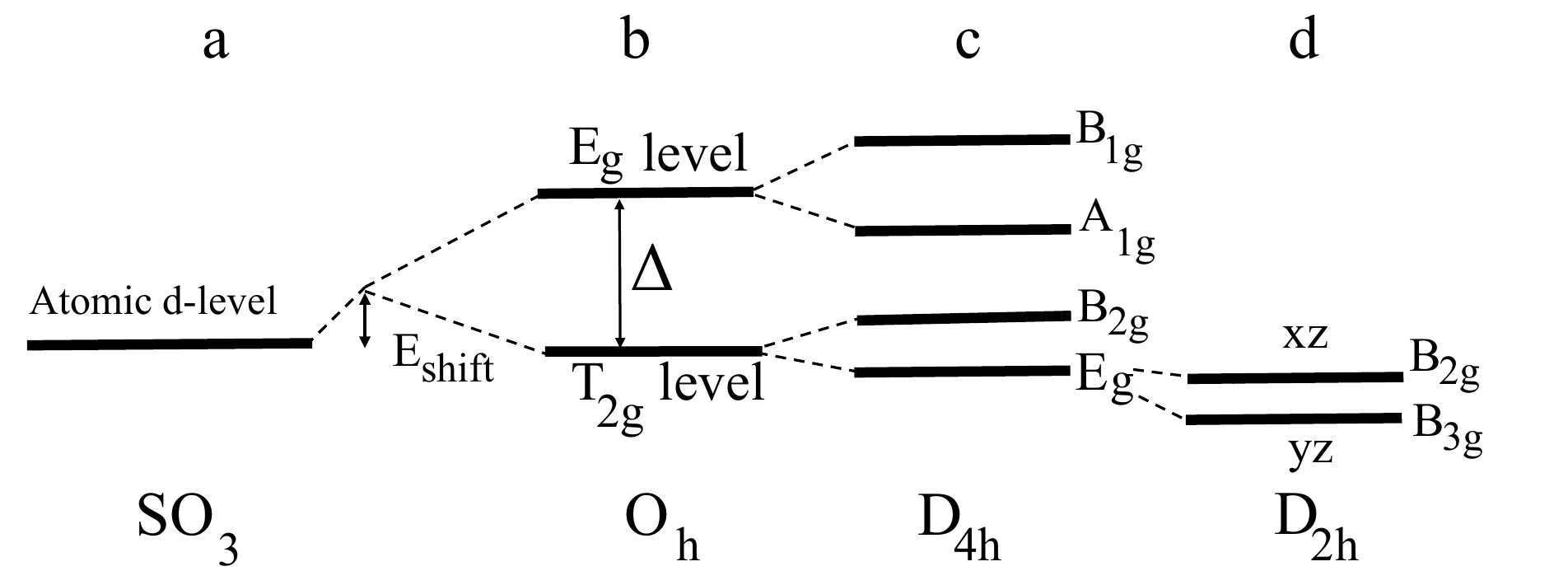}
\caption{Scheme of the  energy levels for a $d^1$-electron configuration  (a) under octahedral symmetry (b), tetragonal elongation along the axis $z$ (c) and rectangular distortion in the $xy$-plane (d). The distortions away from the octahedral geometry are driven by the Jahn-Teller phenomenon. $\Delta$ is the so-called crystal field splitting and $E_{\text{shift}}$ is an energy shift. The angular dependent orbital components resulting at the end of the symmetry lowering (d) can be computed using symmetry arguments\cite{DP_Group} and are given in the figure as polynomials. Their radial part is subject to a numerical solution of the radial Schr\"odinger equation.
\label{Fig:JT}
}
\end{center}
\end{figure}

A less obvious but both conceptually and
technologically very much relevant energy is the magnetic anisotropy energy.
By this energy contribution one designates the change of the total energy that
is observed  when the spin magnetic moment is \textit{rotated} with respect to
the crystallographic axes. Notice that this is a different rotation from the one
of Sec.~4.1: there, one spin was rotated with respect to its \emph{neighbours},
here all spins are rotated together with respect to the \emph{lattice}. The
exchange interaction is blind to the second rotation, and one has to look for
another source of energy. One source of magnetic anisotropy is the spin-orbit coupling.\\
We aim at discussing this component of the magnetic anisotropy energy by studying the simple example of the rotational energy of the spin of a $d$-electron (a in Fig.~\ref{Fig:JT}) embedded into an octahedral crystal field (b). This might be the case of a transition metal ion embedded into a molecule with octahedral geometry\cite{DP_Group}. The resulting octahedral symmetry is, typically, lowered to $D_{4h}$ by an elongation along the $z$-axis (c) and, successively, to $D_{2h}$ by a rectangular distortion in the $xy$-plane (d), via the Jahn-Teller type effect that, finally, produces a non-degenerate orbital state, see (d) in Fig.~\ref{Fig:JT}.
\noindent Let us, for the time being, stop at the situation of an elongation along the $z$-axis that produces a doubly degenerate $E_g$-ground state. The angular parts of the orbital wave functions building basis states in the space $E_g$ are
\begin{eqnarray}
i\cdot \sqrt{\frac{1}{2}} \left( Y_2^{- 1} + Y_2^1 \right)&\doteq& d_{yz}\nonumber\\
\sqrt{\frac{1}{2}} \left( Y_2^{- 1} - Y_2^1 \right)&\doteq& d_{xz}
\end{eqnarray}
We are interested in computing the energy of the electron with these orbital basis states and with its spin state pointing along an arbitrary direction, specified, e.g., by the spherical angles $\theta,\phi$. For constructing such a spin state we rotate the spin basis state $\uparrow$ with the rotation matrix
\begin{equation}
\label{Eq:su2}
\left(
\begin{array}{cc}
\cos\frac{\varphi }{2}-i\cdot n\sin\frac{\varphi }{2}&(-i\cdot l -m)\sin\frac{\varphi }{2}\\
(-i\cdot l +m)\sin\frac{\varphi }{2}&\cos\frac{\varphi }{2}+i\cdot n\sin\frac{\varphi }{2}
\end{array}
\right)
\end{equation}
describing the rotation by an angle $\varphi$ about an axis defined by the unit vector $(l,m,n)$. We perform first the rotation about $z$ by an angle $\phi$, followed by a rotation by an angle $\theta$ about the $y$-axis. The composition of the resulting two matrices produces the coordinates of the spin state ``up'' with respect to a new $z'$-axis identified by the spherical angles $\theta,\phi$:
\begin{eqnarray}
\label{Eq:bstates}
\begin{pmatrix}
e^{-i\cdot \frac{\phi}{2}}&0\\
0&e^{i\cdot \frac{\phi}{2}}
\end{pmatrix}\cdot \begin{pmatrix}\cos\frac{\theta }{2}&-\sin\frac{\theta}{2}\\
\sin\frac{\theta}{2}&\cos\frac{\theta}{2}\end{pmatrix}\begin{pmatrix}1\\
0\end{pmatrix}=\begin{pmatrix}e^{-i\cdot \frac{\phi}{2}}\cdot \cos\frac{\theta }{2}\\e^{i\cdot \frac{\phi}{2}}\cdot \sin\frac{\theta}{2}\end{pmatrix}
\end{eqnarray}
i.e.
\begin{equation}
Y^{+'} = e^{-i\cdot \frac{\phi}{2}}\cdot \cos\frac{\theta }{2}Y^+ + e^{i\cdot \frac{\phi}{2}}\cdot \sin\frac{\theta}{2}Y^-
\end{equation}
The component of the Hamilton operator relevant for introducing a dependence of
the energy eigenvalues on the direction of the spin of the electron is the
spin-orbit coupling operator $A_{LS}\cdot \vec L\cdot \vec S$ (Ref.~\cite{DP_QM}, p.~109-111).
This operator originates from the Dirac equation and is, accordingly, a relativistic correction to Schr\"odinger quantum mechanics. $A_{LS}$ is a parameter related to some average over the radial wave function of the radially dependent part of the spin-orbit coupling operator. The order of magnitude of $A_{LS}$ is such that the corrections introduced by this operator are, typically, much smaller than the crystal field splitting or the Jahn-Teller distortion splitting, so that this operator can be considered as a small perturbation.
To first order perturbation theory, the energy eigenvalues of the Hamilton operator within the space spanned by the basis states $\{d_{yz}Y^{+'}, d_{xz}Y^{+'}\}$ are the eigenvalues of the
$2\times 2$ matrix
\begin{eqnarray}
\begin{pmatrix}
E_{E_g}+A_{LS}\cdot <d_{yz}Y^{+'}\vert\vec L\cdot \vec S\vert d_{yz}Y^{+'}>&A_{LS}\cdot <d_{yz}Y^{+'}\vert\vec L\cdot \vec S\vert d_{xz}Y^{+'}>\\
A_{LS}\cdot <d_{xz}Y^{+'}\vert\vec L\cdot \vec S\vert d_{yz}Y^{+'}>& E_{E_g}+A_{LS}\cdot <d_{xz}Y^{+'}\vert\vec L\cdot \vec S\vert d_{xz}Y^{+'}>
\end{pmatrix}
\end{eqnarray}
For computing the matrix elements, one often simplifies the algebra by writing
\begin{equation}
\vec L\cdot \vec S=L_+\cdot S_- + L_-\cdot S_+ + L_z\cdot S_z
\end{equation}
using the operators $I_{\pm}\dot= \sqrt{\frac{1}{2}}[I_x\pm i\cdot I_y]$ ($I$ standing for either ``L'' or ``S'') and the corresponding
computation rules
\begin{eqnarray}
L_+Y_2^1=\sqrt{2}\cdot Y_2^2\quad\quad L_+Y_2^0 = \sqrt{3}Y_2^1 \quad\quad  L_+Y_2^{-1} = \sqrt{3}Y_2^0\quad\quad L_+Y_2^{-2}= \sqrt{2}Y_2^{-1}\nonumber\\
L_-Y_2^2=\sqrt{2}\cdot Y_2^1\quad \quad L_-Y_2^1 = \sqrt{3}Y_2^0 \quad \quad L_-Y_2^{0} = \sqrt{3}Y_2^{-1}\quad\quad L_-Y_2^{-1}= \sqrt{2}Y_2^{-2}
\end{eqnarray}
\begin{equation}
S_+Y^{-\frac{1}{2}} = \frac{1}{\sqrt{2}}\cdot Y^{\frac{1}{2}}\quad\quad S_-Y^{\frac{1}{2}}= \frac{1}{\sqrt{2}}\cdot Y^{-\frac{1}{2}}
\end{equation}
and
\begin{equation}
L_z Y_l^m= m\cdot Y_l^m\quad\quad S_zY^\pm = \pm\frac{1}{2}\cdot Y^\pm
\end{equation}
Using these computation rules, one obtains the Hamilton matrix
\begin{eqnarray}
\begin{pmatrix}
E_{E_g}&\frac{i}{2}\cdot A_{LS}\cdot \cos \theta\\
-\frac{i}{2}\cdot A_{LS}\cdot \cos \theta & E_{E_g}
\end{pmatrix}
\end{eqnarray}
Only the $L_zS_z$ term survives: $L_z d_{xz}= i\,d_{yz}$, so that the orbital
matrix element equals $i$, while the rotated spin state contributes
$<Y^{+'}\vert S_z\vert Y^{+'}> = \frac{1}{2}\cos\theta$. Its eigenvalues are
\begin{equation}
\label{Eq:first_order}
E_{E_g}\pm \frac{1}{2}\cdot A_{LS}\cdot \vert \cos \theta\vert
\end{equation}
The presence of the $\cos\theta$-correction  introduces a magnetic
anisotropy, produced at the atomic level by the local crystal field.\\
A more realistic situation arises when the rectangular distortion in the $xy$-plane establishes a non-degenerate orbital ground state with symmetry $yz$. To first order perturbation theory, the correction produced by the spin-orbit coupling operator amounts to $A_{LS}\cdot<d_{yz}Y^{+'}\vert\vec L\cdot \vec S \vert d_{yz}Y^{+'}>$ and vanishes because of the vanishing orbital matrix element. For better estimating the correction introduced by the spin-orbit interaction, one must include excited states in the perturbational approach, i.e. one must go to second order perturbation theory. The next excited state is the one with $xz$-symmetry. If this state is included, one obtains the new ground state as the lowest of the eigenvalues of the $2\times 2$ matrix
\begin{eqnarray}
\begin{pmatrix}
E_{B_{3g}}+A_{LS}\cdot <d_{yz}Y^{+'}\vert\vec L\cdot \vec S\vert d_{yz}Y^{+'}>&A_{LS}\cdot <d_{yz}Y^{+'}\vert\vec L\cdot \vec S\vert d_{xz}Y^{+'}>\\
A_{LS}\cdot <d_{xz}Y^{+'}\vert\vec L\cdot \vec S\vert d_{yz}Y^{+'}>& E_{B_{2g}}+A_{LS}\cdot <d_{xz}Y^{+'}\vert\vec L\cdot \vec S\vert d_{xz}Y^{+'}>
\end{pmatrix}
\end{eqnarray}
The new ground state energy amounts to
\begin{equation}
 E_{B_{3g}}-\frac{A_{LS}^2}{4}\frac{\cos^2\theta}{E_{B_{2g}}-E_{B_{3g}}}\pm {\cal O}(\cos^4 \theta)
\end{equation}
This result highlights the fact that magnetic anisotropy from spin-orbit coupling  arises, in physically relevant situations, in second order perturbation theory at the earliest and produces corrections of the order of $\frac{A_{LS}^2}{E_1-E_0}$, $E_1$ being the energy of some relevant excited state close to the ground state $E_0$. Accordingly, magnetic anisotropy, produced by spin-orbit interaction in conjunction with the symmetry breaking originating from the local crystal field, is, in general, a very small correction to the total energy per atom.\\
One can generalize the single-site magnetic anisotropy computed above to a crystal by thinking of it as entering the free energy of a body as some powers of the direction cosines. In a bulk crystal with cubic axes, for instance, the single-site (or ``single-ion'') magnetic anisotropy  entails the fourth power of the direction cosines\cite{Gay,GR}. The strength of their coefficients is proportional to the fourth power of $A_{LS}$ and amounts to a few $\mu$eV\cite{Gay,Bruno} per atom. At the surfaces of a bulk crystal or in ultrathin films, the translational symmetry along the normal direction is broken and this introduces a magnetic anisotropy term that can be written as the second power of the direction cosine\cite{Gay,Bruno}, $-\lambda\cdot \cos^2\theta(\vec x)$,
$\theta$ being the angle of the spin vector with respect to the direction perpendicular to the film (taken to be the $z$-direction henceforth). $\lambda$ is proportional to the second power of $A_{LS}$ and is of the order of meV per atom\cite{Gay,GR}. $\lambda$ can be positive or negative: a positive value of the coupling constant $\lambda$ favors the spin vectors at the site $\vec x$ to point along $\pm z$, while a negative value favors an in-plane direction. The existence of this anisotropy was postulated by N\'eel long time ago\cite{Neel}, but a computation of the coupling constant $\lambda$  was achieved, for the first time from first principles, about 30 years later by Gay and
Richter\cite{Gay}. The value of $\lambda$ for the monolayer of Fe is, for instance, about $0.3$ meV per surface unit cell\cite{Gay,Shi}.\\
Notice the two powers of the direction cosine side by side: in the bulk the
leading term is of fourth order and worth a few $\mu$eV, at a surface it is of
second order and worth a fraction of a meV. Breaking the translational symmetry
along one direction thus buys two orders of magnitude of anisotropy. The
occurrence of the $\cos^2\theta$ single-ion magnetic anisotropy is one of
the most striking phenomena identified within the subject of magnetism of 3d
transition metal films\cite{Bruno}. It leads to the possibility that the ground
state of an ultrathin film has a perpendicular magnetization. This special
state of matter and its relevant topological aspects will be dealt with in
Chapter 11.

\section{Magnetic anisotropy: dipole-dipole magnetostatic interaction.}
There is a further source of magnetic anisotropy: it resides in the
magnetostatic dipole-dipole interaction between the magnetic moments. This
interaction originates from Maxwell classical magnetostatics and is well
described, accordingly, by Maxwell equations. It can be computed quite
precisely, using formulas for a continuum distribution of magnetic dipoles.
The general equations governing the dipolar interaction are obtained in Chapter 5. One important result of these general equations is
that the dipolar energy contains a single-site term that can be written as
\begin{equation}
\Omega\cdot \cos^2\theta(\vec x)
\end{equation}
$\Omega$ is an energy per bulk unit cell and is a measure of the dipolar coupling constant:
\begin{equation}
\Omega\doteq \frac{\mu_0}{2}\cdot M_0^2\cdot a^3
\end{equation}
$M_0$ being the absolute value of the magnetization vector (see Chapter 5 for a discussion of the origin of this constant). For transition metal atoms $\Omega$ is of the order of meV. Its precise value depends on various parameters that describe the magnetic moments in the unit cell.
Notice that the coefficient of the $\cos^2\theta$ term arising from the dipolar
energy is positive, i.e. it favors an in-plane direction for the magnetic
moment to point at. Because of this property, it competes with the
N\'eel anisotropy whenever the value of the $\lambda$-coupling constant is positive.
This competition is one of the most important fundamental aspects of ultrathin film magnetism.\\
It is worth closing the Chapter by putting the three numbers next to each other,
per atom: the interatomic exchange coupling of Sec.~4.1 is some tens of meV, the
N\'eel anisotropy $\lambda$ of Sec.~4.3 is a fraction of a meV, and the dipolar
constant $\Omega$ of this Section is again of the order of a meV. The exchange
interaction is thus in no danger of being overruled -- the moments will be
aligned with each other in any case -- but the two anisotropies are of the same
order and enter the energy balance with opposite signs. Which of the two wins is
decided by quantities as delicate as one atomic layer of thickness or a few
degrees of temperature, and this is precisely what makes the reorientation
transition of Chapter 8 and the stripe patterns of Chapter 11 observable in
ultrathin films.

\setcounter{section}{0} 
\renewcommand{\thesection}{\thechapter.\Alph{section}} 
\setcounter{equation}{0}
\renewcommand{\theequation}{\thesection\arabic{equation}}
\section*{Appendices}

\section{Delta-function model of the RKKY interaction.}
\label{App:RKKY}
We introduce in a 1d free electron gas a perturbing potential localized at the
origin and look for the total charge density produced by it at a location $x$.
As we are interested here in the behavior of the charge (or spin) density far
away from the location of the perturbing potential, we do not use the more
realistic but also more cumbersome traditional potential well with finite
width.  Instead, in order to simplify the mathematics, we consider a
Dirac-Delta like perturbation potential of strength $-J_{sd}$ and width $a_0$
located at the origin of a one dimensional solid filled with a free electron gas: $V(x) = -J_{sd}\cdot a_0\cdot \delta(x)$.
The system is contained within a segment extending from $-L/2$ to $+L/2$ along the $x$-axis. We refer to the segment with $x<0$ as the left-hand side $l$ and to the segment with $x>0$ as the right-hand side $r$. We solve the Schr\"odinger equation
\begin{equation}
[\frac{-\hbar ^2}{2m}\frac{\partial^2}{\partial x^2} -J_{sd}\cdot a_0\cdot \delta(x)]\psi(x) = E\psi(x)
\end{equation}
under the boundary conditions for a $\delta$ like potential\cite{Wikidelta}
\begin{equation}
\psi_l(0) = \psi_r(0)
\end{equation}
and
\begin{equation}
-\frac{\hbar^2}{2m}\Big[\psi_r'(0)-\psi_l'(0)\Big] - J_{sd}\cdot a_0\psi(0)  =   0
\end{equation}
In the two regions $l$ and $r$ the respective solutions have to fulfil the
Schr\"odinger equation
\begin{equation}
\frac{-\hbar ^2}{2m}\frac{\partial^2 \psi(x)}{\partial x^2} = E\cdot \psi(x)
\end{equation}
We distinguish two cases: $E<0$ and $E>0$.\\
For $E<0$ the equation away from the singularity has two solutions, one growing exponentially toward $\pm \infty $, the other decaying exponentially toward $\pm \infty $. This last solution is the only physical one as it has a finite norm and produces a \textbf{bound} state with energy amounting to
\begin{equation}
E_b = -\frac{\hbar ^2 \kappa^2}{2m}
\end{equation}
with $\kappa \doteq \frac{J_{sd}\cdot a_0\cdot m}{\hbar ^2}$ and wave function
\begin{equation}
\psi_l^b(x) = \sqrt{\kappa}e^{\kappa x}\quad \psi_r^b(x) = \sqrt{\kappa}e^{-\kappa x}
\end{equation}
In the range $E>0$ we have free electrons moving left and right, and the solutions in each range $l,r$, under periodic boundary conditions at $\pm L/2$, are
\begin{eqnarray*}
E = \frac{\hbar ^2 k^2}{2m};\quad k = \frac{2\pi}{L}\cdot n&&\\
\psi_l(x) = A_l\sqrt{\frac{2}{L}}\sin kx + B_l\sqrt{\frac{2}{L}} \cos kx;&\quad& \psi_r (x)= A_r\sqrt{\frac{2}{L}}\sin kx + B_r\sqrt{\frac{2}{L}} \cos kx
\end{eqnarray*}
both  basis functions being normalized to $1/2$ in the range $x\in [-L/2,0]$.
The boundary conditions at $x=0$ read
\begin{equation}
B_l=B_r;\quad k\cdot (A_l - A_r)-2\cdot \kappa B_l = 0
\end{equation}
There are two classes of wave functions fulfilling these conditions.
One class has
\begin{eqnarray*}
B_l=0;\quad A_l=A_r\Rightarrow&& \\
\psi_l^u(x) = \sqrt{\frac{2}{L}}\sin kx;&\quad& \psi_r^u(x) = \sqrt{\frac{2}{L}}\sin kx
\end{eqnarray*}
where the total wave function is normalized to 1 over the segment with length $L$.\\
The second class has $B_l=B_r\doteq B\not=0$ and must be even under change of sign, so that $A_r=-A_l=A$ and $B=\frac{A k}{\kappa}$. This type of wave function reads
\begin{eqnarray*}
\psi_l^g(x) = A\left[\sqrt{\frac{2}{L}} \sin kx + \frac{k}{\kappa}\sqrt{\frac{2}{L}} \cos kx\right]\\
\psi_r^g(x) = A\left[-\sqrt{\frac{2}{L}}\sin kx + \frac{k}{\kappa}\sqrt{\frac{2}{L}} \cos kx\right]
\end{eqnarray*}
$A$ must be chosen so that the entire wave function is normalized to 1 in the range $x\in [-L/2,L/2]$. This means
\begin{equation}
A = \frac{\kappa}{\sqrt{\kappa^2 + k^2}}
\end{equation}
and
\begin{eqnarray*}
\psi_l^g(x) = \frac{1}{\sqrt{\kappa^2 + k^2}}\left[\sqrt{\frac{2}{L}} \kappa\cdot \sin kx + \sqrt{\frac{2}{L}} k\cdot\cos kx\right]\\
\psi_r^g(x) = \frac{1}{\sqrt{\kappa^2 + k^2}}\left[-\sqrt{\frac{2}{L}}\kappa\cdot\sin kx + \sqrt{\frac{2}{L}}k\cdot \cos kx\right]
\end{eqnarray*}
Using these wave functions, we compute the total charge density
at a point, e.g. $x\geq 0$:
\begin{eqnarray*}
\rho(x)&=&\kappa e^{-2\kappa x}+ \frac{2}{L}\cdot \frac{L}{2\pi}\cdot
\int_0^{k_F} dk \Bigl[ \frac{(\kappa \sin kx - k\cos kx)^2}{\kappa^2 + k^2} + \sin^2 kx\Bigr]\\
&=& \kappa e^{-2\kappa x}+\frac{1}{\pi}\cdot
\int_0^{k_F} dk \Bigl[ 1-\frac{\kappa^2\cos 2 k x + \kappa k\sin 2 k x}{\kappa^2 + k^2} \Bigr]\\
&=&\kappa e^{-2\kappa x}+\frac{k_F}{\pi}-\frac{1}{\pi}
\int_0^{\infty} dk \Bigl[\frac{\kappa^2\cos 2 k x + \kappa k\sin 2 k x}{\kappa^2 + k^2} \Bigr]\\
&+&\frac{1}{\pi}\int_{k_F}^{\infty} dk \Bigl[\frac{\kappa^2\cos 2 k x
+ \kappa k\sin 2 k x}{\kappa^2 + k^2} \Bigr]
\end{eqnarray*}
One can prove by complex integration that
\begin{eqnarray*}
\frac{\kappa\!\cdot\!i}{2} e^{-2 \kappa x}\Bigl[E_1(-2 \kappa x\!+\! i 2 k x)\!-\!E_1(-2 \kappa x- i 2 k x)\Bigr]\! = \!\int_{k}^{\infty}\!dk\! \frac{\kappa^2\cos 2 k x + \kappa k\sin 2 k x}{\kappa^2 + k^2}
\end{eqnarray*}
$E_1(z)$ being the exponential integral (Ref.~\cite{AS}, p.~228)
\begin{equation}
E_1(z) = \int_z^\infty dt\cdot\frac{e^{-t}}{t} \quad \mid \arg z\mid <\pi
\end{equation}
($t$ parametrizes a path in the complex plane along which the integral is taken). To prove this identity we perform the integration along the path $-2 \kappa x + i t$, $t\geq 2 k x$ and take into account that $E_1(\bar z) = \bar E_1(z)$. We then have
\begin{eqnarray*}
\frac{\kappa\!\cdot\!i}{2} e^{-2 \kappa x}\Bigl[E_1(-2 \kappa x\!+\! i 2 k x)\!-\!E_1(-2 \kappa x- i 2 k x)\Bigr]\! =\\
\frac{\kappa\!\cdot\!i}{2} e^{-2 \kappa x}\Bigl[\int_{2kx}^\infty \frac{e^{2 \kappa x -it}}{-2\kappa x +it}i\cdot dt + \int_{2kx}^\infty \frac{e^{2 \kappa x +it}}{-2\kappa x -it}i\cdot dt\Bigr]=\\
-\frac{\kappa}{2}\int_{2kx}^\infty\frac{(\cos t -i\sin t)(-2\kappa x-it)+
(\cos t+i\sin t)(-2\kappa x+it)}{4 \kappa^2 + t^2}=\\
\kappa\int_{2 k x}^\infty \frac{2 \kappa x \cos t + t\sin t}{4\kappa^2 + t^2}=\\
\int_{k}^{\infty}\!du \frac{\kappa^2\cos 2 u x + \kappa u\sin 2 u x}{\kappa^2 + u^2}\quad QED
\end{eqnarray*}
(with $u\doteq\frac{t}{2x}$).
We can now write
\begin{eqnarray*}
\rho(x) &=&\kappa e^{-2\kappa x}+ \frac{k_F}{\pi}+\frac{\kappa\!\cdot\!i}{2\pi} e^{-2 \kappa x}\\
&\cdot&\Bigl[E_1(-2 \kappa x + 2 i k_Fx)- E_1(-2 \kappa x - 2 i k_F x)-E_1(-2 \kappa x + i0)+ E_1(-2 \kappa x-i0)\Bigr]
\end{eqnarray*}
Taking into account that
\begin{displaymath}
E_1(-x \pm i0) = -E_i(x)\mp i\pi
\end{displaymath}
we obtain
\begin{eqnarray*}
\rho(x) &=&
\kappa e^{-2\kappa x}+ \frac{k_F}{\pi}+\frac{\kappa\!\cdot\!i}{2\pi} e^{-2 \kappa x}\\
&\cdot &\Bigl[E_1(-2 \kappa x + 2 i k_Fx)- E_1(-2 \kappa x - 2 i k_F x)+ 2\pi i\Bigr]\\
&=&\frac{k_F}{\pi}+\frac{\kappa\!\cdot\!i}{2\pi} e^{-2 \kappa x}
\cdot\Bigl[E_1(-2 \kappa x + 2 i k_Fx)- E_1(-2 \kappa x - 2 i k_F x)\Bigr]
\end{eqnarray*}
Notice that the contribution of the bound state to the total charge density cancels out with one component of the contribution originating from the free electron states. This is in line with similar analytical studies of the same problem\cite{Balt}. The next figure plots the charge density as a function of the variable $x$ for characteristic values of $\kappa$ and $k_F$, given the relation
\begin{equation}
\label{Eq:kappa}
\frac{\kappa}{k_F} = \frac{\pi}{4}\cdot \frac{J_{sd}}{E_F}\cdot \frac{a_0}{a}
\cdot n
\end{equation}
\begin{figure}[H]
\begin{center}
\includegraphics[width=0.5\textwidth]{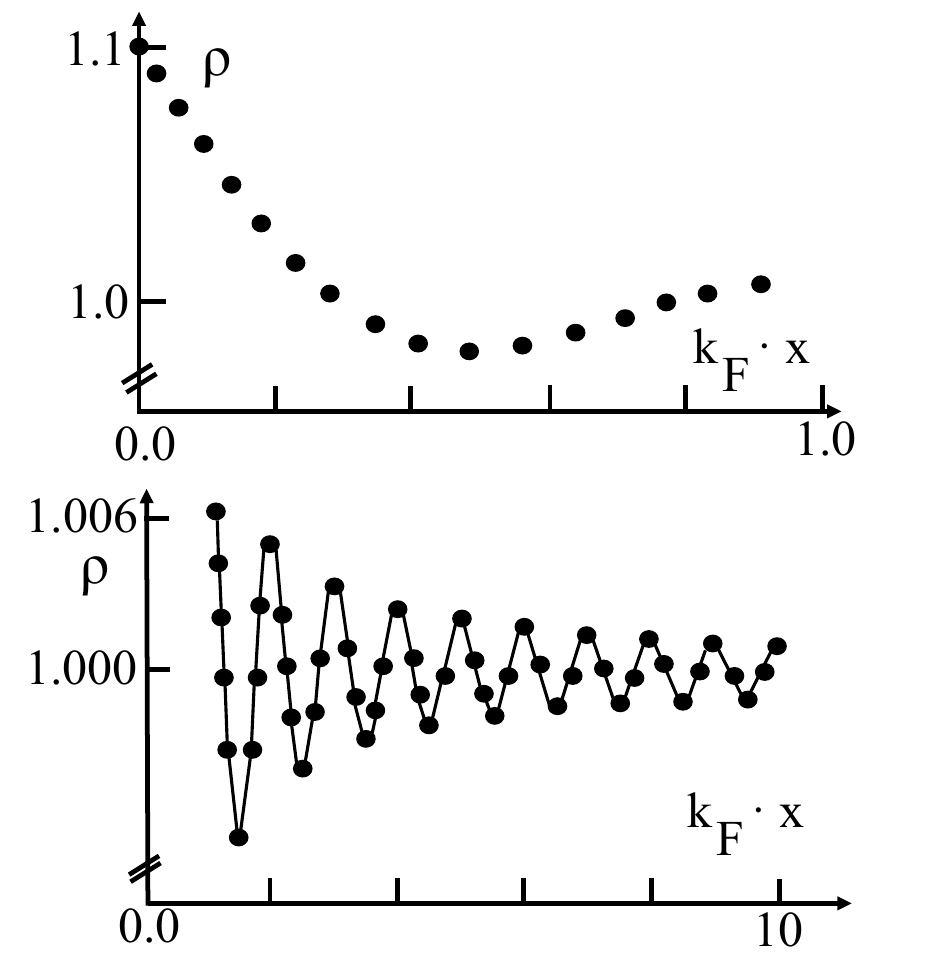}
\caption{Plot of the graph of $\rho(k_F\cdot x)$ at $k_F=\pi$ and $\frac{\kappa}{k_F}= 0.2$.}
\label{Fig:RKKY}
\end{center}
\end{figure}
Some particular limits are worked out now.
For small $x$ we have ($E_1(z)\approx -\gamma -\ln z$)
\begin{displaymath}
\rho(x)\approx \frac{k_F}{\pi}+ \frac{\kappa}{\pi}\left(\pi-\arctan \frac{k_F}{\kappa}\right)-{\cal O}(x)\approx \frac{k_F}{\pi} + \frac{\kappa}{2}-{\cal O}(x)
\end{displaymath}
For large $z$ we have
\begin{displaymath}
E_1(z)\approx \frac{e^{-z}}{z}
\end{displaymath}
and accordingly
\begin{eqnarray}
\frac{e^{2\kappa x-2ik_Fx}}{-2\kappa x + 2 ik_Fx}-\frac{e^{2\kappa x+2ik_Fx}}{-2\kappa x - 2 ik_Fx}=\frac{4 i e^{2\kappa x}}{4 \kappa^2 x^2 + k_F^2x^2}\Bigl[ \kappa x \sin 2k_F x-k_F x \cos 2 k_F x\Bigr]
\end{eqnarray}
and in the limit of large $x$ (and small $\kappa$) we obtain
\begin{equation}
\rho(x) \approx \frac{k_F}{\pi} + {\cal O}\left(\frac{\kappa \cos 2 k_F x}{k_F x}\right)
\end{equation}
This result underlines the formation of a Friedel-like oscillation\cite{Friedel} of the charge density away from a localized perturbation. The
wavelength of the oscillation  is $\frac{\pi}{k_F}$. The oscillation decays as
the inverse of the distance from the perturbation. This decay holds for a
strictly one dimensional problem. Different dimensionalities and geometries give rise to a different power for the decay. In 3D, for instance, one finds that
the decay from a point localized perturbation is like the inverse third power of
the distance from the disturbance. In the situation of two dimensional epitaxial
multilayers, the power of the decay is ``2''\cite{BrunoJsd}, which is the decay
quoted for $J(d_{Cu})$ in Sec.~4.2.

\chapter{Energy functionals and Landau-Lifshitz equation in transition metals ultrathin films.}
\label{Chap:LL}

\renewcommand{\thesection}{\thechapter.\arabic{section}}
\renewcommand{\theequation}{\thechapter.\arabic{equation}}

In the next Chapters, our focus will be on the finite temperature
properties of ultrathin ferromagnetic films. Ultrathin films of transition metal
atoms are a many electron system, i.e. they are ruled by a quantum mechanical
many-body operator. As such, according to the
lore of statistical physics, their magnetic properties are, formally,
computable by the so called partition function of the many body quantum
mechanical operator. This partition function is, actually, computed by summing
over \textbf{all} possible eigenstates of the Hamilton operator, but finding
them is a task that has not been solved yet for concrete condensed matter
systems. A paper by Landau\cite{Landau_32} indicates a practical computational
scheme that could replace the full quantum mechanical approach. This scheme
consists in replacing the original Hamilton operator, defined over quantum mechanical spin operators, with a suitable functional ${\cal F}$ of a
classical (so called order parameter) field $m(\vec r)$ (scalar or vectorial) defined over the continuous coordinates $\vec r$ -- a so called \textbf{Landau functional} ${\cal F}[m(\vec r)]$. The Landau functional is often designated  as  ``the effective Hamiltonian'', so that one encounters often the corresponding designation ${\cal H}[m(\vec r)]$. The functional is constructed with the aim of describing the free energy of those elementary excitations that matter most for determining the finite temperature behavior. The law of thermodynamics then states that the probability of the realization of a given spin configuration (excitation) $m(\vec r)$ is
\begin{equation}
\label{Eq:Boltzmann}
\frac{e^{-\beta{\cal F}[m(\vec r)]}}{\sum_{m(\vec r)}e^{-\beta{\cal F}[m(\vec r)]}}
\end{equation}
($\beta\doteq\frac{1}{k_B T}$). In Eq.~\eqref{Eq:Boltzmann}, the quantity $\sum_{m(\vec r)}e^{-\beta{\cal F}[m(\vec r)]}$ -- formally the ``sum'' over all configurations -- is the classical partition function that replaces the quantum mechanical one. We will learn in the next Chapter how the ``sum'' is, mathematically speaking, actually performed. The scope of this Chapter is to find those Landau functionals that properly describe the interactions symbolically represented by the coupling parameters
$J,\lambda,\Omega$ and to verify them by some kind of experimental findings.\\
An energy functional, as it is used in the literature, expresses the total free energy of a configuration as an integral over the volume of some classical spin distribution $\vec m(\vec r)$. The use of classical vectors is not immediately obvious. In fact, in atoms spins are strictly quantized and the transition from quantized atomic spins to a classical vector field in metallic ferromagnets is not necessarily univocal. In addition, one deals with magnetic moments that are not exactly residing on a lattice point but are somewhat spread out over finite
distances, given e.g. by the size of the $d$-orbitals. The moments are built by
the contribution of a few electrons and it is not a priori clear that these
electrons can be ``bundled'' into a classical vector that represents all of them. A further difficulty: the ferromagnetic coupling between them is not a direct type exchange coupling between atomic like orbitals but is mediated, in a
complex manner, by the surrounding free electron like orbitals via the RKKY
mechanism. This coupling can therefore be described only at the conceptual
level of some kind of ``model'' rather than at the level of first principles
using a precise, many-electron Schr\"odinger equation. It adds to the
difficulty of finding a proper functional that total energy ab-initio
calculations are effective in distinguishing between a ferromagnetic and a
paramagnetic ground state, but are not yet so advanced as to be able to compute
reliably the energy of one local magnetic moment ``rotated'' with respect to
the environment. One therefore must aim at finding, in the best of circumstances, a suitable ``model'' for the energy functional, a kind of suitable ``caricature'' of a microscopic, not yet completely known many-electron Hamilton operator.\\
One problem intrinsic to epitaxial layers of transition metals is that their
thickness is in the subnanometer to few nanometer range. The numerous monoatomic
steps that appear in the deposit produce thickness fluctuations that are
non-negligible. There is no model that takes these defects into account and
the discussion of epitaxial films must be conducted on the basis of energy
functionals established onto a perfectly smooth medium.\\
The Chapter is organized as follows. Section~5.1 constructs the exchange energy
functional and the exchange stiffness $A$. Section~5.2 introduces the
Landau-Lifshitz equation of motion, whose eigenmodes -- the spin waves -- turn
$A$ into a measurable quantity. Section~5.3 collects the remaining functionals,
those of the magnetic anisotropies, of the Zeeman energy and of the dipolar
interaction. Section~5.4 puts the two halves together: it applies the
Landau-Lifshitz equation to an ultrathin film in the two field geometries that
are actually used in the laboratory, and thereby provides the experimental
route to $\lambda$ and $\Omega$; it closes with a Table of typical values.
Appendix~\ref{App:magstat} derives the dipolar functional from the equations of
magnetostatics.

\section{The exchange energy functional.}
In the hierarchy of interactions, the exchange interaction is certainly the most relevant one. Among the authors that have undertaken the task of describing the exchange energy  for a ferromagnetic transition metal, there is a considerable agreement that the ``Hamilton operator'' best describing the exchange energy is, paradoxically, not a quantum mechanical one. Rather, the
energy of the ``rotation'' of a spin at the site $i$ with respect to a spin $j$
in the neighborhood is given by $-J_{ij}\cdot S^2\cdot \vec n_i\cdot \vec n_j$,
where $\vec n$ is a classical unit vector that can
point in any spatial direction\cite{Small}. The vectors $\vec n$ are used to
describe a spin vector $\vec S\dot=S\cdot \vec n$, with $S$ (the spin at the lattice site) being in units of
$\hbar$. The corresponding magnetic moment vector is given by $-g\cdot \mu_B\cdot S\cdot \vec n$, $g$ being the $g$-factor. The corresponding magnetization is given in units of $M_0 = \frac{g\cdot \mu_B\cdot S}{a^3}$, $a$ being the lattice constant of a lattice that we assume to be a simple cubic one, for simplicity. For transition metals, one uses $g\approx 2$. $S$ is not necessarily a half integer but is a real valued number that allows one to express the computed or experimental magnetic moment ($S\approx 1.1$ for bulk Fe, for instance). $i$ and $j$ need not be nearest neighbours: one finds that the exchange coupling, given by the energy $J_{ij}$, extends beyond the nearest neighbours\cite{Small}. For itinerant electrons, one uses $S^2$ rather than $S\cdot(S+1)$ to express the square of the spin vector. One problem with this description is that the actual value of $J_{ij}$ depends on the average alignment of the spins surrounding the couple $i,j$\cite{Luchini}, so that the generalization of the energy of a couple to a large set of spins with the classical Heisenberg Hamiltonian\cite{Small,Wang}
\begin{equation}
\label{Eq:Heis_lattice}
{\cal H}_J = -\frac{1}{2}\cdot \sum_{i\not=j} J_{ij}\vec S_i\cdot \vec S_j
\end{equation}
can be rigorously justified only in the case of small deviations from the
totally ferromagnetic state. In this case, the classical Heisenberg Hamiltonian -- a kind of ``functional'' of a field $\vec n$ that ``lives'' on
a lattice -- is a proper rendering of the Taylor expansion of the total energy to
second order about the ferromagnetic ground state\cite{Wang,Small}. The
comparison with experiment foresees that one should extract from the computed $J_{ij}$ an effective value that stands for the coupling of two
neighbouring spins, so that the exchange energy functional that best describes excitations of the transition metal ferromagnetic state is the classical Heisenberg Hamiltonian on a lattice, with $j$ being restricted to the nearest neighbours ($j=i+nn$) and $J$ being an averaged exchange coupling parameter:
\begin{equation}
{\cal H}_J\approx -\frac{J}{2}\cdot S^2\cdot \sum_{i,j=i+nn}\vec n_i\cdot \vec n_j
\end{equation}
A further approximated version of this energy functional foresees that the
lattice is replaced by a continuum of sites $\vec r$. This approximation is
based on the fact that only small deviations from the ferromagnetic ground state
are typically considered, so that a noticeable change of the spin direction is
only achieved over large distances. On such distances, the lattice constant
might be considered to be small enough for a transition to a continuous spin distribution to be physically meaningful. The algebra leading to a functional of $\vec n(\vec r)$ proceeds as follows. We start by writing
\begin{equation}
-\vec n_i\cdot \vec n_j\! =\! \frac{1}{2}(\vec n_i\!-\!\vec n_j)^2\!-\!\frac{\vec n_i^2}{2}\!-\!\frac{\vec n_j^2}{2}
\end{equation}
We assume a simple cubic lattice and write the six sites $i+nn$ as
\begin{equation}
\big(\vec r\! + \!(\pm a,0,0)\big)\quad \big(\vec r \!+\! (0,\pm a,0)\big) \quad \big(\vec r\! + \!(0,0,\pm a)\big)
\end{equation}
with $a$ being the distance to the nearest neighbor. With this choice, we obtain, for each coordinate $p$,
\begin{eqnarray}
n_p\big(\vec r\! +\! (\pm a,0,0)\big)\approx n_p(\vec r)\! +\! \frac{\partial n_p(\vec r)}{\partial x}\cdot (\pm a)\nonumber\\
n_p\big(\vec r\! + \!(0,\pm a,0)\big)\approx n_p(\vec r)\! +\! \frac{\partial n_p(\vec r)}{\partial y}\cdot (\pm a)\nonumber\\
n_p\big(\vec r \!+ \!(0,0,\pm a)\big)\approx n_p(\vec r)\! + \!\frac{\partial n_p(\vec r)}{\partial z}\cdot (\pm a)
\end{eqnarray}
and
\begin{equation}
\big(n_p(\vec r)\! -\!n_p(\vec r\!+\!a(\pm 1,0,0))\big)^2 \!\approx \!\big(\frac{\partial n_p(\vec r)}{\partial x}\cdot a\big)^2
\end{equation}
A similar expression holds true for the remaining nearest neighbours along $y$ and $z$, the partial derivative being taken with respect to the appropriate coordinate. Performing the sum over all neighbours and over all coordinates we obtain
\begin{equation}
\sum_{j}\frac{1}{2}\left(\vec n_i\!-\!\vec n_j\right)^2\approx \sum_{p=x,y,z}\big(\vec \nabla n_p(\vec r)\cdot a\big)^2
\end{equation}
For the transition to the continuum version of the Hamiltonian,
$\sum_i...$ is replaced by $\frac{1}{a^3}\int d^3x...$. The total exchange energy, originally defined on a lattice, can be written as an energy functional over the field $\{\vec n(\vec r)\}$
\begin{eqnarray}
{\cal H}_J[\vec n(\vec r)]=\frac{1}{2}\cdot J\cdot S^2\cdot \frac{1}{a} \cdot\!\int d^3x\!\cdot\!\Big[\!(\vec \nabla n_x)^2\!+\!(\vec \nabla n_y)^2\! +\! (\vec \nabla n_z)^2\Big]- \frac{J\cdot S^2\!\cdot z}{2}\cdot \frac{V}{a^3}
\end{eqnarray}
$V$ being the volume of the ferromagnetic medium and $z$ the number of nearest
neighbours ($z=6$ for the simple cubic lattice).
To be in line with the literature\cite{Diop} about the practical determination of the exchange integral, we use
\begin{equation}
\label{Eq:A}
A\dot=\frac{1}{2}\cdot J\cdot S^2\cdot \frac{1}{a}
\end{equation}
the letter ``$A$'' designating the so called ``exchange stiffness''\cite{Diop}, and write the spatially variable component of the exchange energy functional (also called the ``elastic energy functional'') as
\begin{equation}
{\cal H}_A= A\cdot\!\int d^3x\!\cdot\!\Big[\!(\vec \nabla n_x)^2\!+\!(\vec \nabla n_y)^2\! +\! (\vec \nabla n_z)^2\Big]
\end{equation}
More compactly and in two equivalent writings:
\begin{eqnarray}
\label{Eq:HA}
{\cal H}_A&=& A\cdot\!\int d^3x\!\cdot\!\sum_{\nu=x,y,z} \big (\vec \nabla n_\nu(\vec r)\big)^2\nonumber\\
& = & A\cdot\!\int d^3x\!\cdot\!\sum_{\nu=x,y,z}\big(\partial_\nu \vec n(\vec r)\big)^2
\end{eqnarray}

\section{The Landau-Lifshitz (LL) equation of motion and the parameter \texorpdfstring{$A$}{A}.}
In this section, we introduce a general method used in research on
ferromagnetism, the equation of motion of Landau and Lifshitz\cite{LL}. For the
sake of illustration, we apply this method to determining the parameter
$A$. This method consists in measuring the so called spin wave stiffness, which is related to the parameter $A$. The spin waves are the eigenmodes of the equation of motion for the vector $\vec S$ (the so called Landau-Lifshitz equation) and represent the fundamental excitations from the ferromagnetic ground state.\\
The equation of motion for $\vec S$ can be constructed by the correspondence principle. In classical mechanics, the angular momentum vector obeys the equation
\begin{equation}
\dot {\vec L} = \vec D
\end{equation}
whereby $\vec D$ is defined as the negative of the functional derivative\cite{Function} of the total energy with respect to a rotational axis $\vec {\delta\varphi}$\cite{Landaumech}. Owing to the correspondence principle, we can write
\begin{equation}
\hbar\cdot\frac{\partial \vec S(\vec r)}{\partial t}= -\frac{\delta {\cal H}[\vec S(\vec r)]}{\vec {\delta \varphi}(\vec r)}
\end{equation}
where ${\cal H}[\vec S]$ is some suitable functional of the field $\vec S$. Recalling that ${\cal H}[\vec S]$ is some integral over the volume of a function of $\vec S$ (and possibly its gradient), we can rewrite the equation of motion more explicitly as
\begin{equation}
\label{Eq:LLfunctional}
\hbar\cdot\frac{1}{a^3}\int dV\,\frac{\partial \vec S(\vec r)}{\partial t}\cdot \vec{\delta\varphi}(\vec r)= -\Big({\cal H}[\vec S+\vec{\delta S}]-{\cal H}[\vec S]\Big)
\end{equation}
This last expression is a good starting point for putting the technology of the ``functional derivative'' into practice. In this section, ${\cal H}$ is the exchange functional and we use this equation for determining the parameter $A$.\\
A first approximation consists in allowing only for ``transversal'' excitations from the ground state, where the spin at $\vec r$ might be rotated away from the ground state direction but its length is not changed.
In addition, for studying the long-wavelength eigenmodes of this equation, we introduce a linearized version of it that only allows small deviations from the ferromagnetic state. This approximation is defined by setting $n_z\approx 1$ and $n_x,n_y<< 1$ in ${\cal H}_A[\vec n]$. This approximation is called the harmonic or spin wave approximation of ${\cal H}_A[\vec n]$ and the corresponding energy functional writes
\begin{equation}
{\cal H}_{A,@n_z\approx 1}= A\cdot \int dV\cdot \!\big[\!(\vec \nabla n_x)^2\!+\!(\vec \nabla n_y)^2\!\big]
\end{equation}
The LL equation then writes
\begin{equation}
\hbar\cdot S\cdot \frac{1}{a^3}\int dV \frac{\partial\vec n(\vec r)}{\partial t}\cdot \vec {\delta\varphi}(\vec r)= -\Big({\cal H}_{A,@n_z\approx 1}[\vec n + \vec{\delta \varphi}\times \vec n]-{\cal H}_{A,@n_z\approx 1}[\vec n]\Big)
\end{equation}
whereby we have used the important property that for a rotational axis $\vec {\delta \varphi}$ one has
\begin{equation}
\label{Eq:rot}
\vec {\delta n}= \vec{\delta \varphi}\times \vec n
\end{equation}
Both sides of this equation are now defined over the local field
$\vec n(\vec r)$. We seek now the equation of motion for $\vec n( \vec r)$. We find it explicitly for e.g. the $x$-component. For this purpose, we use
\begin{equation}
\label{Eq:deltan_x}
\vec {\delta\varphi} = \Big(\delta \varphi,0,0\Big)\quad\quad
\delta {\vec n} = \Big(0,-n_z\cdot \delta \varphi,+n_y\cdot \delta \varphi\Big)
\end{equation}
so that
\begin{eqnarray}
\hbar\cdot S\cdot \int \frac{dV}{a^3}\cdot \frac{\partial n_x}{\partial t}\cdot \delta \varphi &= &+2\cdot A\cdot a^3\cdot \int \frac{dV}{a^3}\vec \nabla n_y(\vec r)\cdot \vec \nabla \delta \varphi\nonumber\\
&=&-2\cdot A\cdot a^3\cdot \int \frac{dV}{a^3}\triangle n_y(\vec r)\cdot \delta \varphi
\end{eqnarray}
to first order in $\delta\varphi$. The last line is obtained by partial integration. By comparing the integrand of the left-hand side with the
integrand of the right-hand side we obtain the sought-for equation of motion for the component $n_x(\vec r)$:
\begin{equation}
\frac{\partial}{\partial t}n_x(\vec r,t) = -\frac{2\cdot A\cdot a^3}{\hbar\cdot S}\cdot \triangle n_y(\vec r,t)
\end{equation}
Similarly, we obtain
\begin{equation}
\frac{\partial}{\partial t}n_y(\vec r,t) = +\frac{2\cdot A\cdot a^3}{\hbar\cdot S}\cdot \triangle n_x(\vec r,t)
\end{equation}
and, owing to the harmonic approximation,
\begin{equation}
\frac{\partial}{\partial t}n_z(\vec r,t) =0
\end{equation}
The eigenmode Ansatz solution (``spin wave Ansatz'') for these coupled differential equations reads
\begin{eqnarray}
n_x(t,\vec r) &=& a_x\cdot e^{i(\omega\cdot t-\vec k\cdot \vec r)}\nonumber\\
n_y(t,\vec r) &=& a_y\cdot e^{i(\omega\cdot t-\vec k\cdot \vec r)}
\end{eqnarray}
Inserting this Ansatz produces a system of coupled homogeneous algebraic equations for the sought for amplitudes $a_x,a_y$. This system has a non trivial solution when the determinantal equation is fulfilled:
\begin{equation}
\text{det}\begin{pmatrix}
\hbar \cdot S\cdot (i\cdot \omega)&-2\cdot A\cdot a^3\cdot (\vec k)^2\\
+2\cdot A\cdot a^3\cdot (\vec k)^2& \hbar \cdot S\cdot (i\cdot \omega)\end{pmatrix}=0
\end{equation}
The solution of this equation is the dispersion relation for the spin wave excitations:
\begin{equation}
\hbar\cdot \omega = \pm \frac{2\cdot A\cdot a^3}{S}\cdot (\vec k)^2
\end{equation}
(the $\pm$-sign refers to waves traveling in $\pm \vec k$ direction). This relation is important because it establishes a relation between the so called spin wave stiffness parameter $D$\cite{Small}
\begin{equation}
D\dot= \frac{\hbar\cdot \omega}{\vec k^2}
\end{equation}
and the sought-for parameter $A$ (or $J\cdot S$):
\begin{equation}
\label{Eq:D}
D= \frac{2\cdot A\cdot a^3}{S}=J\cdot S\cdot a^2
\end{equation}
A table summarizing the values of $D$ and $A$ reported in the literature can be
found in Ref.~\cite{Diop}. For Fe, for instance, the computed value for $D$ is
$D\approx 350\ \text{meV}\cdot \text{\AA}^2$ (see \cite{Small} and the references
therein). With $S\approx 1.1$ and $a = 2.87$ \AA, one obtains $J\cdot S\approx
42$ meV, i.e. $J\approx 38$ meV.
The main experimental method
used to determine $D$ is the inelastic scattering of neutrons. Neutrons have a
magnetic moment that can interact with the spins of a ferromagnet. By this
interaction, spin waves can be excited or annihilated and the energy loss or
gain acquired by the neutrons gives directly the energy of the excited spin
wave. In addition, by determining the momentum transfer, the dispersion
relation can be measured over the entire Brillouin zone\cite{Shirane}.

\section{The functionals for magnetic anisotropies, Zeeman energy and dipolar energy.}
We introduce now further important components of the energy functional in
transition metals ultrathin magnetic films, considered as a continuum medium of
matter with an underlying lattice constant $a$ and spin distribution described by
the classical unit vector field $\vec n(\vec r)$. The single site N\'eel magnetic
anisotropy writes, in terms of the field $\vec n$, as
\begin{equation}
-\lambda\cdot n^2_z
\end{equation}
As this two-fold anisotropy originates from the breaking of translational symmetry perpendicular to the film, it is located mainly at the two surfaces bounding the thin film and is only weakly dependent on the film thickness\cite{GR}. Accordingly, for the N\'eel-type magnetic anisotropy, one can envisage an energy functional that writes
\begin{equation}
\label{Eq:Hlambda}
{\cal H}_\lambda[\vec n]= -\frac{\lambda}{\frac {d}{a}}\cdot \frac{1}{a^3}\cdot \int d^3x \cdot n_z^2(\vec r)
\end{equation}
$\lambda$ being the N\'eel magnetic anisotropy constant per surface unit cell (or per atom).\\
The ultrathin film may contain magnetic single-ion anisotropies originating from the lattice\cite{GR}. In a cubic lattice, for instance, their energy functional writes\cite{GR}
\begin{equation}
{\cal H}_\Lambda[\vec n]= \Lambda\cdot \frac{1}{a^3}\int d^3x\cdot \Big[n_x^2(\vec r)\cdot n_y^2(\vec r) + n_x^2(\vec r)\cdot n_z^2(\vec r) + n_y^2(\vec r)\cdot n_z^2(\vec r)\Big]
\end{equation}
$\Lambda$ is the coupling constant of the cubic anisotropy per atom.\\
The Zeeman energy functional writes
\begin{equation}
{\cal H}_B[\vec n]= g\cdot\mu_B\cdot S\cdot \frac{1}{a^3}\cdot \int d^3x \cdot \vec n(\vec r)\cdot \vec B(\vec r)
\end{equation}
We have included the general situation where $\vec B$ also depends on the coordinates $\vec r$.\\
The dipolar energy is less obvious. In Appendix~\ref{App:magstat} we will obtain the essential formulas applicable to ultrathin epitaxial overlayers. We will show that the essential energy characterising the dipolar interaction per atom is given by the coupling constant
\begin{equation}
\label{Eq:Omega}
\Omega\dot=\frac{\mu_0}{2}\cdot M_0^2\cdot a^3
\end{equation}
For the limited purpose of this Chapter, which is understanding the key experimental tool for determining the various coupling constants, one can use a simplified dipolar energy functional involving only the local component of the total dipolar energy:
\begin{equation}
\label{Eq:HOmega}
{\cal H}_\Omega[\vec n]\approx \Omega\cdot\frac{1}{a^3}\cdot \int d^3x\cdot n_z^2(\vec r)
\end{equation}
As $\Omega$ is \textbf{always} positive, the dipolar interaction favors the magnetization being oriented in the film plane.

\section{Application of the LL equation to ultrathin films.}
\begin{figure}[H]
\begin{center}
\includegraphics[width=0.45\textwidth]{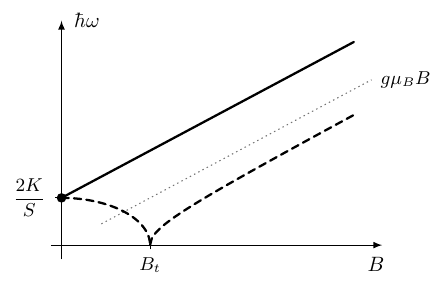}
\caption{Spin wave frequency as a function of the applied magnetic field for an
ultrathin film with perpendicular spontaneous magnetization, drawn from
Eqs.~\eqref{Eq:omega_perp}, \eqref{Eq:omega_inplane} and
\eqref{Eq:omega_tilted} with $K=\lambda\frac{a}{d}-\Omega$. Continuous line:
field applied along the normal, Eq.~\eqref{Eq:omega_perp}; its intercept at
$B=0$ (full circle) measures $2K/S$. Dashed lines: field applied in the plane.
Below the threshold field $B_t$ the magnetization is tilted and the frequency
follows Eq.~\eqref{Eq:omega_tilted}, starting at $B=0$ from the same value
$2K/S$ -- at zero field the two geometries cannot differ -- and vanishing at
$B_t$; above $B_t$ the film is saturated in the plane and the frequency follows
Eq.~\eqref{Eq:omega_inplane}, rising from zero at $B_t$ and approaching the
asymptote $g\mu_B B$ (dotted). The perpendicular branch lies above that
asymptote by $2K/S$, so that the two solid and dashed branches remain parallel
at large field rather than merging.}
\label{Fig:SW}
\end{center}
\end{figure}

A possible path used in thin film technology to determine the parameters of
the various interactions entails exciting spin waves in some specific
configurations of spontaneous magnetization. For this purpose, one uses light\cite{Brillouin}, an external field with periodic time dependence\cite{FMR},
or the technology of pulsed precessional motion\cite{Knowles}. These experimental methods aim at detecting the frequency of the excited spin
waves, which is then related, by solving the LL equation, to the
sought-for parameters. For the sake of illustration, we study the situation of an ultrathin film with spontaneous magnetization along the film normal
(here $z$). We study first the solutions of the Landau-Lifshitz equation for a magnetic field applied along the $z$-axis. Our convention is
that the magnetic field is applied along the negative $z$-axis, so that we seek
those solutions with positive $n_z$. Spin waves excited by
light or by time dependent magnetic fields typically have $\vec k\approx 0$, so that the exchange (elastic) Hamiltonian can be neglected. The relevant Hamiltonian writes
\begin{equation}
\label{Eq:Hperp}
\int\frac{dV}{a^3} \Big[\big[-\lambda\cdot \frac{a}{d}+\Omega\big]\cdot n_z^2-g\cdot \mu_B\cdot S\cdot B\cdot n_z\Big]
\end{equation}
We compute explicitly e.g. the equation of motion for the $n_x$-component. We use $\delta n_z = +n_y\cdot \delta\varphi$ from Eq.~\eqref{Eq:deltan_x} to write
\begin{eqnarray}
\hbar\cdot \dot n_x = \Big[\frac{2\cdot \big[\lambda\cdot \frac{a}{d}-\Omega\big]}{S}\cdot n_z+ g\cdot\mu_B\cdot B\Big]\cdot n_y
\end{eqnarray}
and, similarly,
\begin{eqnarray}
\hbar\cdot \dot n_y &=& -\Big[\frac{2\cdot \big[\lambda\cdot\frac{a}{d}-\Omega\big]}{S}\cdot n_z+  g\cdot\mu_B\cdot B\Big]\cdot n_x\nonumber\\
\dot n_z &=& 0
\end{eqnarray}
We seek a solution with $n_z\approx 1$. In this situation we obtain, by solving the determinantal equation,
\begin{equation}
\label{Eq:omega_perp}
\hbar\cdot \omega = \frac{2\cdot \big[\lambda\cdot\frac{a}{d}-\Omega\big]}{S} + g\cdot\mu_B\cdot B
\end{equation}
For this applied field geometry, the sought-for parameter $\lambda\cdot\frac{a}{d}-\Omega$ can be read out in a plot of $\hbar\cdot\omega$ vs $B$ from the intercept with the vertical axis of a straight line.\\
In an alternative experiment, the magnetic field is applied in the plane of the film. This produces a rotation of the magnetization away from the perpendicular direction and toward the in-plane direction specified by the magnetic field. We apply a field in the negative $x$-direction, and we seek those solutions of the Landau-Lifshitz equations with $n_x\approx +1$, i.e. we consider the film in the saturated in-plane state. In this situation, we expect the spin waves to be small rotations about the $x$-axis, governed by the free energy density
\begin{equation}
-\big[\lambda\cdot\frac{a}{d}-\Omega\big]\cdot \frac{n_z^2}{a^3}  -\frac{g\cdot \mu_B\cdot S\cdot B}{a^3}\cdot n_x
\end{equation}
and their Landau-Lifshitz equation reads
\begin{eqnarray}
\hbar\cdot \dot n_x &=& \frac{2\cdot(\lambda\cdot\frac{a}{d}-\Omega)}{S}\cdot n_y\cdot n_z\approx 0\nonumber\\
\hbar\cdot \dot n_y &=& -\frac{2\cdot(\lambda\cdot\frac{a}{d}-\Omega)}{S}\cdot n_x\cdot n_z+g\cdot \mu_B\cdot B\cdot n_z\approx n_z\cdot\Big(g\cdot \mu_B\cdot B-\frac{2\cdot(\lambda\cdot\frac{a}{d}-\Omega)}{S}\Big)\nonumber\\
\hbar \cdot \dot n_z &=& -n_y\cdot g\cdot \mu_B\cdot B
\end{eqnarray}
The solution of the determinantal equation writes
\begin{equation}
\label{Eq:omega_inplane}
\hbar\omega = \sqrt{g\cdot\mu_B\cdot B\cdot\Big(g\cdot\mu_B\cdot B-\frac{2\cdot(\lambda\cdot\frac{a}{d}-\Omega)}{S}\Big)}
\end{equation}
This solution is real, and therefore physically relevant, when $B$ is larger
than the threshold field at which the spin waves ``go soft'', i.e. at which
their frequency approaches zero. The threshold field reads
\begin{equation}
\label{Eq:Bt}
B_t= \frac{2\cdot\big(\lambda\cdot\frac{a}{d}-\Omega\big)}{g\cdot\mu_B\cdot S}
\end{equation}
Its determination is a further method for obtaining the parameter $\lambda\cdot\frac{a}{d}-\Omega$: the mode softens as $B$ approaches $B_t$ from above. When the applied field is much larger than this threshold field the frequency approaches, asymptotically,
\begin{equation}
\hbar\omega\rightarrow g\cdot\mu_B\cdot B
\end{equation}
Below $B_t$ the film is no longer saturated in the plane: the magnetization
sits at an intermediate angle fixed by the balance between the Zeeman and the
anisotropy energy. The corresponding branch -- the left-hand dashed branch of
Fig.~\ref{Fig:SW}, which reaches zero at $B_t$ from below -- follows from the
same Landau-Lifshitz equation, linearized about that tilted equilibrium. We
carry out this calculation now. It is convenient to abbreviate the effective
perpendicular anisotropy energy per atom and the Zeeman energy per atom at
saturation as
\begin{equation}
\label{Eq:Kh}
K\dot=\lambda\cdot\frac{a}{d}-\Omega\quad\quad h\dot= g\cdot\mu_B\cdot S\cdot B
\end{equation}
so that the energy per atom of a uniform configuration $\vec n$, i.e. the
integrand of Eq.~\eqref{Eq:Hperp} multiplied by $a^3$, reads $-K\cdot
n_z^2-h\cdot n_x$.\\
\textbf{The equilibrium angle.} We let the magnetization tilt in the plane
spanned by the film normal and the applied field, $\vec n_0 =
(\sin\vartheta,0,\cos\vartheta)$, $\vartheta$ being the angle from the normal.
The energy per atom
\begin{equation}
E(\vartheta)= -K\cdot \cos^2\vartheta - h\cdot \sin\vartheta
\end{equation}
is stationary when
\begin{equation}
\frac{dE}{d\vartheta}= \cos\vartheta\cdot\big(2\cdot K\cdot \sin\vartheta - h\big)=0
\end{equation}
Besides the saturated solution $\cos\vartheta = 0$ treated above, there is the
tilted solution
\begin{equation}
\label{Eq:tilt}
\sin\vartheta = \frac{h}{2\cdot K}=\frac{B}{B_t}
\end{equation}
with $B_t$ given by Eq.~\eqref{Eq:Bt}. The magnetization thus leaves the normal
as soon as a field is applied and arrives in the film plane exactly at $B_t$,
where the two solutions merge.\\
\textbf{The energy to second order.} We describe small deviations from $\vec
n_0$ in the local orthonormal frame
\begin{equation}
\vec e_1 = (\cos\vartheta,0,-\sin\vartheta)\quad \vec e_2 = (0,1,0)\quad \vec e_3 = \vec n_0
\end{equation}
which is right-handed ($\vec e_1\times \vec e_2 = \vec e_3$). Writing
\begin{equation}
\vec n = \vec e_3\cdot\sqrt{1-u^2-v^2} + u\cdot \vec e_1 + v\cdot \vec e_2
\end{equation}
with $u,v<<1$, so that the length of $\vec n$ is preserved, and expanding
$-K n_z^2-h n_x$ to second order in $u,v$, the terms linear in $u$ cancel by
virtue of the equilibrium condition~\eqref{Eq:tilt}, and one is left with
\begin{equation}
\label{Eq:Equadratic}
E = E(\vartheta) + K\cdot \cos^2\vartheta\cdot u^2 + K\cdot v^2
\end{equation}
Both coefficients are positive, i.e. the tilted solution is a minimum, as long
as $\vartheta$ is given by Eq.~\eqref{Eq:tilt}.\\
\textbf{The precession.} Rotations about $\vec e_1$ and about $\vec e_2$ move
the vector $\vec n$ according to $\vec{\delta n}=\vec{\delta\varphi}\times \vec
n$, i.e. $\delta v = -\delta\varphi_1$ and $\delta u = +\delta\varphi_2$. The
Landau-Lifshitz equation~\eqref{Eq:LLfunctional}, applied to these two rotations
and to the energy~\eqref{Eq:Equadratic}, gives the pair
\begin{eqnarray}
\hbar\cdot S\cdot \dot u &=& +\frac{\partial E}{\partial v}= +2\cdot K\cdot v\nonumber\\
\hbar\cdot S\cdot \dot v &=& -\frac{\partial E}{\partial u}= -2\cdot K\cdot \cos^2\vartheta\cdot u
\end{eqnarray}
Eliminating $v$ produces a harmonic oscillation of $u$ with
$(\hbar\cdot S\cdot\omega)^2 = 4\cdot K^2\cdot \cos^2\vartheta$, i.e., using
Eq.~\eqref{Eq:tilt} to express $\cos\vartheta$ through the applied field,
\begin{equation}
\label{Eq:omega_tilted}
\hbar\omega = \frac{2\cdot K}{S}\cdot\sqrt{1-\Big(\frac{B}{B_t}\Big)^2}= g\cdot\mu_B\cdot\sqrt{B_t^2-B^2}
\end{equation}
This is the left-hand dashed branch of Fig.~\ref{Fig:SW}. Two checks are worth
making. At $B=B_t$ it vanishes, meeting the saturated branch
Eq.~\eqref{Eq:omega_inplane} at the same point and from the other side: the
softening at $B_t$ is therefore approached from both directions, which is what
makes $B_t$ such a convenient experimental quantity. At $B=0$ it returns
$\hbar\omega = 2K/S$, i.e. exactly the intercept~\eqref{Eq:omega_perp} of the
perpendicular geometry at zero field -- as it must, since at zero field there is
nothing to distinguish the two experiments.
The next figure summarizes the results we have just achieved. The continuous line refers to a magnetic field applied perpendicular to the plane, along which direction the spontaneous magnetization resides\cite{Brillouin}. The dashed line, instead, refers to a magnetic field applied in the plane of the film\cite{FMR}.
\noindent Graphs similar to those sketched in Fig.~\ref{Fig:SW} also appear in a situation where the magnetization is e.g. along a certain direction in the plane of the film and the free energy contains terms originating from in-plane magnetic anisotropies\cite{Knowles,Pini,SSC,FMR,Grad}. One observes that the spin wave energy also softens to almost zero, as at a specific field strength the competition between Zeeman energy and anisotropy energy renders the system almost isotropic. In practice, the experimental graph corresponding to the dashed curve undergoes a minimum rather than a cusp.\\
We conclude the section with a Table that summarizes values for the parameters that are representative of epitaxial ultrathin films.
\begin{table}[H]
\begin{center}
\begin{tabular}{|c|c|c|c|c|c|}
\hline
Table 1&$J\!\cdot\!S^2$&$\lambda$  &$\Omega$  &$\Lambda$&$\mu_B\cdot B$\\
&[meV]&[meV]&[meV]&[meV]&[meV]\\
\hline
$^1$&46\cite{Small}&&&&\\
\hline
$^2$&36\cite{Small}&&&&\\
\hline
$^3$&&0.38\cite{Gay}&&&\\
\hline
$^4$&&&$0.3$\cite{Gay}&&\\
\hline
$^5$&&&$0.28$&&\\
\hline
$^6$&&$0.7\!-\!1.1$\cite{Shi}&&&\\
\hline
$^{7}$&&$\approx\!1\!$\cite{FMR}&&&\\
\hline
$^8$&&&&$0.0026$\cite{Grad}&\\
\hline
$^9$&&&&0.0023\cite{SSC}&\\
\hline
$^{10}$&&&&&0.058@$1$T\\
\hline
\end{tabular}
\caption*{Summary of typical experimental and theoretical values for the parameters of ultrathin films of transition metals.\\
{\small $^1$: Theoretical values for bulk bcc Fe. Realistic computations find values for the exchange interaction between atoms which are further away than nearest neighbors. The value we use here is obtained by simplifying Eq. 4.10 of Ref.~\cite{Small} for the spin wave stiffness constant to $D\approx S\cdot J\cdot a^2$. An effective value for $J$ is then obtained using the values of Table 1 in Ref.~\cite{Small}. $S\approx 1.1$\cite{Small}.\\
$^2$: The experimental value is obtained from the experimental value for $D$ quoted in Ref.~\cite{Small}.\\
$^3$: Per unit surface cell.\\
$^4$: Per Fe atom.\\
$^5$: Per bulk unit cell. Fe atoms occupying a bcc lattice, i.e. 2 atoms in the unit cell, each carrying a magnetic moment of $2.2 \mu_B$ ($g\approx 2$\cite{Small}, $S\approx 1.1$\cite{Small}). $\mu_0 = 4\pi\cdot 10^{-7} \cdot \frac{T\cdot m}{A}$, $\mu_B = 5.8\cdot 10^{-2}\cdot \frac{\text{meV}}{T}$,
$a= 2.87\cdot 10^{-10}\cdot m$.\\
$^6$: Per unit surface cell.\\
$^7$: 3 ML Fe on Cu(100), room temperature (estimated from Table 1 in Ref.~\cite{FMR}).\\
$^8$: Fe/W(110), per bulk unit cell.\\
$^9$: $\approx 1.8$ ML fcc Co/Cu(110)fcc, per bulk unit cell.\\
 }
 }
\end{center}
\end{table}

\setcounter{section}{0} 
\renewcommand{\thesection}{\thechapter.\Alph{section}} 
\setcounter{equation}{0}
\renewcommand{\theequation}{\thesection\arabic{equation}}
\section*{Appendices}

\section{The magnetostatic equations for the dipolar interaction.}
\label{App:magstat}
The fundamental elements for describing the magnetostatic energy of a body are the Maxwell equations of magnetostatics:
\begin{equation}
\vec \nabla \cdot \vec B = 0 \quad ;\quad \vec \nabla \times \vec B = \mu_0 \cdot \vec \nabla \times \vec M
\end{equation}
Using the relation between the technical magnetic field $\vec H$ and the
physical magnetic field $\vec B$, i.e.
\begin{equation}
\vec B = \mu_0\cdot (\vec H + \vec M)
\end{equation}
one can rewrite the equations of magnetostatics in terms of the technical field
$\vec H$:
\begin{equation}
\vec \nabla \cdot \vec H = -\mu_0\cdot \vec \nabla \cdot \vec M \quad ;\quad \vec \nabla \times \vec H = 0
\end{equation}
The general equation of magnetostatics that represents the starting point for the computation is the total magnetostatic self-energy of a continuous distribution of permanent magnetization $\vec M(\vec r)$, which can be written, according to\cite{Jak}, as
\begin{eqnarray}
 {\cal H}_M[\vec M]&=&-\frac{\mu_0}{2}\cdot \int d^3x \cdot \vec M(\vec r)\cdot \vec H(\vec r) -\frac{\mu_0}{2}\cdot \int d^3x \cdot \vec M^2(\vec r)
 \end{eqnarray}
We have a ``slab'' model for the ultrathin films in mind, describing a
film as a continuum medium with coordinates $\vec r$. Laterally, the film has
an area $L^2$ (coordinates $\vec \rho=(x,y)$) while along the vertical direction ($z$) the thickness $d$ is finite ($d\!<\!<\!L$). The slab hosts a continuous spin distribution. We require that the spin distribution has a strictly two-dimensional spatial spin degree of freedom, i.e. the spin distribution is rigid along $z$ and can only assume some profile along the $xy$-plane. This requirement is strictly true only if the film thickness is small enough that a rotation of the spin along the vertical direction costs too much exchange energy. We recall, finally, that in metallic ferromagnets such as Fe, the spins at the lattice sites can be considered as classical vectors\cite{Small}, i.e. the spin field can be written in terms of a dimensionless classical vector field
$\vec n(\vec r)$ with unit length and related to the magnetization vector by the relations $\vec S(\vec r) = S\cdot \vec n(\vec r)$ and $\vec M(\vec r) = -\frac{g\cdot \mu_B\cdot S\cdot \vec n(\vec r)}{a^3}$.
We also use, for the absolute value of the magnetization, the labeling
$\frac{g\cdot \mu_B\cdot S}{a^3}\doteq M_0$
and introduce a characteristic dipolar energy per atom by defining the parameter
\begin{equation}
\Omega\dot=\frac{\mu_0}{2}\cdot M_0^2\cdot a^3
\end{equation}
We now work out a formula for ${\cal H}_M[\vec M]$ that is specific to the slab model of an ultrathin film. For achieving this task, we consider the solution of the Maxwell equation for the field $\vec H(\vec r)$:
\begin{equation}
\vec H(\vec r) = \frac{1}{4\pi}\cdot \vec \nabla \int d^3x'\frac{\vec \nabla' \cdot \vec M(\vec r')}{\vert \vec r-\vec r'\vert}
\end{equation}
We use the identity of vector analysis
\begin{equation}
-\vec M(\vec r')\cdot \vec \nabla' \frac{1}{\mid \vec r -\vec r' \mid}= \frac{\vec \nabla^{'}\cdot \vec M(\vec r')}{\mid \vec r -\vec r' \mid} + \vec \nabla^{'}\cdot (\vec M(\vec r')\cdot \frac{1}{\mid \vec r -\vec r' \mid})
\end{equation}
to redirect the divergence of the magnetization toward the gradient of the scalar $\frac{1}{\mid \vec r -\vec r' \mid}$. The divergence term on the right is rendered vanishing by transforming it via Gauss' law into a surface integral and letting the surface go to infinity, where the magnetization (well behaved and localized) is vanishing. After these transformations, the field $\vec H$ writes, explicitly,
\begin{equation}
\begin{pmatrix}
H_x(\vec r)\\
H_y(\vec r)\\
H_z(\vec r)
\end{pmatrix}
 = -\frac{1}{4\pi}\cdot \int d^3x'\begin{pmatrix}\frac{\partial }{\partial x}\frac{\partial }{\partial x'}\frac{1}{\mid \vec r -\vec r' \mid}&\frac{\partial }{\partial x}\frac{\partial }{\partial y'}\frac{1}{\mid \vec r -\vec r' \mid}&\frac{\partial }{\partial x}\frac{\partial }{\partial z'}\frac{1}{\mid \vec r -\vec r' \mid}\\
\frac{\partial }{\partial y}\frac{\partial }{\partial x'}\frac{1}{\mid \vec r -\vec r' \mid}&\frac{\partial }{\partial y}\frac{\partial }{\partial y'}\frac{1}{\mid \vec r -\vec r' \mid}&\frac{\partial }{\partial y}\frac{\partial }{\partial z'}\frac{1}{\mid \vec r -\vec r' \mid}\\
\frac{\partial }{\partial z}\frac{\partial }{\partial x'}\frac{1}{\mid \vec r -\vec r' \mid}&\frac{\partial }{\partial z}\frac{\partial }{\partial y'}\frac{1}{\mid \vec r -\vec r' \mid}&\frac{\partial }{\partial z}\frac{\partial }{\partial z'}\frac{1}{\mid \vec r -\vec r' \mid}\end{pmatrix}
\begin{pmatrix}
M_x(\vec r')\\
M_y(\vec r')\\
M_z(\vec r')
\end{pmatrix}
\end{equation}
\noindent We insert this last equation into ${\cal H}_M[\vec M]$ and obtain an equivalent relation
\begin{eqnarray}
 {\cal H}_M[\vec M]&=& +\frac{\Omega}{4\pi}\cdot \frac{1}{a^3}\cdot \sum_{i,j=x,y,z} \int d^3x \int d^3x'\cdot n_i(\vec r)\cdot n_j(\vec r')\cdot \frac{\partial}{\partial i}\frac{\partial}{\partial j'}\frac{1}{\vert \vec r -\vec r' \vert}\nonumber\\
&-& \frac{\mu_0}{2}\cdot \int d^3x \cdot \vec M^2(\vec r)
\end{eqnarray}
The first term is the sum of 9 integrals. In our slab model, the magnetization does not depend on the $z$-coordinate, so
that the integrals over $z$ and $z'$ can be performed. Four of the integrals vanish, namely those that contain the derivative
according to $z$ or $z'$ only once\footnote{To be computed is an integral of the type
\begin{equation}
\int_0^d dz\int_0^d dz' \frac{\partial}{\partial z} \frac{1}{\sqrt{(\vec \rho-\vec \rho')^2 + (z-z')^2}}
\end{equation}
The integral over $z$ gives
\begin{equation}
\frac{1}{\sqrt{(\vec \rho-\vec \rho')^2 + (d-z')^2}}- \frac{1}{\sqrt{(\vec \rho-\vec \rho')^2 + (z')^2}}
\end{equation}
The integrals of these two functions over $z'$ cancel out exactly.
}.
This vanishing produces a split of the functional into two separate sectors: the vertical one and the in-plane one. The vertical sector writes
\begin{eqnarray}
{\cal H}_M[n_z]&=& +\frac{\Omega}{4\pi}\cdot \frac{1}{a^3}\cdot \int d^2\rho \int d^2\rho' \cdot n_z(\vec \rho)\cdot n_z(\vec \rho')\nonumber\\
&\cdot &\int_0^d dz\int_0^d dz' \frac{\partial}{\partial z}\frac{\partial}{\partial z'}\frac{1}{\sqrt{(\vec \rho-\vec \rho')^2 + (z-z')^2}}
\end{eqnarray}
The in-plane sector writes
\begin{eqnarray}
{\cal H}_M[n_x,n_y]&=& +\frac{\Omega}{4\pi}\cdot \frac{1}{a^3}\cdot \sum_{i,j=x,y} \int d^2\rho \int d^2\rho'\cdot n_i(\vec \rho)\cdot n_j(\vec \rho')\cdot\nonumber\\
&\cdot &  \frac{\partial}{\partial i}\frac{\partial}{\partial j'}\int_0^d dz\int_0^d dz'\frac{1}{\sqrt{(\vec \rho-\vec \rho')^2 + (z-z')^2}}
\end{eqnarray}
One can use the identity
\begin{equation}
\Big[\frac{\partial}{\partial x}\frac{\partial}{\partial x'}+\frac{\partial}{\partial y}\frac{\partial}{\partial y'}+\frac{\partial}{\partial z}\frac{\partial}{\partial z'}\Big]\frac{1}{\sqrt{(\vec r-\vec r')^2}}= 4\pi\delta(\vec r-\vec r')
\end{equation}
to rewrite the energy in the vertical sector as
\begin{eqnarray}
{\cal H}_M[n_z]&=& \Omega\cdot \frac{d}{a}\cdot \frac{1}{a^2}\cdot \int d^2\rho \cdot n^2_z(\vec \rho)\nonumber\\
&-&\frac{\Omega}{4\pi}\cdot \frac{1}{a^3}\int d^2\rho \int d^2\rho' n_z(\vec \rho)\cdot n_z(\vec \rho')\cdot \big[\frac{\partial}{\partial x}\frac{\partial}{\partial x'}+\frac{\partial}{\partial y}\frac{\partial}{\partial y'}\big]\nonumber\\
&&\int_0^d dz\int_0^d dz'\frac{1}{\sqrt{(\vec \rho-\vec \rho')^2 + (z-z')^2}}
\end{eqnarray}
This way of writing allows a local dipolar energy term
\begin{equation}
\Omega\cdot \frac{d}{a}\cdot \frac{1}{a^2}\cdot \int d^2\rho \cdot n^2_z(\vec \rho)
\end{equation}
to emerge. Formally, this term writes exactly as the functional arising from the single-atom N\'eel magnetic anisotropy, Eq.~\eqref{Eq:Hlambda}, the only remarkable difference being that it is always positive.\\
We provide now some more technical aspects about the kernel
\begin{equation}
\label{Eq:kernel}
\int_0^d dz\int_0^d dz'\frac{1}{\sqrt{(\vec \rho-\vec \rho')^2 + (z-z')^2}}
\end{equation}
appearing in all integrals for the non-local component of the dipolar interaction.\\
\textbf{A.} It can be computed exactly. Writing $\varrho\dot=\vert\vec
\rho-\vec\rho'\vert$ and substituting $u=z-z'$, the double integral reduces to
$2\int_0^d du\,(d-u)/\sqrt{\varrho^2+u^2}$ and gives
\begin{equation}
2\cdot d\cdot \ln\frac{d+\sqrt{d^2 + \varrho^2}}{\varrho} -2\cdot \sqrt{d^2 + \varrho^2} + 2\cdot \varrho
\end{equation}
This result allows the variables $z$ and $z'$ to be eliminated from the non-local terms, which become integrals over the in-plane variables $(x,y)$ only.\\
\textbf{B.}
A further possible useful representation of this kernel is obtained by inserting the identity
\begin{equation}
\frac{1}{\sqrt{(x-x')^2+(y-y')^2+(z-z')^2}}\!=\! \frac{1}{2\pi}\int d^2k\frac{e^{i\mathbf k\cdot(\mathbf x-\mathbf x')}\cdot e^{-k\cdot \vert z-z'\vert}}{k}
\end{equation}
and performing the integral
\begin{equation}
\int_0^d dz \int_0^d dz'e^{-k\cdot \vert z-z'\vert}= 2\cdot\Big(\frac{d}{k}+\frac{e^{-k\cdot d}-1}{k^2}\Big)
\end{equation}
so that the kernel writes
\begin{eqnarray}
\int_0^d\! dz\int_0^d\! dz'\frac{1}{\sqrt{(\vec \rho-\vec \rho')^2\! +\!(z-z')^2}}\!=\!\frac{1}{2\pi}\cdot \int d^2k\cdot \underbrace{\big[2\cdot(\frac{d}{k^2}+\frac{e^{-k\cdot d}-1}{k^3})\big]}_{\dot=f(k)}\cdot e^{i\vec k\cdot(\vec \rho-\vec \rho')}
\end{eqnarray}
Later, for simplicity of writing, we will use the kernel
\begin{equation}
G(\vec \rho-\vec \rho')\dot=\frac{1}{2\pi}\cdot \int d^2k\cdot f(k)\cdot e^{i\vec k\cdot(\vec \rho-\vec \rho')}
\end{equation}
within the dipolar functional.\\
\textbf{C.} We now give an estimate of the size of the non-local terms compared to the local one. Suppose we have an element with thickness $d<<L$, $L$ being the lateral size of this element. We would like to know how the various dipolar terms scale with $d$ and $L$. We fill the local term with a uniform
perpendicular spin distribution and ascertain that it scales as
$\frac{d}{a}\cdot (\frac{L}{a})^2$, i.e. as the thickness multiplied by the area
of the element. For the non-local terms, we use again a uniform spin
distribution. By redirecting the in-plane derivatives to the spin fields we
observe that the integrands are only non-vanishing at the boundaries of the
element. We assume that the boundary has a thickness $w>d$ and proceed to perform
an integral that resembles the one of Eq. 10 in the ``Supplemental Material''
to Ref.~\cite{Argentina}. This integral describes the self-energy of the effective
``charges'' appearing at the boundary of the element in virtue of the in-plane
derivatives of the spin fields. The evaluation of this integral (Eq. 14 of the SM to
Ref.~\cite{Argentina}) shows that it scales as
$(\frac{d}{a})^2\cdot \frac{L}{a}$, i.e. the relative strength of the non-local
terms to the local one is of the order of $\frac{d}{L}<<1$. This provides an
explanation of why we have taken only the local term
of the dipolar interaction into account when discussing the LL equation in ultrathin films with uniform magnetization distribution.

\chapter{Ferromagnetism of transition metal ultrathin films at finite temperatures: the Non Linear Sigma Model.}
\label{Chap:Mermin}
\renewcommand{\thesection}{\thechapter.\arabic{section}}
\renewcommand{\theequation}{\thechapter.\arabic{equation}}

Provided ferromagnetic order appears in the ground state of a spin system, the question arises about what happens to it when the temperature is ``switched on''. For answering this question in ultrathin films we inspect some general properties of the effective Hamiltonians we have constructed in Chapter 5.\\
\paragraph{The exchange functional.}
This functional
\begin{equation}
\label{Eq:HA_rot}
{\cal H}_A[\vec n]= A\cdot\!\int d^3x\!\cdot\!\Big[\!(\vec \nabla n_x)^2\!+\!(\vec \nabla n_y)^2\! +\! (\vec \nabla n_z)^2\Big]
= A\cdot\!\int d^3x\cdot \sum_{\nu}\sum_i \big(\partial_i n_\nu(\vec x)\big)^2
\end{equation}
describes the elastic (exchange) energy the system has to pay when the field $\vec n(\vec x)$ has some non-uniformities. The index $\nu=x,y,z$ labels the components of the spin field and the index $i$ the spatial coordinates. One symmetry of this Hamiltonian is essential to understand its finite temperature properties: the Hamiltonian does not change when \textbf{all} the spins
are rotated by the same amount. We prove now this statement.\\
\textbf{The statement.} Let $R$ be a rotation of the three-dimensional spin
space, i.e. a real $3\times 3$ matrix obeying $R^{T}R=\mathds 1$, or in
components
\begin{equation}
\label{Eq:orthogonal}
\sum_\mu R_{\mu\nu}\cdot R_{\mu\lambda}= \delta_{\nu\lambda}
\end{equation}
and let the rotated configuration be
\begin{equation}
\label{Eq:rotated}
n'_\mu(\vec x)= \sum_\nu R_{\mu\nu}\cdot n_\nu(\vec x)
\end{equation}
the \textbf{same} matrix $R$ acting at every point $\vec x$. We claim that
${\cal H}_A[\vec n\,']={\cal H}_A[\vec n]$.\\
\textbf{The rotated configuration is an admissible one.} The field $\vec n$ is
constrained to unit length, so we must first check that $\vec n\,'$ still is.
Using Eq.~\eqref{Eq:orthogonal},
\begin{equation}
\sum_\mu \big(n'_\mu\big)^2 = \sum_\mu\sum_{\nu,\lambda} R_{\mu\nu}R_{\mu\lambda}\,n_\nu n_\lambda= \sum_{\nu,\lambda}\delta_{\nu\lambda}\,n_\nu n_\lambda = \sum_\nu \big(n_\nu\big)^2 = 1
\end{equation}
\textbf{The integrand is invariant point by point.} Because $R$ does not depend
on $\vec x$, it commutes with the derivative:
\begin{equation}
\partial_i n'_\mu(\vec x)= \sum_\nu R_{\mu\nu}\cdot \partial_i n_\nu(\vec x)
\end{equation}
The quantity to be examined is therefore of exactly the same form as the one
just computed, with $n_\nu$ replaced by $\partial_i n_\nu$:
\begin{eqnarray}
\sum_\mu\sum_i\big(\partial_i n'_\mu\big)^2 &=& \sum_i\sum_{\nu,\lambda}\Big(\sum_\mu R_{\mu\nu}R_{\mu\lambda}\Big)\big(\partial_i n_\nu\big)\big(\partial_i n_\lambda\big)\nonumber\\
&=&\sum_i\sum_{\nu,\lambda}\delta_{\nu\lambda}\big(\partial_i n_\nu\big)\big(\partial_i n_\lambda\big)= \sum_\nu\sum_i \big(\partial_i n_\nu\big)^2
\end{eqnarray}
The integrand of Eq.~\eqref{Eq:HA_rot} is thus not merely equal after
integration: it is unchanged at every single point $\vec x$. Integrating,
${\cal H}_A[\vec n\,']={\cal H}_A[\vec n]$, which is the claim. Notice that only
the orthogonality~\eqref{Eq:orthogonal} was used, so that the invariance holds
for the full group $O(3)$, reflections included; its continuous part, the group
$SO(3)$ of proper rotations, is the one that matters in what follows.\\
The second symmetry emerges when ${\cal H}_A$ is computed for the slab model. As the spin configuration is unchanged along the vertical direction, we can integrate ${\cal H}_A$ along $z$ from $0$ to $d$. The result is the exchange Hamiltonian for the slab model of ultrathin films:
\begin{equation}
\label{Eq:HAd}
{\cal H}_{Ad}= A\cdot\!d\!\cdot\int d^2x\!\cdot\Big[\!\big(\vec \nabla n_x(\vec \rho)\big)^2\!+\!\big(\vec \nabla n_y(\vec \rho)\big)^2\! +\big(\vec \nabla n_z(\vec \rho)\big)^2\!\Big]
\end{equation}
The integral in ${\cal H}_{Ad}$ runs over the coordinates $\vec \rho=(x,y)$ and the gradients are two-dimensional gradients as well. In other words, ${\cal H}_{Ad}$ is effectively a two-dimensional Hamiltonian.\\
These two properties -- isotropy and two-dimensionality -- drive the slab model into the domain covered by a famous theorem of mathematical physics: the Mermin-Wagner theorem\cite{Mermin}. This theorem states that long range order is impossible in two-dimensional systems with isotropic interaction, in the thermodynamic limit of infinitely extended systems and at finite temperatures. The content of this theorem was actually predicted long before the proof by Mermin and Wagner, in a paper by Landau\cite{Landau_29}. One of the aims of this Chapter is to ``prove'' this theorem with heuristic arguments, to render it accessible to a student with a standard knowledge of statistical mechanics.\\
\paragraph{The functionals for $\lambda,\Omega$ and $\vec B$.}
Owing to the fact that the spin configuration is independent of $z$, one can define functionals of these interactions for the slab model as well:
\begin{eqnarray}
\label{Eq:slabfun}
{\cal H}_{\lambda}&=& -\lambda\cdot \frac{1}{a^2}\cdot\int d^2x \,n_z^2(\vec \rho)\nonumber\\
{\cal H}_{B}&=& g\,\mu_B\,S\,\frac{d}{a}\cdot\frac{1}{a^2}\cdot \vec B\cdot \int d^2x \,\vec n(\vec \rho)
\nonumber\\
{\cal H}_{\Omega}&\approx & + \Omega\cdot \frac{d}{a}\cdot\frac{1}{a^2}\,\int d^2x \,n_z^2(\vec \rho)
\end{eqnarray}
These interactions ``live'' on an effectively two-dimensional system but \textbf{none} of them shares the rotational invariance of the exchange functional:
the N\'eel anisotropy $-\lambda n_z^2$ and the local dipolar term
$\Omega n_z^2$ single out an axis, the Zeeman term $-g\mu_B S\vec B\cdot \vec n$
singles out a direction. They break the symmetry~\eqref{Eq:rotated} explicitly,
and this is why -- however small they are -- they will be able to restore what
the exchange interaction alone cannot sustain\cite{Poki}. It is a further aim of this Chapter to give heuristic arguments for the restoring of long range order by minute symmetry breaking interactions.

\section{The non-linear sigma model of the two-dimensional Heisenberg ferromagnet and the absence of spontaneous magnetization.}
\subsection{The NLSM.}
The finite temperature properties of any system are described by the
partition function. We have in mind, for the sake of illustration, the classical Heisenberg model on a lattice and write the partition function as
\begin{equation}
{\cal Z}(J,\beta)=\sum_{\{\vec S_i,\vec S_j@\vert S_i\vert^2=1\}} e^{-\beta\cdot \big[-\frac{J}{2}\sum_{<i,j>}\vec S_i\cdot \vec S_j\big]}
\end{equation}
$\beta\doteq\frac{1}{k_B\cdot T}$. The partition function involves the sum over all possible configurations $(\vec
S_i,\vec S_j)$ (the ``phase space'' of the spin degree of freedom). On the lattice, however, it has not been computed exactly yet\footnote{It is, however, a good starting point of an approximate method of computation known as ``Monte Carlo'' simulation\cite{Monte}.}. The sum over a discrete distribution is difficult to perform analytically and the constraint $\vert S_{i,j}\vert^2=1$ introduces a non-linearity that makes the problem even more difficult. Under these circumstances, one prefers to translate the Heisenberg Hamiltonian onto a
continuum of sites and with the assumption that the lattice constant is small.
This leads us back to the classical Heisenberg functional over the field $\vec
n(\vec x)$, ${\cal H}_A[\vec n]$ and, referring to the slab, ${\cal H}_{Ad}[\vec n]$. The variables $\vec n(\vec x)$ are vectors with real components, subject to the constraint that $\vec n^2(\vec x)=1$.
There are an infinite number of such configurations and they are not countable,
so that, in practice, the ``sum'' over such configurations needs to be defined
and performed in a proper way. The proper way is called ``functional integration''. It was
suggested in 1958 by L.D. Landau (see Ref.~\cite{Landau_V_1}, p.478-480) as a technology relevant for performing the ``sum'' over non-countable degrees of freedom. It is better illustrated by carrying out some practical examples. We will illustrate it on the task of computing the partition function for ${\cal H}_{Ad}[\vec n]$. This example has also an added value, because the sum over all $\vec n$ configurations must include the constraint that $\vec n$ is a vector of length ``1''. There are various approaches to how this constraint is taken into account. One of these approaches is the so-called Non Linear Sigma Model (NLSM)\cite{Dupuis}. A further one is the Renormalization Group approach by A.M. Polyakov\cite{Pol}. The first approach is, mathematically speaking, the subject of this Section. The Polyakov approach will be the subject of the next Chapter.\\
For the purpose of defining the NLSM, we continue, for the sake of simplicity, thinking of $\vec x$ as being restricted to a discrete lattice and write, formally,
\begin{equation}
{\cal Z}(Ad,\beta)\dot= \int D'[\vec n(\vec x)]\cdot  e^{-\beta\cdot {\cal H}_{Ad}[\vec n(\vec x)]}
\end{equation}
The symbol $\int D' [\vec n(\vec x)]$ means ``functional integration'' and replaces the sum over configurations on a discrete lattice with the so called functional integral over the various continuous distributions
$[\vec n(\vec x)]$. The volume element $D'[\vec n(\vec x)]$ amounts to
\begin{equation}
D'[\vec  n(\vec x)]\equiv \Pi_{\vec x_i} d'n_x(\vec x_i)\cdot d'n_y(\vec x_i)\cdot d'n_z(\vec x_i)
\end{equation}
The ``prime'' attached to the symbol ``$D$'' reminds us that only those configurations are allowed for which $\vec n^2(\vec x_i)=1$. One possible way of implementing this constraint leads to the NLSM effective functional and consists in replacing $n_z(\vec
x_i)$ with $\sqrt{1-n_x^2-n_y^2}$ in both the exponent and the volume element.
Because of the condition $n_x^2+n_y^2+n_z^2 = 1$, the volume element
$d'n_x(\vec x)\cdot d'n_y(\vec x)\cdot d'n_z(\vec x)$ is the surface element of a unit sphere, i.e.
\begin{eqnarray}
d'n_x(\vec x_i)\cdot d'n_y(\vec x_i)\cdot d'n_z(\vec x_i)= \frac{dn_x(\vec x_i)\cdot dn_y(\vec x_i)}{\cos\vartheta} = \frac{dn_x(\vec x_i)\cdot dn_y(\vec x_i)}{\sqrt{1-n^2_x(\vec x_i)-n^2_y(\vec x_i)}}
\end{eqnarray}
$\vartheta$ being the angle between $\vec n$ and the $z$-axis, so that $\cos\vartheta=n_z$. We use an identity to rewrite the term coming from the measure:
\begin{equation}
\frac{1}{\Pi_{\vec x_i} \sqrt{1-n^2_x(\vec x_i)-n^2_y(\vec x_i)}}= e^{\left(-\ln \Pi_{\vec x_i}\sqrt{1-n^2_x(\vec x_i)-n^2_y(\vec x_i)}\right)}
\end{equation}
and write the partition function formally as
\begin{eqnarray}
{\cal Z}(Ad,\beta)=\int \Pi_{\vec x_i} dn_x(\vec x_i)\cdot dn_y(\vec x_i)\cdot e^{\left(-\beta {\cal L}[\vec n(\vec x_i)]\right)}
\end{eqnarray}
The generalization of ${\cal L}[\vec n(\vec x_i)]$ to the space of continuous variables $\vec x$ is an effective Hamiltonian (often called ``Landau free energy functional'' or ``Lagrangian'') that maps the Heisenberg Hamiltonian, including the constraint $\vec n^2=1$, into the so called \textbf{Non-Linear-Sigma-Model} (NLSM) effective Hamiltonian
\begin{eqnarray}
\label{Eq:NLSM}
{\cal L}_{Ad,\beta}[\vec n]&=& A\cdot\!d\!\cdot\int d^2x\!\cdot\Big[\!(\vec \nabla n_x)^2\!+\!(\vec \nabla n_y)^2\! +\! \left(\vec \nabla \sqrt{1-n_x^2-n_y^2}\right)^2\Big]\!\nonumber\\
&+&\frac{1}{2\beta}\cdot \frac{d}{a}\cdot \int \frac{d^2x}{a^2}\cdot \ln (1-n_x^2-n_y^2)
\end{eqnarray}
One uses\cite{Dupuis} the letter ``${\cal L}$'' rather than
``${\cal H}$'' to signify that the functional ``${\cal L}$'' can contain elements which are not obvious in ``${\cal H}$'', such as a temperature dependent component (in this specific case originating from the measure). Regarding the thermodynamic properties of the system, the partition function computed over the effective NLSM Lagrangian ${\cal L}_{Ad,\beta}[\vec n]$ should be entirely equivalent to the partition function computed over the original Heisenberg ferromagnet (see Ref.~\cite{Dupuis} for an accurate discussion of this point).

\subsection{The absence of spontaneous magnetization.}
\label{Sec:nomagn}
We continue the discussion and use the NLSM Lagrangian to ``prove'' the absence of spontaneous magnetization in a slab described by the NLSM effective Lagrangian. In fact, it turns out that
the absence of spontaneous magnetization is already apparent within the \textbf{harmonic} approximation of the NLSM. This approximation assumes only spin configurations with small deviations from the ground state. Assume that the ground state is a ferromagnetic one with spins pointing e.g. along $z$, i.e. $n_x=n_y=0$\footnote{Owing to the isotropic nature of this model, $z$ need not be along the normal of the two-dimensional system.}; this means that $n_x,n_y$ should be considered to be small and only the lowest powers of $n_x,n_y$ are taken into account in ${\cal L}_{Ad,\beta}$. The harmonic approximant of ${\cal L}_{Ad,\beta}$ writes, accordingly,
\begin{equation}
\label{Eq:Lharm}
{\cal L}_{Ad,\text{harm}}= A\cdot\!d\!\cdot \int d^2x\!\cdot\!\Big[\!(\vec \nabla n_x)^2\!+\!(\vec \nabla n_y)^2\Big]
\end{equation}
The next step is to define the spontaneous spin polarization $\cal P$. A suitable definition is
\begin{equation}
{\cal P} \dot= <n_z>_{{\cal L}_{Ad,\text{harm}}}\doteq<\cos\theta>_{{\cal L}_{Ad,\text{harm}}}
\end{equation}
$\theta$ being the angle between the vector $\vec n$ and the $z$-axis. The symbol
\begin{equation}
<...>_{{\cal L}_{Ad,\text{harm}}}
\end{equation}
denotes the operation of a statistical average. It involves ``functional integration'', and its explicit computation is therefore suitable to illustrate this technology, see Appendix~\ref{App:Gauss}. One can translate this definition
to the fields $n_x$ and $n_y$:
\begin{eqnarray}
{\cal P}&=&<\sqrt{1-n_x^2-n_y^2}>_{{\cal L}_{Ad,\text{harm}}}
\approx 1-\frac{1}{2}\Big(<n_x^2>_{{\cal L}_{Ad,\text{harm}}}+<n_y^2>_{{\cal L}_{Ad,\text{harm}}}\Big)
\end{eqnarray}
Computing explicitly $<n_x^2>_{{\cal L}_{Ad,\text{harm}}}$ and $<n_y^2>_{{\cal L}_{Ad,\text{harm}}}$ (see Appendix~\ref{App:App1}) we obtain a most remarkable result
\begin{eqnarray}
\label{Eq:P2D}
{\cal P} \approx 1-\frac{1}{4\pi\cdot\beta\cdot A\cdot d}\cdot \ln\frac{L}{a}
\end{eqnarray}
with $L$ being the lateral size of the slab. Formally, ${\cal P}$ vanishes when the film reaches a lateral size $L_c$,
\begin{equation}
\label{Eq:L_c}
L_c\doteq a\cdot e^{4\pi\cdot \beta\cdot A\cdot d}
\end{equation}
Our interpretation of this result is the following. We assign to $L_c$ the
meaning of a cross-over length: regions of the film with spatial dimensions
well below $L_c$ have a spontaneous magnetization. When spatial lengths well
above $L_c$ are sampled, they show no spontaneous magnetization. This is in
line with the Mermin-Wagner theorem, which sets the spontaneous magnetization to
vanish in the limit $L\rightarrow \infty$. The result of Eq.~\eqref{Eq:L_c} is
more telling, in the sense that it organizes the two-dimensional NLSM into spin blocks of highly correlated spins. Notice that, when the length scale
$L_c$ is approached, $n_x$ and $n_y$ are no longer ``small'' and further terms
of the Landau functional must be taken into account for estimating the decay of
the spontaneous magnetization with a better precision. A scenario where the
spontaneous magnetization approaches zero smoothly in the limit $L\rightarrow
\infty$ is the most likely result. We point out that one has not succeeded in
computing the ``true'' vanishing behaviour of the spontaneous magnetization so
far.\\
A special limit of the two-dimensional NLSM is more tractable. It appears when one of the in-plane components is set identically to zero (say, $n_y$). This limiting situation is known as the 2D \textbf{planar} model, where the spins only have two components, here $n_z$ and $n_x$. In this situation, one can compute ${\cal P}$ more accurately\cite{Poki} (see Appendix~\ref{App:App2}), and find
\begin{equation}
\label{Eq:PXZ}
{\cal P}\approx \Big(\frac{L}{a}\Big)^{-\frac{1}{8\pi\cdot \beta\cdot A\cdot d}}
\end{equation}
In the 2D planar model, the spontaneous magnetization is found to vanish,
smoothly, in the thermodynamic limit $L\rightarrow\infty$, following a power law. It was one of the most remarkable discoveries in condensed matter of the past 40 years that the isotropic, 2D planar model contains topological excitations such as vortices and anti-vortices\cite{Kost} that produce a special type of phase transition at a finite temperature (the so called Berezinskii-Kosterlitz-Thouless temperature), consisting in the unbinding of vortex-antivortex pairs. The vanishing of the spontaneous magnetization is, however, not affected by this transition.

\section{Symmetry breaking interactions and the spontaneous magnetization in the two-dimensional NLSM.}
The experimental observation of a spontaneous magnetization, for both in-plane\cite{Vaterlaus} and perpendicular orientation (see Figs.~7, 11, 12, 13 of Ref.~\cite{Michele}
for results on Fe on Cu(100) and Fe on Ag(100)), suggests that, in practical
ultrathin films, the NLSM, as discussed in the previous
Section, needs to be modified in order to host the observed macroscopic
ferromagnetic behaviour. A possible route to install a spontaneous
magnetization is to allow for symmetry breaking interactions.
The simplest functional containing symmetry breaking interactions in ultrathin films writes
\begin{equation}
\label{Eq:Lsb}
{\cal L}_\lambda[n_x,n_y]\doteq\big(\lambda-\Omega\cdot \frac{d}{a}\big)\int \frac{d^2x}{a^2}\cdot \big(n_x^2(\vec x)+ n_y^2(\vec x)\big)
\end{equation}
It is the sum of the first and the third of the functionals~\eqref{Eq:slabfun},
rewritten with the help of $n_z^2=1-n_x^2-n_y^2$ and with the constant term
dropped. For the sake of making an example, we let this term favor the perpendicular $z$ direction by setting $\big(\lambda-\Omega\frac{d}{a}\big)>0$.
The adding of a symmetry breaking interaction profoundly modifies the result for the spontaneous spin polarization, which writes (Appendix~\ref{App:App3})
\begin{equation}
\label{Eq:Psb}
{\cal P} \approx 1-\frac{1}{4\pi\cdot\beta\cdot A\cdot d}\cdot \ln\sqrt{\frac{4\pi\cdot A\cdot d}{\big(\lambda-\Omega\cdot \frac{d}{a}\big)}}
\end{equation}
The symmetry breaking interactions introduce a non-vanishing spontaneous magnetization at low enough temperatures. Notice that the logarithm of the system size $L$ of Eq.~\eqref{Eq:P2D} has been replaced by the logarithm of a quantity that does not depend on $L$ at all. If one pushes the spin wave approximation up to large enough temperatures, the equation ${\cal P}(\beta)=0$ introduces a characteristic temperature at which the spontaneous magnetization, sustained by the symmetry breaking interactions, vanishes:
\begin{equation}
\frac{1}{\beta_c} = \frac{4\pi\cdot A\cdot d}{\ln\sqrt{\frac{4\pi\cdot A\cdot d}{\big(\lambda-\Omega\cdot \frac{d}{a}\big)}}}
\end{equation}
Most remarkably, except for the $\ln$ correction, this characteristic temperature is of the order of $A\cdot d$, i.e. it is mainly determined by the interatomic exchange interaction. This is a notion that we are comfortable with from the Mean Field Approximation of the Heisenberg Hamiltonian, which (as it is well known) predicts the Curie temperature to be of the order of the interatomic exchange for any system, independently of its dimensionality\cite{AM}.

\setcounter{section}{0} 
\renewcommand{\thesection}{\thechapter.\Alph{section}} 
\setcounter{equation}{0}
\renewcommand{\theequation}{\thesection\arabic{equation}}
\section*{Appendices}

\section{The statistical average for a Gaussian distributed field.}
\label{App:Gauss}
\setcounter{equation}{0}
The situation is one where the system is described by an order parameter field $m(\vec x)$. For mathematical simplicity, we assume that the field is a scalar one and we let the vector $\vec x$ span a volume $L^D$ in a $D$-dimensional space.
The computation of the statistical average is often performed in Fourier space, so that we introduce, at this point, some concepts from the theory of Fourier series that we need for practical computations.
We assume that the field can be expressed as a Fourier sum
\begin{equation}
m(\vec  x)= \frac{1}{\sqrt{N^D}}\cdot \sum_{\vec  k} m_{\vec k}\cdot e^{i\cdot \vec  k\cdot \vec  x}
\end{equation}
$N^D$ is the number of unit cells in the system, with the relation
$L^D=N^D\cdot a^D$, $a$ being the lattice constant of a hypercubic lattice, and
$\vec k$ the reciprocal lattice vectors
$\vec  k = \frac{2\pi}{N\cdot a}\cdot (n_1,n_2,...,n_D)$,
$n_i\in {\cal Z}$.
The values of $\vec k$ are assumed to occupy the first Brillouin zone and, because of periodic boundary conditions, they are distributed on a fine-meshed lattice in a (hyper)cube with volume $(\frac{2\pi}{a})^D$
\footnote{We need some mathematical relations of Fourier expansions.\\
-- The orthonormality relation:
\[\int_V d^Dx\cdot e^{i\vec k\cdot x}\cdot e^{i\vec k'\cdot x} = N^D\cdot a^D\cdot \delta_{\vec k,-\vec k'}\]
-- The inverse Fourier transform:
\[m_{\vec k} = \frac{1}{\sqrt{N^D}}\cdot \frac{1}{a^D}\int_V d^Dx\, m(\vec x) \cdot e^{-i\cdot \vec k\cdot \vec x}\]
-- The second orthonormality relation:
\[\sum_{\vec k} e^{i\vec k\cdot x}\cdot e^{-i\vec k\cdot x'} = N^D\cdot a^D \cdot \delta(\vec x-\vec x')\]
-- When the sums over $\vec k$ are transformed into integrals (it is often easier to perform ``integrals'' than ``sums'') the translation key is
\[\sum_{\vec k} \rightarrow \frac{N^D\cdot a^D}{(2\pi)^D}\int d^D k\]
(the prefactor takes into account the density of $\vec k$-vectors). The second orthonormality relation then reads, accordingly,
\[\frac{1}{(2\pi)^D}\int d^Dk\cdot e^{i\vec k\cdot x}\cdot e^{-i\vec k\cdot x'} =\delta(\vec x-\vec x')\]
which is the Fourier transform of the Dirac delta function as known in the conventional literature.}.
The Fourier transformed field is specified on the fine-meshed set within the first Brillouin zone. In virtue of the choice we have made for the Fourier sum, the field $m_{\vec k}$ has complex values, i.e.
\begin{equation}
m_{\vec k} = m'_{\vec k}+\imath\cdot m''_{\vec k}
\end{equation}
We require that the field $m(\vec x)$ is real, so that $m_{-\vec k}$ is not an independent degree of freedom, as
\begin{equation}
m_{-\vec k} = \overline {m_{\vec k}}
\end{equation}
The Lagrangian (or \textbf{Landau functional}) ${\cal L}[m(\vec x)]$ of the field is the integral over the volume of some Lagrange density that contains the various interactions and, possibly, further terms arising from the entropy. Formally, inserting the Fourier transform of the field into the Lagrangian, one produces the Fourier transform of the Lagrangian, which associates to a distribution $[m_{\vec k}]$ a scalar ${\cal L}[m_{\vec k}]$. According to Landau\cite{Landau_32,Poki} the probability that the distribution $[m_{\vec k}]$ is realized in the system at equilibrium is given by
\begin{equation}
W[m_{\vec k}]= \frac{e^{-\beta\cdot {\cal L}[m_{\vec k}]}}{{\cal Z}(\beta)}
\end{equation}
where
\begin{equation}
{\cal Z}(\beta)=\sum_{[m_{\vec k}]}e^{-\beta\cdot {\cal L}[m_{\vec k}]}
\end{equation}
is the partition function of the system. This is the key principle for computations in equilibrium statistical mechanics. $[m_{\vec k}]$ builds an infinite, non-countable set, so that one has to specify how to ``sum'' over all states. This operation is, technically, known as functional integration:
\begin{equation}
\sum_{[m_{\vec k}]}\rightarrow \int D[m_{\vec k}]\doteq \int \Pi_{\vec k}dm'_{\vec k}\cdot dm''_{\vec k}
\end{equation}
We proceed now to apply this principle to a Lagrangian that often results in statistical physics:
\begin{equation}
\label{Eq:Lgauss}
{\cal L}[m_{\vec k}] = \sum_{\vec k}\epsilon(k)\cdot \Big[m'^2_{\vec k}+ m''^2_{\vec k}\Big]
\end{equation}
$\epsilon(k)$ is a scalar function of $k\doteq \vert\vec k\vert$ and is called the \textbf{spectrum of fluctuations}. It contains the various coupling constants that characterize the interactions. The Lagrangian depends quadratically on the fields $m'_{\vec k}$ and $m''_{\vec k}$: the sets $[m'_{\vec k}]$ and $[m''_{\vec k}]$ are said to be \textbf{Gaussian distributed fields}\footnote{As the $m_{\vec k}$ are Gaussian distributed, so is any linear combination of them, such as the real space field $m(\vec x)$\cite{Stat}.}.
As an example, we compute an important relation for Gaussian distributed fields:
\begin{equation}
\label{Eq:mkmh}
<m_{\vec k}\cdot \overline{m_{\vec h}}>_{{\cal L}[m_{\vec k}]}= \delta_{\vec h,-\vec k}\cdot \frac{1}{\beta\cdot \epsilon(k)}
\end{equation}
$<...>_{{\cal L}[m_{\vec k}]}$ is the statistical average over the Lagrangian and corresponds to the thermodynamic expectation value (for the remainder of this Appendix we will write simply $<...>$). We show this relation step by step. The initial step is the computation of $<m'^2_{\vec k}>$:
\begin{eqnarray}
<m'^2_{\vec k}> = \frac{\int D[m_{\vec q}]\cdot m'^2_{\vec k}\cdot e^{-\beta\cdot {\cal L}[m_{\vec q}]}}{\int D[m_{\vec q}]\cdot e^{-\beta\cdot {\cal L}[m_{\vec q}]}}
\end{eqnarray}
The exponential function in the integrand writes
\begin{equation}
e^{-\beta\cdot {\cal L}[m_{\vec q}]}=\Pi_{\vec q}e^{-\beta\cdot \epsilon(q)\cdot m'^2_{\vec q}}\cdot \Pi_{\vec q}e^{-\beta\epsilon(q)\cdot m''^2_{\vec q}}
\end{equation}
This factorization allows one to write
\begin{eqnarray}
<m'^2_{\vec k}>\!=\!\frac{\int dm'_{\vec k}\cdot m'^2_{\vec k}\cdot e^{-\beta\cdot \epsilon(k)\cdot m'^2_{\vec k}}}{\int d m'_{\vec k}\cdot e^{-\beta\cdot \epsilon(k)\cdot m'^2_{\vec k}}}\!\cdot\!\underbrace{\Pi_{\vec q\not=\vec k}\frac{\int dm'_{\vec q}\cdot e^{-\beta\cdot \epsilon(q)\cdot m'^2_{\vec q}}}{\int d m'_{\vec q}\cdot e^{-\beta\cdot \epsilon(q)\cdot m'^2_{\vec q}}}}_{1}
\!\cdot\!\underbrace{\Pi_{\vec q}\frac{\int dm''_{\vec q}\cdot e^{-\beta\cdot \epsilon(q)\cdot m''^2_{\vec q}}}{\int d m''_{\vec q}\cdot e^{-\beta\cdot \epsilon(q)\cdot m''^2_{\vec q}}}}_{1}\nonumber\\
\end{eqnarray}
The Gaussian integrals are rendered analytically computable by extending the integration limits to infinity. This introduces an unphysical range for $m_{\vec k}$ but the error that arises is negligible. The evaluation of Gaussian integrals is given e.g. in Ref.~\cite{Wein}:
\begin{eqnarray}
\int_{-\infty}^\infty dx\cdot x^n\cdot e^{-a\cdot x^2} &=&2\cdot
\frac{(n-1)!!}{2^{\frac{n}{2}+1}\cdot a^{\frac{n}{2}}}\cdot
 \sqrt{\frac{\pi}{a}}
\quad n\ \text{even}, n>0
\nonumber\\
&=& \sqrt{\frac{\pi}{a}}\quad n=0\nonumber\\
&=& 0\quad n\ \text{odd}
\end{eqnarray}
$(n-1)!!\doteq (n-1)(n-3)\cdots 1$ being the double factorial. We obtain therefore
\begin{equation}
<m'^2_{\vec k}>= \frac{1}{2}\cdot \frac{1}{\beta\cdot \epsilon(k)}
\end{equation}
In a similar way we obtain, for the imaginary component,
\begin{equation}
<m''^2_{\vec k}>= \frac{1}{2}\cdot \frac{1}{\beta\cdot \epsilon(k)}
\end{equation}
so that we have shown that
\begin{equation}
<m_{\vec k}\cdot \overline{m_{\vec k}}>_{{\cal L}[m_{\vec k}]}= \frac{1}{\beta\cdot \epsilon(k)}
\end{equation}
for a scalar, complex field $m_{\vec k}$. We notice that the computation of the expression $<m_{\vec k}\cdot m_{\vec h}>$ with $\vec h\not=-\vec k$
would involve Gaussian integrals with odd $n$ and would therefore vanish. We have therefore demonstrated the original relation~\eqref{Eq:mkmh} completely.\\
We use this relation for computing $<m^2(\vec x)>$, in order to make contact with Sec.~\ref{Sec:nomagn}:
\begin{eqnarray}
<m(\vec x)^2>&=&\frac{a^D}{L^D}\sum_{\vec k,\vec h}e^{i(\vec k+\vec h)\cdot \vec x}\cdot <m_{\vec k}\cdot m_{\vec h}>=\frac{a^D}{L^D}\sum_{\vec k}\frac{1}{\beta\epsilon(k)}
\end{eqnarray}
The sum over $\vec k$ is transformed into an integral over the first Brillouin zone by the transition
\begin{equation}
\sum_{\vec k}\longrightarrow \frac{L^D}{(2\pi)^D}\int d^Dk
\end{equation}
Recalling that the fields at $-\vec k$ are not independent degrees of freedom, we only need to integrate over half the Brillouin zone. We, however, continue integrating over the full Brillouin zone but take this restraint into account by using
\begin{equation}
<m_{\vec k}\cdot \overline{m_{\vec h}}>_{{\cal L}[m_{\vec k}]}= \delta_{\vec h,-\vec k}\cdot \frac{1}{2\cdot \beta\cdot \epsilon(k)}
\end{equation}
We obtain
\begin{equation}
<m^2(\vec x)>=\frac{a^D}{(2\pi)^D}\int_{\text{1st B.Z.}}d^Dk\cdot \frac{1}{2\cdot \beta\cdot \epsilon(k)}
\end{equation}
It is also customary, for quantitative computations, to replace the hypercubic first Brillouin zone with a $D$-dimensional sphere of the same volume. The volume element in spherical coordinates writes
\begin{equation}
d^Dk = k^{D-1}dk\cdot d\Omega_{D-1}
\end{equation}
where $d\Omega_{D-1}$ is the differential solid angle. As the spectrum of fluctuations only depends on $k$, one can perform the integral of the angular part, $\int d\Omega_{D-1}$, which yields the surface area of a $D$-dimensional unit sphere:
\begin{equation}
\frac{2\cdot \pi^{\frac{D}{2}}}{\Gamma(\frac{D}{2})}
\end{equation}
The radius of the spherical Brillouin zone amounts to
\begin{equation}
k_{BZ}= \frac{2\pi}{a}\cdot \Big[\frac{D\cdot \Gamma(\frac{D}{2})}{2\cdot \pi^{\frac{D}{2}}}\Big]^{\frac{1}{D}}
\end{equation}
which in two dimensions is $k_{BZ}=2\sqrt{\pi}/a$, the value used in the Appendices that follow. Accordingly,
\begin{equation}
<m^2(\vec x)>=\frac{a^D}{(2\pi)^D}\cdot \frac{2\cdot \pi^{\frac{D}{2}}}{\Gamma(\frac{D}{2})}\cdot \int_{\text{Spherical BZ}} k^{D-1}\cdot dk\cdot \frac{1}{2\cdot \beta\cdot \epsilon(k)}
\label{Eq:msquare}
\end{equation}

\section{Application 1: the harmonic NLSM.}
\label{App:App1}
\setcounter{equation}{0}
The Lagrangian is the harmonic functional of Eq.~\eqref{Eq:Lharm},
\begin{equation}
{\cal L}[n_x,n_y] = A\cdot\!d\!\cdot \int d^2x\!\cdot\!\Big[\!(\vec \nabla n_x)^2\!+\!(\vec \nabla n_y)^2\Big]
\end{equation}
The Fourier transform of this Lagrangian writes
\begin{eqnarray}
{\cal L}[n^x_{\vec k},n^y_{\vec k}] &=& \frac{a^2}{L^2}\cdot A\cdot d\cdot (\imath )^2\cdot \sum_{\vec k,\vec h} \vec k\cdot\vec h \cdot \big[n^x_{\vec k}\cdot n^x_{\vec h} + n^y_{\vec k}\cdot n^y_{\vec h}\big]\cdot\underbrace{\int d^2x\, e^{i\cdot \vec k\cdot \vec x}\cdot e^{i\cdot \vec h\cdot \vec x}}_{\delta_{\vec k,-\vec h}}\nonumber\\
&=&\sum_{\vec k}\big(A\cdot d\cdot a^2\cdot k^2\big)\cdot \big[\vert n^x_{\vec k}\vert^2 + \vert n^y_{\vec k}\vert^2\big]
\end{eqnarray}
i.e. the spectrum of fluctuations of each of the two in-plane components is $\epsilon(k)=A\cdot d\cdot a^2\cdot k^2$ and, accordingly, referring to Eq.~\eqref{Eq:msquare} with $D=2$,
\begin{equation}
<n_x^2(\vec x)>=\int_0^{\frac{2\sqrt{\pi}}{a}}dk\cdot \frac{1}{4\pi\cdot \beta\cdot A\cdot d\cdot k}
\end{equation}
The integral on the right hand side is divergent at the lower integration limit (one speaks of an ``infrared divergence''). We therefore need a suitable strategy to manage this divergence. The usual strategy is to allow only spin waves with $k\geq\frac{2\sqrt{\pi}}{L}$, i.e. those whose wavelength fits into a system of lateral size $L$. The resulting statistical averages then give the physical properties of the system on a scale $L$. The thermodynamic limit $L\rightarrow \infty$ cannot be explored by this strategy, but only speculated at. With this restriction we obtain, for instance,
\begin{equation}
<n_x^2(\vec x)>=\int_{\frac{2\sqrt{\pi}}{L}}^{\frac{2\sqrt{\pi}}{a}}dk\cdot \frac{1}{4\pi\cdot\beta\cdot A\cdot d\cdot k}= \frac{1}{4\pi\cdot\beta\cdot A\cdot d}\cdot \ln\frac{L}{a}
\end{equation}
and the same for $<n_y^2>$, which inserted into the definition of ${\cal P}$ gives Eq.~\eqref{Eq:P2D}.

\section{Application 2: the XZ-model in the harmonic approximation.}
\label{App:App2}
\setcounter{equation}{0}
When the $n_y$-component in the harmonic Lagrangian of the NLSM is set to vanish exactly, one obtains the Lagrangian of the non-linear XZ-model. This Lagrangian can be written using the polar coordinate representation
$n_x(\vec x)= \sin \theta(\vec x)$, $n_z(\vec x)=\cos\theta(\vec x)$. The two gradient terms of the NLSM then combine into a single one,
\begin{eqnarray}
{\cal L}[\theta] &=&A\cdot d\cdot \int d^2x \cdot \Big[\big(\vec \nabla \sin\theta\big)^2+\big(\vec \nabla \cos\theta\big)^2\Big]\nonumber\\
&=& A\cdot d\cdot \int d^2x \cdot \big(\cos^2\theta+\sin^2\theta\big)\cdot(\vec \nabla\theta)^2 = A\cdot d\cdot \int d^2x \cdot (\vec \nabla\theta)^2\nonumber\\
&=& \sum_{\vec k}\big[A\cdot d\cdot a^2\cdot k^2\big]\cdot \vert \theta_{\vec k}\vert^2
\end{eqnarray}
The field $\theta_{\vec k}$ is, accordingly, Gaussian distributed, with the same spectrum of fluctuations as in Appendix~\ref{App:App1}, so that $<\theta^2(\vec x)>=\frac{1}{4\pi\beta A d}\ln\frac{L}{a}$. The one approximation still contained in the word ``harmonic'' is now a different and more interesting one: $\theta$ is an angle, defined modulo $2\pi$, and treating it as a variable running over the whole real axis discards precisely those configurations in which $\theta$ winds by a multiple of $2\pi$ around a point -- the vortices and antivortices met in Sec.~\ref{Sec:nomagn}.
For computing the spontaneous spin polarization in this model we rewrite
\begin{equation}
{\cal P}= <\cos\theta>
\end{equation}
as
\begin{equation}
{\cal P}=<e^{i\theta}>
\end{equation}
by adding a term odd in $\theta$ and therefore with vanishing statistical average. We then compute $<e^{i\cdot \theta}>$ using a remarkable theorem on Gaussian distributed quantities $n_i$\footnote{See e.g. Ref.~\cite{Landau_V_1}, p.~338.}
\begin{equation}
<e^{\sum_i\alpha_i\cdot n_i}>= e^{\frac{1}{2}\sum_{i,j}\alpha_i\cdot \alpha_j\cdot <n_i\cdot n_j>}
\end{equation}
which, applied to the present case, writes
\begin{eqnarray}
{\cal P}&=&<e^{i\theta(\vec x)}>=e^{-\frac{1}{2}<\theta^2(\vec x)>}= \Big(\frac{L}{a}\Big)^{-\frac{1}{8\pi\cdot \beta\cdot A\cdot d}}
\end{eqnarray}
which is Eq.~\eqref{Eq:PXZ}. Notice that this result did not require the expansion of ${\cal P}$ to lowest order in the deviations, which is why it is more accurate than Eq.~\eqref{Eq:P2D}.

\section{Application 3: the 2D NLSM with symmetry breaking interactions.}
\label{App:App3}
\setcounter{equation}{0}
We use now a spectrum of fluctuations that originates from including the symmetry breaking interactions of Eq.~\eqref{Eq:Lsb}:
\begin{equation}
\epsilon(k) = A\cdot d\cdot a^2\cdot k^2 +\big(\lambda-\Omega\cdot \frac{d}{a}\big)
\end{equation}
Accordingly,
\begin{equation}
<n^2_x(\vec x)>=\frac{a^2}{4\pi\cdot \beta}\int_0^{\frac{2\sqrt{\pi}}{a}}dk\cdot k\cdot \frac{1}{A\cdot d\cdot a^2\cdot k^2 + \big(\lambda-\Omega\cdot \frac{d}{a}\big)}
\end{equation}
and the same expression for $<n^2_y>$. This is an elementary integral, and since $\lambda-\Omega\frac{d}{a}<<4\pi A d$ it gives
\begin{equation}
<n^2_x(\vec x)>=\frac{1}{8\pi\cdot\beta\cdot A\cdot d}\cdot \ln\frac{4\pi\cdot A\cdot d+\big(\lambda-\Omega\cdot \frac{d}{a}\big)}{\big(\lambda-\Omega\cdot \frac{d}{a}\big)}\approx \frac{1}{4\pi\cdot\beta\cdot A\cdot d}\cdot \ln\sqrt{\frac{4\pi\cdot A\cdot d}{\big(\lambda-\Omega\cdot \frac{d}{a}\big)}}
\end{equation}
which leads to
\begin{equation}
{\cal P} \approx 1-\frac{1}{4\pi\cdot\beta\cdot A\cdot d}\cdot \ln\sqrt{\frac{4\pi\cdot A\cdot d}{\big(\lambda-\Omega\cdot \frac{d}{a}\big)}}
\end{equation}
i.e. Eq.~\eqref{Eq:Psb}. The infrared divergence of Appendix~\ref{App:App1} has disappeared: the symmetry breaking constant, however small, cuts the spectrum of fluctuations off at small $k$, and the size $L$ of the system has dropped out of the result.

\chapter{Finite Temperature ferromagnetism of transition metal ultrathin films: the Polyakov Renormalization Group (RG).}
\label{Chap:Polyakov}
\renewcommand{\thesection}{\thechapter.\arabic{section}}
\renewcommand{\theequation}{\thechapter.\arabic{equation}}

In 1975\cite{Pol} Polyakov constructed a Renormalization group approach to the
Heisenberg model that avoided introducing explicitly the non-linear terms
produced by the constraint $\vec n^2(\vec x)=1$ in the Lagrangian. Polyakov's
argument goes beyond the content of the Mermin-Wagner theorem and introduces a
novel development in the physics of two-dimensional or nearly two-dimensional
systems. Regarding the specific situation of transition metal ultrathin films, the Polyakov RG provides a natural framework to relate the symmetry breaking interactions occurring in these systems with the observation of a Curie temperature.

\section{The Renormalization group.}
The RG is a computational method of physics. It was discovered in the 1950's for particle physics\cite{WikiRG,Stuck,Gell,Bog} and rediscovered for the solution of problems of statistical physics and condensed matter physics by K. Wilson\cite{Nobel}.\\
To understand the RG, we observe that the main source of information about the
thermodynamic properties of a system is the Mean Field Approximation. We will,
in Chapter 10, discuss in detail the MFA of the Heisenberg model as a source of
qualitative results regarding the phase transition in ultrathin films. For the
purpose of introducing the RG, we summarize here the relevant elements of the MFA
approach\cite{MFA}. The MFA, as used in many fields of science, is an approximate approach based on an inequality by Bogoliubov\cite{Feynman}. It
replaces the Hamiltonian containing two-body interactions\footnote{We use a
``spin language'' we are acquainted with to describe the interaction as
$S_i\cdot S_j$, $S_i$ respectively $S_j$ being the spins at the lattice sites ``i'' and ``j''. $[S_i]$, then, defines an ``order parameter field'' on the lattice.} with a single particle trial Hamiltonian, parametrized in such a way as to minimize a trial free energy. The essential element when constructing the trial free energy is the setting of the so called \textbf{correlation function}
\begin{eqnarray}
G(i-j)&\doteq &\big<\big(S_i-<S_i>\big)\cdot \big(S_j-<S_j>\big)\big>
\end{eqnarray}
to be vanishing. For the correlation function, one expects, typically (up to a power law factor that is not important for the next considerations)
\begin{equation}
G(i-j)\propto a\cdot e^{-\frac{\vert i-j\vert}{\xi}}
\end{equation}
with $\xi$ being the so called \textbf{correlation length}. The correlation
function (and, with it, the correlation length) are fundamental quantities in
statistical physics. The correlation length, for instance, states that when the
spin at the site ``i'' is flipped, only spins residing within linear spatial
sizes specified by the correlation length will change their spin states in
response to the spin flip at ``i''. A vanishing correlation function is
interpreted as the correlation length being of the order of the lattice constant.
Accordingly, the spin field described by the MFA is associated with a high
degree of spatial order described by the equilibrium value. An important
consequence of the small correlation length entailed by the MFA is that systems
with much larger correlation lengths (such as those poised at a second order
phase transition) cannot be properly described by the MFA: ``spin waves''
(planar component of the Fourier transform of the order parameter) with
wavelength between $a$ and $\xi$ exist that produce fluctuations of the
order parameter field away from the thermal average, up to spatial length
scales of the order of $\xi$. Exactly at this point the RG sets in. The
renormalization group method provides a mathematical
tool that ``averages out'' these fluctuations, leaving behind an order parameter
field defined on spatial scales larger than $\xi$. The long scale field is more ordered than the original one and can be, possibly, managed by an MFA.
The mathematical tool that we are now going to introduce is the operation of
``smoothing''.
\paragraph{The smoothing operation of the RG.} The aim of statistical physics is the computation of the Gibbs free energy
\begin{equation}
G(T,B)=-\frac{1}{\beta}\cdot \ln \int D[m(\vec x)] e^{-\beta\cdot {\cal L}[m(\vec x)]}
\end{equation}
${\cal L}[m(\vec x)]$ is a Landau free energy functional that contains the coupling constants and thermodynamic variables such as the temperature $T$ and the magnetic field $B$. The ``sum'' over all possible configurations of the field $m(\vec x)$ (represented
by the functional integration) is the partition function. Its logarithm leads to
a thermodynamic potential, which is called the ``Gibbs free energy'' as it
contains the thermodynamic variables $T,B$. By means of known relations from the
subject of ``Thermodynamics'', most macroscopic properties of the system can be
obtained from $G(T,B)$. The simplest approximate calculation of the partition
function goes back to Landau\cite{Landau_V_1}. Within Landau theory, the
partition function is calculated by replacing the functional integration with
the maximum value of the integrand. This is equivalent to stating that the
equilibrium distribution $m(\vec x)$ is the result of the variational problem
\begin{equation}
\frac{\delta {\cal L}[m(\vec x)]}{\delta m(\vec x)}=0
\end{equation}
which typically leads to Euler-Lagrange equations. The Landau approximation of
the partition function is, accordingly, a MFA. We will later carry out
the Landau approximation when studying the ferromagnetic phase transition in
ultrathin films. The Landau approximation of the partition function neglects the
fluctuations of the order parameter field. These fluctuations are approximately
taken into account by the ``smoothing'' operation. When defining the smoothing
operation, one has in mind that the original field $m(\vec x)$ can be expressed as a Fourier sum
\begin{equation}
m(\vec  x)= \frac{1}{\sqrt{N^D}}\cdot \sum_{\vec  k} m_{\vec k}\cdot e^{i\cdot \vec  k\cdot \vec  x}
\end{equation}
$N^D$ is the number of unit cells in the system, with the relation
$L^D=N^D\cdot a^D$, $a$ being the lattice constant of a hypercubic lattice, and
$\vec k$ the reciprocal lattice vectors
$\vec  k = \frac{2\pi}{N\cdot a}\cdot (n_1,n_2,...,n_D)$,
$n_i\in {\cal Z}$.
The values of $\vec k$ are assumed to occupy the first Brillouin zone and, because of periodic boundary conditions, they are distributed on a fine-meshed lattice in a (hyper)cube with volume $(\frac{2\pi}{a})^D$
\footnote{We need some mathematical relations of Fourier expansions.\\
-- The orthonormality relation:
\[\int_V d^Dx\cdot e^{i\vec k\cdot x}\cdot e^{i\vec k'\cdot x} = N^D\cdot a^D\cdot \delta_{\vec k,-\vec k'}\]
-- The inverse Fourier transform:
\[m_{\vec k} = \frac{1}{\sqrt{N^D}}\cdot \frac{1}{a^D}\int_V d^Dx\, m(\vec x) \cdot e^{-i\cdot \vec k\cdot \vec x}\]
-- The second orthonormality relation:
\[\sum_{\vec k} e^{i\vec k\cdot x}\cdot e^{-i\vec k\cdot x'} = N^D\cdot a^D \cdot \delta(\vec x-\vec x')\]
-- When the sums over $\vec k$ are transformed into integrals (it is often easier to perform ``integrals'' than ``sums'') the translation key is
\[\sum_{\vec k} \rightarrow \frac{N^D\cdot a^D}{(2\pi)^D}\int d^D k\]
(the prefactor takes into account the density of $\vec k$-vectors). The second orthonormality relation then reads, accordingly,
\[\frac{1}{(2\pi)^D}\int d^Dk\cdot e^{i\vec k\cdot x}\cdot e^{-i\vec k\cdot x'} =\delta(\vec x-\vec x')\]
which is the Fourier transform of the Dirac delta function as known in the conventional literature.}.
It is also customary, for quantitative computations, to replace the hypercubic first Brillouin zone with a $D$-dimensional sphere of the same volume. The radius of the spherical Brillouin zone amounts to
\begin{equation}
\label{Eq:kBZ}
k_{BZ}= \frac{2\pi}{a}\cdot \Big[\frac{D\cdot \Gamma(\frac{D}{2})}{2\cdot \pi^{\frac{D}{2}}}\Big]^{\frac{1}{D}}
\end{equation}
The field $m(\vec  x)$ is considered as consisting of two components:
\begin{eqnarray}
S(\vec  x)&=&\frac{1}{\sqrt{N^D}}\cdot \sum _{0\leq\vert \vec  k\vert <
k_L} S(\vec  k)\cdot e^{i\cdot \vec  k\cdot \vec  x}\nonumber\\
\sigma(\vec  x)&=& \frac{1}{\sqrt{N^D}}\cdot \sum _{k_L<\vert \vec  q\vert <k_{BZ}} \sigma(\vec  q)\cdot e^{i\cdot \vec  q\cdot \vec  x}
\end{eqnarray}
with
\begin{equation}
k_L\doteq \frac{k_{BZ}}{\frac{L}{a}}
\end{equation}
DEFINITION. Given is a Landau functional ${\cal L}[S,\sigma]$. The smoothing operation (also called ``partial sum'' or ``partial trace'') is defined as
\begin{equation}
{\cal Z}[S]\doteq\int D[\sigma(\vec  x)] e^{-\beta\cdot {\cal L}[S(\vec  x),\sigma(\vec  x)]}
\end{equation}
COMMENTS.\\
1. This operation is said to ``average out'' the short wavelength fluctuations (the ``spin wave''-like fluctuations) and to ``eliminate'' them from the problem.\\
2. It leaves behind a functional that contains only the ``long wavelength'' field $S(\vec  x)$. One interprets the quantity left behind from the smoothing
operation as the partition functional ${\cal Z}[S]$ of the field $S(\vec
x)$. This partition functional is used to define the Landau functional for the field $S$ by means of the\\
DEFINITION. The Landau functional for the long spatial scales ${\cal L}[S]$ is defined as
\begin{equation}
{\cal L}[S]\doteq -\frac{1}{\beta}\ln {\cal Z}[S]
\end{equation}
3. One also finds, in the literature, the definition
\begin{equation}
e^{-\beta\cdot {\cal L}[S(\vec x)]}\doteq\int D[\sigma(\vec  x)] e^{-\beta\cdot {\cal L}[S(\vec  x),\sigma(\vec  x)]}
\end{equation}
Both definitions lead to the same ${\cal L}[S(\vec x)]$.\\
For most Lagrangians the smoothing operation cannot be performed exactly. For a perturbational computation of ${\cal L}[S(\vec x)]$, the original Landau functional is supposed to have the following structure:
\begin{equation}
{\cal L}[S,\sigma]= {\cal L}_0[\sigma]+ {\cal L}_1[S,\sigma]
\end{equation}
${\cal L}_0[\sigma]$ is considered to be the functional in leading order in $\sigma$.
How the two components -- the one containing the short wavelength components
and the one containing the long wavelength components -- are combined to build
the component ${\cal L}_1[S,\sigma]$ is irrelevant when general principles are
discussed. However, for the purpose of using a perturbational version of the
smoothing operation, ${\cal L}_1[S,\sigma]$ is supposed to be ``small'' in the
field $\sigma$ with respect to ${\cal L}_0[\sigma]$. We now compute:
\begin{eqnarray}
{\cal L}[S(\vec x)]&=&-\frac{1}{\beta}\ln \int D[\sigma(\vec  x)] e^{-\beta\cdot {\cal L}[S(\vec  x),\sigma(\vec  x)]}\nonumber\\
&\approx& -\frac{1}{\beta}\cdot \ln \int D[\sigma] e^{-\beta\cdot {\cal L}_0[\sigma]}\Big[1+\Big(-\beta\cdot {\cal L}_1[S,\sigma]+ \frac{\beta^2}{2}\cdot {\cal L}_1[S,\sigma]^2\Big)\Big]\nonumber\\
&=& -\frac{1}{\beta}\cdot \ln\Big[Z_0(T,B)\cdot\Big(1-\beta\cdot\frac{\int D[\sigma] e^{-\beta\cdot {\cal L}_0[\sigma]}\cdot {\cal L}_1[S,\sigma]}{Z_0(T,B)}\nonumber\\
&+& \frac{\beta^2}{2}\cdot\frac{\int D[\sigma] e^{-\beta\cdot {\cal L}_0[\sigma]}\cdot {\cal L}_1[S,\sigma]^2 }{Z_0(T,B)}\Big)\Big]\nonumber\\
&\approx & G_0(T,B)\nonumber\\
& + &\frac{\int D[\sigma] e^{-\beta\cdot {\cal L}_0[\sigma]}\cdot {\cal L}_1[S,\sigma]}{Z_0(T,B)}\nonumber\\
&+& \frac{\beta}{2}\cdot \Big[\frac{\int D[\sigma] e^{-\beta\cdot {\cal L}_0[\sigma]}\cdot {\cal L}_1[S,\sigma]}{Z_0(T,B)}\Big]^2-\frac{\beta}{2}\cdot\frac{\int D[\sigma] e^{-\beta\cdot {\cal L}_0[\sigma]}\cdot {\cal L}_1[S,\sigma]^2}{Z_0(T,B)}
\end{eqnarray}
COMMENTS.\\
Within this perturbational approach of the smoothing operation, the functional
${\cal L}[S(\vec x)]$ consists of three components:\\
I. The first, originating from 0th order perturbation theory, is $G_0(T,B)$. It does not contain the field $S(\vec x)$ and, accordingly, does not contribute to the physical properties of the system at large spatial scales.\\
II. The second component originates from 1st order perturbation theory and is the functional average of ${\cal L}_1[S,\sigma]$ over the field $\sigma$. We introduce a simple notation to indicate this averaging operation:
\begin{equation}
\frac{\int D[\sigma] e^{-\beta\cdot {\cal L}_0[\sigma]}\cdot {\cal L}_1[S,\sigma]}{Z_0(T,B)}\doteq <{\cal L}_1[S,\sigma]>_\sigma
\end{equation}
III. The third component originates from second order perturbation theory and writes
\begin{eqnarray}
+\underbrace{\frac{\beta}{2}\cdot \Big[\frac{\int D[\sigma] e^{-\beta\cdot {\cal L}_0[\sigma]}\cdot {\cal L}_1[S,\sigma]}{Z_0(T,B)}\Big]^2}_{\doteq \frac{\beta}{2}\cdot <{\cal L}_1[S,\sigma]>^2_\sigma}
\underbrace{-\frac{\beta}{2}\cdot \frac{\int D[\sigma] e^{-\beta\cdot {\cal L}_0[\sigma]}\cdot {\cal L}_1[S,\sigma]^2}{Z_0(T,B)}}_{\doteq -\frac{\beta}{2}<{\cal L}^2_1[S,\sigma]>_\sigma}
\end{eqnarray}
In summary, up to second order perturbation theory
\begin{eqnarray}
{\cal L}[S(\vec x)]&\approx & <{\cal L}_1[S,\sigma]>_\sigma\nonumber\\
&-& \frac{\beta}{2}\cdot \Big[<{\cal L}^2_1[S,\sigma]>_\sigma-<{\cal L}_1[S,\sigma]>^2_\sigma\Big]
\end{eqnarray}
COMMENTS.\\
1. The main output of the renormalization group algorithm is the Landau functional for the large scale field $S(\vec x)$, defined at the spatial scale $L$.\\
2. The question arises up to which length ``$L$'' the smoothing operation must be performed. This final length was set by Wilson to be the correlation length $\xi$.\\
3. Suppose that we have obtained a smoothed functional ${\cal L}[S]$. How can we use it? It belongs to the lore of the renormalization group that the smoothed functional ${\cal L}[S]$ is more suitable to e.g. a standard Mean Field Approximation than the original functional ${\cal L}[m]$.\\
4. DEFINITION. The Landau functional ${\cal L}[m]$ is said to be renormalizable if ${\cal L}[m]$ and ${\cal L}[S]$ are formally identical.\\
5. Although ${\cal L}[S]$ might turn out to have the
same structure as ${\cal L}[m]$, the various coupling constants appearing in ${\cal L}[S]$ might have been ``changed'' (more precisely: renormalized) by the smoothing process. This might lead to profound changes with respect to the results of the mean field approximation applied to ${\cal L}[m]$.\\
6. Bogoliubov and Shirkov\cite{Bog} and Gell-Mann and Low\cite{Gell} have described a way of improving the perturbational result. In their algorithm\footnote{This aspect is very technical and will be illustrated on the basis of the concrete examples that we will work out.}, the smoothing operation is used as a way of obtaining \textbf{recursion relations} for the coupling constants. These recursion relations are then used for constructing a set of differential equations -- the so called Gell-Mann-Low equations -- that the various coupling constants must obey. The differential equations are integrated between ``$a$'' and ``$\xi$''.\\
The experimental situation in ultrathin film magnetism allows one to identify three
phenomena that can be accounted for by the renormalization group algorithm: the
existence of a phase transition in perpendicularly (and in-plane)
magnetized two-dimensional ferromagnets (the subject of this Chapter), the reorientation transition in perpendicularly magnetized films (the subject of Chapter 8) and the Ising character of the ferromagnetic-to-paramagnetic
transition (the subject of Chapter 9).

\section{The correlation function in the NLSM and in the NLSM with symmetry breaking interactions.}
\label{Sec:corr}
We assume a spectrum of excitations that behaves as
\begin{equation}
\label{Eq:gap}
A\cdot d\cdot a^2\cdot \vec k^2 + \Big(\lambda-\Omega\cdot \frac{d}{a}\Big)
\end{equation}
with $\lambda-\Omega\cdot \frac{d}{a}>0$.
With this spectrum of excitations, the system has a spontaneous magnetization
along the vertical direction. Along the in-plane direction the spontaneous magnetization is vanishing. We insert the Fourier transforms
\begin{equation}
n_x(\vec x) = \frac{a}{L}\cdot\sum_{\vec k}e^{i\cdot \vec k\cdot \vec x}\cdot n_{\vec k}
\end{equation}
and
\begin{equation}
n_x(\vec x') = \frac{a}{L}\cdot\sum_{\vec k}e^{i\cdot \vec k\cdot \vec x'}\cdot n_{\vec k}
\end{equation}
to compute the correlation function for the field $\vec n_\perp= (n_x,n_y)$ as
\begin{eqnarray}
G_\perp(\vec x, \vec x') &\doteq & <\vec n_\perp(\vec x)\cdot \vec n_\perp(\vec x')>\nonumber\\
&=& <n_x(\vec x)\cdot n_x(\vec x')> + <n_y(\vec x)\cdot n_y(\vec x')>\nonumber\\
&=& 2\cdot <n_x(\vec x)\cdot n_x(\vec x')>
\end{eqnarray}
because the $n_x$ and $n_y$ components have the same correlator. With
$r\doteq \vert \vec x-\vec x'\vert$,
\begin{eqnarray}
<n_x(\vec x)\cdot n_x(\vec x')>&=& \frac{a^2}{L^2}\sum_{\vec k,\vec k'}<n_{\vec k}\cdot n_{\vec k'}>\cdot e^{i\vec k\vec x+ i\vec k'\vec x'}\nonumber\\
&=&\frac{a^2}{L^2}\sum_{\vec k,\vec k'}\delta_{\vec k,-\vec k'}<\vert n_{\vec k}\vert^2>\cdot e^{i\vec k\vec x+ i\vec k'\vec x'}\nonumber\\
&=& \frac{a^2}{L^2}\sum_{\vec k}<\vert n_{\vec k}\vert^2>\cdot e^{i\vec k(\vec x-\vec x')}\nonumber\\
&=&\frac{1}{(2\pi)^2}\cdot a^2\cdot \int d^2k\cdot <\vert n_{\vec k}\vert^2>\cdot e^{i\vec k(\vec x-\vec x')}\nonumber\\
&=& \frac{1}{2\pi}\cdot a^2\cdot \int_0^{\infty} k\cdot dk\cdot \frac{1}{2\cdot \beta\cdot \big[A\cdot d\cdot a^2\cdot k^2+ \big(\lambda-\Omega\cdot \frac{d}{a}\big)\big]}\nonumber\\
&\cdot & \underbrace {\frac{1}{2\pi}\cdot \int d\varphi\, e^{ik\cdot r \cdot \cos\varphi}}_{J_0(k\cdot r)}\nonumber\\
&=& \frac{1}{4\pi\cdot \beta\cdot A\cdot d}\cdot \int_0^{\infty}\frac{dk\cdot k\cdot J_0(k\cdot r)}{k^2+ \frac{\lambda-\Omega\cdot \frac{d}{a}}{A\cdot d\cdot a^2}}\nonumber\\
&=&\frac{1}{4\pi\cdot \beta\cdot A\cdot d}\cdot K_0\Big(\frac{r}{\xi}\Big)
\end{eqnarray}
$K_0(x)$ is the McDonald function of 0th order.
From its asymptotic behaviour, we obtain, for large values of $r$,
\begin{equation}
<n_x(\vec x)\cdot n_x(\vec x')>\sim \frac{1}{4\pi\cdot \beta\cdot A\cdot d}\cdot \sqrt{\frac{\pi\xi}{2r}}\cdot e^{-\frac{r}{\xi}}
\end{equation}
with
\begin{equation}
\label{Eq:xi}
\xi\doteq a\cdot \sqrt{\frac{A\cdot d}{\lambda-\Omega\cdot \frac{d}{a}}}
\end{equation}
being the correlation length of the transversal components.
If one were to consider the isotropic 2D Heisenberg model as being the limiting case
of the gapped NLSM when $\lambda-\Omega\cdot \frac{d}{a}$ tends to zero, one would
ascertain that the correlation length of the isotropic model tends to infinity.
Because of the infinite correlation length, the exponential tends to 1 and the question arises about the functional dependence of the correlation function on $r$, left behind in the isotropic model after the exponential has been dropped. The question about the correlation function in the isotropic 2D Heisenberg model is still unanswered. A possible attempt at this is to compute $G(r)$ for the isotropic model as
\begin{equation}
\frac{1}{4\pi\cdot \beta\cdot A\cdot d}\int_0^{\infty}\frac{dk}{k}\cdot J_0(k\cdot r)
\end{equation}
This integral is divergent. To extract the divergent component one computes
\begin{equation}
\frac{1}{4\pi\cdot \beta\cdot A\cdot d}\int_{\frac{2\sqrt{\pi}}{L}}^{\infty}\frac{dk}{k}\cdot J_0(k\cdot r)
\end{equation}
in the limit of large $L$. Using the small argument expansion of the integral,
$\int_\epsilon^\infty \frac{dk}{k}J_0(kr)= -\ln\frac{\epsilon\, r}{2}-\gamma_E+{\cal O}(\epsilon^2r^2)$
with $\gamma_E$ the Euler constant, one finds that
\begin{equation}
\label{Eq:Gisotropic}
\frac{1}{4\pi\cdot \beta\cdot A\cdot d}\int_{\frac{2\sqrt{\pi}}{L}}^{\infty}\frac{dk}{k}\cdot
J_0(k\cdot r)\approx -\frac{1}{4\pi\cdot \beta\cdot A\cdot d}\cdot \ln\frac{r}{L}
\end{equation}
up to an additive constant.
This result suggests that the correlation function, for points at distances much smaller than the lateral size of the system, decays logarithmically. However, as $L$ cannot be eliminated from the problem, the result we have just obtained is not rigorous.\\
The situation is different when one in-plane component (let us say the $n_y$ component) is set to vanish identically, thus producing the cross-over from the 2D isotropic Heisenberg model to the 2D planar model with only two spin components. In this case we can use the exact relationship $n_x=\sin\theta$ ($\theta$ being the angle between the spin vector and the $z$-direction) to compute the correlation function in a more rigorous way. We compute:
\begin{eqnarray}
<n_x(\vec x)\cdot n_x(\vec x')>&=& <\sin\theta(\vec x)\cdot \sin\theta(\vec x')>\nonumber\\
&=& -\frac{1}{4}\cdot <\Big(e^{i\theta(\vec x)}-e^{-i\theta(\vec x)}\Big)\cdot
\Big(e^{i\theta(\vec x')}-e^{-i\theta(\vec x')}\Big)>\nonumber\\
&\approx & -\frac{1}{4}\cdot <\Big(e^{i\cdot n_x(\vec x)}-e^{-i\cdot n_x(\vec x)}\Big)\cdot
\Big(e^{i\cdot n_x(\vec x')}-e^{-i\cdot n_x(\vec x')}\Big)>\nonumber\\
&=&  -\frac{1}{4}\cdot\Big(<e^{i\cdot n_x(\vec x)+i\cdot n_x(\vec x')}>+ <e^{-i\cdot n_x(\vec x)-i\cdot n_x(\vec x')}>\nonumber\\
&-& <e^{i\cdot n_x(\vec x)-i\cdot n_x(\vec x')}>- <e^{-i\cdot n_x(\vec x)+i\cdot n_x(\vec x')}>\Big)\nonumber\\
&=&-\frac{1}{4}\cdot\Big(2\cdot e^{-\frac{1}{2}<(n_x(\vec x)+ n_x(\vec x'))^2>}-2\cdot e^{-\frac{1}{2}<(n_x(\vec x)- n_x(\vec x'))^2>}\Big)\nonumber\\
&=& -\frac{1}{2}\cdot e^{-<n^2_x(\vec x)>}\cdot \Big(e^{-<n_x(\vec x)\cdot n_x(\vec x')>}- e^{+<n_x(\vec x)\cdot n_x(\vec x')>}\Big)\nonumber\\
&=& -\frac{1}{2}\cdot e^{-\frac{1}{4\pi\cdot\beta\cdot A\cdot d}\cdot \ln\frac{L}{a}}\cdot \Big(e^{\frac{1}{4\pi\cdot\beta\cdot A\cdot d}\cdot \ln\frac{r}{L}}-
e^{-\frac{1}{4\pi\cdot\beta\cdot A\cdot d}\cdot \ln\frac{r}{L}}\Big)\nonumber\\
&\underbrace{\longrightarrow}_{L\rightarrow\infty}& \frac{1}{2}\Big(\frac{r}{a}\Big)^{-\frac{1}{4\pi\cdot\beta\cdot A\cdot d}}
\end{eqnarray}
Notice that $L$ has been eliminated, so that this expression holds true in the thermodynamic limit. The 2D planar model with only two components produces a power law correlation function\cite{Poki}.

\section{The isotropic 2D-Heisenberg ferromagnet.}
\label{Sec:isotropic}
The 2D Landau functional\cite{Pol} for a $p$-component\footnote{We introduce a $p$-component field: $p=2$ corresponds to the $XY$-model, $p=3$ to the Heisenberg model.} field $\vec n(\vec \rho)$ writes
\begin{equation}
{\cal L}_A[\vec n]= \Gamma\cdot \int d^2\rho \sum_{\nu=x,y} \big(\partial_\nu \vec n\big)^2
\end{equation}
The coupling constant $\Gamma$ for an ultrathin slab amounts to $A\cdot d$.
The field $\vec n(\vec x)$ is constrained by the condition $(\vec n)^2 = 1$. The renormalization protocol, applied to this model Hamiltonian by A.M. Polyakov\cite{Pol}, foresees introducing an ad-hoc orthonormal basis set
\begin{equation}
\{\vec n_0, \vec e_1,...,\vec e_{p-1}\}\quad \vec n_0\cdot \vec e_a =0
\end{equation}
of basis vectors and parametrizing $\vec n$ according to
\begin{equation}
\label{Eq:param}
\vec n = \vec n_0\cdot\sqrt{1-\vec \phi^2}+\vec \phi;\quad\vec\phi(\vec \rho) \doteq \sum_{a=1}^{p-1}\phi_a\cdot \vec e_a(\vec \rho)
\end{equation}
This parametrization ensures that $\vec n_0^2=\vec n^2$. Within the lore of RG, one has in mind a field $\vec n_0$ that varies over spatial scales larger than $L$, i.e. its Fourier components contain the vectors $0\leq \vert \vec k\vert \leq k_L$. The transversal field $\vec \phi$, instead, is defined over
the components $k_L\leq \vert \vec q\vert \leq k_{BZ}$. The
plane waves entering the Fourier expansion of the field $\vec \phi$ are the
so-called \textbf{spin waves} and represent small deviations from the
ferromagnetic ground state. The effective exchange Hamiltonian contains other
types of excitations (for instance, the so-called \textbf{skyrmions} -- local
topological, ``vortex-like'' excitations\cite{Skir}) which are not taken into account by the plane wave expansion.\\
The result of the RG protocol is a renormalized coupling constant for the effective Lagrangian of the field $\vec n_0$, ${\cal L}[\vec n_0]$. ${\cal L}[\vec n_0]$ is, formally, identical to the original Lagrangian ${\cal L}[\vec n]$. Let us designate the original coupling constant as $\Gamma_a$, i.e. the coupling constant at the shortest scale $L=a$, and the renormalized coupling constant at the scale $L$ as $\Gamma_L$. Then one obtains
\begin{equation}
\label{Eq:pertbody}
\Gamma_L =  \Gamma_a\cdot \Big(1-\frac{(p-2)}{4\pi\cdot \beta\cdot \Gamma_a}\cdot \ln \frac{L}{a}\Big)
\end{equation}
COMMENTS.\\
1. The Gell-Mann-Low equations producing the renormalized coupling constant $\Gamma_L$ must be integrated between the length ``$a$'' and
the correlation length, i.e. the renormalization stops at the
correlation length. In the specific situation of the isotropic model, the correlation length grows as the lateral size of the system. In practice, however, for $p>2$ and at any finite temperature $T$, there is a characteristic length
\begin{equation}
\label{Eq:Lc}
L_c= a\cdot e^{\frac{4\pi\cdot\beta\cdot A\cdot d}{(p-2)}}
\end{equation}
at which the exchange interaction vanishes. Within this length, the spins are
strongly correlated and can be considered to build a ``spin block''. Over
larger distances, the isotropic Heisenberg ferromagnet consists of blocks with
no exchange coupling between them. The isotropic 2D Heisenberg magnet is a kind
of ``superparamagnet'', with fluctuations continuously changing the macroscopic
spin configurations while keeping the spins within the spin blocks almost
aligned. The scenario of the exponentially large length scale $L_c$ appears at
any temperature, i.e. the Polyakov renormalization entails that the isotropic Heisenberg ferromagnet has no phase transition. In view of these considerations, one can state that the Polyakov renormalization goes beyond the Mermin-Wagner theorem, providing a clear picture about why long range
order should vanish in the two-dimensional Heisenberg ferromagnet.\\
2. For $p=2$: this is the so-called two-dimensional $XY$ model, $XY$ meaning that
the spin only has two components. In this case, $\Gamma$ is independent of the
length scale and the spins, even on large scales, have a finite stiffness against
fluctuations. The physical properties of the planar model were worked out in depth by Kosterlitz and Thouless\cite{PP,Kost}.\\
3. It is the lore of the renormalization group algorithm that,
by solving the Gell-Mann-Low equation constructed using the first order
perturbation theory, one obtains the summation of all ``leading logarithms'' of
the type $(\frac{1}{4\pi\cdot\beta\cdot \Gamma}\cdot \ln L)^m$ appearing in the
successive orders of perturbation theory\cite{Bog}. For this reason, the Gell-Mann-Low algorithm produces an improvement of the perturbational result (by coincidence, in the specific case of the Polyakov renormalization of $\Gamma$, the result of the Gell-Mann-Low equation coincides with the perturbational result). However, there are further, non-leading logarithms that usually contribute to the renormalization of the coupling constant. When $\Gamma$ is small, these non-leading logarithms might play a role in softening the decrease of $\Gamma$. These non-leading logarithms have not yet been computed, so that we do not know exactly whether $\Gamma$ will vanish at $L\rightarrow \infty$ or whether it will level off at some finite value, thus allowing some kind of order.\\
4. Within Polyakov RG, one can also compute that the magnetization along a certain direction $z$ renormalizes according to
\begin{equation}
\label{Eq:nzren}
n_z(\vec x, L) = n_z(\vec x,a)\cdot \Big(1-(p-2)\frac{1}{4\pi\cdot \beta\cdot \Gamma_a}\cdot \ln \frac{L}{a}\Big)^{\frac{(p-1)}{2\cdot (p-2)}}
\end{equation}
Accordingly, for a given $L$, it vanishes exactly at the same temperature at which $\Gamma_L$ vanishes.\\
5. In the argument presented above, the spatial length scale $L$ (the lateral size of the system) cannot be eliminated and the thermodynamic limit $L\rightarrow \infty$ is ill defined. There is a formal method that fixes the transition temperature from the ordered, ferromagnetic phase to the disordered phase at exactly $T=0$, in line with the Mermin-Wagner theorem. The method is that of performing the computation of the coupling constant at dimensions $D>2$ and
taking the result to the limit $D\rightarrow 2$\cite{Pol,Brezin}. In fact, in dimensions $D>2$, one finds that
\begin{equation}
\Gamma(L\rightarrow \infty)= \Gamma_a\cdot\Big(1-(p-2)\cdot \frac{k_B\cdot T}{4\pi\cdot \Gamma_a}\cdot \frac{1}{D-2}\Big)
\end{equation}
One can therefore define a temperature at which the coupling constant and the magnetization vanish:
\begin{equation}
k_B\cdot T_C  = \frac{4\pi\cdot \Gamma_a}{(p-2)}\cdot (D-2)
\end{equation}
$T_C$ vanishes in the limit $D\rightarrow 2$.\\

\section{The critical temperature in the gapped Heisenberg ferromagnet.}
There is a fundamental difference between the 2D isotropic Heisenberg model and
the Heisenberg model augmented by a small effective anisotropic term that
consists of a single ion perpendicular anisotropy and the short range component
of the dipolar interaction. These additional interactions introduce a gap into
the spectrum of fluctuations. The gap is responsible for the occurrence of a
finite spontaneous magnetization that is absent in the isotropic model. In
addition, in virtue of this gap, a finite correlation length enters the problem.
In the process of going to the larger spatial scales over which the field $\vec n_0$ is defined, the value of the gap is also renormalized by the spin waves with wavelength smaller than the correlation length. If we let, for the moment, this renormalization aside and use the bare value for the gap, the correlation length is the one of Eq.~\eqref{Eq:xi},
\begin{equation}
\xi= a\cdot \sqrt{\frac{A\cdot d}{\lambda-\Omega\cdot \frac{d}{a}}}
\end{equation}
At this length, the renormalization of the coupling constant $\Gamma$ stops and the renormalized value to be used for the writing of the Landau functional of the field $\vec n_0$ writes
\begin{equation}
\Gamma(\lambda,\Omega,d)= \Gamma_a\cdot \Big(1-\frac{(p-2)}{4\pi\cdot \beta\cdot \Gamma_a} \cdot\ln \sqrt{\frac{A\cdot d}{\lambda-\Omega\cdot \frac{d}{a}}}\Big)
\end{equation}
The renormalized value vanishes, formally, at a characteristic temperature given by (we use, for clarity, the original designation $\Gamma=A\cdot d$)
\begin{equation}
\label{Eq:Tc}
k_B\cdot T_c= \frac{8\pi\cdot A\cdot d}{(p-2)\cdot \ln \frac{A\cdot d}{\lambda-\Omega\cdot \frac{d}{a}}}
\end{equation}
At the same temperature, the renormalized spontaneous magnetization also
vanishes. It is therefore natural to identify this temperature as the Curie
temperature of the gapped 2D Heisenberg ferromagnetic slab. Notice that, apart from the logarithmic correction, the Curie temperature computed is of the order of $A\cdot d$ -- the same order of magnitude one computes using the Mean Field Approximation of the gapped Heisenberg ferromagnet.

\setcounter{section}{0} 
\renewcommand{\thesection}{\thechapter.\Alph{section}} 
\setcounter{equation}{0}
\renewcommand{\theequation}{\thesection\arabic{equation}}
\section*{Appendices}

\section{Polyakov Renormalization group method: technical aspects.}
\label{App:Poly}
\setcounter{equation}{0}
The 2D Landau functional for a $p$-component field $\vec n(\vec \rho)$ writes
\begin{equation}
{\cal L}_A[\vec n]= A\cdot d\cdot \int d^2\rho \sum_{\nu=x,y}
\big(\partial_\nu \vec n\cdot \partial_\nu \vec n\big)
\end{equation}
The field $\vec n(\vec x)$ is constrained by the condition $(\vec n)^2 = 1$. The renormalization protocol, applied to this model Hamiltonian by A.M. Polyakov\cite{Pol}, foresees introducing an ad-hoc orthonormal basis set
\begin{equation}
\{\vec n_0, \vec e_1,...,\vec e_{p-1}\}\quad \vec n_0\cdot \vec e_a =0
\end{equation}
of basis vectors and parametrizing $\vec n$ according to
\begin{equation}
\vec n = \vec n_0\cdot\sqrt{1-\vec \phi^2}+\vec \phi;\quad\vec\phi(\vec \rho) \doteq \sum_{a=1}^{p-1}\phi_a\cdot \vec e_a(\vec \rho)
\end{equation}
This parametrization ensures that $\vec n_0^2=\vec n^2$. Within the lore of RG, one has in mind a field $\vec n_0$ that varies over spatial scales larger than $L$, i.e. its Fourier components contain the vectors $0\leq \vert \vec k\vert \leq k_L$. The transversal field $\vec \phi$, instead, is defined over
the components $k_L\leq \vert \vec q\vert \leq k_{BZ}$. The
plane waves entering the Fourier expansion of the field $\vec \phi$ are the
so-called \textbf{spin waves} and represent small deviations from the
ferromagnetic ground state. The effective exchange Hamiltonian contains other
types of excitations (for instance, the so-called \textbf{skyrmions} -- local
topological, ``vortex-like'' excitations\cite{Skir}) which are not taken into account by the plane wave expansion. Using this special parametrization, the Landau functional rewrites
\begin{eqnarray}
{\cal L}_A[\vec n_0(\vec \rho),\vec \phi(\vec \rho)] &=&
A\cdot d\cdot \int d^2\rho\Big[\partial_\nu \left(\vec n_0(\vec \rho)\sqrt{1-\vec \phi^2} +\vec \phi(\vec \rho)\right)\Big]\nonumber\\
&\cdot& \Big[\partial_\nu \left(\vec n_0(\vec \rho)\sqrt{1-\vec \phi^2} +\vec \phi(\vec \rho)\right)\Big]
\end{eqnarray}
(we use the convention of summing over two equal indices). When resolving the square, a term containing $\vec n_0\cdot\vec \phi$ appears, that vanishes because of the orthogonality of the vectors. The remaining terms can be summarized to give
\begin{eqnarray}
{\cal L}[\vec n_0,\vec \phi] =
A\cdot\! d\!\cdot\! \int\! d^2\rho \!&\cdot& \Big[\partial_\nu \vec n_0\cdot \partial_\nu \vec n_0\cdot \big(1-(\vec \phi)^2\big)\nonumber\\
&+& 2\cdot \partial_\nu \vec n_0\cdot \partial_\nu \vec \phi\cdot \sqrt{(1-(\vec \phi)^2)}\nonumber\\
&+& \partial_\nu \vec \phi\cdot \partial_\nu \vec \phi\Big]
\end{eqnarray}
The first term has the sought for structure in the field $\vec n_0$.\\
We evaluate further the middle summand, i.e. the cross product
\begin{equation}
\partial_\nu \vec n_0\cdot \vec e_a\cdot \partial_\nu \phi_a + \phi_a\cdot\partial_\nu \vec n_0\cdot \partial_\nu\vec e_a
\end{equation}
$\partial_\nu \phi_a$ is odd in the field $\phi_a$ and will therefore turn out to vanish when the average over $\vec\phi$ is performed. The remaining one
\begin{equation}
\phi_a\cdot\partial_\nu \vec n_0\cdot \partial_\nu \vec e_a
\end{equation}
requires a closer look. Without loss of generality, we can set
\begin{equation}
\partial_\nu \vec n_0 = c_{\nu b}\cdot \vec e_b
\end{equation}
(see the sketch).
\begin{figure}[H]
\begin{center}
\includegraphics[width=0.5\textwidth]{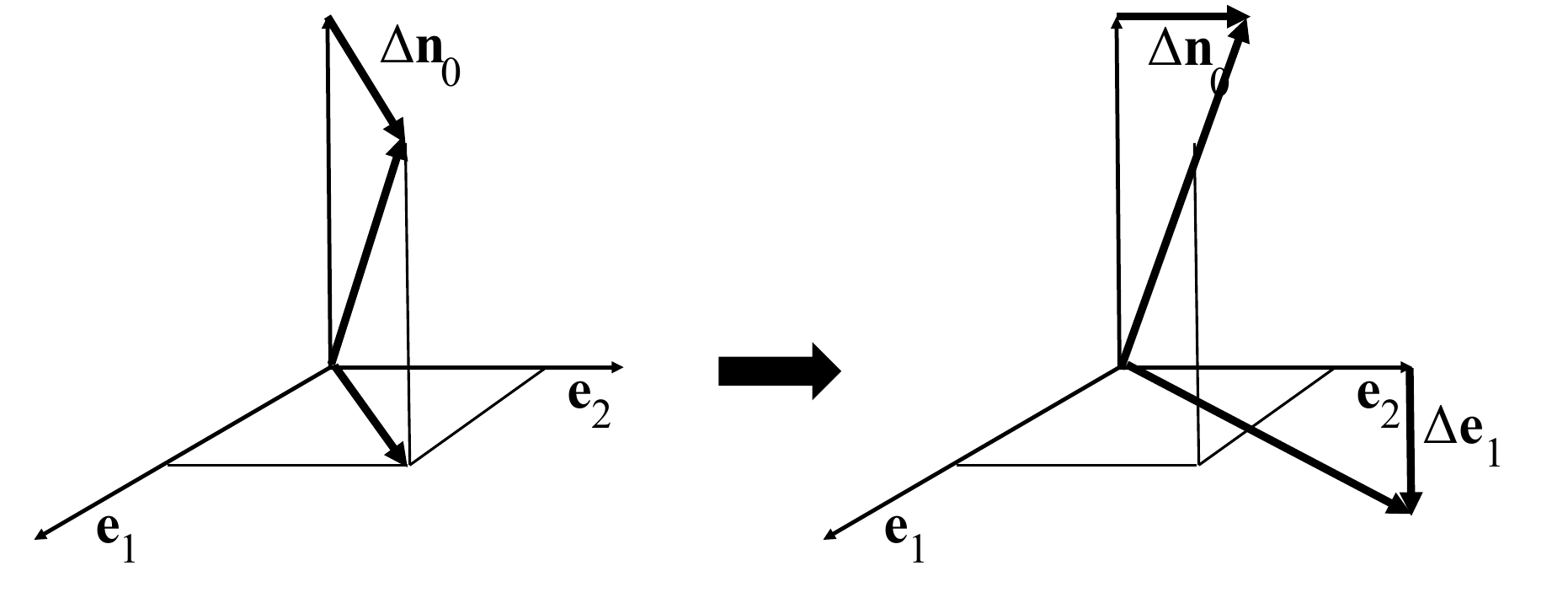}
\end{center}
\caption{Sketch of the geometrical issues in the Polyakov RG of the elastic Heisenberg Lagrangian.}
\label{Fig:Geometry}
\end{figure}
We expect that, for a given $\nu$, the set $\vec e_a$ can be chosen so that
\begin{equation}
\partial_\nu \vec e_a = - c_{\nu a}\cdot \vec n_0
\end{equation}
The same set, however, cannot fulfill this condition for a different $\nu$. In general, therefore,
\begin{equation}
\partial_\nu \vec e_a = - c_{\nu a}\cdot \vec n_0+ f^b_{\nu,a}\cdot \vec e_b
\end{equation}
By choosing the set $\{\vec e_a\}$ so that the coefficients are proportional to the second derivatives with respect to $\vec n_0$, we can continue assuming
\begin{equation}
\partial_\nu \vec e_a = - c_{\nu a}\cdot \vec n_0
\end{equation}
Accordingly,
\begin{equation}
\phi_a\cdot\partial_\nu \vec n_0\cdot \partial_\nu\vec e_a\approx c_{\nu b}\cdot\vec e_b\cdot \phi_a\cdot (- c_{\nu a})\cdot \vec n_0 =0
\end{equation}
in virtue of the orthogonality of the vectors $\vec n_0$ and $\vec e_a$.\\
We now turn to the third term and find
\begin{eqnarray}
\partial_\nu \vec \phi\cdot \partial_\nu \vec \phi&=&\partial_\nu(\phi_a\vec e_a)\cdot \partial_\nu(\phi_b\vec e_b)\nonumber\\
&=& c_{\nu a}\cdot c_{\nu b}\cdot \phi_a\cdot\phi_b + \partial_\nu\phi_a\cdot \partial_\nu\phi_a
\end{eqnarray}
so that, summarizing, the exchange functional in the fields $\vec n_0,\vec \phi$ writes
\begin{eqnarray}
{\cal L}_A[\vec n_0,\vec \phi] &=&
A\cdot\! d\!\cdot\! \int\! d^2\rho\!\cdot \partial_\nu\phi_a\cdot \partial_\nu\phi_a\nonumber\\
&+&A\cdot\! d\!\cdot\! \int\! d^2\rho\!\cdot\Big[\partial_\nu \vec n_0\cdot \partial_\nu \vec n_0\cdot \big(1-(\vec \phi)^2\big)+ c_{\nu a}\cdot c_{\nu b}\cdot \phi_a\cdot\phi_b\Big]\nonumber\\
\end{eqnarray}
To first order perturbation theory, the functional for the field $\vec n_0$ writes
\begin{eqnarray}
{\cal L}_A[\vec n_0] &=&
A\cdot\! d\!\cdot\! \int\! d^2\rho\!\cdot\Big[\partial_\nu \vec n_0\cdot \partial_\nu \vec n_0\cdot \big(1-<(\vec \phi)^2>_{\vec \phi}\big)\nonumber\\
&+& <c_{\nu a}\cdot c_{\nu b}\cdot \phi_a\cdot\phi_b>_{\vec \phi}\Big]
\end{eqnarray}
The symbol $<...>_{\vec \phi}$ indicates that the functional represented by the dots is averaged over the transversal, small field $\vec \phi$ according to the functional average
\begin{equation}
<...>_{\vec \phi}= \frac{\int D(\phi_1)\cdot D(\phi_2)...\cdot D(\phi_{p-1})\cdot e^{-\beta\cdot A\cdot\! d\!\cdot\! \int\! d^2\rho\!\cdot \partial_\nu\phi_a\cdot \partial_\nu\phi_a}\cdot (...)}{\int D(\phi_1)\cdot D(\phi_2)...\cdot D(\phi_{p-1})\cdot e^{-\beta\cdot A\cdot\! d\!\cdot\! \int\! d^2\rho\!\cdot \partial_\nu\phi_a\cdot \partial_\nu\phi_a}}
\end{equation}
The averaging over the field $\vec \phi$ is ruled by the following relations:
\begin{eqnarray}
<\phi_a(\vec \rho)\phi_b(\vec \rho)>_{\vec \phi}&=& \delta_{ab}\cdot <\phi(\vec \rho)\phi(\vec \rho)>_{\vec \phi}\nonumber\\
<\phi(\vec \rho)\phi(\vec \rho)>_{\vec \phi}&=& \frac{a^2}{4\pi^2}\cdot
\int_{\frac{2\sqrt{\pi}}{L}}^\frac{2\sqrt{\pi}}{a} d^2k\cdot
\frac{1}{\epsilon(k)}\nonumber\\
<\vec \phi^2>_{\vec \phi}&=&(p-1)\cdot <\phi_a(\vec \rho)\phi_a(\vec \rho)>_{\vec \phi}
\end{eqnarray}
Taking into account that
\begin{equation}
<c_{\nu a}\cdot c_{\nu b}\cdot \phi_a\cdot\phi_b>_{\vec \phi}= \partial_\nu \vec n_0\cdot \partial_\nu \vec n_0\cdot \frac{1}{p-1}\cdot <\vec \phi(\vec \rho)^2>_{\vec \phi}
\end{equation}
Polyakov\cite{Pol} obtained the perturbative effective exchange Hamiltonian for the field $\vec n_0$ as
\begin{eqnarray}
{\cal L}_A[\vec n_0]&=& A\cdot d\cdot
\Big(1-(p-2)\cdot\frac{a^2}{4\pi^2}\cdot
\int_{\frac{2\sqrt{\pi}}{L}}^\frac{2\sqrt{\pi}}{a} d^2k\cdot
\frac{1}{\epsilon(k)}\Big)\cdot \int d^2\rho \sum_{\nu=x,y}\partial_\nu \vec n_0\cdot \partial_\nu \vec n_0\nonumber\\
\end{eqnarray}
The net result of averaging out spin waves is that the long wavelength field is described by the same Heisenberg effective Landau functional, albeit with a renormalized coupling constant (for simplicity of writing, we call $A\cdot d$ the ``$\Gamma$''-coupling constant)
\begin{equation}
\label{Eq:pert}
\Gamma_L =  \Gamma_a\cdot \Big(1-(p-2)\cdot \frac{a^2}{4\pi^2}\cdot
\int_{\frac{2\sqrt{\pi}}{L}}^\frac{2\sqrt{\pi}}{a} d^2k\cdot
\frac{1}{\epsilon(k)}\Big)
\end{equation}
$\Gamma_a$ is the coupling constant at the length scale $L=a$.
\paragraph{The Gell-Mann-Low equations.}
Eq.~\eqref{Eq:pert} is a perturbational result, connecting the coupling constant at the scale ``$a$'' with the coupling constant at the scale ``$L$''. Within the lore of the renormalization group algorithm\cite{Bog}, the perturbational Eq.~\eqref{Eq:pert} can be used as a \textbf{recursion relation} for building the so called Gell-Mann-Low differential equations. For this purpose, the recursion relation is written as
\begin{eqnarray}
\Gamma(L+dL)-\Gamma(L)&=& \Gamma(L)\cdot \Big(-(p-2)\cdot \frac{a^2}{4\pi^2}\cdot \int_{\frac{2\sqrt{\pi}}{L+dL}}^\frac{2\sqrt{\pi}}{L} d^2k\cdot
\frac{1}{\epsilon(k)}\Big)
\end{eqnarray}
In the specific case
\begin{equation}
\epsilon(k) = 2\cdot \beta\cdot \Gamma(L)\cdot a^2\cdot k^2
\end{equation}
so that
\begin{equation}
\frac{a^2}{4\pi^2}\cdot \int_{\frac{2\sqrt{\pi}}{L+dL}}^\frac{2\sqrt{\pi}}{L} d^2k\cdot \frac{1}{\epsilon(k)}= \frac{1}{4\pi\cdot \beta\cdot \Gamma(L)}\cdot \ln \frac{L+dL}{L}
\end{equation}
and
\begin{equation}
\Gamma(L+dL)-\Gamma(L)= \Gamma(L)\cdot \Big(-(p-2)\frac{1}{4\pi\cdot \beta\cdot \Gamma(L)}\cdot \ln \frac{L+dL}{L}\Big)
\end{equation}
The standard version of the Gell-Mann-Low equation uses the variable
\[\ln \frac{L}{a}\equiv \zeta\]
so that the Gell-Mann-Low equation for the coupling constant $\Gamma(L)$ writes
\begin{equation}
\frac{d\Gamma}{d\zeta}= -\frac{p-2}{4\pi\beta}
\end{equation}
to be solved with the initial condition $\Gamma(\zeta=0) = \Gamma_a$.
The solution of this differential equation (now in the variable $L$) is
\begin{equation}
\Gamma(\xi)= \int_a^\xi \frac{dL'}{L'}\cdot\Big(-\frac{(p-2)}{4\pi\cdot \beta}\Big)+\Gamma_a
\end{equation}
i.e.
\begin{equation}
\Gamma(\xi)= \Gamma_a\cdot \Big(1-\frac{(p-2)}{4\pi\cdot \beta\cdot \Gamma_a} \cdot\ln \frac{\xi}{a}\Big)
\end{equation}
\paragraph{The renormalization of the magnetization.}
To find how the magnetization renormalizes, we add a magnetic field functional by applying a magnetic field along a given direction
($\vec B = \vec t\cdot B$)
\[\underbrace{\frac{g\cdot\mu_B\cdot S\cdot d}{a^3}\cdot B}_{b}\cdot \int_S n(\vec x)\,d^2x\]
$n(\vec x)\equiv \vec n(\vec x)\cdot \vec t$.
We compute the transformation properties for the coupling coefficient $b$:
\begin{eqnarray}
<n>_{\vec \phi} &=& n_{0}\Big(1-\frac{1}{2}<\vec \phi^2>_{\vec \phi}\Big)\nonumber\\
&=&n_{0}\Big(1-\frac{p-1}{2}\cdot\frac{1}{4\pi\beta\Gamma_a}\cdot\ln \frac{L}{a}\Big)
\end{eqnarray}
and
\[\frac{d\ln b(\zeta)}{d\zeta}= -\frac{(p-1)}{2}\cdot \frac{1}{4\pi\beta\Gamma_a\cdot \left(1-(p-2)\cdot \frac{1}{4\pi\beta \Gamma_a}\cdot \zeta\right)}\]
to be solved with the initial condition $b(\zeta=0) = b$.
The solution of this equation is
\begin{equation}
b(\xi) = b\cdot \Big(1-(p-2)\frac{1}{4\pi\beta\cdot \Gamma_a}\cdot \ln\frac{\xi}{a}\Big)^{\frac{(p-1)}{2\cdot (p-2)}}
\end{equation}
By Polyakov renormalization, the length of $\vec n$ is fixed and the applied magnetic field renormalizes. In practical situations, the magnetic field is fixed and the component of $\vec n$ along $\vec t$ renormalizes, accordingly,
as
\begin{equation}
n(L) = n(a)\cdot \Big(1-(p-2)\frac{1}{4\pi\cdot \beta\cdot \Gamma_a}\cdot \ln\frac{L}{a}\Big)^{\frac{(p-1)}{2\cdot (p-2)}}
\end{equation}
This transformation property is also valid if we let the magnetic field go to 0.
\paragraph{The Polyakov renormalization in dimension $D=2+\epsilon$.}
In the argument presented above, the spatial length scale $L$ (the lateral size
of the system) cannot be eliminated and the thermodynamic limit $L\rightarrow
\infty$ is ill defined. There is a formal method that fixes the transition
temperature from the ordered, ferromagnetic phase to the disordered phase at
exactly $T=0$, in line with the Mermin-Wagner theorem. The method is that of
performing the computation of the coupling constant at dimensions $D>2$ and
taking the result to the limit $D\rightarrow 2$.\\
The transformation property for the coupling constant $\Gamma$ must be computed from the recursion relation
\begin{equation}
\Gamma(L_2) = \Gamma(L_1)\Big(1-\frac{(p-2)}{(p-1)}\cdot <\vec \phi^2>_T\Big)
\end{equation}
In a general dimension $D$, there are several geometrical issues that must be taken into account for computing $<\vec \phi^2>_T$.
For example, the infinitesimal element becomes
\begin{equation}
d^Dq \doteq d\Omega^D\cdot q^{D-1}\cdot dq
\end{equation}
and
\begin{eqnarray}
\int d\Omega^D = D\cdot \frac{\pi^{\frac{D}{2}}}{\Gamma(1+\frac{D}{2})}
\end{eqnarray}
We approximate the square Brillouin zone by a sphere. Its radius amounts to
\begin{equation}
k_{BZ}= \frac{2\pi}{a}\cdot \Big[\frac{D\cdot \Gamma(\frac{D}{2})}{2\cdot \pi^{\frac{D}{2}}}\Big]^{\frac{1}{D}}
\end{equation}
By defining $\int \frac{d\Omega^D}{(2\pi)^D}\equiv C(D)$ we obtain
\begin{equation}
k_{BZ} = \frac{1}{a}\cdot\left(\frac{D}{C(D)}\right)^{\frac{1}{D}}
\end{equation}
The integral in $q$ space is between a sphere with radius
\[q_2\equiv \frac{k_{BZ}}{L_2/a}\]
and a sphere with radius
\[q_1\equiv \frac{k_{BZ}}{L_1/a}\]
i.e.
\begin{equation}
<\vec \phi^2>_T= (p-1)\cdot \frac{1}{2\beta\Gamma(L_1)}\cdot C(D)\cdot \int_{q_2}^{q_1} dq\cdot q^{D-3}
\end{equation}
The Gell-Mann-Low equation follows from
\begin{eqnarray}
\Gamma(L'+dL') &=& \Gamma(L')\cdot \Big(1-(p-2)\cdot \frac{1}{2\beta\cdot\Gamma(L')}\cdot C(D)\cdot q^{D-3}\vert_{\frac{k_{BZ}}{L'/a}}\cdot k_{BZ}\cdot \frac{dL'}{L'^2}\Big)\nonumber\\
\frac{d\Gamma}{dL'} &=& -\frac{1}{2}(p-2)\cdot (k_B\cdot T)\cdot \underbrace{D\cdot \left(\frac{C(D)}{D}\right)^{\frac{2}{D}}}_{(2\pi)^{-1} @ D=2}\cdot \frac{1}{L'^{D-1}}
\end{eqnarray}
to be solved with the initial condition
\[\Gamma(L'=a) = \Gamma_a\]
The solution writes
\[\Gamma(L\rightarrow \infty)= \Gamma_a\cdot\left(1-(p-2)\cdot \frac{k_B\cdot T}{4\pi\cdot \Gamma_a}\cdot \frac{1}{D-2}\right)\]
One can define a temperature at which the coupling constant and the magnetization vanish:
\begin{equation}
k_B\cdot T_C  = \frac{4\pi\cdot \Gamma_a}{(p-2)}\cdot (D-2)
\end{equation}
$T_C$ vanishes in the limit $D\rightarrow 2$.

\chapter{A slab model of the reorientation transition.}
\label{Chap:reorientation}

\renewcommand{\thesection}{\thechapter.\arabic{section}}
\renewcommand{\theequation}{\thechapter.\arabic{equation}}

\begin{figure}[H]
\begin{center}
\includegraphics[width=0.35\textwidth]{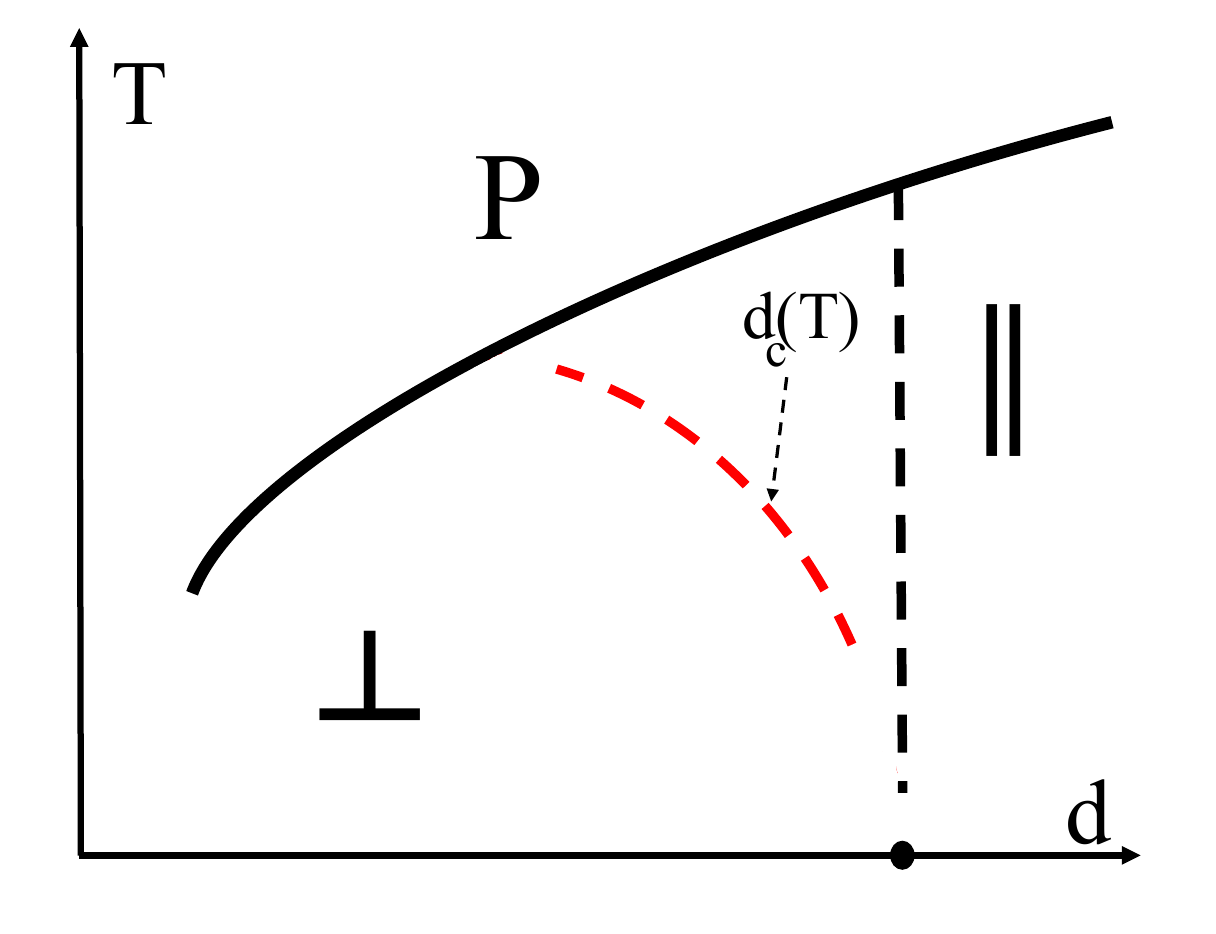}
\end{center}
\caption{Sketch of the transition lines in the temperature (T)-thickness (d) parameter plane, as expected from experimental\cite{Qiu} results and theoretical works\cite{Politi,Shi}.}
\label{Fig:Reo}
\end{figure}

In 1954, Louis N\'eel\cite{Neel} suggested
that the breaking of the translational symmetry perpendicular to the film plane
introduces, in thin magnetic layers, a spin anisotropy energy that could favor the alignment of the magnetization vector perpendicular to the film plane. At that time, it was not possible to compute the strength of magnetic anisotropies from first principles. This situation changed when monolayers of 3d metals became computable by the density functional approach\cite{Freeman}. In 1986, Gay and Richter presented the first self-consistent electronic structure calculation of the spin anisotropy of a ferromagnetic overlayer\cite{GR}. They predicted a N\'eel spin anisotropy energy of about 0.4 meV per atom in the monolayer of Fe. As the strength of the classical dipole
interaction, favoring the in-plane spin orientation, is about 0.3 meV per
atom, the energy balance ought to produce a ground state with perpendicular
magnetization. The perpendicular magnetization was indeed inferred or observed by several experimental works on ultrathin magnetic films, see e.g. Refs.~\cite{Carbone,Pescia_Fe_Cu,Stampa}. These very same experiments reported the vanishing of the perpendicular magnetization with increasing thickness\cite{Carbone} or temperature\cite{Pescia_Fe_Cu,Stampa}. This behavior was interpreted\cite{PP,Tet} as a sign of a possible reorientation transition from perpendicular to
in-plane magnetization. The reorientation transition was experimentally
observed by Pappas et al.\cite{Pappas}. Transition lines in the
temperature-thickness parameter plane were later obtained experimentally\cite{Qiu} and theoretically\cite{Politi,Shi} and revealed that the perpendicular
state could also turn paramagnetic without an intermediate in-plane phase,
depending on the film thickness.\\
The reorientation transition was explained on the basis of the Polyakov renormalization scheme of the coarse grained Heisenberg model, augmented by a N\'eel type anisotropy and the dipole-dipole interaction\cite{Pescia_Fe_Cu,Tet,Abanov,Politi}. These computations and the Monte Carlo simulations of the same model\cite{Shi} differ in the detail but their results can be cast into a conceptually simple statement. The Landau free
energy per surface unit cell contains a term quadratic in the cosine of
$\vartheta$, the angle between the coarse grained magnetization vector and the
film normal:
\begin{equation}
\label{Eq:aniso}
\big[-\lambda(T)+\Omega(T)\!\cdot\!\frac{d}{a}\big]\cdot \cos^2\vartheta
\end{equation}
The coefficient $\lambda(T)$ is the strength of the N\'eel anisotropy and, in some
circumstances\cite{GR}, is positive at $T\!=\!0$ (for the one monolayer of Fe, e.g., $\lambda(T=0)=0.38$ meV). $\Omega(T)$ is the strength of the dipolar energy per surface unit cell and is always positive ($\approx + 0.3$ meV for the one monolayer of Fe). $d$ is the film thickness: it enters linearly the dipolar contribution to the free energy per unit cell, as in the ground state calculations\cite{GR}. The coefficient $\lambda$, instead, is a surface property and is unaffected by the film thickness\cite{GR}. Accordingly, at $T=0$ a critical thickness $d_c=a\cdot\!\frac{\lambda(T=0)}{\Omega(T=0)}$ results at which the quadratic term vanishes\cite{Politi,Shi}. At $d_c$, the magnetization turns from a perpendicular state at $d\!<\!d_c$ to an in-plane state at $d\!>\!d_c$. The question is about how $d_c$ changes at finite temperatures. The works of Refs.~\cite{Pescia_Fe_Cu,Tet,Abanov,Politi,Shi} find that, at finite temperatures, the short wavelength spin wave fluctuations modify the balance between the coefficients of the quadratic term. $d_c$ is found to \textbf{decrease smoothly} with temperature. In Fig.~\ref{Fig:Reo}, a possible phase diagram in the $T-d$ parameter space is drawn. The thick continuous line marks the transition from the ferromagnetic to the paramagnetic state (P, the line drawn freely from\cite{RF}), at typically a few 100 K. The thin lines indicate the transition from perpendicular ($\perp$) to parallel
($\parallel$) magnetization. Along the hypothetical vertical dashed line, the balance between the coefficients of the anisotropic term does not change with temperature. Along the continuous red line (the reorientation transition line $d_c(T)$, drawn freely from \cite{Politi,Shi,Qiu}) the balance is affected by fluctuations ($d_c$, in ultrathin films of Fe, is typically less than $1$ nm). The curvature of the continuous red line allows experiments that work at fixed thickness and variable temperature\cite{Pappas} to observe the reorientation transition before the transition to the paramagnetic state takes place.
\noindent In this Chapter, we apply the RG protocol to an ultrathin slab with the aim of computing the approximate topology of the $d_c(T)$ transition line.

\paragraph{The anisotropic functional.}
The anisotropic component of the total functional writes
\begin{eqnarray}
\label{Eq:LDeltaOmega}
{\cal L}(\Delta,\Omega)&=& -\Delta\cdot \frac{1}{a^2}\int d^2\rho\cdot n_z^2\nonumber\\
&-&\frac{\Omega}{4\pi}\cdot \frac{1}{a^3}\int d^2\rho \int d^2\rho' n_z(\vec \rho)\cdot n_z(\vec \rho')\nonumber\\
&\cdot &\big[\frac{\partial}{\partial x}\frac{\partial}{\partial x'}+\frac{\partial}{\partial y}\frac{\partial}{\partial y'}\big]G(\vec \rho-\vec \rho')\nonumber\\
&+& \frac{\Omega}{4\pi}\cdot \frac{1}{a^3}\cdot \sum_{i,j=x,y} \int d^2\rho \int d^2\rho'\cdot n_i(\vec \rho)\cdot n_j(\vec \rho')\cdot\nonumber\\
&\cdot &  \frac{\partial}{\partial i}\frac{\partial}{\partial j'}G(\vec \rho-\vec \rho')
\end{eqnarray}
with
\begin{equation}
\label{Eq:G}
G(\vec \rho-\vec \rho')\dot=\frac{1}{2\pi}\cdot \int d^2k\cdot f(k)\cdot e^{i\vec k\cdot(\vec \rho-\vec \rho')}
\end{equation}
The first line is the local anisotropy of Eq.~\eqref{Eq:aniso}: it collects the
N\'eel anisotropy and the local part of the dipolar interaction, so that
\begin{equation}
\label{Eq:Delta}
\Delta \doteq \lambda-\Omega\cdot\frac{d}{a}
\end{equation}
The remaining two lines are the non-local part of the dipolar functional
established in Chapter 5.

\paragraph{The renormalization of the ${\cal L}(\Delta)$-functional.}
The kernel of the ${\cal L}(\Delta)$-functional writes, in the Polyakov parametrization,
\begin{equation}
n_z^2 \!=\! n^2_{0_z}\big(1-(\vec\phi)^2\big) \!+\! \sum_{a,b} e_{a_z}\cdot
e_{b_z}\cdot \phi_a\cdot \phi_b \!+\! 2\cdot n_{0_z}\sqrt{1- (\vec \phi)^2}\cdot \sum_a\phi_a\cdot e_{a_z}
\end{equation}
and the effective kernel for the field $n_{0_z}$ writes, to first order,
\begin{equation}
n^2_{0_z}\big(1-< (\vec\phi)^2>_{\phi}\big)\!+\! \sum_{a,b} e_{a_z}\cdot
e_{b_z}\cdot <\phi_a\cdot \phi_b>_{\phi} \!+\! 2\cdot n_{0_z}\cdot \sum_a e_{a_z}\underbrace{<\sqrt{1-(\vec \phi)^2}\cdot \phi_a>_{\phi}}_{0,\text{uneven power}}
\end{equation}
Recall that
\begin{equation}
<\phi_a\phi_b(\vec x)>_{\phi}= \delta_{ab}\cdot \frac{1}{p-1}<(\vec \phi)^2>_{\phi}
\end{equation}
so that the effective kernel simplifies to
\begin{equation}
n^2_{0_z}\big(1-<(\vec \phi)^2>_{\phi}\big)+ \frac{1}{p-1}<(\vec \phi)^2>_{\phi}\cdot \sum_{a} e_{a_z}\cdot e_{a_z}
\end{equation}
We need to evaluate
\[\sum_{a} e_{a_z}(\vec x)\cdot e_{a_z}(\vec x)\]
The triple of vectors $(\vec n_0, \vec e_1,\vec e_2)$ arises from the triple of vectors $(\vec e_x,\vec e_y,\vec e_z)$ by an orthogonal matrix transformation:
\begin{equation}
\begin{pmatrix}\vec n_0\\ \vec e_1\\\vec e_2\end{pmatrix}= \begin{pmatrix}n_{0_x}&n_{0_y}&n_{0_z}\\
e_{1_x}&e_{1_y}&e_{1_z}\\
e_{2_x}&e_{2_y}&e_{2_z}\end{pmatrix}\begin{pmatrix}\vec e_x\\ \vec e_y\\\vec e_z\end{pmatrix}
\end{equation}
The orthogonality of this matrix means that multiplying the transpose of this matrix with the matrix itself gives the unit matrix:
\begin{equation}
\begin{pmatrix}n_{0_x}&e_{1_x}&e_{2_x}\\
n_{0_y}&e_{1_y}&e_{2_y}\\
n_{0_z}&e_{1_z}&e_{2_z}
\end{pmatrix}\begin{pmatrix}n_{0_x}&n_{0_y}&n_{0_z}\\
e_{1_x}&e_{1_y}&e_{1_z}\\
e_{2_x}&e_{2_y}&e_{2_z}\end{pmatrix}= \begin{pmatrix}1&0&0\\0&1&0\\0&0&1\end{pmatrix}
\end{equation}
From the matrix multiplication one can read out the relation, generalized to a $p$-component field,
\begin{equation}
\label{Eq:ortho}
\sum_a e_{u_a}\cdot e_{v_a} = \delta_{uv}-n_{0_u}\cdot n_{0_v}\quad\quad\text{in particular}\quad \sum_a e^2_{a_z} = 1-n^2_{0_z}
\end{equation}
and the effective kernel writes
\begin{equation}
n^2_{0_z}\big(1-<{\vec \phi}^2>_{\phi}\big)\!+\!\frac{1}{p-1}<(\vec \phi)^2>_{\phi}\cdot (1-n^2_{0_z})
\end{equation}
Neglecting those components that do not contain the field $n_{0_z}$ we obtain for the kernel of ${\cal L}_0(\Delta)$
\begin{equation}
 n^2_{0_z}\cdot \big(1-\frac{p}{p-1}\cdot <{\vec \phi}^2>_{\phi}\big)
\end{equation}
We recall that
\begin{equation}
<{\vec \phi}^2>_{\phi}= (p-1)\cdot <\phi^2>_{\phi}
\end{equation}
and that
\begin{equation}
\label{Eq:corrloc}
<\phi(\vec \rho)\phi(\vec \rho)>_{\vec \phi}= \frac{a^2}{4\pi^2}\int d^2k\,\frac{1}{\epsilon(k)}
\end{equation}
$\epsilon(k)$ is the spectrum of fluctuations. We assume that the symmetry breaking interactions are weak so that we can set
\begin{equation}
\label{Eq:eps}
\epsilon(k) = 2\cdot \beta\cdot A\cdot d\cdot a^2\cdot k^2
\end{equation}
Accordingly, ${\cal L}_0(\Delta)[n_{0_z}]$ writes
\begin{equation}
{\cal L}_0(\Delta)= -\Delta\cdot \Big(1-p\cdot \frac{1}{4\pi\cdot \beta\cdot A\cdot d}\cdot \ln \frac{L}{a}\Big)\cdot \frac{1}{a^2}\int d^2\rho\, n^2_{0_z}
\end{equation}
We observe that ${\cal L}_0(\Delta)[n_{0_z}]$ is formally identical to ${\cal L}_0(\Delta)[n_{z}]$, except for the coupling constant $\Delta$, which obeys the recursion relation (we use again $A\cdot d\doteq \Gamma$)
\begin{equation}
\Delta(L) = \Delta(a)\cdot \Big(1-p\cdot \frac{1}{4\pi\cdot \beta\cdot \Gamma}\cdot \ln \frac{L}{a}\Big)
\end{equation}
The recursion relation is translated into a Gell-Mann-Low equation by the rule
\begin{eqnarray}
\Delta(L+dL)& = & \Delta(L)\cdot \Big(1-p\cdot \frac{1}{4\pi\cdot \beta\cdot \Gamma(L)}\cdot \ln \frac{L+dL}{L}\Big)
\end{eqnarray}
or, using the variable $\zeta\doteq \ln \frac{L}{a}$,
\begin{eqnarray}
\frac{d \ln\Delta(\zeta)}{d\zeta}& = & \frac{-p}{4\pi\cdot \beta\cdot \Gamma(a)\cdot \big(1-\frac{p-2}{4\pi\beta\Gamma(a)}\cdot \zeta\big)}
\end{eqnarray}
with solution
\begin{equation}
\label{Eq:Deltasol}
\Delta(\zeta)= \Delta(a)\cdot \big(Z(\zeta)\big)^{\frac{p}{p-2}}
\end{equation}
where we have introduced
\begin{eqnarray}
\label{Eq:Z}
Z(\zeta) &\doteq & 1- \frac{p-2}{4\pi\beta\Gamma(a)}\cdot \zeta\nonumber\\
Z(L)&=& 1- \frac{p-2}{4\pi\beta\Gamma(a)}\cdot \ln\frac{L}{a}
\end{eqnarray}

\paragraph{The renormalization of ${\cal L}_\Omega[\vec n]$.}
We need to compute the transformation properties of the non-local products
\begin{equation}
n_u(\vec \rho)\cdot n_v(\vec \rho')
\end{equation}
which are common to all components of the dipolar functional. Inserting the Polyakov parametrization we obtain (linear terms in $\phi_a$ and $\phi_b$ are dropped as their average will turn out to vanish)
\begin{eqnarray}
n_u(\vec \rho)\cdot n_v(\vec \rho') &=& n_{u_0}(\vec \rho)\cdot n_{v_0}(\vec \rho')\underbrace{\left(\sqrt{1-\vec \phi^2(\vec \rho)}\right)\cdot \left(\sqrt{1-\vec \phi^2(\vec \rho')}\right)}_{1-\frac{1}{2}\cdot \vec \phi^2(\vec \rho)-\frac{1}{2}\cdot \vec \phi^2(\vec \rho')}\nonumber\\
&+& \sum_{a,b} e_{u_a}(\vec \rho)\cdot e_{v_b}(\vec\rho')\cdot \phi_a(\vec \rho)\cdot \phi_b(\vec \rho')
\end{eqnarray}
The smoothing operation applied to these components leads to
\begin{eqnarray}
<n_u(\vec \rho)\cdot n_v(\vec \rho')>_{\vec \phi} &=& n_{u_0}(\vec \rho)\cdot n_{v_0}(\vec \rho')\Big(1-<\vec \phi^2(\vec \rho)>_{\vec \phi}\Big)\nonumber\\
&+&<\phi(\vec \rho)\cdot \phi(\vec \rho')>_{\vec \phi}\cdot \sum_{a} e_{u_a}(\vec \rho)\cdot e_{v_a}(\vec \rho')
\end{eqnarray}
We have replaced $<\phi_a\cdot\phi_a>$ with a multiplicative prefactor $<\phi(\vec \rho)\cdot \phi(\vec \rho')>_{\vec \phi}$ as we expect the smoothing result for the correlation function to be independent of $a$. We observe the appearance of one non-local product
\begin{equation}
n_{u_0}(\vec \rho)\cdot n_{v_0}(\vec \rho')\Big(1-<\vec \phi^2(\vec \rho)>_{\vec \phi}\Big)
\end{equation}
that produces the components ``A'' and ``B'' of the total dipolar functional:\\
Component ``A'' writes
\begin{eqnarray}
&-&\frac{\Omega}{4\pi}\cdot \frac{1}{a^3}\cdot \big(1-<\vec \phi^2(\vec \rho)>_{\vec \phi}\big)\int d^2\rho \int d^2\rho' n_{0_z}(\vec \rho)\cdot n_{0_z}(\vec \rho')\nonumber\\
&\cdot &\big[\frac{\partial}{\partial x}\frac{\partial}{\partial x'}+\frac{\partial}{\partial y}\frac{\partial}{\partial y'}\big]G(\vec \rho-\vec \rho')
\end{eqnarray}
Component ``B'' writes
\begin{eqnarray}
&+&\frac{\Omega}{4\pi}\cdot \frac{1}{a^3}\cdot \big(1-<\vec \phi^2(\vec \rho)>_{\vec \phi}\big)\cdot \sum_{i,j=x,y} \int d^2\rho \int d^2\rho'\cdot n_{0_i}(\vec \rho)\cdot n_{0_j}(\vec \rho')\cdot\nonumber\\
&\cdot &  \frac{\partial}{\partial i}\frac{\partial}{\partial j'}G(\vec \rho-\vec \rho')
\end{eqnarray}
Components ``A'' and ``B'' carry formally the same functional dependence as the
Lagrangian of the field $\vec n$. The coupling constant is renormalized by the
spin waves with wavelength between $a$ and $L$, according to
\begin{equation}
\Omega(L) = \Omega(a)\cdot \Big(1-<\vec \phi^2(\vec \rho)>_{\vec \phi}\Big)
\end{equation}
We use the spectrum of fluctuations $\epsilon(k) = 2\cdot \beta\cdot \Gamma\cdot a^2\cdot k^2$ to evaluate $<\vec \phi^2(\vec \rho)>$ and obtain
\begin{equation}
\Omega(L) = \Omega(a)\cdot \Big(1-(p-1)\cdot\frac{1}{4\pi\cdot \beta\cdot \Gamma}\cdot \ln\frac{L}{a}\Big)
\end{equation}
which is transformed into a Gell-Mann-Low equation by the rule
\begin{equation}
\Omega(L+dL) = \Omega(L)\cdot \Big(1-(p-1)\cdot\frac{1}{4\pi\cdot \beta\cdot \Gamma(L)}\cdot \ln\frac{L+dL}{L}\Big)
\end{equation}
i.e.
\begin{equation}
\frac{d\ln\Omega(\zeta)}{d\zeta} = -(p-1)\cdot \frac{1}{4\pi\beta\cdot \Gamma(\zeta)}
\end{equation}
with solution
\begin{equation}
\label{Eq:Omegasol}
\Omega(\zeta)=\Omega(a)\cdot Z^{\frac{p-1}{p-2}}(\zeta)
\end{equation}
The renormalization stops at the correlation length $\xi$; we will discuss it later. It is convenient to introduce the corresponding value of the RG variable,
\begin{equation}
\label{Eq:zetaxi}
\zeta_\xi \doteq \ln\frac{\xi}{a}
\end{equation}
The smoothing operation produces a further non-local product:
\begin{equation}
\sum_{a} e_{u_a}(\vec \rho)\cdot e_{v_a}(\vec \rho')\cdot <\phi_a(\vec \rho)\cdot \phi_a(\vec \rho')>_{\vec \phi}
\end{equation}
This product can only be dealt with using some kind of approximation (see Ref.~\cite{Politicomment} for a possible method).
A possible strategy to deal with this non-local term is to approximate
\begin{equation}
e_{u_a}(\vec \rho)\cdot e_{v_a}(\vec \rho')
\end{equation}
by
\begin{equation}
 e_{u_a}(\vec \rho)\cdot e_{v_a}(\vec \rho)
\end{equation}
By this approximation, only the product of the transversal vectors of fluctuations is rendered local. The amplitude of the fluctuations, instead, is kept non-local. One then uses the orthogonality relation~\eqref{Eq:ortho} to write
\begin{equation}
\sum_a e_{u_a}(\vec \rho)\cdot e_{v_a}(\vec \rho) = \delta_{uv}-n_{u_0}(\vec \rho)\cdot n_{v_0}(\vec \rho)
\end{equation}
and, dropping the field independent contribution,
\begin{eqnarray}
\sum_{a} e_{u_a}(\vec \rho)\cdot e_{v_a}(\vec \rho')\cdot <\phi_a(\vec \rho)\cdot \phi_a(\vec \rho')>_{\vec \phi}=-n_{u_0}(\vec \rho)\cdot n_{v_0}(\vec \rho)\cdot <\phi(\vec \rho)\cdot \phi(\vec \rho')>_{\vec \phi}
\end{eqnarray}
For performing the smoothing operation, we use the formula
\begin{equation}
\label{Eq:corrnonloc}
<\phi(\vec \rho)\phi(\vec \rho')>_{\vec \phi}= \frac{a^2}{4\pi^2}\int d^2k\,\frac{1}{\epsilon(k)}\cdot e^{i\cdot \vec k\cdot(\vec \rho-\vec \rho')}
\end{equation}
Component ``C'' writes
\begin{eqnarray}
&+&\frac{\Omega}{4\pi}\cdot \frac{1}{a^3}\int d^2\rho \int d^2\rho' n_{0_z}(\vec \rho)\cdot n_{0_z}(\vec \rho)\nonumber\\
&\cdot & \frac{a^2}{4\pi^2}\int d^2q\,\frac{1}{\epsilon(q)}\cdot e^{i\cdot \vec q\cdot(\vec \rho-\vec \rho')}\nonumber\\
&\cdot &\big[\frac{\partial}{\partial x}\frac{\partial}{\partial x'}+\frac{\partial}{\partial y}\frac{\partial}{\partial y'}\big]G(\vec \rho-\vec \rho')
\end{eqnarray}
Component ``D'' writes
\begin{eqnarray}
&-&\frac{\Omega}{4\pi}\cdot \frac{1}{a^3}\cdot \sum_{i,j=x,y} \int d^2\rho \int d^2\rho'\cdot n_{0_i}(\vec \rho)\cdot n_{0_j}(\vec \rho)\cdot\nonumber\\
&\cdot & \frac{a^2}{4\pi^2}\int d^2q\,\frac{1}{\epsilon(q)}\cdot e^{i\cdot \vec q\cdot(\vec \rho-\vec \rho')}\nonumber\\
&\cdot &  \frac{\partial}{\partial i}\frac{\partial}{\partial j'}G(\vec \rho-\vec \rho')
\end{eqnarray}
The functionals ``C'' and ``D'' contain
the product of components of the field $\vec n_0$ at the same in-plane
coordinate, i.e. the two components are local in the field. This is a
consequence of our approximation and violates, in principle, the
renormalizability of the dipolar functional. We will show how to deal with the appearance of local terms, produced by the smoothing operation itself.\\
We evaluate first component ``C''. We use now
\begin{equation}
G(\vec \rho-\vec \rho')=\frac{1}{2\pi}\cdot \int d^2k\cdot f(k)\cdot e^{i\vec k\cdot(\vec \rho-\vec \rho')}
\end{equation}
We notice that the derivatives with respect to $x$
and $x'$, applied to $G(\vec \rho-\vec \rho')$, lead to a multiplicative factor $i\cdot k_x\cdot (-i)\cdot k_x$. The derivatives with respect to $y$ and $y'$ add a similar factor $k_y^2$. The net result is a multiplicative factor $k^2$, so that ``C'' writes
\begin{eqnarray}
&+&\frac{\Omega}{4\pi}\cdot \frac{1}{a^3}\int d^2\rho \int d^2\rho' n^2_{0_z}(\vec \rho)\nonumber\\
&\cdot & \frac{a^2}{4\pi^2}\int d^2q\,\frac{1}{\epsilon(q)}\cdot e^{i\cdot \vec q\cdot(\vec \rho-\vec \rho')}\nonumber\\
&\cdot &\frac{1}{2\pi}\cdot \int d^2k\cdot k^2\cdot f(k)\cdot e^{i\vec k\cdot(\vec \rho-\vec \rho')}
\end{eqnarray}
As the product of the fields only depends on $\vec \rho$, the integral over $\vec \rho'$ can be performed:
\begin{equation}
\int d^2\rho'\, e^{i\cdot \vec q\cdot(\vec \rho-\vec \rho')}\cdot e^{i\vec k\cdot(\vec \rho-\vec \rho')} = (2\pi)^2\cdot \delta(\vec k + \vec q)
\end{equation}
Next we have
\begin{eqnarray}
\int d^2q\int d^2k \frac{1}{\epsilon(q)}\cdot k^2\cdot f(k)\cdot \delta(\vec k + \vec q)\!
=\! \int d^2q \frac{1}{\epsilon(q)}\cdot q^2\cdot f(q)
\end{eqnarray}
i.e. the delta function ``synchronizes'' the $q$-vector arising from the smoothing with the $k$-vector originating from the kernel of the dipolar interaction.
Gathering all the remaining pieces, ``C'' writes
\begin{eqnarray}
&+&\frac{\Omega}{4\pi}\cdot \frac{a^2}{(2\pi)^2}\cdot (2\pi)^2\cdot \frac{1}{2\pi}\cdot\frac{1}{a}\int d^2q \frac{1}{\epsilon(q)}\cdot q^2\cdot f(q)\nonumber\\
&\cdot &\frac{1}{a^2}\int d^2\rho\cdot  n^2_{0_z}(\vec \rho)
\end{eqnarray}
The smoothing operation has generated a new term which contributes to the local N\'eel anisotropy functional. Using $\epsilon(q)= 2\cdot\beta\cdot \Gamma\cdot a^2\cdot q^2$ and $f(q)=2\big(\frac{d}{q^2}+\frac{e^{-q d}-1}{q^3}\big)$, the coefficient -- let us call it $A_\perp$ -- of this term writes
\begin{equation}
A_\perp = +\Omega\cdot \frac{1}{4\pi\cdot \beta \cdot\Gamma}\cdot\big[\frac{1}{2\pi}\cdot \int_0^{2\pi}d\omega\big]\cdot \frac{1}{a}\int dq  \big[\frac{d}{q} + \frac{e^{-q\cdot d}-1}{q^2}\big]
\end{equation}
We have evidenced the integral over the angle $\omega$ in the $\vec q$-plane because this integral produces a different result in component ``D''. We also need to specify the domain of integration for the variable $q$. For building the differential equation that transforms $A_\perp$ from the scale $L$ to the scale $L+dL$ we use the limits of integration $\frac{k_{BZ}\cdot a}{L+dL}$ and $\frac{k_{BZ}\cdot a}{L}$. Assuming that $L$ is larger than $d$, one can also replace
\begin{equation}
\big[\frac{d}{q} + \frac{e^{-q\cdot d}-1}{q^2}\big]
\end{equation}
with the lowest term in its Taylor expansion, which amounts to $\frac{d^2}{2}$.
Using $k_{BZ} = \frac{2\cdot \sqrt{\pi}}{a}$, the integral over $q$ gives then
$d^2\cdot \sqrt{\pi}\cdot \frac{dL}{L^2}$, so that the differential equation for $A_\perp(L)$ writes
\begin{equation}
\frac{d A_\perp(L)}{dL}=+\Omega(L)\cdot \frac{d}{a} \cdot \frac{1}{4\pi\cdot \beta \cdot\Gamma(L)}\cdot\big[\sqrt{\pi}\cdot \frac{1}{2\pi}\cdot
\int_0^{2\pi}d\omega\big]\cdot \frac{d}{a}\cdot \frac{a}{L^2}
\end{equation}
In the variable $\zeta$ the differential equation writes
\begin{equation}
\frac{d A_\perp(\zeta)}{d\zeta}=+\Omega(\zeta)\cdot \frac{d}{a} \cdot \frac{1}{4\pi\cdot \beta \cdot\Gamma(\zeta)}\cdot\big[\sqrt{\pi}\cdot \frac{1}{2\pi}\cdot
\int_0^{2\pi}d\omega\big]\cdot \frac{d}{a}\cdot e^{-\zeta}
\end{equation}
to be solved with the initial condition
\begin{equation}
A_\perp(\zeta=0) = 0
\end{equation}
Component ``D'' involves the monomials $n^2_{0_x}$, $n^2_{0_y}$, $n_{0_x}\cdot n_{0_y}$ and $n_{0_y}\cdot n_{0_x}$. The in-plane derivatives are redirected in $\vec k$ space and produce the four products $k_x^2=k^2\cdot \cos^2\omega$, $k_y^2=k^2\cdot \sin^2\omega$, $k_x\cdot k_y=k_y\cdot k_x=k^2\cdot \cos\omega\cdot \sin\omega$. When the angular integration in $\vec q$ space is considered, one obtains the following multiplicative numbers:
\begin{equation}
\frac{1}{2\pi}\int_0^{2\pi} \cos^2 \omega\cdot d\omega \quad \frac{1}{2\pi}\int_0^{2\pi} \sin^2 \omega\cdot d\omega \quad \frac{1}{2\pi}\int_0^{2\pi} \cos \omega\cdot \sin\omega\, d\omega
\end{equation}
The coefficient of the dipolar functional containing the monomials $n_{0_y}\cdot n_{0_x}$ vanishes by integration over the angular variable.
The coefficients of the dipolar functional (let us call them $A_\parallel$) containing the monomials $n^2_{0_x}$ and $n^2_{0_y}$ obey the differential equation
\begin{equation}
\frac{d A_\parallel(\zeta)}{d\zeta}=+\Omega(\zeta)\cdot \frac{d}{a} \cdot \frac{1}{4\pi\cdot \beta \cdot\Gamma(\zeta)}\cdot\big[\sqrt{\pi}\cdot\frac{1}{2\pi}\int_0^{2\pi} \cos^2 \omega\cdot d\omega\big]\cdot \frac{d}{a}\cdot e^{-\zeta}
\end{equation}
to be solved with the initial condition
\begin{equation}
A_\parallel(\zeta=0) = 0
\end{equation}
We insert, for $p=3$,
\begin{equation}
\Omega(\zeta) = \Omega(a)\cdot Z^2(\zeta)\quad \Gamma(\zeta) = \Gamma(a)\cdot Z(\zeta)
\end{equation}
and perform the angular integration. The differential equations write
\begin{eqnarray}
\frac{d A_\perp(\zeta)}{d\zeta}&=&+\Omega(a)\cdot \frac{d}{a} \cdot \frac{1}{4\pi\cdot \beta \cdot\Gamma(a)}\cdot\sqrt{\pi}\cdot \frac{d}{a}\cdot Z(\zeta)\cdot e^{-\zeta}\nonumber\\
\frac{d A_\parallel(\zeta)}{d\zeta}&=&+\Omega(a)\cdot \frac{d}{a} \cdot \frac{1}{4\pi\cdot \beta \cdot\Gamma(a)}\cdot\frac{\sqrt{\pi}}{2}\cdot \frac{d}{a}\cdot Z(\zeta)\cdot e^{-\zeta}
\end{eqnarray}
The integration of these equations leads to an elementary integral:
\begin{equation}
\int_0^{\zeta_\xi} d\zeta'\,Z(\zeta')\cdot e^{-\zeta'}= 1-K+ e^{-\zeta_\xi}\cdot \Big(-1+K+K\cdot \zeta_\xi\Big)\quad\quad K\doteq \frac{1}{4\pi\beta\Gamma(a)}
\end{equation}
and the solutions
\begin{eqnarray}
A_\perp(\zeta_\xi)&=&+\Omega(a)\cdot \frac{d}{a} \cdot K\cdot\sqrt{\pi}\cdot \frac{d}{a}\cdot \Big[1-K+ e^{-\zeta_\xi}\cdot \Big(-1+K+K\cdot \zeta_\xi\Big)\Big]\nonumber\\
A_\parallel(\zeta_\xi)&=&+\Omega(a)\cdot \frac{d}{a} \cdot K\cdot\frac{\sqrt{\pi}}{2}\cdot \frac{d}{a}\cdot \Big[1-K+ e^{-\zeta_\xi}\cdot \Big(-1+K+K\cdot \zeta_\xi\Big)\Big]
\end{eqnarray}
In the Mean Field Approximation of the anisotropic functional in the field $\vec n_0$, one obtains the Gibbs free energy density by inserting into the functional the spin configuration
\begin{equation}
\vec n_0 (\vec \rho) = \big(\sin\vartheta,0,\cos \vartheta\big)
\end{equation}
When inserted into components ``A'' and ``B'', and after redirecting the
derivatives, one observes that such a spatially uniform spin distribution
renders the contribution of the components ``A'' and ``B'' to the MFA Gibbs
free energy vanishing in the limit of an in-plane infinite slab. Using
$n^2_{0_x}+n^2_{0_y}=1-\cos^2\vartheta$, so that the in-plane term ``D''
contributes to the coefficient of $\cos^2\vartheta$ with the opposite sign to
its own, the Gibbs free energy density then writes
\begin{equation}
\label{Eq:GibbsMFA}
\Big[-\Delta(\zeta_\xi)+ A_\perp(\zeta_\xi)+ A_\parallel(\zeta_\xi)\Big]\cdot \cos^2\vartheta
\end{equation}
A transition line in the $d-T$ parameter space is defined by the vanishing of the coefficient of $\cos^2\vartheta$.
When only those terms of the coefficient are considered that are linear in $\frac{1}{\beta}$, one obtains an equation of the type
\begin{equation}
\label{Eq:quad}
{\cal A}\cdot \Big(\frac{d}{a}\Big)^2 + {\cal B}\cdot \frac{d}{a} + {\cal C} = 0
\end{equation}
With the constants
\begin{equation}
K= \frac{1}{4\pi\cdot \beta \cdot\Gamma(a)}\quad M\doteq 1- 3\cdot K\cdot \zeta_\xi
\end{equation}
($M$ is $Z^{\frac{p}{p-2}}=Z^3$ of Eq.~\eqref{Eq:Deltasol} linearized in $K$, at $p=3$)
the coefficients ${\cal A}$, ${\cal B}$ and ${\cal C}$ write
\begin{equation}
{\cal A}= \Omega(a)\cdot K\cdot \frac{3\sqrt{\pi}}{2}\quad {\cal B} = \Omega(a)\cdot M \quad {\cal C}= -\lambda\cdot M
\end{equation}
${\cal B}$ and ${\cal C}$ follow from $-\Delta(\zeta_\xi)=-\big(\lambda-\Omega(a)\frac{d}{a}\big)M$, i.e. from the definition~\eqref{Eq:Delta} of $\Delta$, and ${\cal A}$ from $A_\perp+A_\parallel$, whose bracket has been set to $1$ to this order.
The solutions of the equation write
\begin{equation}
\Big(\frac{d}{a}\Big)_{1,2} = \frac{-\Omega(a)\cdot M\pm\sqrt{(\Omega(a)\cdot M)^2 + 4\big(\Omega(a)\cdot K\cdot \frac{3\sqrt{\pi}}{2}\big)\cdot \lambda\cdot M}}{2\cdot \Omega(a)\cdot K \cdot \frac{3\sqrt{\pi}}{2}}
\end{equation}
The first solution ($+$) has a Taylor expansion for small $K$
\begin{equation}
\label{Eq:dcexp}
\Big(\frac{d}{a}\Big)_{1}= \frac{\lambda}{\Omega(a)}-\frac{3\sqrt{\pi}\lambda^2}{2\Omega^2(a)}\cdot K + \frac{9\sqrt{\pi}\lambda^2}{2\Omega^2(a)}\cdot\Big(\frac{\sqrt{\pi}\cdot \lambda}{\Omega(a)}-\zeta_\xi\Big)\cdot K^2 + {\cal O}(K^3)
\end{equation}
that gives the exact value $\frac{\lambda}{\Omega(a)}$ for $K=0$. Viewed in the $T-d$ parameter plane, the transition line decreases linearly with temperature from the ground state value of $\frac{d}{a}$. Provided the correlation length is large enough, so that $\zeta_\xi>\sqrt{\pi}\lambda/\Omega(a)$ (a limit which is also intrinsic to the Polyakov RG), the curvature is negative, in accordance with the topology expected from the experimental results\cite{Qiu} and the Monte Carlo simulations\cite{Shi}.

\chapter{Phase transitions of the second order in epitaxial ultrathin films.}
\label{Chap:2Dphase}
\renewcommand{\thesection}{\thechapter.\arabic{section}}
\renewcommand{\theequation}{\thechapter.\arabic{equation}}

Phase transitions of the second order were systematically studied in a seminal
paper by Landau\cite{Landau_phase}. This paper established a fundamental
knowledge on this topic that provided the basis for the research on
experimental and theoretical phase transition in matter. The subject was also
strongly influenced by a remarkable  result of statistical physics, the exact
solution of the 2D Ising model\cite{Ising} by Onsager\cite{Ons}. The exact
solution showed both the originality and the limits of the Landau theory of
second order phase transitions and started a long search for a more compelling
picture, in line with experimental results. This search culminated in the
formulation of the scaling hypothesis\cite{Scale,Poki} in the 1960's and the
subsequent masterpiece on the topic, Wilson's renormalization
group\cite{Wilson}. The renormalization group method provided the long
sought-for correction to the Landau theory. Before approaching the scaling
hypothesis and Wilson's algorithm, we therefore give a short review of Landau
method.
\section{Landau functionals and Landau theory.}
One starts with an Hamilton operator describing the system -- typically a quantum mechanical one. Let us think, to be specific, of the Heisenberg operator\footnote{We use, for operators,  calligraphic symbols, as we use for functionals. The context will clarify which quantity we are dealing with.} with spin $S$, defined onto a lattice:
\begin{equation}
{\cal H} = -\frac{1}{2}\cdot \sum_{<i,j>}J_{ij} \vec S_{i}\cdot \vec S_{j} +
g\cdot \mu_B\cdot \sum_i \vec B_i\cdot \vec S_{i}
\end{equation}
The degrees of freedom are the quantum mechanical vector operators $\vec S_{i}$ (here to be understood as being dimensionless operators with eigenvalues $S_z = -S,...,+S$). We seek to compute the Gibb's free energy for the system described by this operator, defined as
\begin{equation}
G({\cal H})\doteq -\beta^{-1}\ln \textbf{Tr}\big(e^{-\beta\cdot {\cal H}}\big)
\end{equation}
The problem is that the partition function
\begin{equation}
\textbf{Tr}\big(e^{-\beta\cdot {\cal H}_I}\big)
\end{equation}
of the original operator can only be computed, for most cases, with some
approximations. One possible approximation is the so called Mean Field
Approximation. The MFA allows to simulate the original operator  by means of an
effective \textbf{classical} Lagrangian (or, ''Landau functional''). The
''Landau functional'' is defined on a set of parameters -- the so called order
parameter field -- associated to each site $i$ of the lattice or, to a further
approximation, to each coordinate $\vec x$ of an Euclidean space. The Landau
functional is a completely different object than the original quantum
Hamiltonian, but its partition function is believed to be a ''good''
approximation of the partition function of the original Hamiltonian.\footnote{We include more details about how to construct Landau functionals and how to use them in statistical physics in Appendix \ref{App:MFA_2D}.} One possible Landau functional that has been used to simulate the original Heisenberg Hamiltonian is the one appearing in the Non Linear Sigma Model\cite{Dupuis}. A further Landau functional that is often used as a replacement of the Heisenberg and Ising model to study, from a quite general point of view, a potential phase transition in any system is the one referred to, in the literature, as ''Landau-Ginzburg-Wilson Hamiltonian''. For a slab of thickness $d$ with underlying simple cubic lattice (number of nearest neighbours: $z=6$) one finds (Appendix \ref{App:MFA_2D})
\begin{eqnarray}
{\cal L}[\vec m]\approx\!&\!+\!&\!A\cdot d\cdot \int d^2x\cdot
\Big(\sum_{\alpha=x,y}\Big[(\frac{\partial m_x}{\partial \alpha})^2 + (\frac{\partial m_y}{\partial \alpha})^2+(\frac{\partial m_z}{\partial \alpha})^2\Big]\Big)\nonumber\\
&+& \!\frac{d}{a}\cdot \frac{1}{a^2}\cdot\int d^2x\!\cdot\Big[\frac{r}{2}\cdot \vec m(\vec x)^2+\frac{u}{4}\cdot \Big(\big(\vec m(\vec x)\big)^2\Big)^2+\mu\cdot \vec B(\vec x)\cdot \vec m(\vec x)\Big]\nonumber\\
&\pm& {\cal O}(\vec m(\vec x)^6)
\end{eqnarray}
The coefficients $A,r,u,\mu$ can be computed within the Mean Field Approximation from the microscopic coupling constants:
\begin{eqnarray}
A&\doteq &\frac{J}{2}\cdot S^2\cdot \frac{1}{a}\nonumber\\
r&\doteq & 3\cdot \frac{S}{(S+1)}\cdot \Big(\frac{1}{\beta}-\frac{1}{\beta_c}\Big)\nonumber\\
u&\doteq &\frac{1}{\beta}\cdot \frac{9}{10}\cdot \frac{S\cdot (S+1)\cdot (2S^2+2S+1)}{(S+1)^4}\nonumber\\
\mu &\doteq & g\cdot\mu_B\cdot S\nonumber\\
(\beta_c)^{-1} &=& 4\cdot A\cdot a
\end{eqnarray}
COMMENTS.\\
1. The degrees of freedom are represented, in this functional, by the order parameter vector field $\vec m(\vec x)$ --  a classical vector with real numbers as components. In contrast to the non-linear sigma model of the Heisenberg Hamiltonian, $\vec m$ does not need to have unit length. In fact, as a consequence of the approximation used in obtaining the functional, the components need to be considered ''small''.\\
2. The coefficients of the gradient-component and of the Zeeman energy are
directly related to the parameters entering the original quantum Hamiltonian.
The appearance of a temperature $\beta^{-1}$ (in the coefficient $r$) and of a
further parameter $u$ in the functional is related to the variational method used to obtain the Lagrangian starting from the original Hamiltonian. In this method, the Lagrangian is seen as a ''trial Free energy functional'' that associate to a given distribution $[\vec m(\vec x)]$ a scalar that depends on the magnetic field and on the temperature.\\
3. A striking property of these coefficients is the linear vanishing of $r$ at a temperature $T_c$. This establish the special point $(T_c,0)$ in the $T-B$-phase diagram, called the Curie point. Typical values of $T_c$ are $1043$ K for bulk Fe, $1395$ K for bulk Co, $629$ K for bulk Ni and $289$ K for bulk Gd. The significance of this point will be elucidated in the next comment.\\
4. Based on the variational principle underlying the MFA approximation, Landau\cite{Landau_29} formulated a principle according to which, among the possible configurations $[\vec m(\vec x)]$, the equilibrium one $[\vec m_B(\vec x)]$ is the one that solve the Euler equation associated with the functional (set the functional derivative to zero).
Once $[\vec m_B(\vec x)]$ has been computed, one can build a possible thermodynamic potential by the rule
\begin{equation}
G\big(\beta,\{\vec B(\vec x)\}\big)\doteq {\cal L}[[\vec m_B(\vec x)]]
\end{equation}
\underline{Example.}\\
We seek a state of equilibrium which can be achieved with a uniform magnetic field. We suppose that the equilibrium spin distribution is also uniform within the slab. The Landau functional relevant for computing such a state writes
\begin{eqnarray}
{\cal L}[\vec m]&\approx & \frac{1}{a^2}\cdot \frac{d}{a}\cdot \int d^2x\Big[ \frac{r}{2}\cdot \vec m(\vec x)^2+\frac{u}{4}(\vec m^2)^2+\mu\cdot \vec B\cdot  \vec m(\vec x)\Big]
\end{eqnarray}
To implement the functional derivative we insert
\begin{equation}
\vec m = \vec m_B + \vec \epsilon
\end{equation}
and compute (to first order in $\vec \epsilon$)
\begin{eqnarray}
\delta {\cal L}&\doteq & {\cal L}[\vec m]- {\cal L}[\vec m_B]\nonumber\\
&=&\!\frac{1}{a^2}\cdot \frac{d}{a}\cdot \int d^2x \cdot \Big[r\cdot \vec m_B + u\cdot \vec m_B^2\cdot \vec m_B+ \mu\cdot \vec B\Big]\cdot \vec \epsilon =0
\end{eqnarray}
The equality to zero must hold for any $\vec \epsilon$ so that the integrand itself must vanish. This leads to the Euler equation (also called ''equation of state'') for the sought for equilibrium $\vec m_B$
\begin{equation}
-\mu\cdot \vec B = \vec m_B\cdot \big(r+ u\cdot\vec m_B^2\big)
\label{Eq:state}
\end{equation}
The solutions $\vec m_0$ of this equation in $\vec B=0$ are vectors of length
\begin{equation}
\sqrt{\frac{-r}{u}}\propto \big (T_c-T)^\frac{1}{2}
\end{equation}
whose tips are distributed onto a sphere. These solutions build the set of possible spontaneous magnetizations of the system. The spherical topology is a direct consequence of the symmetry group of the Heisenberg Hamiltonian being $SO_3$ (the continuous group of rotations in space). The square-root dependence of $\vert \vec m_0\vert$ on $(T_c-T)$ predicts a non-analytic vanishing of the spontaneous magnetization at the Curie temperature $T_c$ via a power-law with a so called critical exponent $\beta = \frac{1}{2}$. Such non-analytic behaviour, with characteristic critical exponents, is also predicted by this variational principle for further thermodynamic quantities. We will indeed observe critical exponents in measurements on ultrathin magnetic transition metal films, although their value will differ from that predicted within this variational principle.\\
5. One can use the solution $\vec m_B$ to compute the MFA $g(r,B,u)$. For positive $r$ and in a region of the parameter space $r, \vec B$ where $\vec m_B$ is small enough, the approximate solution of the Euler equation writes
\begin{equation}
\vec m_B\approx -\frac{\mu\cdot \vec B}{r}
\end{equation}
and
\begin{equation}
g(r>0,\vec B) = \ell\big(r,\vec m_B(\vec B)\big)\approx \frac{r}{2}\cdot \big(-\frac{\mu\cdot \vec B}{r}\big)^2 +\mu\cdot \vec B\cdot \frac{-\mu\vec B}{r}= -\frac{1}{2}\cdot \frac{(\mu\cdot B)^2}{r}
\end{equation}
(see the interval $r>0$ in the next figure).
For negative $r$ we approximate the solution of the equation of state with\footnote{The factor $\frac{1}{2}$ introduced for $r<0$ in $\frac{\mu\cdot B}{2\cdot r}$ is known from the computation of the so called susceptibility: at $m^2=-\frac{r}{u}$ one has $\frac{\partial (-\mu B)}{\partial m}= r+3um^2=-2r$.}
\begin{equation}
m_B = -\sqrt{\frac{-r}{u}} + \frac{\mu\cdot B}{2\cdot r}
\end{equation}
$g(r<0,B)$ computes from
\begin{equation}
\frac{r}{2}\cdot \big( -\sqrt{\frac{-r}{u}} + \frac{\mu\cdot B}{2\cdot r}\big)^2 + \frac{u}{4}\cdot \big( -\sqrt{\frac{-r}{u}} + \frac{\mu\cdot B}{2\cdot r}\big)^4 + \mu\cdot B\cdot \big(-\sqrt{\frac{-r}{u}} + \frac{\mu\cdot B}{2\cdot r}\big)
\end{equation}
and amounts, expanding in powers of $B$, to
\begin{equation}
g(r<0,B)\approx \big(-\frac{r^{2}}{4u}\big)-\mu B\sqrt{-\frac{r}{u}}+\frac{\mu ^{2}B^{2}}{4r}+\frac{\mu ^{3}B^{3}}{8r^{2}}\sqrt{-\frac{u}{r}}+\frac{\mu ^{4}B^{4}u}{64r^{4}}+....
\end{equation}
A sketch of $g(r,B)$ is provided by the next figure.
\begin{figure}[H]
\centering
\resizebox{0.6\textwidth}{!}
{
\begin{tikzpicture}
    \begin{axis}[
       xlabel=$r$, ylabel=$b$, zlabel=$g$,
       xlabel style={font=\huge},
    ylabel style={font=\huge},
    zlabel style={font=\huge},
        view={30}{30}, 
        colormap/viridis, 
        colorbar, 
        colorbar style={
            ylabel={}, 
        yticklabel style={/pgf/number format/fixed} 
        }
    ]

    \addplot3[
        surf, shader=interp,
        domain=1:4,      
        domain y=-5:5,   
        samples=30,
    ]
    {-0.5*y^2/x};        
     \addplot3[
        surf, shader=interp,
        domain=-4:-1,    
        domain y=-5:5,   
        samples=30,
    ]
    {-x^2/4-abs(y)*sqrt(-x)};  
    \end{axis}
   \end{tikzpicture}
   }
   \caption{Sketch of $g(r,b)$ for $\vert r\vert$ sufficiently far from the critical point (the empty interval $\vert r\vert<1$ in the figure). Computed with $u=1$ and $\mu=1$; $g,r,b$ are then in units of $u$. The cusp along $b=0$ for $r<0$ is the signature of the spontaneous magnetization: $\frac{\partial g}{\partial b}$ jumps by $2\sqrt{-r/u}$ across it.}
   \label{Fig:g}
   \end{figure}
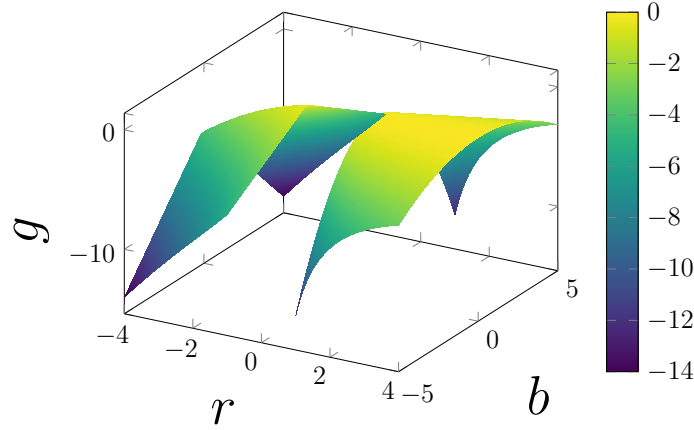
\noindent 6. Data collapsing.
The equation for the length of the magnetization vector in a uniform magnetic field writes
\begin{equation}
-\mu\cdot B = m\cdot r+ u\cdot m^3
\end{equation}
Written in this way, it is a relation between three thermodynamic variables:
the temperature $r$, the magnetic field $B$ and the spin polarization $m$. In
practice, one measures the magnetization as a function of the temperature and
plots the resulting set of data in a graphic for different applied magnetic
fields. One observes an opening up of the set of curves that is expression of
the relation between them imposed by the equation of state. There is, however, a subtle symmetry of the equation of state that allows the possibility of the collapsing of the experimental data onto one single curve. One possible expression of this symmetry is called ''Griffiths scaling law'', after the author that suggested it on the basis of general considerations\cite{Griff}. To let this symmetry emerge, suppose to divide the equation of state by $m^3$. One obtains\footnote{According to our convention, $m$ is negative for positive applied magnetic fields.}
\begin{equation}
-\frac{\mu\cdot B}{m^3} = \frac{r}{m^2}+ u
\label{Eq:Griff_MFA}
\end{equation}
This relation reads: the equation of state relates the variable $\frac{B}{m^3}$ to the \textbf{single} variable $\frac{r}{m^2}$. In other words: when the experimental data are plotted using these special variables, they collapse onto one single curve. The graph of this curve (the so called ''scaling function'') is given, within MFA Landau theory, by the relation
\begin{equation}
h(x) = x+u
\end{equation}
\begin{figure}[H]
\centering
\begin{tikzpicture}
\begin{axis}[
    name=linpanel,
    width=5.2cm, height=4.4cm, scale only axis,
    axis lines = left,
    xlabel = {$x=\frac{r}{m^2}$},
    ylabel = {$h=-\frac{\mu B}{m^3}$},
    xmin=-1.4, xmax=5, ymin=-0.5, ymax=6,
    xtick={-1,0,1,2,3,4,5}, ytick={0,2,4,6},
    label style={font=\small}, tick label style={font=\small},
    title={(a) linear}, title style={font=\small},
]
\addplot [domain=-1.4:5, samples=2, color=black, line width=1.0pt] {x+1};
\draw[dashed, black!55] (axis cs:-1.4,1) -- (axis cs:0,1);
\node[font=\footnotesize, anchor=west, black!55] at (axis cs:0.15,0.8) {$h(0)=u$};
\end{axis}
\begin{axis}[
    name=logpanel, at={(linpanel.south east)}, anchor=south west, xshift=1.9cm,
    width=5.2cm, height=4.4cm, scale only axis,
    xmode=log, ymode=log,
    xlabel = {$x=\frac{r}{m^2}$},
    ylabel = {$h=-\frac{\mu B}{m^3}$},
    xmin=1e-1, xmax=1e4, ymin=5e-1, ymax=1e6,
    xtick={1e-1,1e1,1e3}, ytick={1e0,1e2,1e4,1e6},
    label style={font=\small}, tick label style={font=\small},
    title={(b) double logarithmic}, title style={font=\small},
]
\addplot [domain=1e-1:1e4, samples=60, color=black, line width=1.0pt] {x+1};
\addplot [dashed, black!55, domain=1e1:1e4, samples=2, line width=0.7pt] {0.35*x^1.75};
\node[font=\footnotesize, anchor=north east, rotate=27] at (axis cs:4e3,4e3) {slope $\gamma_{\text{MFA}}=1$};
\node[font=\footnotesize, anchor=south east, rotate=40, black!55] at (axis cs:4e2,1e5) {slope $\frac{7}{4}$};
\end{axis}
\end{tikzpicture}
\caption{The scaling function in Griffiths' variables, as Landau theory predicts it: the straight line $h=x+u$, here with $u=1$. (a) linear; (b) double logarithmic, where its large-$x$ slope, which is the critical exponent $\gamma$, is seen to be $1$. The dashed line of slope $\frac{7}{4}$ is what the experiment gives (Fig. \ref{Fig:H}).}
\label{Fig:Griff_MFA}
\end{figure}
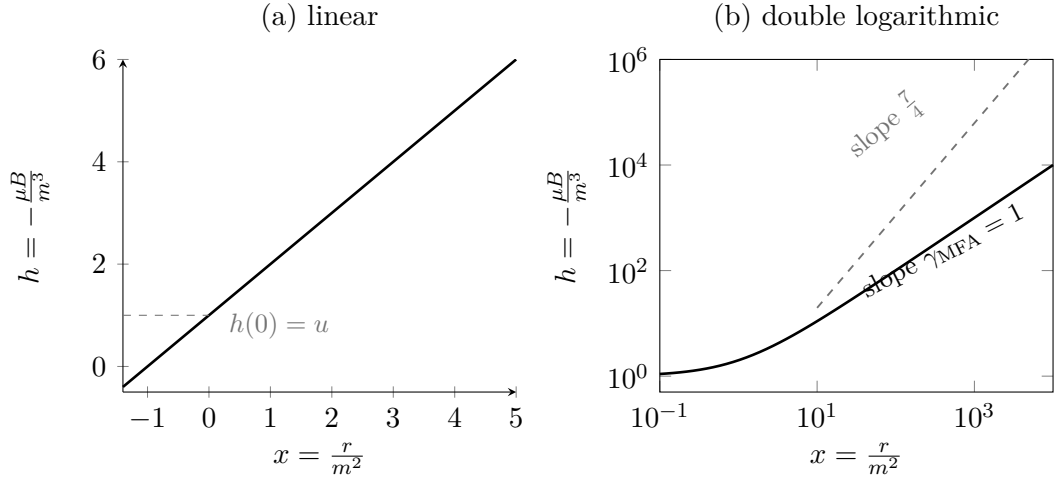
As analyzed by Griffiths and other authors\cite{Poki,Stan,Griff,Wid}, such a collapsing (''scaling'') is related to a fundamental property of systems in the vicinity of the second order transition point that we will discuss in the next section in depth\footnote{The  word ''scaling'', used as synonym to ''data collapsing'', originates within this more comprehensive phenomenological approach discovered by Griffiths et al\cite{Poki,Stan,Griff,Wid}}.\\
\noindent 7. There is a further set of special variables that collapses the experimental data. This further set was proposed by Milo\v{s}evi\'c and Stanley\cite{Stan,Domb}. Divide the equation of state by $B^{\frac{2}{3}}\cdot m$ to obtain
\begin{equation}
-\mu\cdot \frac{B^{\frac{1}{3}}}{m}= \big(\frac{r}{B^{\frac{2}{3}}}\big) + u\cdot \big(\frac{m}{B^{\frac{1}{3}}}\big)^2
\label{Eq:MS_MFA}
\end{equation}
This relation can be read as: the equation of state relates the special variable
$\big(\frac{r}{B^{\frac{2}{3}}}\big)$ to the special variable $\big(\frac{m}{B^{\frac{1}{3}}}\big)$. Plotted using these variables, the experimental data should collapse onto one single curve. Writing
\begin{equation}
x\doteq \frac{r}{B^{\frac{2}{3}}}\quad\quad y\doteq \frac{\vert m\vert}{B^{\frac{1}{3}}}
\end{equation}
equation (\ref{Eq:MS_MFA}) becomes
\begin{equation}
x = \frac{\mu}{y}-u\cdot y^2
\end{equation}
which is the scaling function in implicit form. Its two limits are the two
critical exponents: $y\rightarrow \sqrt{-x/u}\propto \vert x\vert^{\beta}$ with
$\beta=\frac{1}{2}$ for $x\rightarrow -\infty$, and $y\rightarrow \mu/x\propto
x^{-\gamma}$ with $\gamma=1$ for $x\rightarrow +\infty$. At $x=0$, i.e. exactly
at $T_c$, $y=(\mu/u)^{\frac{1}{3}}$, which is the exponent $\delta=3$.
\begin{figure}[H]
\centering
\begin{tikzpicture}
\begin{axis}[
    width=7.4cm, height=5.0cm, scale only axis,
    axis lines = left,
    xlabel = {$x=\frac{r}{B^{2/3}}$},
    ylabel = {$y=\frac{\vert m\vert}{B^{1/3}}$},
    xmin=-9, xmax=20, ymin=0, ymax=3.2,
    xtick={-5,0,5,10,15,20}, ytick={0,1,2,3},
    label style={font=\small}, tick label style={font=\small},
]
\addplot [domain=0.05:3, samples=250, variable=\t, color=black,
          line width=1.0pt] ({1/t - t^2},{t});
\draw[dashed, black!45] (axis cs:0,0) -- (axis cs:0,1);
\node[font=\footnotesize, anchor=south west, black!55] at (axis cs:0.3,0.95)
      {$y(0)=\big(\frac{\mu}{u}\big)^{1/3}$};
\node[font=\footnotesize, anchor=south west] at (axis cs:-4.3,2.55)
      {$y\sim\vert x\vert^{\beta},\ \beta=\frac{1}{2}$};
\node[font=\footnotesize, anchor=south west] at (axis cs:8,0.25)
      {$y\sim x^{-\gamma},\ \gamma=1$};
\end{axis}
\end{tikzpicture}
\caption{The scaling function in the variables suggested by Milo\v{s}evi\'c and Stanley, as Landau theory predicts it ($\mu=u=1$). The measured counterpart of this curve is Fig. \ref{Fig:ST}: same variables, but with $\beta=\frac{1}{8}$ and $\gamma=\frac{7}{4}$ in place of $\frac{1}{2}$ and $1$.}
\label{Fig:MS_MFA}
\end{figure}
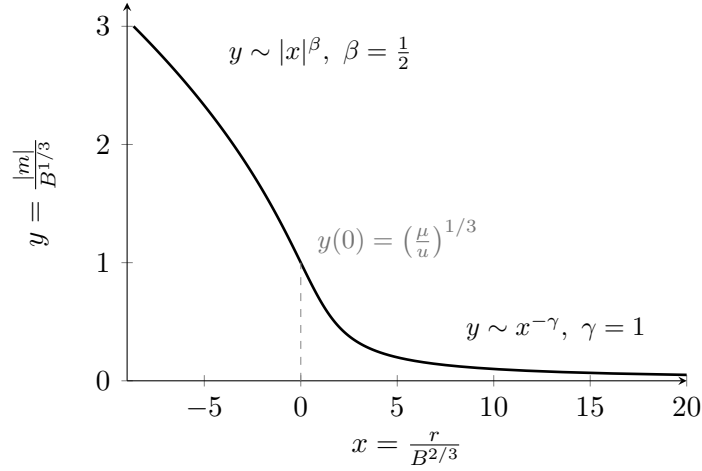
\noindent 8. The use of the solutions of this Euler equation for building the thermodynamic potential might lead to inaccuracies, as this approach neglects other, possibly important, spin configurations. A non-variational interpretation of the Landau functional that takes into account (almost) all spin configurations was also proposed by Landau\cite{Landau_32,Poki,Landau_V_1}. This principle states that the probability of  realizing a given spin configuration is proportional to
\begin{equation}
e^{-\beta\cdot {\cal L}[\vec m]}
\end{equation}
Accordingly, the partition function is the sum (functional integral) of such exponentials over all configurations of the order parameter field.\\
\underline{Example.}
The spontaneous spin polarization vectors lying on a sphere are an example of degeneracy. Because of this degeneracy, going from one solution to another does not require going over any free energy  barrier. Thus, as a
consequence of degeneracy, there emerges a type of excited states in which the
local ground state changes very gradually over space, so as to form a wave of
very long spatial length. Such an excited state is locally one of the equilibrium states and its free energy can be infinitely close to the actual minimum of the Landau functional, provided its wavelength is very long. This type of excitations allowed by the continuous symmetry of the problem are called ''Goldstone'' excitations.\\
Once the system has been kicked in one of the excited states, to which minimum will it decay afterwards? Will it decay into the minimum it originates from? Or to a completely different minimum? As there is nothing to prevent a minimum decaying into a totally different minimum, an accurate analysis of the stability of the spontaneous magnetization against Goldstone excitations is required.\\
Let us assume the slab has acquired a state of spontaneous magnetization, which we set along the $z$-axis without loss of generality. We introduce an excitation by adding a spatial dependent transversal vector $(m_x,m_y,0)$. The Landau functional per slab volume unit cell writes (considering only the lowest powers of $m_x^2$ and $m_y^2$)
\begin{eqnarray}
{\ell}(m_x,m_y,m_B)&=& A\cdot a^3\cdot \big[(\vec \nabla m_x)^2 +
(\vec \nabla m_y)^2\big]\nonumber\\
\!&+&\!\Big[\frac{r}{2}\!\cdot\! (m_x^2\!+\!m_y^2\!+\!m_B^2)\!+\!\frac{u}{4}\cdot m_B^4+\frac{u}{2}\cdot\! m_B^2\cdot(m_x^2+m_y^2)+\mu\cdot B\cdot m_B\Big]\nonumber\\
\end{eqnarray}
and, using the equation of state (\ref{Eq:state}) in the form
$r+u\cdot m_B^2 = -\mu\cdot \frac{B}{m_B}$,
\begin{eqnarray}
\delta \ell(m_x,m_y)&\doteq &\ell(m_x,m_y,m_B)-\ell(0,0,m_B)\nonumber\\
&=&A\cdot a^3\cdot \Big[(\vec \nabla m_x)^2 +(\vec \nabla m_y)^2\Big]- \frac{1}{2}(m_x^2+m_y^2)\cdot\mu\cdot\frac{B}{m_B}
\end{eqnarray}
The probability for the realization of the fluctuation $(m_x,m_y)$ is
\begin{equation}
\frac{e^{-\beta\cdot \frac{d}{a}\cdot \frac{1}{a^2}\int d^2\rho\cdot \delta \ell\big(m_x(\vec \rho),m_y(\vec \rho)\big)}}{Z}
\end{equation}
with
\begin{equation}
Z=\int D\vec m(\vec \rho)e^{-\beta\cdot \frac{d}{a}\cdot \frac{1}{a^2}\int d^2\rho\cdot \delta \ell\big(m_x(\vec \rho),m_y(\vec \rho)\big)}
\end{equation}
Using the functional Gaussian integration technology we compute, using
\begin{equation}
<\vert m_{\vec k}\vert^2>_T= \frac{1}{\beta}\cdot\frac{1}{ A\cdot d\cdot a^2\cdot k^2 -  \mu\cdot\frac{B}{m_B}}
\end{equation}
the mean of the square of the fluctuations:
\begin{equation}
<m_x^2+m_y^2>_T = \frac{2}{\beta}\cdot a^2\cdot \int \frac{d^2k}{(2\pi)^2}\cdot\frac{1}{ A\cdot d\cdot a^2\cdot k^2 -  \mu\cdot\frac{B}{m_B}}
\end{equation}
(the integral is over the first Brillouin zone).
The integral diverges logarithmically in the limit $B\rightarrow 0$ if the spontaneous magnetization is finite. This means that the spontaneous magnetization is unstable against transversal fluctuations. Mathematically speaking, one can avoid this divergence by letting $m_0\rightarrow 0$ at any finite temperature -- but this is the content of the Mermin-Wagner theorem of Chapter 6\footnote{In a seminal paper\cite{Landau_29}, Landau pointed out that long range order is not possible in truly two dimensional systems.}. The divergence of the integral occurring in the thermodynamic limit is due to the lower integral range approaching $0$ as the volume of the sample approaches infinity and is referred to in the literature as ''infrared divergence''. This infrared divergence plays a key role in determining physical properties of any 2d-system, even if interactions breaking the exact continuous symmetry occur in real systems. In the context of this Chapter, this instability invalidates the Landau theory for two-dimensional isotropic systems.
\section{The scaling hypothesis and experimental results on ultrathin films.}
The scaling hypothesis\cite{Griff,Stan,Wid,Scale} arose from the following
situation. The Mean Field Landau theory did provide a deep understanding of phase
transitions but its quantitative predictions, for instance those relating to  the precise value of the critical exponents and to the precise graph of the  scaling functions, were not supported by
experiments. In particular, it predicted a phase transition in any space
dimension and independently of the number of components of the order parameter.
The exact solution of the 2D-Ising model by Onsager\cite{Ons}, instead,
provided exact values for the critical exponents in a 2D system that did not match the results of the MFA Landau
theory. This situation turned out to supply a set of theoretical and
experimental data that did not have a fundamental and comprehensive explanation.
In the middle of the 1960's, several authors tried to find such a comprehensive theory. Their effort produced a phenomenological  Ansatz to phase transitions called ''the scaling hypothesis'' \cite{Stan}. This hypothesis is based on a symmetry of the MFA Landau theory that remained hidden for about 30 years.\\
Consider the mathematical expressions for the $g(r,b)$ that we have obtained within the MFA variational principle in the previous Subsection:
\begin{eqnarray}
g(r>0,b)&\approx & -\frac{1}{2}\cdot \frac{b^2}{r}\nonumber\\
g(r<0,b)&\approx & -\frac{r^2}{4\cdot u}\pm b\cdot \sqrt{-\frac{r}{u}}
\end{eqnarray}
Multiply each variable by a ''scaling factor'' $L^{\Delta_x}$, $L$ being some number and $\Delta_x$ being a so called ''scaling dimension'' of the variable
$x$. It is not immediately obvious but it can be easily demonstrated that one can choose one of the scaling dimensions so that $g(r,b)$ itself, in  both ranges of $r$, is multiplied by the same scaling factor, $L^{\Delta_g}$:
the first line scales as $L^{2\Delta_b-\Delta_r}$; in the second line the two
terms scale as $L^{2\Delta_r-\Delta_u}$ and $L^{\Delta_b+\frac{\Delta_r}{2}-\frac{\Delta_u}{2}}$, and both become equal to the
first one for the single choice
\begin{equation}
\Delta_u = 3\cdot \Delta_r-2\cdot \Delta_b\quad\quad\text{giving}\quad\quad \Delta_g = 2\cdot \Delta_b-\Delta_r
\end{equation}
This apparently strange symmetry states that, mathematically speaking, the function $g(r,b)$ should be considered to be an homogeneous function of its variables, i.e.
\begin{eqnarray}
g\Big(r\cdot L^{\Delta_r},b\cdot L^{\Delta_b},u\cdot L^{\Delta_u}\Big)= L^{\Delta_g}\cdot g(r,b,u)
\end{eqnarray}
Notice that $u$ has to be scaled along with $r$ and $b$. If instead one insists
that $u$ be left alone, $\Delta_u=0$, the two relations above fix everything:
$\Delta_b=\frac{3}{2}\Delta_r$ and $\Delta_g=2\Delta_r$, from which the Landau
exponents $\beta=\frac{1}{2}$ and $\delta=3$ follow by the two formulae
(\ref{Eq:beta}) and (\ref{Eq:delta}) below. That the Landau exponents come out
of the homogeneity property alone is the reason why that property, and not the
Landau expressions themselves, was retained.\\
In the 1960's, several works\cite{Poki,Griff,Stan,Wid} decided that the
mathematical expression originating from Landau MFA should be rejected but its
symmetry should be kept as a phenomenological approach to obtaining the proper
thermodynamics of a system poised at a second order phase transition. The
scaling hypothesis states, therefore, that thermodynamic quantities such as
the thermodynamic potential or the average magnetization are, close to the
critical point, homogeneous functions of their variables. This mathematical
property produces a set of predictions that, formally, leads to non-Landau
critical exponents and ''data collapsing'' into scaling functions. The scaling hypothesis was supported e.g. by the spin block construction of L. Kadanoff\cite{Kad} but it is the Renormalization group algorithm propagated for statistical physics by K. Wilson that allows to find a more rigorous proof of the scaling hypothesis, a clear explanation for the meaning of the scale factor ''$L$'' and a practical computation of the value of the scaling dimensions. For the
sake of being practical we anticipate the results of the scaling hypothesis as
formulated by B. Widom\cite{Wid} and compare them with experimental findings on epitaxial transition metal ultrathin films. We leave the ''hard'' work of finding the meaning of $L$ and of computing the scaling exponents for the Landau-Ginzburg-Wilson scalar Hamiltonian to the next Chapter.\\
\paragraph{Widom's scaling.}
The formulation of the scaling hypothesis along the lines developed by B. Widom provide a useful way of accessing experimental data. Widom's formulation is also mathematically neat. Take e.g. a magnetic system poised to undergo a second order phase transition at the transition point $r=B=0$. In this situation, the Gibb's free energy is a function of the variables $r$ and $B$, their small values being an expression of the system being in the vicinity of the critical point $r=B=0$. Widom's scaling consists in making an assumption about how $G$ depends on the variables $r$ and $B$: it assumes that the thermodynamic potential is an \textbf{homogeneous} function of these variables, i.e.
\begin{equation}
G\Big(r,B\Big)= L^{-\Delta_G}\cdot G\Big(r\cdot L^{\Delta_r}, B\cdot L^{\Delta_B}\Big)
\end{equation}
This equation states that when the independent variables $r,B$ are substituted with themselves multiplied by so called \textbf{scaling factors} $L^{\Delta_r}$ respectively $L^{\Delta_B}$, the thermodynamic potential is also multiplied by a scaling  factor, $L^{-\Delta_G}$ (the ''minus sign'' is for the sake of convention). This transformation property is what one calls, technically speaking, the ''homogeneity'' property of the function $G(r,B)$. The homogeneity relation is supposed to hold for any $L$. In this relation,
$L$ is a number that defines, together with the so called \textbf{scaling dimensions} $\Delta_x$, the scaling factors specific to the thermodynamic quantity. We will show by using the renormalization group that $L$ is some length in units of the lattice constant. $L$ will turn out to be much larger than ''1'' and can be as large as the correlation length of the system, i.e. $1<<L\leq \xi$. At the time Widom and Kadanoff published their scaling hypothesis, there was neither a way of computing this homogeneity property from general principles nor a way of computing the scaling dimensions. Their value could be obtained empirically from experimental results that were interpreted imposing the property of homogeneity\cite{Stan}.\\
In the following, we will cast Widom's scaling using the language proper of a magnetic system poised at the second order phase transition. Let us, for instance, translate the homogeneity relation of the Gibb's potential into an homogeneity relation of the order parameter $m$, known to be related to $G(r,B)$ by the equation
\begin{equation}
\frac{g\cdot\mu_B\cdot S}{a^3}\cdot m=\frac{\partial G}{\partial B}
\end{equation}
Assuming Widom's scaling for $G(r,B)$ we obtain
\begin{equation}
m(r,B) = L^{-\Delta_G}\cdot \frac{\partial G}{\partial B_L}\cdot \frac{\partial B_L}{\partial B}= L^{-\Delta_G+\Delta_B}\cdot m\Big(r\cdot L^{\Delta_r}, B\cdot L^{\Delta_B}\Big)
\end{equation}
One can write this relation as
\begin{equation}
m\Big(r\cdot L^{\Delta_r},B\cdot L^{\Delta_B}\Big)= m(r,B)\cdot L^{\Delta_G-\Delta_B}
\label{Eq:hom_m}
\end{equation}
This relation reads: when the independent variables are multiplied by the
scaling factors, the scaling factors themselves must cancel out in the argument
of the function. The only result of inserting the scaling factors is that the
function as a whole is multiplied by some scaling factor, characteristic of the
function itself. Relations of homogeneity such as these require that they are
satisfied for any $L$. In order for this to occur, the independent variables in
the function can only appear as a special combination, which we are now trying
to find. For this purpose, we set the argument of the function $m$ as being, tentatively,
\begin{equation}
r^{p}\cdot B^{q}
\end{equation}
with $p$ and $q$ to be determined so that
\begin{equation}
\Delta_r\cdot p + \Delta_B\cdot q = 0
\end{equation}
This last equation ensures that the scaling factors cancel out in the argument of the function $m$ so that the functional equation can hold true for any $L$.
One solution is e.g. $p=1$ and $q= -\frac{\Delta_r}{\Delta_B}$,  resulting in $m$ being a function of one single variable
\begin{equation}
\frac{r}{B ^{\frac{\Delta_r}{\Delta_B}}}
\end{equation}
The functional equation allows the function $m$ to be multiplied by the scaling factor $L^{\Delta_G-\Delta_B}$ when $L$ is inserted. This degree of freedom translates into the possibility of multiplying $m\left(\frac{r}{B^{\frac{\Delta_r}{\Delta_B}}}\right)$ with a suitable variable, such as
\begin{equation}
B^{\frac{\Delta_G-\Delta_B}{\Delta_B}}
\end{equation}
A solution of the functional equation that takes all degrees of freedom into account writes, accordingly,
\begin{equation}
m(r,B) = B^{\frac{\Delta_G-\Delta_B}{\Delta_B}}\cdot m\left(\frac{r}{B^{\frac{\Delta_r}{\Delta_B}}}\right)
\end{equation}
This result implies that plotting the quantity
\begin{equation}
\frac{m(r,B)}{B^{\frac{\Delta_G-\Delta_B}{\Delta_B}}}\quad \text{versus}\quad \frac{r}{B^{\frac{\Delta_r}{\Delta_B}}}
\end{equation}
should produce the \textbf{collapsing} of the $m(r,B)$ experimental data onto
one single graph. The exponents appearing here are the ones that the original
proposals of the equation of state\cite{Stan,Domb} wrote with the symbols
inherited from Landau theory:
\begin{equation}
\frac{\Delta_G-\Delta_B}{\Delta_B}= \frac{1}{\delta}\quad \frac{\Delta_r}{\Delta_B}= \frac{1}{\beta\cdot \delta}
\end{equation}
so that the two axes are $\frac{m}{B^{1/\delta}}$ and
$\frac{r}{B^{1/(\beta\delta)}}$ -- the variables of Milo\v{s}evi\'c and Stanley
of COMMENT 7, for which Landau theory predicts the curve of
Fig. \ref{Fig:MS_MFA}.
The collapsing of the experimental data using these variables is shown e.g. in Ref.\cite{Back_Nature} or Ref.\cite{Saratz}.
\begin{figure}[H]
\centering
\includegraphics[width=0.95\textwidth]{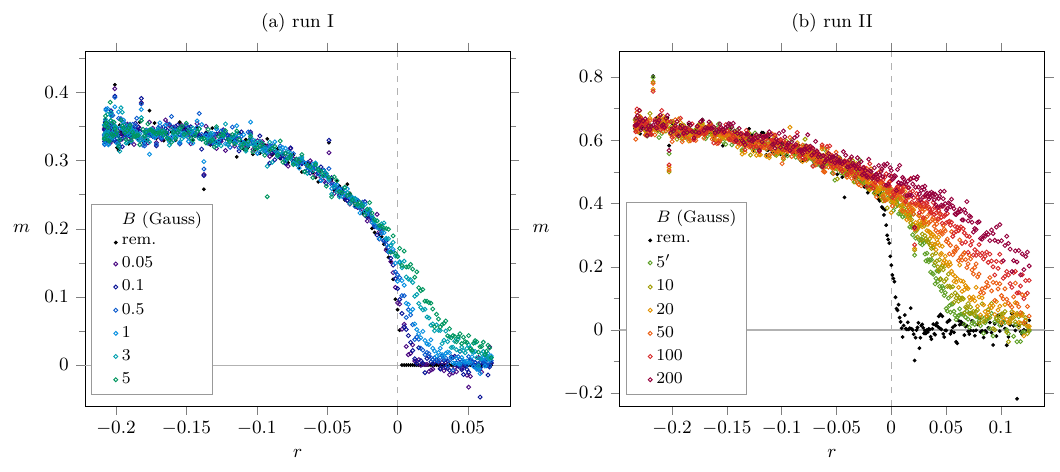}
\caption{Experimental $m(r)$ data at different applied magnetic fields, $r=\frac{T-T_c}{T_c}$, $m$ in arbitrary units, the field in Gauss given by the colour code of the legend. The black points are the remanence, i.e. $B=0$; they are the only ones that vanish identically above the transition. The two panels are two measurement runs, whose arbitrary units differ by a factor $1.88$; the field of $5$ G was measured in both, and the second one is denoted $5'$ in the collapsing plots that follow. The experimental data were obtained for an in-plane magnetized Fe film on W(110), with thickness of about 1.7 ML. For more details see Ref.\cite{Back_Nature}.}
\label{Fig:Raw}
\end{figure}
\begin{figure}[H]
\centering
\includegraphics[width=0.95\textwidth]{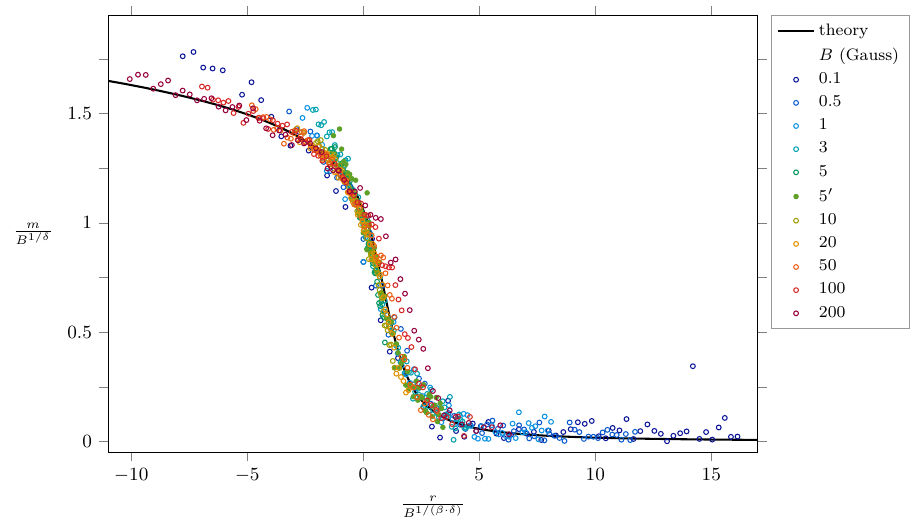}
\caption{Collapsing of the experimental data of Fig.\ref{Fig:Raw} in the variables of Milo\v{s}evi\'c and Stanley, $\frac{m}{B^{1/\delta}}$ against $\frac{r}{B^{1/(\beta\delta)}}$, with the 2D Ising values $\beta=\frac{1}{8}$ and $\delta=15$. The continuous line is the theoretical curve; the data are the eleven fields of Fig.\ref{Fig:Raw}, not connected. Measured on the theoretical curve, the two tails give $\frac{d\ln y}{d\ln\vert x\vert}=+0.124$ for $x<-3$ and $-1.747$ for $x>+3$, i.e. $\beta=\frac{1}{8}$ and $\gamma=\frac{7}{4}$. This is the experimental counterpart of Fig.\ref{Fig:MS_MFA}, where the same two tails carry the Landau values $\frac{1}{2}$ and $1$.}
\label{Fig:ST}
\end{figure}
A further possible solution of the functional equation is
\begin{equation}
m(r,B)= r^{\frac{\Delta_G-\Delta_B}{\Delta_r}}\cdot m\left(\frac{B}{r^{\frac{\Delta_B}{\Delta_r}}}\right)
\end{equation}
The Griffiths\cite{Griff} variables used for the collapsing plots e.g. in Ref. \cite{Back_Nature,Saratz} are obtained by reading the homogeneity relation
(\ref{Eq:hom_m}) the other way round, i.e. as a statement on $B(m,r)$. Since
(\ref{Eq:hom_m}) says that $m$ acquires the factor $L^{\Delta_m}$ with
$\Delta_m\doteq \Delta_G-\Delta_B$, the inverse function obeys
\begin{equation}
B\Big(m\cdot L^{\Delta_m}, r\cdot L^{\Delta_r}\Big)= B(m,r)\cdot L^{\Delta_B}
\end{equation}
It is solved as before, by using the freedom in $L$: choose $L$ so that
$m\cdot L^{\Delta_m}=1$, i.e. $L= m^{-\frac{1}{\Delta_m}}$. Then
\begin{equation}
\frac{B(m,r)}{m^{\frac{\Delta_B}{\Delta_m}}}= B\left(1,\frac{r}{m^{\frac{\Delta_r}{\Delta_m}}}\right)\doteq h\left(\frac{r}{m^{\frac{1}{\beta}}}\right)
\end{equation}
where the two exponents are, with the definitions above,
\begin{equation}
\frac{\Delta_B}{\Delta_m}=\frac{\Delta_B}{\Delta_G-\Delta_B}= \delta\quad\quad \frac{\Delta_r}{\Delta_m}=\frac{\Delta_r}{\Delta_G-\Delta_B}=\frac{1}{\beta}
\end{equation}
The two axes are therefore $\frac{B}{m^{\delta}}$ and $\frac{r}{m^{1/\beta}}$,
which for the Landau exponents $\delta=3$, $\frac{1}{\beta}=2$ are exactly the
two variables of equation (\ref{Eq:Griff_MFA}) and of Fig. \ref{Fig:Griff_MFA}.
A short computation with the two exponents shows that the large-$x$ slope of
$h(x)$ in a double logarithmic plot is $\gamma$ itself: Landau theory gives $1$,
the measurement gives $\frac{7}{4}$.
\begin{figure}[H]
\centering
\includegraphics[width=0.95\textwidth]{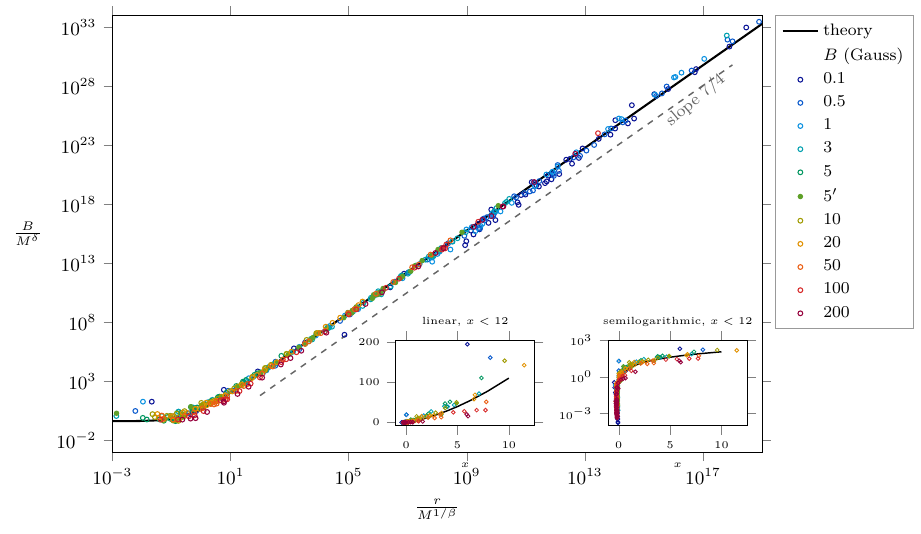}
\caption{Collapsing of the experimental data of Fig.\ref{Fig:Raw} in the canonical variables proposed by Griffiths\cite{Griff}: $\frac{B}{M^{\delta}}$ against $\frac{r}{M^{1/\beta}}$, with $\beta=\frac{1}{8}$, $\delta=15$. Main panel: double logarithmic, branch $x>0$; the collapse holds over twelve decades and the measured slope is $1.7500$, i.e. $\gamma=\frac{7}{4}$, against the Landau value $1$ of Fig.\ref{Fig:Griff_MFA}(b). Insets: the region $x<12$, linear and semilogarithmic, which carries the branch $x<0$. The continuous line is the theoretical curve, the circles are the eleven fields.}
\label{Fig:H}
\end{figure}
\paragraph{Particular cases.} Some particular cases of the scaling equation are also forthcoming in the literature. Here we shortly introduce them.
The  scaling equation for $B=0$ writes, for instance,
\begin{equation}
m(r,0) = L^{\Delta_B-\Delta_G}\cdot m\left (r\cdot L^{\Delta_r},0\right)
\end{equation}
and has solution
\begin{equation}
m(r) \propto r^{\frac{\Delta_G-\Delta_B}{\Delta_r}}\doteq r^{\beta}
\end{equation}
This result shows that the standard critical exponent $\beta$ relates to scaling dimensions of $r$ and $B$ according to
\begin{equation}
\beta = \frac{\Delta_G-\Delta_B}{\Delta_r}
\label{Eq:beta}
\end{equation}
The  scaling equation for $r=0$ writes:
\begin{equation}
m(0,B) = L^{\Delta_B-\Delta_G}\cdot m\left (0,B\cdot L^{\Delta_B}\right)
\end{equation}
and has solution
\begin{equation}
m(B)\propto B^{\frac{\Delta_G-\Delta_B}{\Delta_B}}\doteq B^{\frac{1}{\delta}}
\end{equation}
This result shows that the standard critical exponent $\delta$ relates to scaling dimensions of $r$ and $B$ according to
\begin{equation}
\delta= \frac{\Delta_B}{\Delta_G-\Delta_B}
\label{Eq:delta}
\end{equation}
We point out that there is a strategy for obtaining the collapsed form of the equation of state which is not based on  eliminating $L$ from the scaling hypothesis. We start again with the homogeneity relation
\begin{equation}
m\Big(r\cdot L^{\Delta_r},B\cdot L^{\Delta_B}\Big)= m(r,B)\cdot L^{\Delta_G-\Delta_B}
\end{equation}
As this relation is valid for any $L$, it is certainly also valid for one specific length, which we choose so that e.g. the first variable on the left-hand side is equal one, i.e.
\begin{equation}
L\propto r^{-\frac{1}{\Delta_r}}
\end{equation}
In this way, the left-hand side is a function of one single variable, producing the sought-for collapsed form of the equation of state. This length is defined to be the correlation length and this relation is also written as
\begin{equation}
\xi\propto r^{-\nu}
\end{equation}
(implying $\nu= \frac{1}{\Delta_r}$).
\paragraph{What the experiment gives.} The three figures above are one and the
same set of measurements, plotted three times. Fig. \ref{Fig:Raw} is the raw
material: eleven fields between $0.1$ and $200$ G, plus the remanence, on a
1.7 ML Fe film on W(110). Fig. \ref{Fig:ST} and Fig. \ref{Fig:H} are the two
collapsed forms, in the two sets of variables of COMMENTS 6 and 7. The
collapse is obtained with
\begin{equation}
\beta = \frac{1}{8}\quad\quad \delta = 15\quad\quad \gamma = \beta\cdot(\delta-1)=\frac{7}{4}
\end{equation}
which are the exact exponents of the 2D Ising model\cite{Ons}, and not with
$\beta=\frac{1}{2}$, $\delta=3$, $\gamma=1$, which is what the Landau theory of
section \ref{Chap:2Dphase}.1 predicts. The two scaling functions of that theory
were drawn, in the same variables, in Fig. \ref{Fig:MS_MFA} and
Fig. \ref{Fig:Griff_MFA}: they have the right shape and the wrong exponents.
Reading the slopes off the measured curves gives $\frac{d\ln y}{d\ln\vert
x\vert}=+0.124$ against $\beta=0.125$, $-1.747$ against $-\gamma=-1.75$ in
Fig. \ref{Fig:ST}, and $1.7500$ against $\gamma=1.75$ over twelve decades in
Fig. \ref{Fig:H}. A film of $1.7$ atomic layers is thus, as far as its critical
behaviour is concerned, a two-dimensional Ising magnet -- the uniaxial
anisotropy of Chapter 4 having reduced the $SO_3$ symmetry of the Heisenberg
Hamiltonian to the $Z_2$ symmetry of the Ising one, and having thereby lifted
the infrared divergence that closes section \ref{Chap:2Dphase}.1. Where these
exponents come from is the subject of the next Chapter.

\setcounter{section}{0} 
\renewcommand{\thesection}{\thechapter.\Alph{section}} 
\setcounter{equation}{0}
\renewcommand{\theequation}{\thesection\arabic{equation}}

\section*{Appendices}
\section{A Mean Field Lagrangian for the quantum Ising model and the classical Heisenberg model.}
\label{App:MFA_2D}
The importance of the Landau theory cannot be underestimated. It provides us
with the essential (albeit not very accurate) results for understanding phase
transitions of the second order. Most important, it introduces a novel concept
in the theory and experiments on the ferromagnetic phase transition: the
magnetization as the \textbf{order parameter}. We build first the Landau functional for the quantum Ising model on a lattice with lattice dependent magnetic field, described by the quantum operator
\begin{equation}
{\cal H}_{I} = -\frac{1}{2}\cdot \sum_{<i,j>}J_{ij} S_{i,z}\cdot S_{j,z} +
g\cdot \mu_B\cdot \sum_i B_i\cdot S_{i,z}
\end{equation}
We refer to this model because it is a minimum model of ferromagnetism
necessary for establishing the essential elements of Landau theory. In a second
step, we generalize the Ising Landau functional to the Heisenberg model.
The problem one faces in statistical physics is that the partition function of the original model can only be computed exactly for a small number of models
\footnote{The partition function of the Ising model, for instance, can only be
computed exactly for one and two dimensional lattices}. As a consequence, one
needs some kind of approximate approach toward computing the partition function
and the associated thermodynamic potential, the Gibb's free energy $G(T,B)$.
One possible approximation is the so called Mean Field Approximation. The MFA
leads to an effective \textbf{classical} Lagrangian (or, ''Landau functional'')
that simulates the original Hamilton operator. One hopes that the Landau
functional is more tractable than the Ising operator, even if the results
obtained from it might be less precise than the result obtained from the
original Hamilton operator\footnote{For example, the Landau functional for the
Ising operator leads to so called ''classical critical exponents'' that do not
precisely agree with the exact results. However, the topology of the phase
diagram and some further practical results from Landau functionals turn out to
be useful for understanding experiments.} For constructing
the Landau functional, we will not follow the original work\cite{Landau_29}, which is mostly based on symmetry considerations. Accordingly, it is of supreme elegance but a more practical approach will provide us with a deeper understanding.\\
Our approach is based on an inequality by Bogoliubov.
Let us assume that one has an Hamiltonian ${\cal H}$ with
\begin{equation}
{\cal H} = {\cal H}_0 + {\cal H}_1
\end{equation}
Let us have then a complete knowledge of the eigenstates of ${\cal H}_0$, so that we are, in principle, capable of computing
its partition function and its Free energy $F({\cal H}_0)$. Let further ${\cal H}_1$ be a ''small'' correction. Bogoliubov inequality states that, for the free energy
\begin{equation}
F({\cal H})\doteq -\beta^{-1}\ln \textbf{Tr}\big(e^{-\beta\cdot {\cal H}}\big)
\end{equation}
the following inequality holds true:
\begin{equation}
F({\cal H})\leq F({\cal H}_0) +  \frac{Tr \big(e^{-\beta\cdot {\cal H}_0}\cdot {\cal H}_1\big)}{Tr\big(e^{-\beta\cdot {\cal H}_0}\big)}
\end{equation}
The last term on the right hand side is the expectation value of ${\cal H}_1$ computed over the set of eigenstates of ${\cal H}_0$. For simplicity of writing we designate this operation with the symbol $<...>_0$, i.e.
\begin{equation}
F({\cal H})\leq F({\cal H}_0)+ <{\cal H}_1>_0
\end{equation}
This inequality can be used to find a possible Landau functional that simulates the quantum Ising ferromagnet. As a first step, one ''guesses'' an Hamilton operator ${\cal H}_0$ that might be suitable for approximating ${\cal H}_I$.
We do this by
means of introducing the single particle operator
\begin{equation}
{\cal H}_0(\phi_i)\doteq g\cdot\mu_B\cdot \sum_i \phi_i\cdot S_{i,z}
\end{equation}
This operator contains the set of parameters $\{\phi_i\}$ that represent an
effective ''mean field'' acting on a site ''i''. That the exchange interaction
could be captured by an effective mean field acting at each site was postulated
originally by Weiss\cite{Neel_Nobel} exactly for the case of ferromagnetism. Later, the idea of a ''mean field'' was translated to further fields of science. In a successive step, the trial Free energy functional
\begin{equation}
{\cal F}_{\text{trial}}[\phi_i]\doteq F\big({\cal H}_0(\phi_i)\big)+ <{\cal H}_{\text{I}}-{\cal H}_0(\phi_i)>_0
\end{equation}
is constructed. ${\cal F}_{\text{trial}}[\phi_i]$ associates to a set of parameters $\{\phi_i\}$ a scalar, i.e. ${\cal F}_{\text{trial}}[\phi_i]$ is a ''functional'' of the ''field'' $[\phi]$.
${\cal F}\big(H_0(\phi_i)\big)$ can be computed exactly:
\begin{eqnarray}
{\cal F}\big({\cal H}_0(\phi)\big) &=& -\frac{1}{\beta}\sum_i\ln\Big[\sum_{x=-S}^{x=S} e^{-\beta \cdot g\cdot \mu_B\cdot \phi_i\cdot x}\Big]\nonumber\\
&=&-\frac{1}{\beta}\cdot\sum_i\cdot \ln \frac{\sinh(2S+1)\cdot \frac{\beta\cdot g\cdot\mu_B\cdot \phi_i}{2}}{\sinh\frac{\beta\cdot g\cdot\mu_B\cdot \phi_i}{2}}
\end{eqnarray}
The average
\begin{eqnarray}
<{\cal H}_{\text{I}}-{\cal H}_0(\phi_i)>_0 &=&-\frac{1}{2}\cdot \sum_{<i,j>}J_{ij}\cdot < S_{i,z}\!\cdot\! S_{j,z}>_0\nonumber\\
&+& g\cdot\mu_B\cdot \sum_i(B_i\!-\!\phi_i)<S_{i,z}>_0
\end{eqnarray}
can be simplified by assuming that the ''fluctuation term'' is small and setting it to zero. This approximation will eventually invalidate some of the results based on the trial functional but, for the moment, it is the one that leads us forward.
\begin{eqnarray}
<\!S_{i,z}\cdot\!S_{j,z}>_0&=&\underbrace{<\Big(S_{i,z}-<S_{i,z}>_0\Big)\cdot \Big(S_{j.z}-<S_{j.z}>_0\Big)>_0}_{\text{set to 0}}\nonumber\\
&+& <S_{i,z}>_0\cdot<S_{j,z}>_0\nonumber\\
&\approx & <S_{i,z}>_0\cdot<S_{j,z}>_0
\end{eqnarray}
We therefore obtain for the ''trial'' Free energy functional
\begin{eqnarray}
{\cal F}_{\text{trial}}[\phi_i]&\approx &-\frac{1}{\beta}\cdot\sum_i\cdot \ln \frac{\sinh(2S+1)\cdot \frac{\beta\cdot g\cdot\mu_B\cdot\phi_i}{2}}{\sinh\frac{\beta\cdot g\cdot\mu_B\cdot \phi_i}{2}}\nonumber\\
&-& \frac{1}{2}\cdot \sum_{<i,j>}J_{ij}\cdot <S_{i,z}>_0\!\cdot\!<S_{j,z}>_0\nonumber\\
&+& g\cdot\mu_B\cdot \sum_i(B_i\!-\!\phi_i)<S_{i,z}>_0
\end{eqnarray}
It is a trial functional: among the degrees of freedom offered by the parameters $\phi$ there might be some which ideally approximate the ''true'' Gibb's thermodynamic potential. The problem of finding the Gibb's Free energy is translated to finding this set of parameters. In words: the classical functional ${\cal F}_{\text{trial}}[\phi_i]$ ''replaces'' the original model.\\
We proceed using a slightly different Ansatz. In fact, Landau based his work on a trial functional that contained a different set of parameters. One can compute $<S_{i,z}>_0$ as the derivative with respect to $\phi_i$ of ${\cal F}\big({\cal H}_0(\phi)\big)$. $<S_{i,z}>_0$ can be expressed as a function of the parameters $\phi_i$:
\begin{eqnarray}
<S_{i,z}>_0=-S\cdot B_S\big(S\cdot\beta\cdot g\cdot\mu_B\cdot\phi_i\big)
\end{eqnarray}
with
\begin{equation}
B_{S}(x)={\frac {2S+1}{2S}}\coth \left({\frac {2S+1}{2S}}\cdot x\right)-{\frac {1}{2S}}\coth \left({\frac {1}{2S}}\cdot x\right)
\end{equation}
being the so called Brillouin function\cite{WikiBrill}\footnote{For example: $B_{\frac{1}{2}}(x)= \tanh (x)$.}
This relation can be inverted to express $\phi_i$ as a function of the new parameter $m_i\doteq \frac{<S_{i,z}>_0}{S}$:
\begin{equation}
S\cdot\beta\cdot g\cdot\mu_B\cdot\phi_i= B_S^{-1}\big(-m_i\big)
\end{equation}
($B_S^{-1}(x)$ being the inverse Brillouin function\cite{InvBrillouin}). By using the inverse function, the trial free Energy  becomes a functional of the parameter set $\{m_i\}$. The set $\{m_i\}$ defines the so called \textbf{order-parameter field} on the lattice. For the S-Ising model, the order parameter field has one component.\\
DEFINITION. The Free energy ${\cal F}_{\text{trial}}$ expressed as a functional of the order parameter field is called the \textbf{Landau (free energy) functional}, indicated with $\cal L$.\\
The Landau functional replacing the quantum $S$-Ising ferromagnet, for instance, writes
\begin{eqnarray}
{\cal L}_{I}[m_i]&=& -\frac{1}{\beta}\cdot \sum_i \ln \frac{\sinh
(2S + 1)\cdot \frac{B_S^{-1}(-m_i)}{2S}}{\sinh \frac{B_S^{-1}(-m_i)}{2S}}\nonumber\\
&-& \frac{1}{2}\cdot S^2\cdot \sum_{<i,j>}J_{ij}\cdot m_i\cdot m_j\nonumber\\
&-&\frac{1}{\beta}\cdot \sum_i B_S^{-1}(-m_i)\cdot m_i\nonumber\\
&+& g\cdot\mu_B\cdot S\cdot \sum_i B_i\cdot m_i
\end{eqnarray}
COMMENTS.\\
1. One important general question pertains the use of Landau functionals in statistical physics. The simplest approach originates from Landau and consists in solving the equation
\begin{equation}
\frac{\partial {\cal L}\left(\beta, B_i,[m_i]\right)}{\partial m_i} = 0
\end{equation}
The original Landau theory\cite{Landau_29} assigned to the solutions  $m_i^{MFA}(\beta,B_i)$ of this equations the significance of an approximate equilibrium distribution at the site $i$. Inserting $m^{MFA}_i(\beta,B_i)$ into ${\cal L}(\beta,B_i,[m_i])$ gives an \textbf{approximate} Gibbs free energy  $G^{MFA}(\beta,[B_i])$. $G^{MFA}(\beta,B_i)$ is the Mean Field Approximation of the Gibb's free energy of the system.\\
It is instructive to show that the equilibrium condition
\begin{equation}
\frac{\partial {\cal L}\left(\beta, B_i,[m_i]\right)}{\partial m_i}\vert_{m_i= m^{MFA}_i(\beta,B_i)}= 0
\end{equation}
is equivalent to the
thermodynamic relation
\begin{equation}
\frac{1}{g\cdot \mu_B\cdot S}\cdot \frac{\partial G^{MFA}(\beta,[B_i])}{\partial B_i}=m_i^{MFA}(\beta,[B_i])
\end{equation}
PROOF:
\begin{eqnarray}
\frac{1}{g\cdot \mu_B\cdot S}\cdot\frac{\partial G^{MFA}(\beta,B_i)}{\partial B_i}
&=& \frac{1}{g\cdot \mu_B\cdot S}\cdot\frac{\partial {\cal L}\Big(\beta,B_i,m^{MFA}_i(\beta,B_i)\Big)}{\partial B_i}
\nonumber\\
&=&\frac{1}{g\cdot \mu_B\cdot S}\cdot\frac{\partial {\cal L}(\beta,B_i,m_i)}{\partial B_i}\vert_{m_i= m^{MFA}_i(\beta,B_i)}\nonumber\\
&+& \frac{1}{g\cdot \mu_B\cdot S}\cdot\underbrace{\frac{\partial {\cal L}(\beta,B_i,m_i)}{\partial m_i}\vert_{m_i= m^{MFA}_i(\beta,B_i)}}_{=0}\nonumber\\
&=& m_i^{MFA}(\beta,[B_i])
\end{eqnarray}
\noindent 2. A further\footnote{More precise but computationally more cumbersome} way of using the Landau functional originates from a paper by Landau again\cite{Landau_32}. Accordingly, the Landau functional is used to find the probability of realizing the distribution $[m_i]$:
\begin{equation}
W[m_i] = Z^{-1}\cdot e^{-\beta\cdot {\cal L}\big(\beta, B_i,[m_i]\big)}
\end{equation}
where $Z$ is the partition function
\begin{equation}
Z(T,B_i)= \sum_{[m_i]} e^{-\beta\cdot {\cal L}\big(\beta, B_i,[m_i]\big)}
\end{equation}
obtained by summing over all possible configurations $[m_i]$. The \textbf{thermodynamic potential} originating from this partition function is the Gibb's free energy
\begin{equation}
G(\beta,[B_i]) = -\frac{1}{\beta}\cdot \ln Z(\beta,B_i)
\end{equation}
In a situation where the applied magnetic field depends on the site, the Gibb's free energy depends also on the site. The thermal average of $m_i$, $\overline{ m_i(\beta,[B_i])}$ is given by the thermodynamic relation
\begin{equation}
\overline{ m_i(\beta,[B_i])}= \frac{1}{g\cdot \mu_B\cdot S}\cdot \frac{\partial G(\beta,[B_i])}{\partial B_i}
\end{equation}
\noindent 3. \textbf{The Weiss field.}
The Mean Field Equation of the S-quantum Ising model writes
\begin{equation}
\frac{\partial {\cal L}_{I}[m_i,\beta,B_i]}{\partial m_i} =0
\end{equation}
The derivative of the first line cancels out partially the derivative of the third line.\footnote{With $y\doteq B_S^{-1}(-m_i)$ one has $\frac{d}{dy}\ln\frac{\sinh\frac{(2S+1)y}{2S}}{\sinh\frac{y}{2S}}=B_S(y)=-m_i$, so that the two derivatives leave exactly $-\frac{1}{\beta}B_S^{-1}(-m_i)$.} Regarding the derivative of the second line: when the sum over all sites was introduced in the original Hamiltonian, the double counting of the sites was corrected by multiplying the sum with a factor $\frac{1}{2}$. When taking the derivative the double counting is mandatory and therefore one must multiply the derivative with a factor of two to counter the original $\frac{1}{2}$ factor. Finally, the remaining equation
\begin{equation}
-S^2\cdot \sum J_{ij}\cdot m_j + g\cdot \mu_B\cdot S\cdot B_i- \frac{1}{\beta}\cdot B_S^{-1}(-m_i)=0
\end{equation}
can be rewritten as
\begin{equation}
m_i = -B_S\Big(-\beta\cdot S^2\cdot \sum J_{ij}\cdot m_j + \beta\cdot g\cdot \mu_B\cdot S\cdot B_i\Big)
\end{equation}
This is a transcendental  implicit algebraic set of coupled equations for the sought for order parameters $\{m_i\}$. When this equation is compared to
\begin{equation}
m_i=-B_S\big(S\cdot\beta\cdot g\cdot\mu_B\cdot\phi\big)
\end{equation}
one can read out the relation
\begin{equation}
\phi= B_i-\frac{S^2\cdot \sum J_{ij}\cdot m_j}{g\cdot\mu_B\cdot S}
\end{equation}
If we recall that the magnetic moment $\mu_j$ on the site $j$ amounts to
$-g\mu_B \cdot S\cdot m_j$ we have the alternative relation
\begin{equation}
\phi= B_i+\frac{1}{(g\cdot\mu_B)^2}\cdot \sum_j J_{ij}\cdot \mu_j
\end{equation}
This relation explains the Weiss postulate\cite{Neel} of an ''exchange field'', provided by the nearest neighbours magnetic moments, as amplifying the applied field at the site $i$.\\
4. We seek to compute ${\cal L}_{I}[m_i,\beta,B_i]$ for small $m_i$.
We use the Taylor expansions\cite{InvBrillouin}
\begin{equation}
B_S^{-1}(x) \approx \frac{3S}{S+1}\cdot x + \frac{9\cdot S\cdot (S+1)\cdot (2S^2+2S+1)}{10\cdot (S+1)^4}\cdot x^3+...
\end{equation}
and
\begin{eqnarray}
-\ln\Big(\frac{\sinh((2S+1)\cdot x)}{\sinh(x)}\Big)&\approx & -\ln(2S+1) - \frac{1}{6}\cdot \big((2S+1)^2- 1\big)\cdot x^2\nonumber\\
&-&\frac{1}{180}\cdot \big(1-(2S+1)^4\big)\cdot x^4 \pm.....
\end{eqnarray}
together with the identity
\begin{equation}
-m_i\cdot m_j = \frac{\big(m_i-m_j\big)^2}{2}- \frac{m_i^2}{2}-\frac{m_j^2}{2}
\end{equation}
to write
\begin{eqnarray}
{\cal L}_I[m_i,\beta,B_i]\approx &+&\frac{J}{4}\cdot S^2\cdot  \sum_{<i,j>} \big(m_i-m_j\big)^2\nonumber\\
&+& \frac{3}{2}\cdot \frac{S}{(S+1)}\cdot \Big(\frac{1}{\beta}-\frac{1}{\beta_c}\Big)\cdot \sum_i m_i^2\nonumber\\
&+&\frac{1}{\beta}\cdot \frac{9}{40}\cdot \frac{S\cdot (S+1)\cdot (2S^2+2S+1)}{(S+1)^4}\sum_i m_i^4\nonumber\\
&+& g\cdot\mu_B\cdot S\cdot \sum_i B_i\cdot m_i\nonumber\\
&\pm& {\cal O}(m_i^6)
\end{eqnarray}
with $\beta_c^{-1} = \frac{z\cdot J\cdot S(S+1)}{3}$.\\
5. \textbf{Exact results for $S=\frac{1}{2}$-quantum Heisenberg model.} For the particular case of $S=\frac{1}{2}$, the functions appearing in the MFA are particularly manageable, so that we are able to make an exact computation of the Landau functional.
As
\begin{equation}
B_{\frac{1}{2}}(x) = \tanh(x)
\end{equation}
we obtain
\begin{equation}
m_i=-\tanh \big(\frac{\beta\cdot g\cdot\mu_B\cdot\phi}{2}\big)
\end{equation}
and
\begin{equation}
\frac{1}{2}\cdot\beta\cdot g\cdot\mu_B\cdot\phi= \tanh^{-1}\big(-m_i\big)
\end{equation}
We therefore obtain the Landau functional
\begin{eqnarray}
{\cal L}_I\big([m_i],\beta,B_i\big)= &-&\frac{1}{\beta}\sum_i\ln \left[2\cdot \cosh\left(\tanh^{-1}m_i\right)\right]\nonumber\\
&-&\frac{1}{2}\sum_{i,j}\frac{J_{ij}}{4}m_i\cdot m_j\nonumber\\
&+&\frac{g\mu_B}{2}\sum_i (B_i+\frac{2}{\beta g\mu_B}\tanh^{-1}m_i)m_i
\end{eqnarray}
which is valid for any $m_i$ in its range of definition.\\\\
An alternative expression for the functional ${\cal L}_{S=\frac{1}{2}}\big([m_i],\beta,B_i\big)$ originates from Bragg and Williams\cite{Bragg}. Using the identities
\begin{equation}
\cosh (\tanh^{-1}m_i) = \frac{1}{\sqrt{(1-m_i^2)}}
\end{equation}
\begin{equation}
\tanh^{-1}m_i = \frac{1}{2}\ln \frac{1+m_i}{1-m_i}
\end{equation}
we obtain
\begin{eqnarray}
{\cal L}\big([m_i],\beta,B_i\big)&=& -\frac{1}{2}\sum_{i,j}\frac{J_{ij}}{4}m_i\cdot m_j\nonumber\\
&+& \frac{1}{2\beta}\sum_i\left[(1+m_i)\ln(1+m_i)+ (1-m_i)\ln(1-m_i)\right]\nonumber\\
&+& \frac{g\mu_B}{2}\sum_i B_im_i
\end{eqnarray}
The Bragg form of the Landau functional displays explicitly the energy and entropic components.
\paragraph{The Landau-Ginzburg-Wilson  Hamiltonian.}
${\cal L}_I\big([m_i],\beta,B_i\big)$ for small $m_i$ can be translated onto an Euclidean space by replacing the sums with integrals. The resulting Landau functional, defined in the continuum space of dimension $D$, is often referred to as the ''Landau-Ginzburg-Wilson Hamiltonian'' and builds the starting point of modern approaches to phase transitions. The Hamiltonian is often given for a medium with $D$-dimensional cubic unit cell and $2$ nearest neighbours along each spatial direction:
\begin{eqnarray}
{\cal L}[m(\vec x),\beta,B(\vec x)]\approx &+&\frac{J}{2}\cdot S^2\cdot a^{2-D}\int d^Dx\Big(\sum_{\alpha=x_1,x_2,x_3,...}\big(\frac{\partial m}{\partial \alpha}\big)^2\Big)
\nonumber\\
&+& \frac{3}{2}\cdot \frac{S}{(S+1)}\cdot \Big(\frac{1}{\beta}-\frac{1}{\beta_c}\Big)\cdot a^{-D}\cdot \int d^Dx \cdot m^2(\vec x)\nonumber\\
&+&\frac{1}{\beta}\cdot \frac{9}{40}\cdot \frac{S\cdot (S+1)\cdot (2S^2+2S+1)}{(S+1)^4}\cdot a^{-D}\nonumber\\
&\cdot & \int d^Dx \cdot  m^4(\vec x)\nonumber\\
&+& g\cdot\mu_B\cdot S\cdot a^{-D}\cdot \int d^Dx \cdot B(\vec x)\cdot m(\vec x)\nonumber\\
&\pm& {\cal O}(m(\vec x)^6)
\end{eqnarray}
with $\beta_c^{-1} = \frac{z\cdot J\cdot S(S+1)}{3}$. By this translation,
$m_i$ becomes the order parameter field $m(\vec x)$.
The Landau-Ginzburg-Wilson Hamiltonian can be specified to a situation of a slab with thickness $d$ and $m(\vec x)$ depending only on the planar coordinates $(x,y)$
\begin{eqnarray}
{\cal L}[m(\vec x),\beta,B(\vec x)]\approx &+&\underbrace{\frac{J}{2}\cdot S^2\cdot \frac{d}{a}}_{\dot= A\cdot d}\cdot \int d^2x\Big(\sum_{\alpha=x,y}\big(\frac{\partial m}{\partial \alpha}\big)^2\Big)
\nonumber\\
&+& \frac{3}{2}\cdot \frac{S}{(S+1)}\cdot \Big(\frac{1}{\beta}-\frac{1}{\beta_c}\Big)\cdot \frac{d}{a}\cdot \int \frac{d^2x}{a^2} \cdot m(\vec x)^2\nonumber\\
&+&\frac{1}{\beta}\cdot \frac{9}{40}\cdot \frac{S\cdot (S+1)\cdot (2S^2+2S+1)}{(S+1)^4}\cdot \frac{d}{a}\cdot \int \frac{d^2x}{a^2}\cdot  m(\vec x)^4\nonumber\\
&+& g\cdot\mu_B\cdot S\cdot \frac{d}{a}\cdot \int \frac{d^2x}{a^2} B(\vec x)\cdot m(\vec x)\nonumber\\
&\pm& {\cal O}(m(\vec x)^6)
\end{eqnarray}
The Landau functional for the $S$-Ising model is often generalized to the Heisenberg ferromagnet by replacing $m(\vec x)$ with the three components order parameter field
\begin{equation}
\Big(m_x(\vec x),m_y(\vec x),m_z(\vec x)\Big)
\end{equation}
and by extending the $Z_2$-invariant polynomials to $O_3$-invariant polynomials. For properly describing the classical Heisenberg ferromagnet (classical meaning that the spin vector operators are replaced by classical vectors of length $S$) one needs the transition to large $S$ in the coefficients of the expansion. With  these adjustments in mind, we write the LGW-Hamiltonian of the Heisenberg ferromagnet for a slab of thickness $d$ as
\begin{eqnarray}
{\cal L}[\vec m,\beta,\vec B]\approx &+&A\cdot d\cdot \int d^2x\cdot\nonumber\\
&\cdot&\Big(\sum_{\alpha=x,y}\Big[(\frac{\partial m_x}{\partial \alpha})^2 + (\frac{\partial m_y}{\partial \alpha})^2+(\frac{\partial m_z}{\partial \alpha})^2\Big]\Big)\nonumber\\
&+& \frac{r}{2\cdot a^2}\cdot \frac{d}{a}\cdot \int d^2x \cdot \vec m(\vec x)^2\nonumber\\
&+&\frac{u}{4\cdot a^2}\cdot \frac{d}{a}\int d^2x \cdot \Big(\big(\vec m(\vec x)\big)^2\Big)^2\nonumber\\
&+&\frac{\mu}{a^2}\cdot \frac{d}{a}\int d^2x\cdot \vec B(\vec x)\cdot \vec m(\vec x)\nonumber\\
&\pm& {\cal O}(\vec m(\vec x)^6)
\end{eqnarray}
The coefficients $A,r,u,\mu$ are defined as follows:
\begin{eqnarray}
A&\doteq &\frac{J}{2}\cdot S^2\cdot \frac{1}{a}\nonumber\\
r&\doteq & 3\cdot \frac{S}{(S+1)}\cdot \Big(\frac{1}{\beta}-\frac{1}{\beta_c}\Big)\nonumber\\
u&\doteq &\frac{1}{\beta}\cdot \frac{9}{10}\cdot \frac{S\cdot (S+1)\cdot (2S^2+2S+1)}{(S+1)^4}\nonumber\\
\mu &\doteq & g\cdot\mu_B\cdot S\nonumber\\
\beta_c^{-1}&=&\frac{z\cdot J\cdot S^2}{3}
\end{eqnarray}
For the simple cubic lattice of the main text, $z=6$, this last line reads
$\beta_c^{-1}=2\cdot J\cdot S^2 = 4\cdot A\cdot a$.
\subsection{Particular solution of the Euler-Lagrange equation.}
The simplest version of the Landau theory foresees that the equilibrium configuration $\overline{m(\vec x)}$ is the solution of the Euler-Lagrange equation, here written explicitly for the Heisenberg model
\begin{equation}
 -2\cdot A\cdot d\cdot \triangle m_x(\vec x)+\frac{r}{a^2}\cdot \frac{d}{a}\cdot m_x(\vec x)+ \frac{u}{a^2}\cdot \frac{d}{a}\cdot m_x^3(\vec x) = -\mu\cdot \frac{d}{a}\cdot \frac{1}{a^2}\cdot B_x(\vec x)
\end{equation}
(and similar for $m_y,m_z$).
We work out some particular solutions of this equation that introduce us to the experimental results.\\
\paragraph{Uniform magnetic field along $z$.}
In this situation, the field $\vec m$ is also uniform and along $z$ (we drop the index 'z' in the following for simplicity). 
The equation of state writes
\begin{equation}
-\mu\cdot B = m\cdot\big(r+ u\cdot m^2\big)
\end{equation}
a) $B=0$. The resulting equation has three solutions
\begin{equation}
 \overline{m_0(T,0)}=0
\end{equation}
which appears at all temperatures, and
\begin{equation}
\overline{m_\pm (T,0)} = \pm\sqrt{\frac{-r}{u}}
\end{equation}
for $r<0$ (i.e. $\frac{1}{\beta}\leq \frac{1}{\beta_c}$). The solution $\overline{m_0}=0$ is a minimum of the Landau functional above $T_c$, while the two solutions with finite \textbf{spontaneous magnetization} are minima of the Landau functional for $\frac{1}{\beta}\leq \frac{1}{\beta_c}$. Notice the power law vanishing of the spontaneous magnetization in zero applied magnetic field
\begin{equation}
\overline{m}\propto \Big(1-\frac{T}{T_C}\Big)^\beta
\end{equation}
with the so called ''critical exponent'' $\beta_{MFA}=\frac{1}{2}$. This law provides a guide to fitting experimental data on the spontaneous magnetization in the vicinity of the transition temperature.\\
b) $r=0$ ($T=T_c$). The solution writes
\begin{equation}
\overline{m(T_c,B)} = \Big(-\frac{\mu\cdot B}{u}\Big)^{\frac{1}{3}}
\end{equation}
i.e.
\begin{equation}
\overline{m(T_c,B)} \propto -\text{sign}(B)\cdot \vert B\vert^{\frac{1}{\delta }}
\end{equation}
with the critical exponent $\delta_{\text{MFA}}=3$. This power law provides a guide to fitting experimental results taken exactly at the transition temperature with  a varying applied magnetic field.\\
c) The susceptibility is defined as
\begin{equation}
\chi(T,B)\doteq \mu_0\cdot \frac{\partial M}{\partial B}
\end{equation}
Within Landau theory, it can be computed exactly for small $B$ in the vicinity of $T_C$.
\begin{eqnarray}
\chi(T,B=0)&=& -\mu_0\cdot \frac{\mu}{a^3}\cdot \frac{\partial m}{\partial B}\vert_{B=0}\nonumber\\
&=& -\mu_0\cdot \frac{\mu}{a^3}\cdot\frac{1}{\frac{\partial B}{\partial m}\vert_{m=\overline{m(r,B=0)}}}
\end{eqnarray}
Above $T_C$ we compute
\begin{equation}
\chi(T>T_C,B=0)= C\cdot \frac{1}{(T-T_C)^\gamma}
\end{equation}
where
\begin{equation}
C= \frac{\mu_0}{a^3}\cdot \frac{(g\mu_B)^2\cdot S(S+1)}{3\cdot k_B}
\end{equation}
is the Curie constant and $\gamma$ is a critical exponent, which amounts to ''1'' within Landau theory. Below $T_C$ we obtain
\begin{equation}
\chi(T<T_C,B=0)= \frac{C}{2}\cdot \frac{1}{(T_C-T)^{\gamma'}}
\end{equation}
$\gamma'$ being also ''1'' within Landau theory.
Characteristic values for ferromagnetica are given in the following table:
\begin{table}[H]
\begin{center}
\begin{tabular}{c|c|c|c|}
& $T_C$[K] & $C$[K] & $\mu [\mu_B]$\\
\hline
Fe & 1043 & 2.22 & 2.22  \\
\hline
Co & 1395 & 2.24 & 1.71\\
\hline
Ni & 629 & 0.588 & 0.605\\
\hline
Gd & 289 & & 7.1\\
\hline
\end{tabular}
\end{center}
\caption{Curie temperature, Curie constant and magnetic moment per atom of the four elemental ferromagnets. Energies can be expressed in units of Kelvin by noting that $1$ meV $\simeq 1.16 \cdot 10^1$ K.}
\label{Tab:Curie}
\end{table}

\paragraph{Local magnetic field applied at the origin.}
An applied field $B(\vec x)=B_0\cdot a^2\cdot \delta(\vec x)$ defines a preferred orientation for the spin polarization at a localized site (taken as the origin of the coordinate system). By solving the Euler-Lagrange
equation of this problem one can compute the spatially resolved response of the
spin polarization $m(\vec x)$. For simplicity, we operate above $T_C$, where
the $m^3$-term is small and we neglect it, rendering the problem mathematically
more simple. The Euler-Lagrange equation writes
\begin{equation}
-2\cdot A\cdot d\cdot a^2\triangle m(\vec x) + r\cdot \frac{d}{a}\cdot m(\vec x)\approx -\mu\cdot \frac{d}{a}\cdot B_0\cdot a^2\cdot \delta(\vec x)
\end{equation}
To solve, we insert
\begin{equation}
m(\vec x)= \sum_{\vec q} m(\vec q)\cdot e^{i\cdot \vec q\cdot \vec x}
\end{equation}
and
\begin{equation}
\delta(\vec x)= \frac{1}{L^2}\sum_{\vec q} e^{i\cdot \vec q\cdot \vec x}
\end{equation}
into the Euler-Lagrange equation
to obtain
\begin{equation}
\Big(2\cdot A\cdot d\cdot a^2 \vec q^2 + r\cdot \frac{d}{a}\Big)\cdot m(\vec q)\approx -\mu\cdot \frac{d}{a}\cdot B_0\frac{a^2}{L^2}
\end{equation}
The solution of this algebraic equation is
\begin{equation}
m(\vec q) = -\mu\cdot \frac{d}{a}\cdot B_0\cdot \frac{a^2}{L^2} \frac{1}{\Big(2\cdot A\cdot d\cdot a^2 \cdot \vec q^2 + r\cdot \frac{d}{a}\Big)}
\end{equation}
We Fourier transform this solution to obtain the Green function of the original Euler-Lagrange equation:
\begin{eqnarray}
G(\vec x) &=& -\mu\cdot \frac{d}{a}\cdot B_0\cdot \frac{a^2}{L^2}\cdot \frac{L^2}{(2\pi)^2}\int d^2q\cdot e^{i\vec q\cdot \vec x}\cdot \frac{1}{\Big(2 A d a^2 \vec q^2 + r\cdot \frac{d}{a}\Big)}\nonumber\\
&=& \frac{-\mu\cdot \frac{d}{a}\cdot B_0}{2\pi\cdot 2\cdot A\cdot d\cdot a^2}\cdot \int_0^{\infty} dq\cdot q\cdot \frac{J_0(\vert \vec q\vert\cdot \vert \vec x\vert )}{q^2+\frac{r\cdot\frac{d}{a}}{2\cdot A\cdot d\cdot a^2}}\nonumber\\
&=&\frac{-\mu\cdot B_0}{4\pi\cdot A\cdot a^3}\cdot K_0(\frac{\vert \vec x\vert}{\xi})
\end{eqnarray}
$K_0$ decays exponentially at large distances from the origin, with the characteristic length $\xi= a\cdot \sqrt{\frac{2\cdot A\cdot a}{r}}$.
$\xi$ is called the \textbf{correlation length}. Within Landau theory, the correlation length diverges upon approaching $T_C$, with the power law
$\xi\propto r^{-\nu}$. The critical exponent $\nu$ amounts, within Landau theory, to $\frac{1}{2}$.

\chapter{Renormalization group algorithm for the $m^4$-Landau-Ginzburg-Wilson Hamiltonian.}
\label{Chap:RG}
\renewcommand{\thesection}{\thechapter.\arabic{section}}
\renewcommand{\theequation}{\thechapter.\arabic{equation}}

The scaling algorithm has turned out to provide precise mathematical
instruments for managing experimental data relating to the phase transition in
epitaxial magnetic films. It also has the
property of supplying a general framework for generalizing mean field results. However, it provides by itself neither a proof of the
homogeneity relation nor values for the scaling dimensions.
One source of quantitative reliable results regarding phase transitions
in two-dimensional systems is certainly the exact solution of the 2D Ising model by Onsager\cite{Ons}. Further reliable quantitative results are reported in numerical works on the same model, such as those quoted in\cite{Domb}. It was
however the application of the renormalization group introduced by K. Wilson
that provided a comprehensive, analytic quantitative analysis of the phase
transition at critical points. In this Chapter, we would like to give a simplified view of the Renormalization group algorithm. Our aim is showing how the homogeneity relation develops from first principles and how values of the scaling dimensions (and the related critical exponents) can be practically computed. We use, to achieve this aim, the $m^4$-model of the phase transition. We emphasize that our approach is not based on precision but on understanding. We hope that a practical view of the concepts underlying the RG treatment of the $m^4$-model will also allow to better understand the Polyakov Renormalization scheme used in Chapter 7. This Chapter is also an ''a posteriori'' justification of the methods used to analyze experimental data on epitaxial films phase transitions in Chapter 9.\\
\section{Computation of scaling dimensions.}
We use the RG algorithm within a perturbational approach. This requires finding the equivalent of the ${\cal L}_0[\sigma]$ and ${\cal L}_1[S,\sigma]$ of Chapter 7 for the $m^4$-model, given by the functional (in $D$-dimensions)
\begin{eqnarray}
{\cal L}[m(\vec x)]&=& \frac{\Gamma}{2}\cdot a^{2-D}\cdot \int d^Dx \big(\vec \nabla m(\vec x)\big)^2\nonumber\\
&+& \frac{r}{2}\cdot a^{-D}\cdot \int d^Dx\, m^2(\vec x)\nonumber\\
&+& \frac{u}{4}\cdot a^{-D}\cdot \int d^Dx\, m^4(\vec x)
\end{eqnarray}
(the Zeeman term is dropped until it is needed, in section \ref{Chap:RG}.3, for
the scaling dimension $\Delta_B$).
The field $m(\vec x)$ is considered as the sum of two components:
\begin{eqnarray}
S(\vec  x)&=&\frac{1}{\sqrt{N^D}}\cdot \sum _{0\leq\vert \vec  k\vert <
k_L} S_{\vec  k}\cdot e^{i\cdot \vec  k\cdot \vec  x}\nonumber\\
\sigma(\vec  x)&=& \frac{1}{\sqrt{N^D}}\cdot \sum _{k_L<\vert \vec  q\vert <k_{BZ}} \sigma_{\vec  q}\cdot e^{i\cdot \vec  q\cdot \vec  x}
\end{eqnarray}
with
\begin{equation}
k_L\doteq \frac{k_{BZ}}{L}
\end{equation}
$L$ being measured in units of the lattice constant $a$ throughout the Chapter.
The two components fulfill useful orthogonality relations, such as (further orthogonality relations will be introduced ad-hoc)
\begin{equation}
\int d^Dx\, S(\vec x)\cdot \sigma(\vec x)=0
\end{equation}
Proof: Insert the Fourier transforms and use
$\int d^Dx\, e^{i(\vec k+\vec q)\cdot \vec x}= a^D\cdot N^D\cdot \delta_{\vec k,-\vec q}$:
\begin{eqnarray}
\int d^Dx\, S(\vec x)\cdot \sigma(\vec x)&=& \frac{1}{N^D}\sum_{\vec k,\vec q}S_{\vec k}\cdot \sigma_{\vec q}\cdot \int d^Dx\, e^{i(\vec k+\vec q)\cdot \vec x}\nonumber\\
&=& a^D\cdot \sum_{\vec k}S_{\vec k}\cdot \sigma_{-\vec k}
\end{eqnarray}
As $\vec k$ and $\vec q$ assume values in different ranges, $\delta_{\vec k,-\vec q} =0$ for every term of the sum and the orthogonality is proven. $\quad$ QED.\\
We next insert $S(\vec x) + \sigma(\vec x)$ into the Landau functional for $m(\vec x)$, resolve the powers of $S(\vec x) + \sigma(\vec x)$ into monomials and use the orthogonality relations for the fields to find
\begin{eqnarray}
{\cal L}[m(\vec x)]&=& a^{-D}\cdot \int d^Dx \Big(\frac{\Gamma\cdot a^2}{2}\cdot \big(\vec \nabla \sigma(\vec x)\big)^2+ \frac{r}{2}\sigma^2(\vec x)\Big)\nonumber\\
&+& a^{-D}\cdot \int d^Dx \Big(\frac{\Gamma\cdot a^2}{2}\big(\vec \nabla S(\vec x)\big)^2+ \frac{r}{2}S^2(\vec x)\Big)\nonumber\\
&+& \frac{u}{4}\cdot a^{-D}\cdot \int d^Dx \cdot \Big(S(\vec x)+ \sigma(\vec x)\Big)^4
\label{Mono}
\end{eqnarray}
From Eq.\ref{Mono} one can read out
\begin{equation}
a^{-D}\cdot \int d^Dx \Big(\frac{\Gamma\cdot a^2}{2}\cdot \big(\vec \nabla \sigma(\vec x)\big)^2+ \frac{r}{2}\sigma^2(\vec x)\Big)
\end{equation}
as being the sought for ${\cal L}_0[\sigma]$ and
\begin{equation}
\frac{u}{4}\cdot a^{-D}\cdot \int d^Dx \cdot \Big(S(\vec x)+ \sigma(\vec x)\Big)^4
\end{equation}
as being the ''interaction'' functional ${\cal L}_1[\sigma,S]$.
The term appearing in the middle line of Eq.\ref{Mono} is quadratic in
$S$ and represents the harmonic contribution ${\cal L}_{\text{harm}}[S]$ to the ${\cal L}[S]$-functional. This term must be added to the smoothed functional, so that the total functional for the field $S(\vec x)$ writes (up to second order perturbation theory)
\begin{eqnarray}
{\cal L}[S]&=& {\cal L}_{\text{harm}}[S]\nonumber\\
&+& <{\cal L}_1[S,\sigma]>_\sigma\quad (\text{1st order})\nonumber\\
&-& \frac{\beta}{2}\cdot \Big[<{\cal L}^2_1[S,\sigma]>_\sigma-<{\cal L}_1[S,\sigma]>^2_\sigma\Big]\quad (\text{2nd order})
\end{eqnarray}
The Fourier transform of the functional ${\cal L}_0$ writes
\begin{equation}
\sum_{\vec q} \Big(\frac{\Gamma}{2}\cdot a^2\cdot \vec q^2 + \frac{r}{2}\Big)\cdot \sigma_{\vec q}\cdot \sigma_{-\vec q}
\end{equation}
i.e. is quadratic in the field $\sigma_{\vec q}$ and, when used for the functional integration implied by the operation $<...>_\sigma$, provides a Gauss average.
\paragraph{1st order.}
Because of the binomial expansion, the kernel of $<{\cal L}_1[S,\sigma]>_\sigma$ contains the monomials\\
-- $<S^4>_\sigma$: this term provides a fourth power term to the $S$-functional that adds up to the harmonic component: the $S$-functional is updated to a $S^4$-model.\\
-- $<\sigma^4>_\sigma$: this term does not contain the field $S$ and provides a non-singular component to  the thermodynamic potential.\\
-- $4\cdot <S^3\cdot\sigma>_\sigma$ and $4\cdot <S\cdot\sigma^3>_\sigma$: the Gauss average of terms containing odd powers of $\sigma$ vanishes exactly.\\
-- $6\cdot<S^2\cdot\sigma^2>_\sigma$. This is the only 1st order term that contributes to the functional of the $S$-field. It provides a further term quadratic in $S$ and, accordingly, it modifies the coupling constant $\frac{r}{2}$, which becomes
\begin{equation}
\frac{r}{2} + 6\cdot\frac{u}{4}\cdot <\sigma^2(x)>_\sigma
\end{equation}
We recall that
\begin{equation}
<\sigma_{\vec q_1}\cdot \sigma_{\vec q_2}>_\sigma= \delta_{\vec q_1+\vec q_2,0}\cdot <\vert \vec \sigma_{q_1}\vert^2>_\sigma= \delta_{\vec q_1+\vec q_2,0}\cdot\frac{1}{\beta\cdot\big(\Gamma\cdot a^2\cdot q^2 +r\big)}
\end{equation}
so that, with
\begin{equation}
C(D)\doteq \frac{1}{(2\pi)^D}\cdot \frac{2\cdot \pi^{\frac{D}{2}}}{\Gamma_E\big(\frac{D}{2}\big)}
\label{Cd}
\end{equation}
($\frac{2\pi^{D/2}}{\Gamma_E(D/2)}$ being the surface of the unit sphere in $D$
dimensions), one obtains
\begin{eqnarray}
<\sigma^2(x)>_\sigma&=&\frac{1}{N^D}\sum_{\vec q}<\vert \vec \sigma_{q}\vert^2>_\sigma\nonumber\\
&=& a^D\cdot C(D)\cdot \int_{k_L}^{k_{BZ}}q^{D-1}\cdot dq\cdot \frac{1}{\beta\cdot\big(\Gamma\cdot a^2\cdot q^2 +r\big)}\nonumber\\
&=& a^D\cdot C(D)\cdot\frac{1}{\beta\cdot \Gamma}\cdot \int_{k_L}^{k_{BZ}}q^{D-1}\cdot dq\cdot \frac{1}{a^2\cdot q^2 +\frac{r}{\Gamma}}\nonumber\\
&\underbrace{=}_{qa\doteq x}&C(D)\cdot\frac{1}{\beta\cdot \Gamma}\cdot\int_{k_L\cdot a}^{k_{BZ}\cdot a}x^{D-1}\cdot dx\cdot \frac{1}{x^2+ \frac{r}{ \Gamma}}
\label{sigma2}
\end{eqnarray}
The kernel of the integral on the right hand side can be expressed as a Taylor series in the variable $\frac{r}{ \Gamma}<<x^2$:
\begin{equation}
\frac{1}{x^2+\frac{r}{\Gamma}}= \frac{1}{x^2}- \frac{\frac{r}{\Gamma}}{x^4} \pm....
\end{equation}
i.e.
\begin{eqnarray}
<\sigma^2(x)>_\sigma&\approx& C(D)\cdot\frac{1}{\beta\cdot \Gamma}\cdot\int_{k_L\cdot a}^{k_{BZ}\cdot a}x^{D-1}\cdot dx\cdot \frac{1}{x^2}\nonumber\\
&-& C(D)\cdot\frac{1}{\beta\cdot \Gamma}\cdot\int_{k_L\cdot a}^{k_{BZ}\cdot a}x^{D-1}\cdot dx \frac{\frac{r}{\Gamma}}{x^4}\nonumber\\
&=& C(D)\cdot\frac{1}{\beta\cdot \Gamma}\cdot\frac{(k_{BZ}\cdot a)^{D-2}}{D-2}\cdot \big(1-\frac{1}{L^{D-2}}\big)\nonumber\\
&-& C(D)\cdot\frac{\frac{r}{\Gamma}}{\beta\cdot  \Gamma}\cdot\int_{k_L\cdot a}^{k_{BZ}\cdot a}x^{D-5}\cdot dx
\end{eqnarray}
The 1st order perturbation theory to lowest order in $\frac{r}{\Gamma}$ foresees that the coupling constant of the $S^2$-component of the functional is modified to
\begin{equation}
\frac{r}{2} + \frac{3}{2}\cdot u\cdot C(D)\cdot\frac{1}{\beta\cdot \Gamma}\cdot\frac{(k_{BZ}\cdot a)^{D-2}}{D-2}\cdot \big(1-\frac{1}{L^{D-2}}\big)
\end{equation}
As $L$ is assumed to be large (the correlation length is bound to diverge close to the transition temperature), the $\frac{1}{L^{D-2}}$ term may be neglected, provided $D>2$. Accordingly, the correction to the coupling coefficient becomes an $L$-independent additive positive constant that effectively  produces a shift of the transition temperature with respect to the original value $T_C$. When considering the second order perturbation theory terms, we consider $r$ replaced by $r^*\propto \big(\frac{1}{\beta}-\frac{1}{\beta^*_C}\big)$ and limit ourselves to the range $D>2$. The term of the kernel linear in $\frac{r}{\Gamma}$, neglected to lowest order, will be reintroduced later because it competes with the second order perturbation expansion terms.\\
\paragraph{Remark: Diagrammatic representation of the perturbation expansion.}
It is instructive to discuss the management of the various perturbation theoretical terms by means of a diagrammatic technology that resembles the Feynman graphs of quantum field theory. We consider again
\begin{equation}
{\cal L}_1[S,\sigma]= \Big<\frac{u}{4}\cdot a^{-D}\cdot \int d^Dx \cdot \Big(S(\vec x)+ \sigma(\vec x)\Big)^4\Big>
\end{equation}
and translate it into Fourier space. The non-vanishing terms read
\begin{eqnarray*}
{\cal L}_1[S,\sigma]&=& \frac{u}{4}\cdot \frac{1}{N^D}\cdot \sum_{\vec k_1,\vec k_2,\vec k_3,\vec k_4} \Big<S_{\vec k_1}\cdot S_{\vec k_2}\cdot S_{\vec k_3}\cdot S_{\vec k_4}\Big>\cdot \delta_{\vec k_1+\vec k_2+\vec k_3+\vec k_4,0}\\
&+& \frac{u}{4}\cdot \frac{6}{N^D}\cdot \sum_{\vec k_1,\vec k_2,\vec q_1,\vec q_2} \Big<S_{\vec k_1}\cdot S_{\vec k_2}\cdot \sigma_{\vec q_1}\cdot \sigma_{\vec q_2}\Big>\cdot \delta_{\vec k_1+\vec k_2+\vec q_1+\vec q_2,0}\\
&+& \frac{u}{4}\cdot \frac{1}{N^D}\cdot \sum_{\vec q_1,\vec q_2,\vec q_3,\vec q_4} \Big<\sigma_{\vec q_1}\cdot \sigma_{\vec q_2}\cdot \sigma_{\vec q_3}\cdot\sigma_{\vec q_4}\Big>\cdot \delta_{\vec q_1+\vec q_2+\vec q_3+\vec q_4,0}
\end{eqnarray*}
The third line is the average of the many-fold product of Gaussian distributed fields. To deal with it one can resort to a nice mathematical theorem regarding Gaussian distributed quantities, known as Isserlis theorem\cite{CS} (in field theory the theorem is known as Wick's theorem). According to this theorem, the average of the product of an even number $n=2m$ of Gaussian distributed quantities $x_1,...,x_{2m}$ can be expressed as the sum of the products of the averages of pairs:
 \begin{equation}
 <x_1\cdot\! x_2\cdot\!....\cdot\! x_{2m}>= \sum <x_{i_1}\cdot\! x_{i_2}>\cdot\! <x_{i_3}\cdot\! x_{i_4}>\cdot ...\cdot <x_{i_{2m-1}}\cdot x_{i_{2m}}>
 \end{equation}
The sum is over all possible ways to group the indices $1,...,2m$ into pairs $(i_1, i_2), . . . , (i_{2m-1}, i_{2m})$.
Wick's theorem for Gaussian distributed quantities, applied to the four-fields average in the third line, generates three two-fields averages:
\begin{eqnarray*}
\Big<\sigma_{\vec q_1}\cdot \sigma_{\vec q_2}\cdot \sigma_{\vec q_3}\cdot\sigma_{\vec q_4}\Big>_\sigma&=&\overbracket{\sigma_{\vec q_1}\cdot \sigma_{\vec q_2}}\cdot \overbracket {\sigma_{\vec q_3}\cdot\sigma_{\vec q_4}}+ \overbracket{\sigma_{\vec q_1}\cdot \overbracket{\sigma_{\vec q_2}\cdot \sigma_{\vec q_3}}\cdot\sigma_{\vec q_4}}+
\overbracket{\sigma_{\vec q_1}\cdot \underline{\sigma_{\vec q_2}}\cdot \sigma_{\vec q_3}}\cdot \underline{\sigma_{\vec q_4}}
\end{eqnarray*}
The fields joined by a cap or underlined are said to be \textbf{contracted}. Contracted fields are averaged over $\sigma$:
\begin{equation}
\overbracket{\sigma_{\vec q_1}\cdot \sigma_{\vec q_2}}\equiv \Big<\sigma_{\vec q_1}\cdot \sigma_{\vec q_2}\Big>
\end{equation}
Following the Wick's theorem, the first order expansion can be rewritten as
\begin{eqnarray*}
{\cal L}_1[S,\sigma]&=& \frac{u}{4}\cdot \frac{1}{N^D}\cdot \sum_{\vec k_1,\vec k_2,\vec k_3,\vec k_4} S_{\vec k_1}\cdot S_{\vec k_2}\cdot S_{\vec k_3}\cdot S_{\vec k_4}\cdot \delta_{\vec k_1+\vec k_2+\vec k_3+\vec k_4,0}\\
&+& \frac{u}{4}\cdot \frac{6}{N^D}\cdot \sum_{\vec k_1,\vec k_2,\vec q_1,\vec q_2} S_{\vec k_1}\cdot S_{\vec k_2}\cdot \overbracket{\sigma_{\vec q_1}\cdot \sigma_{\vec q_2}}\cdot \delta_{\vec k_1+\vec k_2+\vec q_1+\vec q_2,0}\\
&+& \frac{u}{4}\cdot \frac{1}{N^D}\cdot \sum_{\vec q_1,\vec q_2,\vec q_3,\vec q_4}\overbracket{\sigma_{\vec q_1}\cdot \sigma_{\vec q_2}}\cdot \overbracket {\sigma_{\vec q_3}\cdot\sigma_{\vec q_4}}\cdot \delta_{\vec q_1+\vec q_2+\vec q_3+\vec q_4,0}\\
&+& \frac{u}{4}\cdot \frac{1}{N^D}\cdot \sum_{\vec q_1,\vec q_2,\vec q_3,\vec q_4}  \overbracket{\sigma_{\vec q_1}\cdot \overbracket{\sigma_{\vec q_2}\cdot \sigma_{\vec q_3}}\cdot\sigma_{\vec q_4}}\cdot \delta_{\vec q_1+\vec q_2+\vec q_3+\vec q_4,0}\\
&+&\frac{u}{4}\cdot \frac{1}{N^D}\cdot \sum_{\vec q_1,\vec q_2,\vec q_3,\vec q_4} \overbracket{\sigma_{\vec q_1}\cdot \underline{\sigma_{\vec q_2}}\cdot \sigma_{\vec q_3}}\cdot \underline{\sigma_{\vec q_4}}\cdot \delta_{\vec q_1+\vec q_2+\vec q_3+\vec q_4,0}
\end{eqnarray*}
Each member of the sums $\sum_{\vec k,\vec q}$ can be associated with a \textbf{graph}, constructed according to the following rules:\\
\textbf{-- a.} Associate to each vector $\vec k$ and $\vec q$ a line representing the corresponding field $S_{\vec k}$ respectively
$\sigma_{\vec q}$.\\
\textbf{-- b.} The lines must terminate in a ''vertex'', owing to the momentum conservation represented by the Kronecker delta symbol.\\
\textbf{-- c.} The $\vec q$-lines must be \textbf{''contracted''}  because of the $<..>_\sigma$ operation, so that they become lines ''internal'' to the graph, in contrast to the $\vec k$-lines, which are ''external''.  The internal lines $\vec q_i,\vec q_j$ contribute to the vertex the ''propagator''
\begin{equation}
\delta_{\vec q_i+\vec q_j,0}\cdot <\vert \sigma_{\vec q_i}\vert^2>_\sigma
\end{equation}
\textbf{-- d.} One assigns to the vertex a ''magnitude''. The magnitude is the coefficient that comprises the strength of the coupling constant of the perturbation expansion, multiplied by the coefficient of the binomial expansion and the combinatorial number of ways the $q$-fields can be contracted (i.e. the number of topologically equivalent graphs).\\
According to these rules, the first order graphs have the three distinct topologies summarized in the next figure.
\begin{figure}[H]
\begin{center}
\includegraphics[width=0.6\textwidth]{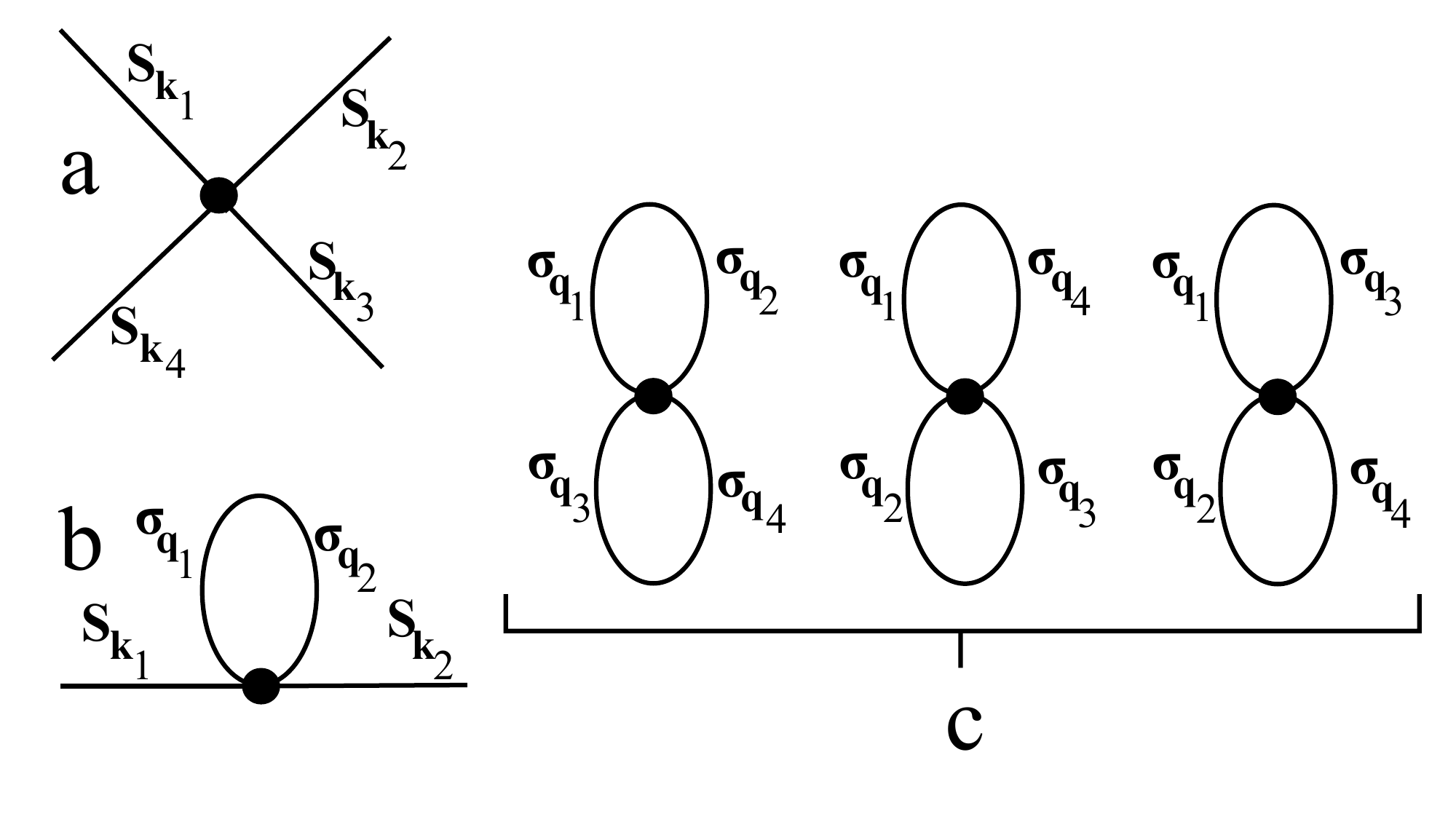}
\caption{a. The graph corresponding to the term
$S_{\vec k_1}\cdot S_{\vec k_2}\cdot S_{\vec k_3}\cdot S_{\vec k_4}\cdot
\delta_{\vec k_1+\vec k_2+\vec k_3+\vec k_4,0}
$
has only external lines. The magnitude of the vertex is $\frac{u}{4}
\cdot 1\cdot 1$.\\
b. The graph corresponding to the term
$S_{\vec k_1}\cdot S_{\vec k_2}\cdot \big<\sigma_{\vec q_1}\cdot \sigma_{\vec q_2}\big>\cdot \delta_{\vec k_1+\vec k_2+\vec q_1+\vec q_2,0}$ has two
external lines and one internal.
The magnitude of the vertex is $\frac{u}{4}\cdot 6 \cdot 1$.\\
c. There are three graphs with identical topology corresponding to the terms
$\big<\sigma_{\vec q_i}\cdot \sigma_{\vec q_j}\big>\cdot \big< \sigma_{\vec q_m}\cdot\sigma_{\vec q_n}\big>\cdot \delta_{\vec q_i+\vec q_j+\vec q_m+\vec q_n,0}$. They have only internal lines. The magnitude of the vertex
is $\frac{u}{4}\cdot 1 \cdot 3$.}
\label{Fig:1storder}
\end{center}
\end{figure}
The perturbational term represented by $a$ has only external lines and contributes a $S^4$-term to the $S$-functional. The perturbational term represented by $b$ has two external lines and therefore contributes a $S^2$-term to the functional. Its coupling amounts to the magnitude of the vertex multiplied by $\frac{1}{N^D}\cdot \sum_{\vec q} \big<\vert \sigma_{\vec q}\vert^2\big>$. The perturbational terms represented by $c$ do not have external lines and therefore do not contribute to the $S$-functional.
\paragraph{2nd order.}
As an introduction, we study the term
\begin{eqnarray}
-\frac{\beta}{2}\cdot (\frac{u}{4})^2\cdot 6^2 \cdot  \Big<\int \frac{d^Dx}{a^D}\cdot S^2(\vec x)\cdot\sigma^2(\vec x)\cdot \int\frac{d^Dx'}{a^D}\cdot S^2(\vec x')\cdot \sigma^2(\vec x')\Big>_\sigma
\end{eqnarray}
which appears in the binomial expansion of
\begin{eqnarray}
-\frac{\beta}{2}\cdot (\frac{u}{4})^2\cdot \Big<\int \frac{d^Dx}{a^D}\cdot \left(S(\vec x)+\sigma(\vec x)\right)^4
\cdot  \int \frac{d^Dx'}{a^D} \left(S(\vec x')+\sigma(\vec x')\right)^4\Big>_\sigma
\end{eqnarray}
To use the diagrammatic technology, we Fourier transform it to
\begin{eqnarray}
-\frac{\beta}{2}\cdot (\frac{u}{4})^2\cdot 6^2\cdot (\frac{1}{N^D})^2&\cdot & \sum_{\vec k_1,\vec k_2,\vec k_3,\vec k_4,\vec q_1,\vec q_2,\vec q_3,\vec q_4}S_{\vec k_1}\cdot S_{\vec k_2}\nonumber\\
&\cdot& \Big<\sigma_{\vec q_1}\cdot \sigma_{\vec q_2}\cdot \sigma_{\vec q_3}\cdot \sigma_{\vec q_4}\Big>_\sigma\nonumber\\
&\cdot & S_{\vec k_3}\cdot S_{\vec k_4}\nonumber\\
&\cdot&\delta_{\vec k_1+ \vec k_2+\vec q_1+\vec q_2,0}\cdot \delta_{\vec k_3+ \vec k_4+\vec q_3+\vec q_4,0}
\end{eqnarray}
Wick's theorem, applied to the four-fields average, generates three two-fields averages:
\begin{eqnarray*}
\Big<\sigma_{\vec q_1}\cdot \sigma_{\vec q_2}\cdot \sigma_{\vec q_3}\cdot\sigma_{\vec q_4}\Big>_\sigma&=&\overbracket{\sigma_{\vec q_1}\cdot \sigma_{\vec q_2}}\cdot \overbracket {\sigma_{\vec q_3}\cdot\sigma_{\vec q_4}}+ \overbracket{\sigma_{\vec q_1}\cdot \overbracket{\sigma_{\vec q_2}\cdot \sigma_{\vec q_3}}\cdot\sigma_{\vec q_4}}+
\overbracket{\sigma_{\vec q_1}\cdot \underline{\sigma_{\vec q_2}}\cdot \sigma_{\vec q_3}}\cdot \underline{\sigma_{\vec q_4}}\\
&=& \delta_{\vec q_1+\vec q_2,0}\cdot \delta_{\vec q_3+\vec q_4,0}\cdot <\vert \vec \sigma_{q_1}\vert^2>_\sigma\cdot <\vert \vec \sigma_{q_3}\vert^2>_\sigma\\
&+&\delta_{\vec q_1+\vec q_4,0}\cdot \delta_{\vec q_2+\vec q_3,0}\cdot <\vert \vec \sigma_{q_1}\vert^2>_\sigma\cdot <\vert \vec \sigma_{q_2}\vert^2>_\sigma\\
&+&\delta_{\vec q_1+\vec q_3,0}\cdot \delta_{\vec q_2+\vec q_4,0}\cdot <\vert \vec \sigma_{q_1}\vert^2>_\sigma\cdot <\vert \vec \sigma_{q_2}\vert^2>_\sigma
\end{eqnarray*}
There are therefore three graphs associated with this term. Each graph has two vertices, following the two Kronecker-delta's appearing in the Fourier transform. The graphs are summarized in the next figure.
\begin{figure}[H]
\begin{center}
\includegraphics[width=0.5\textwidth]{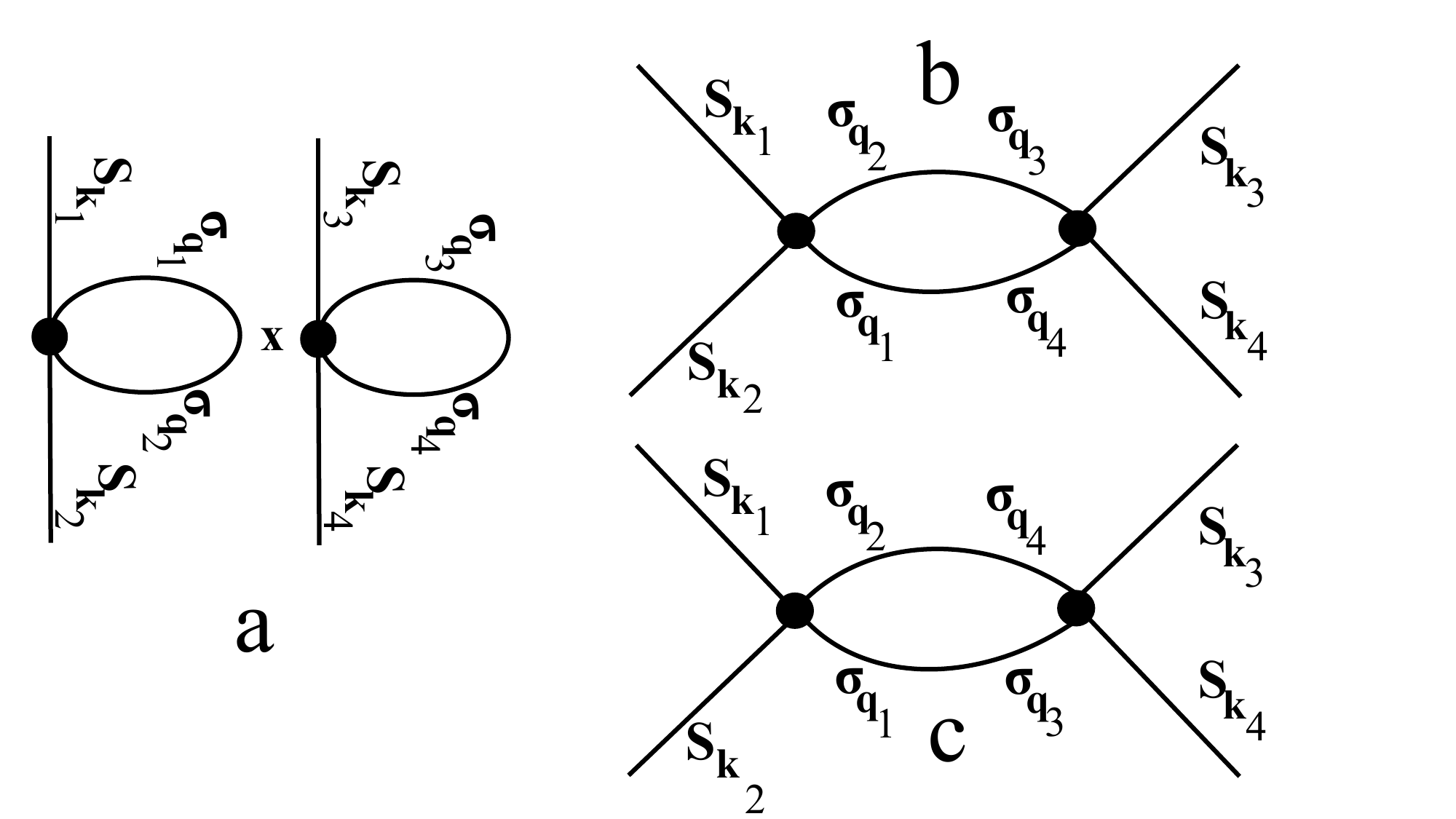}
\caption{Some second order graphs. For a comment see the bulk of the section.}
\label{Fig:2ndorderI}
\end{center}
\end{figure}
The graph of the type in ''a'' on the left is designated as being \textbf{disconnected}: the two vertices are not ''connected'' by any lines. The graphs in ''b'' and ''c'' are connected and topologically equivalent. It is a general result of the perturbation expansion -- the linked-cluster theorem, see e.g.\cite{Rad} -- that the disconnected diagrams in $\big<{\cal L}^2_{1}\big>_\sigma$ are cancelled out exactly by the diagrams in $\big<{\cal L}_{1}\big>^2_\sigma$, which are all disconnected. Accordingly, the second order contribution to the $S$-functional reduces to the connected diagrams in $\big<{\cal L}^2_{1}\big>_\sigma$. In the next figure
\begin{figure}[H]
\begin{center}
\includegraphics[width=0.55\textwidth]{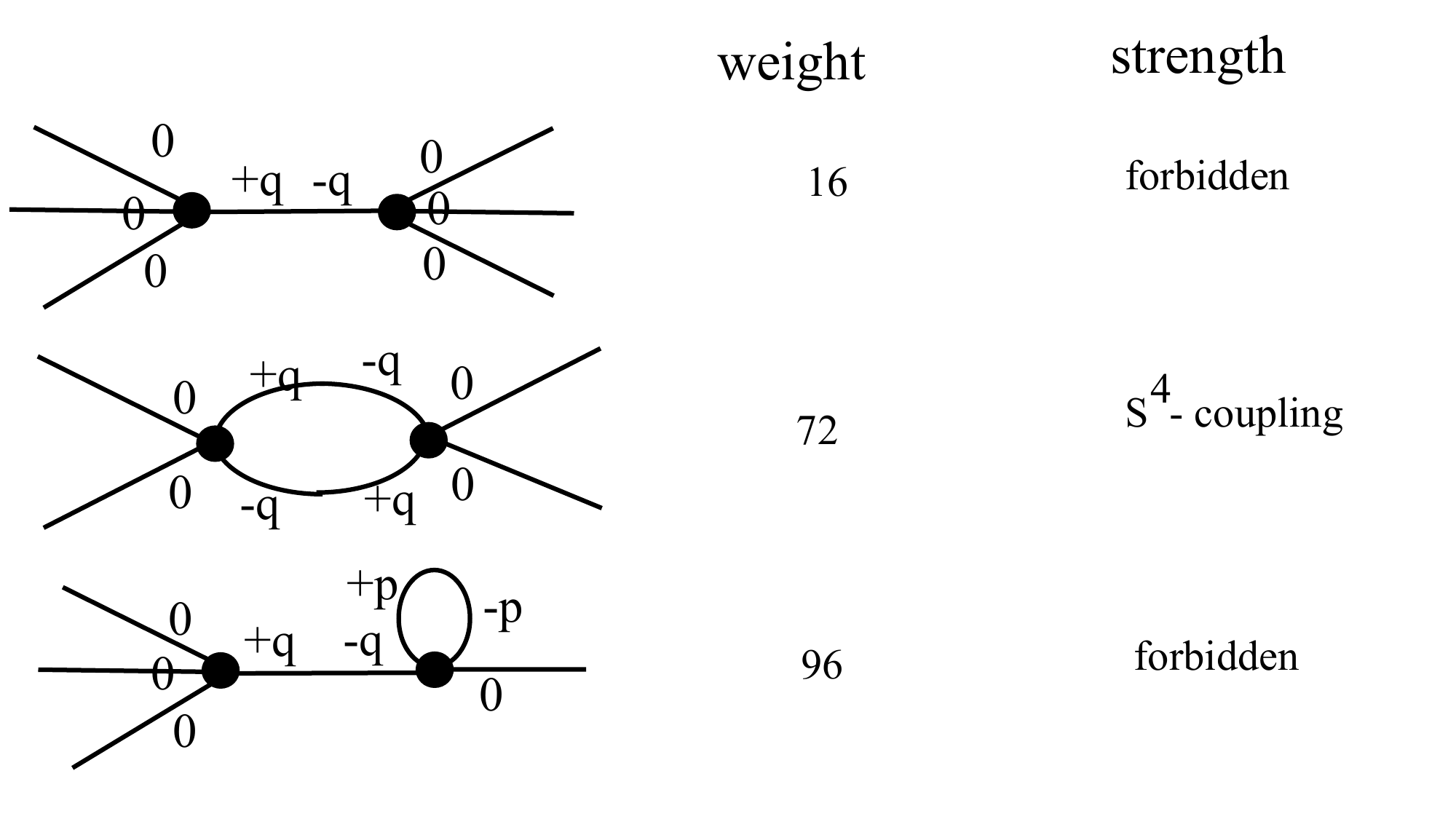}\\[4pt]
\includegraphics[width=0.55\textwidth]{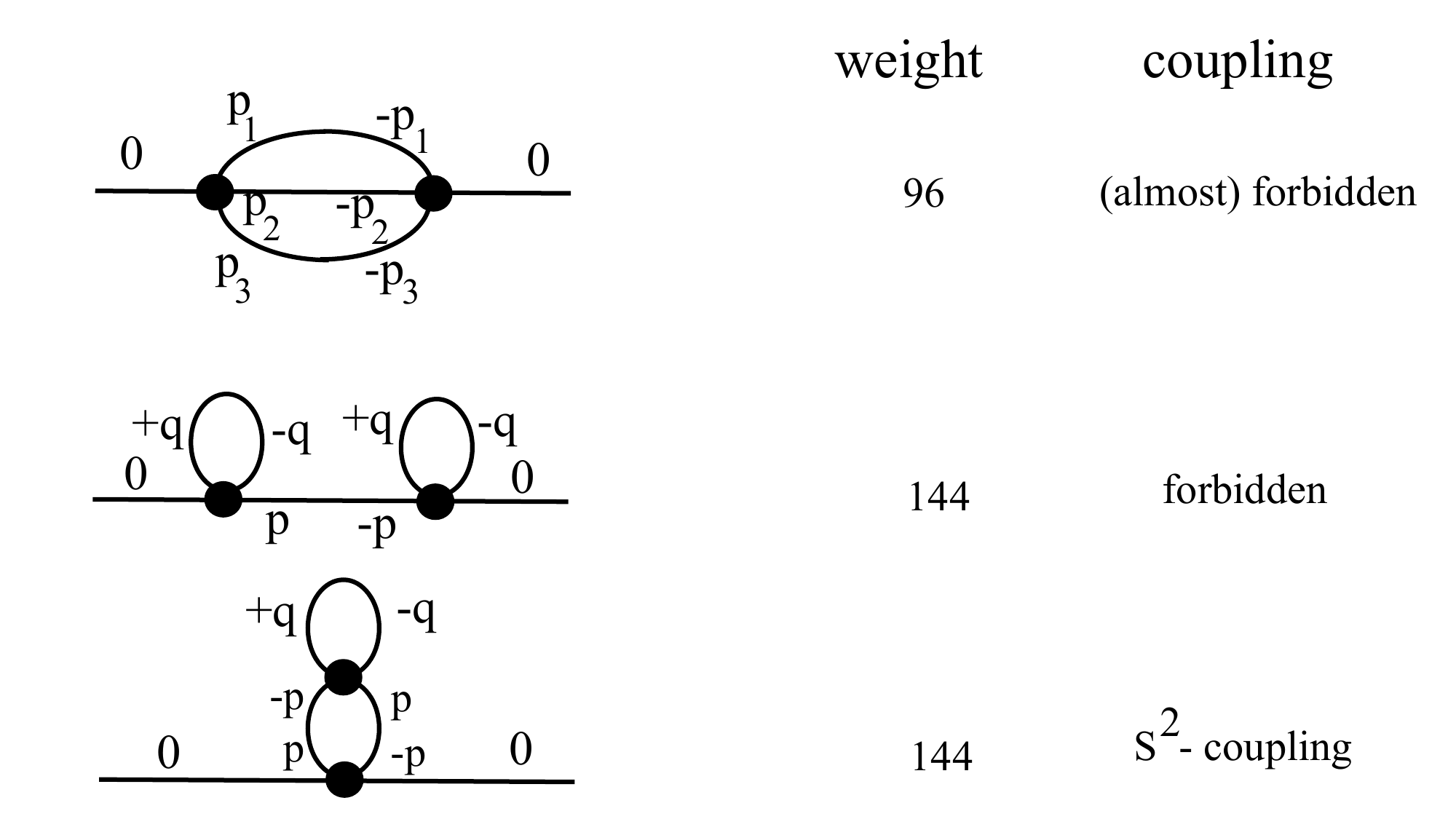}\\[4pt]
\includegraphics[width=0.33\textwidth]{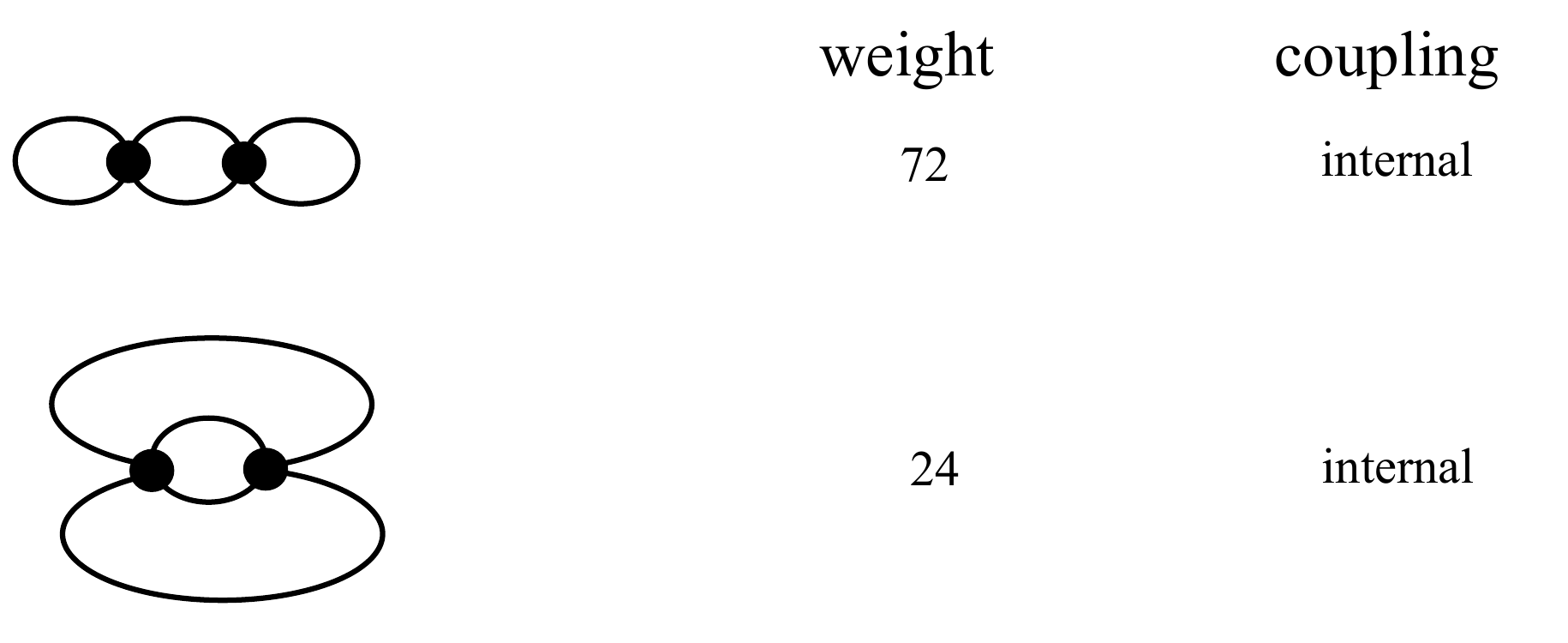}
\caption{Second order connected graphs, their magnitude and strength estimate.}
\label{Fig:2ndorder}
\end{center}
\end{figure}
we summarize the second order connected diagrams, their magnitude and make a
statement about whether they are allowed or not with respect to momentum
conservation (Kronecker-delta). On the left of the figure one finds  the
connected diagrams in  $\big<{\cal L}^2_{1}\big>_\sigma$. In the figure, the
lines are labeled with the momenta rather than with the corresponding fields,
for simplicity. The column in the middle shows the magnitude (or weight) of the
vertices (the numbers in the column headed ''weight'' in the figure must be
multiplied by $-\frac{\beta}{2}\cdot (\frac{u}{4})^2$). The column on the right
gives an estimate of their strength, based on the rule of conservation of
momentum at each vertex. For judging whether the momenta are conserved (diagram
is allowed) or not conserved (diagram is forbidden), one uses the rule that the
momenta of the external lines are small, so that we can set them to ''0'' for a
simplified estimate of momentum conservation (Polyakov-rule). The momenta of
the internal lines, instead, are large and two internal momenta that join a
four line vertex can only realize momentum conservation if they have opposite
sign. In one graph, the coupling is labeled with ''(almost) forbidden''. This is because one side of the vertex has momentum ''0''. The other side has three large momentum vectors, so that one could think, at first glance, that they cannot sum up to ''0''. However, three large momentum vectors can be summed to ''0'', albeit in a geometrically limited way. This is why this graph contributes to higher order of the renormalization procedure.\\
We compute now explicitly the graph with non-forbidden $S^4$-output. This second order correction writes:
\begin{eqnarray}
\frac{1}{4}\cdot \frac{1}{N^D}\cdot \sum_{\vec k_{1,2,3,4}}\Big[...\Big]\cdot S_{\vec k_1}\cdot S_{\vec k_2}\cdot S_{\vec k_3}\cdot S_{\vec k_4}
\end{eqnarray}
with
\begin{eqnarray}
\Big[...\Big]&=& -\frac{\beta}{2}\cdot 36\cdot 8 \cdot(\frac{u(1)}{4})^2\cdot \frac{1}{N^D}\sum_{\vec q_{1,2,3,4}}<\sigma_{\vec q_1}\cdot \sigma_{\vec q_3}>_\sigma\cdot <\sigma_{\vec q_2}\cdot \sigma_{\vec q_4}>_\sigma\nonumber\\
&\cdot &\delta_{\vec k_1+ \vec k_2+\vec q_1+\vec q_2,0}\cdot \delta_{\vec k_3+ \vec k_4+\vec q_3+\vec q_4,0}\nonumber\\
&=& -9\cdot \beta\cdot \frac{1}{\beta^2}\cdot u(1)^2\cdot \frac{1}{N^D}\cdot \sum_{\vec q_{1,2,3,4}} \frac{1}{\Gamma\cdot a^2 q_1^2 + r}\cdot \frac{1}{\Gamma\cdot a^2 q_2^2+r}\nonumber\\
&\cdot&  \delta_{\vec k_1+ \vec k_2+\vec q_1+\vec q_2,0}\cdot \delta_{\vec k_3+ \vec k_4+\vec q_3+\vec q_4,0}\cdot \delta_{\vec q_1+\vec q_3,0}\cdot \delta_{\vec q_2+ \vec q_4,0}\nonumber\\
&=& -9\cdot \beta \cdot \frac{1}{\beta^2}\cdot u(1)^2\cdot \frac{1}{N^D}\cdot \sum_{\vec q_{1,2}} \frac{1}{\Gamma\cdot a^2 q_1^2 + r}\cdot \frac{1}{\Gamma\cdot  a^2 q_2^2+r}\nonumber\\
&\cdot &  \delta_{\vec k_1+ \vec k_2+\vec q_1+\vec q_2,0}\cdot \delta_{\vec k_3+\vec k_4 -\vec q_1-\vec q_2,0}\nonumber\\
&=&-9\cdot \beta \cdot \frac{1}{\beta^2}\cdot u(1)^2\cdot \frac{1}{N^D}\cdot \sum_{\vec q}\frac{1}{\Gamma\cdot  a^2 q^2 + r}\cdot \frac{1}{\Gamma\cdot a^2 (\vec k_3 +\vec k_4-\vec q)^2+r}\nonumber\\
&\cdot & \delta_{\vec k_1+ \vec k_2+\vec k_3+\vec k_4,0}\underbrace{=}_{\text{Polyakov:} \vec k_3=\vec k_4 =0}\nonumber\\
&=& -9\cdot \beta \cdot \frac{1}{\beta^2}\cdot u(1)^2\cdot a^D\cdot C(D)\cdot \int_{k_L}^{k_{BZ}} q^{D-1}dq \frac{1}{\big(\Gamma\cdot  a^2\cdot q^2 + r\big)^2}\nonumber\\
&\cdot & \delta_{\vec k_1+ \vec k_2+\vec k_3+\vec k_4,0}
\end{eqnarray}
The second order $S^4$ term writes, finally,
\begin{eqnarray}
\frac{1}{4}\cdot \frac{1}{N^D}\cdot \sum_{\vec k_{1,2,3,4}}&\cdot& \Big[
-9\cdot \frac{1}{\beta}\cdot u(1)^2\cdot a^D\cdot C(D)\cdot \int_{k_L}^{k_{BZ}} q^{D-1}dq \frac{1}{\big(\Gamma\cdot a^2 q^2 + r\big)^2}\Big]\nonumber\\
&\cdot & S_{\vec k_1}\cdot S_{\vec k_2}\cdot S_{\vec k_3}\cdot S_{\vec k_4}\cdot
\delta_{\vec k_1+ \vec k_2+\vec k_3+\vec k_4,0}
\end{eqnarray}
From this equation we observe that the coupling constant $u$ is changed by the second order perturbation theory and is determined by the recursion relation
\begin{equation}
u(L)= u(1)-9\cdot \frac{1}{\beta}\cdot \Big(\frac{u(1)}{\Gamma}\Big)^2\cdot C(D)\cdot \int_{k_L\cdot a}^{k_{BZ}\cdot a} x^{D-1}\,dx \frac{1}{(x^2 + \frac{r}{\Gamma})^2}
\end{equation}
To lowest order in the integrand we have
\begin{equation}
u(L)= u(1) -9\cdot \frac{1}{\beta}\cdot \Big(\frac{u(1)}{\Gamma}\Big)^2\cdot C(D)\cdot
\int_{k_L\cdot a}^{k_{BZ}\cdot a} dx \cdot x^{D-5}
\end{equation}
We must now go back to the first order terms and look closely at the terms we considered so far in that order. We recall that the first order perturbation theory corrects the term $r$ according to the exact formula (\ref{sigma2})
\begin{equation}
r(L)= r(1)+ 3\cdot u(1)\cdot C(D)\cdot\frac{1}{\beta\cdot \Gamma}\cdot\int_{k_L\cdot a}^{k_{BZ}\cdot a} x^{D-1}\,dx\cdot \frac{1}{x^2+ \frac{r(1)}{ \Gamma}}
\end{equation}
The Taylor expansion of the integrand produces, to lowest order, one term that we did already consider and which was found to merely produce a shift of the Curie temperature. The next order Taylor expansion term instead, we did not consider yet, and it writes
\begin{equation}
r(L) = r(1) - 3\cdot \frac{1}{\beta}\cdot \frac{u(1)\cdot r(1)}{\Gamma^2}\cdot C(D)\cdot\int_{k_L\cdot a}^{k_{BZ}\cdot a}x^{D-5}\cdot dx
\end{equation}
We obtain the Gell-Mann-Low equations by writing the recursion relations as
\begin{eqnarray}
r(L+dL)-r(L) &=& - 3\cdot \frac{1}{\beta}\cdot \frac{u(L)\cdot r(L)}{\Gamma^2}\cdot C(D)\nonumber\\
&\cdot & \Big[\int_{k_{L+dL}\cdot a}^{k_{BZ}\cdot a} dx\cdot x^{D-5}- \int_{k_L\cdot a}^{k_{BZ}\cdot a} dx\cdot x^{D-5}\Big]\nonumber\\
u(L+dL)-u(L) &=& - 9\cdot \frac{1}{\beta}\cdot \frac{u(L)\cdot u(L)}{\Gamma^2}\cdot C(D)\nonumber\\
&\cdot & \Big[\int_{k_{L+dL}\cdot a}^{k_{BZ}\cdot a} dx\cdot x^{D-5}- \int_{k_L\cdot a}^{k_{BZ}\cdot a} dx\cdot x^{D-5}\Big]
\end{eqnarray}
We compute the integrals and find, to lowest order in $\frac{dL}{L}$:
\begin{equation}
\Big[\int_{k_{L+dL}\cdot a}^{k_{BZ}\cdot a} dx\cdot x^{D-5}- \int_{k_L\cdot a}^{k_{BZ}\cdot a} dx\cdot x^{D-5}\Big]= (k_{BZ}\cdot a)^{D-4}\cdot \frac{dL}{L^{D-3}}
\end{equation}
This produces the differential equations for $r(L)$ and $u(L)$:
\begin{eqnarray}
\frac{dr}{dL} &=& -3\cdot \frac{1}{\beta}\cdot\frac{u(L)\cdot r(L)}{\Gamma^2}\cdot C(D)\cdot (k_{BZ}\cdot a)^{D-4}\cdot L^{3-D}\nonumber\\
\frac{du}{dL} &=& -9\cdot \frac{1}{\beta}\cdot \frac{u(L)\cdot u(L)}{\Gamma^2}\cdot C(D)\cdot (k_{BZ}\cdot a)^{D-4}\cdot L^{3-D}
\end{eqnarray}
We solve first the equation for $u$:
\begin{eqnarray*}
\int _{u(1)}^{u(L)} \frac{du}{u^2} &=& -9\cdot \frac{1}{\beta}\cdot (\frac{1}{\Gamma})^2\cdot C(D)\cdot (k_{BZ}\cdot a)^{D-4}\cdot \int_1^L x^{3-D}dx\\
\frac{1}{u(1)}-\frac{1}{u(L)} &=& -9\cdot \frac{1}{\beta}\cdot (\frac{1}{\Gamma})^2\cdot C(D)\cdot (k_{BZ}\cdot a)^{D-4}\cdot \frac{1}{4-D}(L^{4-D}-1)
\end{eqnarray*}
i.e.
\begin{equation}
u(L) = u(1)\cdot \frac{1}{1+9\cdot \frac{1}{\beta}\cdot \frac{u(1)}{\Gamma^2 }\cdot C(D)\cdot (k_{BZ}\cdot a)^{D-4}\frac{1}{4-D}\cdot (L^{4-D}-1)}
\label{uL}
\end{equation}
We insert this exact result for $u(L)$ into the differential equation for $r$ and integrate:
\begin{eqnarray*}
\int_{r(1)}^{r(L)}\frac{dr}{r} &=& -\frac{1}{3}\int_1^L \frac{1}{1+ 9\cdot \frac{u(1)}{\beta\cdot \Gamma^2}\cdot C(D)\cdot (k_{BZ}\cdot a)^{D-4}\frac{1}{4-D}\cdot (x^{4-D}-1)}\\
&\cdot & 9\cdot \frac{u(1)}{\beta\cdot \Gamma^2 }\cdot C(D)\cdot (k_{BZ}\cdot a)^{D-4}\cdot x^{3-D}\cdot dx
\end{eqnarray*}
The integrand is $\frac{1}{3}$ of the logarithmic derivative of the denominator, so that
\begin{equation}\ln\frac{r(L)}{r(1)}= -\frac{1}{3}\ln\left[1+9\cdot \frac{u(1)}{\beta\cdot \Gamma^2}\cdot C(D)\cdot (k_{BZ}\cdot a)^{D-4}\frac{1}{4-D}\cdot (L^{4-D}-1)\right]
\label{rL}
\end{equation}
Relevant for finding the sought for scaling dimensions is the behaviour of these solutions at large values of $L$, which we list first for the situation where the spatial dimension $D$ is less than the threshold value of $4$ (at this threshold value, we need to consider the solutions separately)
\begin{eqnarray}
u(L)&\propto & T_C\cdot L^{D-4}\quad\quad\text{(independently of } u(1))\nonumber\\
r(L)&\propto &r(1)\cdot L^{\frac{D-4}{3}}
\end{eqnarray}
\section{Mean Field Approximation of the smoothed Landau functional.}
The renormalization group equations produce a flow of the coupling constants with the parameter $L$. A natural length at which the renormalization should be set to stop is the correlation length of the system. Why this length? The field $S(\vec x)$ produced by eliminating the excitations with wave-length less than $\xi$ can be viewed as consisting of large spin  blocks. If one excites a spin at the site $i$, the excitation cannot propagate to the spins of the next ''spin block'', which are, on average, too far away. Accordingly, the field $S(\vec x)$ is expected to show little deviations from a uniform spin orientation. This is a situation where its Landau functional can be treated by the mean-field approximation.\\
We apply this principle, for instance, to the Landau functional of the field $S(\vec x)$ resulting from the ''smoothing'' operation:
\begin{eqnarray}
{\cal L}[S]&=&\frac{\Gamma}{2} a^{2-D}\int d^Dx\left(\vec \nabla S\right)^2\nonumber\\
&+& \frac{r(L)}{2}\int \frac{d^Dx}{a^D} \cdot S^2(\vec x)\nonumber\\
&+& \frac{u(L)}{4}\int \frac{d^Dx}{a^D}\cdot S^4(\vec x)
\end{eqnarray}
The mean field result for the correlation length above $T_C$ writes
\begin{equation}
\frac{\xi}{a} \propto \sqrt{\frac{\Gamma}{r(1)\cdot (\frac{\xi}{a})^{\frac{D-4}{3}}}}
\end{equation}
This is an implicit equation for $\frac{\xi}{a}$ with solution
\begin{equation}
\frac{\xi}{a}\propto \Big(\frac{\Gamma}{T-T_C}\Big)^{\frac{3}{2+D}}
\end{equation}
The critical exponent $\beta$ is obtained from
\begin{equation}
S_0\propto \sqrt{-\frac{r(1)}{T_C}\cdot \xi^{\frac{D-4}{3}-(D-4)}}\propto (T_C-T)^{\frac{3D-6}{2D+4}}
\end{equation}
\section{From the Renormalization group to Widom's scaling hypothesis.}
The results from the smoothing operation represent the key to enabling the use of the renormalization group within the framework of Widom's scaling hypothesis. The two equations explicitly show that the coupling constants $u(L)$ and $r(L)$ acquire a scaling factor when the excitations with wave length shorter than $L$ are eliminated from the problem by means of the smoothing operation. A further, more subtle result, is that the Landau functional, defined on the field $S(\vec x)$, is formally identical to the Landau functional defined on the original field $m(\vec x)$, except for the renormalized coupling constants. Why should we care about this symmetry, i.e. about the invariance of the Landau functional? Before answering  this question, we point out that our statement about the Landau functionals ${\cal L}(m)$ and ${\cal L}(S)$ being identical is actually not accurate. The fact is that the fields $S$ and $m$ contain excitations that extend over different wave lengths, so the two fields are not comparable. This ''error'' is amended by a simple ''trick'': the lattice constant ''$a$'', which still appears in the Landau functional ${\cal L}(S)$ after the smoothing operation, is replaced by
\begin{equation}
\frac{a_L}{L}
\end{equation}
$a_L$ being the lattice constant of a new, enlarged lattice, with respect to which $S(\vec x)$ contains excitations with exactly the same wave length as the original field $m(\vec x)$. This transformation is the so called \textbf{dilation} operation.  We will later check that the lattice constant transformation will also lead to multiplicative scaling factors and trivial scaling dimensions. Having defined the two fields onto the same set of excitations is, however, not enough to establish perfect invariance of the Landau functionals. In fact, the smoothing operation has suppressed many wave-length components, so that the length of the field $S(\vec x)$ is considerably reduced. In order to restore a length comparable with the original field $m(\vec x)$ one multiplies the field $S(\vec x)$ with a factor which is chosen to be of the form $L^{\Delta_S}$:
\begin{equation}
S(\vec x)\cdot L^{\Delta_S}\doteq {Z}(\vec x)
\end{equation}
This is the third operation inherent to the renormalization group. Only after these three operations (''smoothing'', ''dilation'' and ''field renormalization'') are completed should one check whether the Landau functional has indeed retained formally its structure, albeit with renormalized coupling constants, which must transform by scaling factors if the homogeneity relations are to be fulfilled.\\
The question is now: what is the benefit of having formally identical ${\cal L}(m)$ and  ${\cal L}(Z)$? The answer is in the use one does of these functionals: ${\cal L}(m)$ enters the functional integral
\begin{equation}
Z(r,u,B)= \int {\cal D}[m(\vec x)]\cdot e^{-\beta\cdot {\cal L}\left(r,u,B,[m(\vec x)]\right)}
\end{equation}
and determines by the rule $G_a(r,u,B)=-\frac{1}{\beta}\ln Z(r,u,B)$ the Gibb's potential per spin of the system with lattice constant $a$. ${\cal L}(Z)$ enters the functional integral
\begin{equation}
Z(r_L,u_L,B_L)= \int {\cal D}[Z(\vec x)]\cdot e^{-\beta\cdot {\cal L}\left(r_L,u_L,B_L,[Z(\vec x)]\right)}
\end{equation}
and determines by the rule $G_{a_L}(r_L,u_L,B_L)=-\frac{1}{\beta}\ln Z(r_L,u_L,B_L)$ the Gibb's potential per spin block of the system with lattice
constant $a_L$. As the structure of the Landau functionals is the same, we expect that the respective thermodynamic potentials have the same dependence on the coupling constants, i.e., taking into account that a ''spin block'' contains $L^D$ spins
\begin{equation}
G_{a_L}(r_L,u_L,b_L)= G(r_L,u_L,b_L) = L^D\cdot G(r,u,B)
\label{Gscale}
\end{equation}
(we have dropped the index ''$_a$'' for simplicity). Suppose now that all coupling constants are renormalized by scaling factors: the previous equation would display the sought for Widom's homogeneity. Comparing (\ref{Gscale}) with the homogeneity relation of Chapter 9 one reads off at once
\begin{equation}
\Delta_G = D
\end{equation}
We have already generated the scaling factor by the operation of smoothing. We now perform the remaining dilation and field rescaling operation to generate the complete set of scaling factors.\\
\paragraph{The dilation operation.}
The smoothed Hamiltonian in $\vec k$ space reads
\begin{eqnarray}
{\cal L}[S(\vec k)]&=&\sum_{\vec k_1,\vec k_2}\frac{\Gamma}{2}\cdot a^2\cdot \vec k_1^2 \cdot S_{\vec k_1}\cdot S_{\vec k_2}\cdot \delta_{\vec k_1+\vec k_2,0}\nonumber\\
&+& \sum_{\vec k_1,\vec k_2} \frac{r(L)}{2}\cdot S_{\vec k_1}\cdot S_{\vec k_2}\delta_{\vec k_1+\vec k_2,0}\nonumber\\
&+& \frac{1}{N^D}\cdot \frac{u(L)}{4}\cdot \sum_{\vec k_1,\vec k_2,\vec k_3,\vec k_4}S_{\vec k_1}\cdot S_{\vec k_2}\cdot S_{\vec k_3}\cdot S_{\vec k_4}\cdot \delta_{\vec k_1+\vec k_2+\vec k_3+\vec k_4,0}
\end{eqnarray}
The sum over $\vec k$ extends originally from $0$ to $\frac{2\pi}{L\cdot a}$. After the transformation $a_L= L\cdot a$ the sum extends from $0$ to $\frac{2\pi}{a_L}$, i.e. it encompasses all $\vec k$ values for a lattice with lattice constant $a_L$.
For instance
\begin{eqnarray*}
\sum_{\vec k_1,\vec k_2}S_{\vec k_1}\cdot S_{\vec k_2}\delta_{\vec k_1+\vec k_2,0}&=& \frac{1}{N^D}\cdot\frac{L^{2D}}{(a_L)^{2D}}\cdot \int d^Dx_1\int d^D x_2\cdot S(\vec x_1)\cdot S(\vec x_2)\nonumber\\
&\cdot & \sum_{\vec k_1,\vec k_2}e^{i\vec k_1\cdot \vec x_1}\cdot e^{i\vec k_2\cdot \vec x_2}\delta_{\vec k_1+\vec k_2,0}
\end{eqnarray*}
We insert
\begin{eqnarray*}
\sum_{\vec k_1,\vec k_2}e^{i\vec k_1\cdot \vec x_1}\cdot e^{i\vec k_2\cdot \vec x_2}\delta_{\vec k_1+\vec k_2,0}&=&
 \sum_{\vec k=0}^{\vec k_{BZ_L}}e^{i\vec k\cdot (\vec x_1-\vec x_2)}= \frac{N^D}{L^D}\cdot a_L^D\cdot \delta (\vec x_1-\vec x_2)
\end{eqnarray*}
and obtain
\[\sum_{\vec k_1,\vec k_2} S_{\vec k_1}\cdot S_{\vec k_2}\delta_{\vec k_1+\vec k_2,0}= \frac{L^D}{a_L^D}\cdot \int d^Dx\cdot S^2(\vec x)\]
The further terms of the Hamiltonian are Fourier transformed in a similar way. The smoothed Hamiltonian for the field $S$ in real space writes, after the dilation operation,
\begin{eqnarray}
{\cal L}[S]&=&\frac{\Gamma}{2} a_L^{2-D}\cdot L^{D-2}\cdot \int d^Dx\left(\vec \nabla S\right)^2\nonumber\\
&+& \frac{r(L)}{2}\cdot \frac{L^D}{a_L^D}\int d^Dx \cdot S^2(\vec x)\nonumber\\
&+& \frac{u(L)}{4}\cdot \frac{L^D}{a_L^D}\int d^Dx \cdot S^4(\vec x)
\label{Ldilated}
\end{eqnarray}
\paragraph{Field rescaling.} We finally must consider that the number of components of the field $S$ is reduced, so that its length has to be restored with the rescaling of the field according to
\[S(\vec x)\cdot L^{\Delta_S}\doteq {Z}(\vec x)\]
The Hamiltonian for the field $Z(\vec x)$ writes, accordingly,
\begin{eqnarray}
{\cal L}[Z]&=&\frac{\Gamma}{2} a_L^{2-D}\cdot L^{D-2}\cdot L^{-2\cdot \Delta_S}\cdot \int d^Dx\left(\vec \nabla Z\right)^2\nonumber\\
&+& \frac{r(L)}{2}\cdot \frac{L^D}{a_L^D}\cdot L^{-2\cdot \Delta_S}\int d^Dx \cdot Z^2(\vec x)\nonumber\\
&+& \frac{u(L)}{4}\cdot \frac{L^D}{a_L^D}\cdot L^{-4\cdot \Delta_S}\int d^Dx \cdot Z^4(\vec x)
\label{Lrescaled}
\end{eqnarray}
We determine $\Delta_S$ by imposing that the scaling dimension of $\Gamma$ remains ''0'' after the dilation and the rescaling of the field. This is equivalent to keeping the energy scale constant throughout the renormalization procedure. This requirement produces
\[\Delta_S = \frac{D-2}{2}\]
Having fixed $\Delta_S$, we can now compute the scaling dimensions arising from dilation, field rescaling \textbf{and} smoothing:
\begin{eqnarray}
\Delta_\Gamma &=& 0\nonumber\\
\Delta_r &=& 2+\frac{D-4}{3}=\frac{D+2}{3}\nonumber\\
\Delta_u &=& 0
\end{eqnarray}
$\Delta_B$ is obtained from the scaling relation $\Delta_B+\Delta_S = D$ and amounts to
\begin{equation}
\Delta_B = \frac{D+2}{2}
\end{equation}
We now plug these scaling dimensions into Widom's scaling relations of Chapter 9 to obtain values for the critical exponents. With $\Delta_G=D$ one has $\Delta_G-\Delta_B=\Delta_S$, so that the formulae of Chapter 9 read
\begin{eqnarray}
\nu &=& \frac{1}{\Delta_r} = \frac{3}{2+D}\nonumber\\
\beta &=& \frac{\Delta_G-\Delta_B}{\Delta_r}=\frac{\Delta_S}{\Delta_r}= \frac{\frac{D-2}{2}}{\frac{D+2}{3}}=\frac{3D-6}{2D+4}\nonumber\\
\delta &=& \frac{\Delta_B}{\Delta_G-\Delta_B}=\frac{\Delta_B}{\Delta_S}=\frac{D+2}{D-2}
\end{eqnarray}
\paragraph{The exponents, and how far they reach.} It is instructive to
evaluate these three formulae, together with $\gamma=\beta\cdot(\delta-1)$,
at the three dimensions of interest, and to compare them with the exact or
best-known values:
\begin{table}[H]
\begin{center}
\begin{tabular}{c|c|c|c|c|}
& $\nu$ & $\beta$ & $\gamma$ & $\delta$\\
\hline
$D=4$, this Chapter & $\frac{1}{2}$ & $\frac{1}{2}$ & $1$ & $3$\\
\hline
$D=4$, Landau & $\frac{1}{2}$ & $\frac{1}{2}$ & $1$ & $3$\\
\hline
$D=3$, this Chapter & $0.600$ & $0.300$ & $1.200$ & $5$\\
\hline
$D=3$, best known & $0.630$ & $0.326$ & $1.237$ & $4.79$\\
\hline
$D=2$, this Chapter & $0.750$ & $0$ & -- & $\infty$\\
\hline
$D=2$, Onsager & $1$ & $\frac{1}{8}$ & $\frac{7}{4}$ & $15$\\
\hline
\end{tabular}
\end{center}
\caption{The critical exponents of the $m^4$-model as computed in this Chapter, against the Landau values, the best known values in three dimensions and the exact two-dimensional values of Onsager\cite{Ons,Domb}.}
\label{Tab:exponents}
\end{table}
\noindent At $D=4$ the exponents are the Landau ones, as they must be: four is
the dimension at which the quartic coupling stops flowing. At $D=3$ the
agreement is remarkable for a calculation of this simplicity -- three per cent
on $\nu$, eight per cent on $\beta$. At $D=2$, on the contrary, the scheme
breaks down: it gives $\beta=0$ and $\delta=\infty$ where Onsager gives
$\frac{1}{8}$ and $15$. Two dimensions is as far from the reference dimension
$4$ as the reference dimension is from the Gaussian regime, and nothing in the
present treatment is expected to survive that distance. What does survive, and
what this Chapter was written for, is the mechanism: the homogeneity relation
of Chapter 9 is not an assumption but a consequence of the flow, the scale $L$
that Widom could not identify is the correlation length, and the scaling
dimensions are computable. The exponents that the films of Chapter 9 actually
display are the two-dimensional Ising ones, and for those the exact solution
remains the reference.
\paragraph{Mean field approximation of the smoothed and dilated Landau functional.}
The Landau functional after smoothing and dilation is (\ref{Ldilated}):
\begin{eqnarray}
{\cal L}[S]&=&\frac{\Gamma}{2} a_L^{2-D}\cdot L^{D-2}\cdot\int d^Dx\left(\vec \nabla S\right)^2\nonumber\\
&+& \frac{r(L)}{2}\cdot \frac{L^D}{a_L^D}\cdot \int d^Dx \cdot S^2(\vec x)\nonumber\\
&+& \frac{u(L)}{4}\cdot \frac{L^D}{a_L^D}\cdot \int d^Dx \cdot S^4(\vec x)
\end{eqnarray}
The new lattice has a correlation length which corresponds to the lattice constant, so that the mean field result for the correlation length above $T_C$ writes
\begin{equation}
1\propto \sqrt{\frac{\Gamma\cdot (\frac{\xi}{a})^{D-2}}{r(1)\cdot (\frac{\xi}{a})^{\frac{D-4}{3}+D}}}
\end{equation}
This is an equation for the sought for $\frac{\xi}{a}$ with solution
\begin{equation}
\frac{\xi}{a}\propto \Big(\frac{\Gamma}{T-T_C}\Big)^{\frac{3}{2+D}}
\end{equation}
as expected. For the spontaneous magnetization $S_0$ we have
\begin{equation}
S_0\propto \sqrt{-r(1)\cdot \xi^{\frac{D-4}{3}-(D-4)}}\propto (T_C-T)^{\frac{3D-6}{4+2D}}
\end{equation}
\paragraph{Mean field approximation of the smoothed, dilated and field rescaled Landau functional.}
The Landau functional after these three renormalization steps is (\ref{Lrescaled}).
The new lattice has a correlation length which corresponds to the lattice constant, so that the mean field result for the correlation length above $T_C$ writes
\begin{equation}
1\propto \sqrt{\frac{\Gamma\cdot (\frac{\xi}{a})^{D-2}}{r(1)\cdot (\frac{\xi}{a})^{\frac{D-4}{3}+D}}}
\end{equation}
(the scaling factors originating from the field rescaling cancel out).
This is an equation for the sought for $\frac{\xi}{a}$ with solution
\begin{equation}
\frac{\xi}{a}\propto \Big(\frac{\Gamma}{T-T_C}\Big)^{\frac{3}{2+D}}
\end{equation}
as expected.
For $Z_0$  we obtain
\begin{equation}
Z_0 \propto \sqrt{\frac{-r(1)\cdot \xi^{\Delta_r}}{u(1)}}\propto \text{const.}
\end{equation}
because $\xi\propto r^{-\nu}=r^{-\frac{1}{\Delta_r}}$, so that $r(1)\cdot \xi^{\Delta_r}$ is a constant
(one could have used this mean field relation as an equation for determining $\xi$). The spontaneous magnetization is given by
\begin{equation}
S_0\propto \frac{1}{\xi^{\Delta_S}}\cdot Z_0 \propto (-r)^{\frac{\Delta_S}{\Delta_r}}=(-r)^{\frac{3D-6}{4+2D}}
\end{equation}
\paragraph{Solution for $D\geq 4$.} For $D>4$ we infer that, for large $L$,
\begin{eqnarray}
u(L) &\rightarrow & \text{const.}\nonumber\\
r(L) &\rightarrow & r(1)
\end{eqnarray}
i.e. the critical exponents are mean field. $D=4$ is a borderline situation.
As
\begin{equation}
\lim_{D\rightarrow 4} \frac{L^{4-D}-1}{4-D} = \ln L
\end{equation}
we obtain
\begin{eqnarray}
u(L) &\propto & \frac{1}{\ln L}\nonumber\\
r(L) &=& r(1)\cdot (\ln L)^{-\frac{1}{3}}
\end{eqnarray}
In $D=4$ the coefficients of the Landau free energy are modified by logarithmic corrections. However, logarithmic corrections do not affect the critical exponents, which remain mean field.
\section{The Brazovskii instability.}
For the purpose of describing the Brazovskii instability\cite{Braz}, we go back to the first order exact recursion relation for $r$:
\begin{equation}
r(q\cdot a)= r(1) + 3\cdot u(1)\cdot \frac{C(D)}{\beta\cdot \Gamma}\cdot \int_{q\cdot a}^{q_{BZ}\cdot a} dq'\cdot \frac{q'^{D-1}}{q'^2 + \frac{r(q'\cdot a)}{\Gamma}}
\end{equation}
By keeping the running coupling constant $r(q'\cdot a)$ in the integrand, one transforms the relation into an integral equation for the sought-for coupling constant. This integral equation can be used, for instance, to find the transition temperature, defined by the condition $r=0$:
\begin{equation}
0 = r(1) + 3\cdot u(L)\cdot \frac{C(D)}{\beta\cdot \Gamma}\cdot\int_q^{q_{BZ}}dq'\cdot q'^{D-3}=r(1) + 3\cdot u(L)\cdot \frac{C(D)}{\beta\cdot \Gamma}\cdot\frac{1}{D-2}\cdot (q_{BZ}^{D-2}-q^{D-2})
\end{equation}
In this relation, $q$ is small and we can set it to zero, so that we obtain, for $D>2$, that $r=0$ fixes the transition temperature to the finite value we have computed in the previous section.\\
The Brazovskii instability appears when the spectrum of fluctuations is set to be minimized not at $q=0$ but at a spherical shell in $\mathbf q$-space, of radius $q_0$. This means that, in the integral equation for
the variable $r(q)$, the denominator is changed from
\begin{equation}
q'^2 + \frac{r}{\Gamma}
\end{equation}
to
\begin{equation}
(q'-q_0)^2 + \frac{r}{\Gamma}
\end{equation}
This transforms the integral equation for $r$ to
\begin{equation}
r(q)= r(1) + 3\cdot u(L)\cdot \frac{C(D)}{\beta\cdot \Gamma}\cdot \int_q^{q_{BZ}}dq'\cdot \frac{q'^{D-1}}{(q'-q_0)^2 + \frac{r}{\Gamma}}
\end{equation}
For any $D\geq 1$ the integrand peaks, for small $r$, at $q'=q_0$. Approximating it by a rectangular function of height $\frac{q_0^{D-1}}{r/\Gamma}$ and full width $2\sqrt{\frac{r}{\Gamma}}$ produces the value
\[\frac{q_0^{D-1}}{\frac{r}{\Gamma}}\cdot 2\sqrt{\frac{r}{\Gamma}}= 2\cdot q_0^{D-1}\cdot \sqrt{\frac{\Gamma}{r}} \]
for the integral. Accordingly, $r$ must satisfy the equation
\[r = r(1) + 6\cdot u(L)\cdot \frac{C(D)}{\beta\cdot \Gamma}\cdot q_0^{D-1}\cdot \sqrt{\frac{\Gamma}{r}}\]
which can be rewritten as
\[r^3-2\cdot r^2\cdot r(1) + r\cdot r(1)^2- \frac{36\cdot u(L)^2\cdot C(D)^2\cdot q_0^{2(D-1)}}{\beta^2\cdot \Gamma}=0\]
A solution $r=0$ requires $\beta\rightarrow \infty$, i.e. $T=0$. Accordingly, the high temperature disordered phase never becomes unstable as $r$ never changes sign at finite temperatures. Any other solution of the model -- such as a modulated phase -- must compete with the disordered phase at any temperature. Should the free energy of the modulated phase become lower than the free energy of the paramagnetic phase, a first order phase transition occurs. See also\cite{Cannas}.

\chapter{Topological aspects of the magnetism of transition metal overlayers.}
\label{Chap:Topo}
\renewcommand{\thesection}{\thechapter.\arabic{section}}
\renewcommand{\theequation}{\thechapter.\arabic{equation}}
Ferromagnetic order, as we have encountered it so far, foresees that, below the
Curie temperature, the graph of the free-energy as a function of the
magnetization has a flat portion\cite{Flat} (path a of Fig.~\ref{Fig:spont}(a)). This
flatness defines a situation in which the magnetization can acquire a value
between zero and a so called ``spontaneous magnetization'' $\pm {M}_s$ without
the free energy changing its value. This ``flatness'' is a property of the
thermodynamic limit, i.e. of infinite bodies. In finite bodies, the free energy
assumes a shape that resembles path b of the same figure, $\pm M_s$ being the
values at which the free energy has minima. Systems that exhibit a spontaneous
magnetization are, for example, the Ising and classical
Heisenberg ferromagnets in three dimensions (3D)\cite{Flat,Jurg}, the 2D-Ising
model or the 2D-planar and classical 2D-Heisenberg models with symmetry breaking single ion interactions\cite{Jose}. These are all quite idealized systems, in which the spins interact purely via the exchange interaction. In real ferromagnetic bodies, however, the dipole-dipole interaction, originating within Maxwell equation of magnetostatics, must be considered alongside the main exchange interaction (of purely quantum mechanical origin) and possible single ion magnetic anisotropies (produced by the spin-orbit interaction). The dipole-dipole interaction is, typically, much weaker than the exchange interaction (by about two orders of magnitude). Yet, it is long-ranged as it decays only with the third power of the distance between two magnetic moments. Because of the dipole-dipole interaction, a theorem, proved by Griffiths\cite{Griff_dipole} for bodies with linear dimension $\Lambda$ approaching infinity along all three spatial dimensions, implies that any non-zero magnetization produces an increase of the free energy. In other words: when the dipole-dipole interaction is taken into account, the graph of the free energy as a function of $M$ has a minimum at $M=0$ at any temperature. Accordingly, in real
ferromagnetic bodies, order can only be local and, globally, the spontaneous
magnetization is exactly vanishing. Griffiths' theorem is therefore at the
origin of what one could define as ''topological disorder'' -- the adjective
''topological'' referring to the order parameter assuming substantially different
values when different regions in space are considered. An example of elementary
''topological excitations'' that promote such a disorder are the so called
''magnetic domains'', studied by Landau\cite{Landau_Domain} and Kittel\cite{Kittel} a long time ago. Topological excitations must be distinguished from an
other type of excitations that occur at finite temperature, the so called ''spin
waves''. In contrast to topological excitations, spin waves promote only minute
deviations of the local magnetization vector from the equilibrium value.
Despite this, spin waves can also destroy global order, as implied, e.g., by the
Polyakov renormalization of the isotropic 2D-Heisenberg ferromagnet of Chapter 7\cite{Mermin}. In this type of disorder, the 2D-character plays an essential role.\\
We point out that Griffiths' no-spontaneous-magnetization rule refers to a
3D-body. In fact, it appears that an important assumption underlying Griffiths'
theorem is the size of the body approaching infinity along all three spatial
dimensions. Recently, a paper has discussed the situation of ferromagnetism in
the presence of exchange, magnetic anisotropy and dipole-dipole interaction but
in an ultrathin film geometry, where only two spatial dimensions are allowed to
increase and the third is assigned a finite, very small thickness\cite{Argentina}. This paper also finds the Griffiths instability in perpendicularly
magnetized ultrathin films. In this Chapter, starting from this instability, we
develop the essential elements of topological disorder in perpendicularly
magnetized ultrathin films. The results of this Chapter reflect themselves in
experimental observations on epitaxial ultrathin 3D transition metals overlayers but could be also relevant for discussing ferromagnetism in the new class of monolayer thin materials obtained by mechanical exfoliation\cite{Gong,Huang,Burch,Li,Novo,Dai}. Mechanically exfoliated samples are known to be
perfectly flat over large distances and have been shown to be vertically
engineerable\cite{Wua}.\\
\section{Topological instability in perpendicularly magnetized ultrathin films.}
\begin{figure}[H]
\begin{center}
\includegraphics[width=0.35\textwidth]{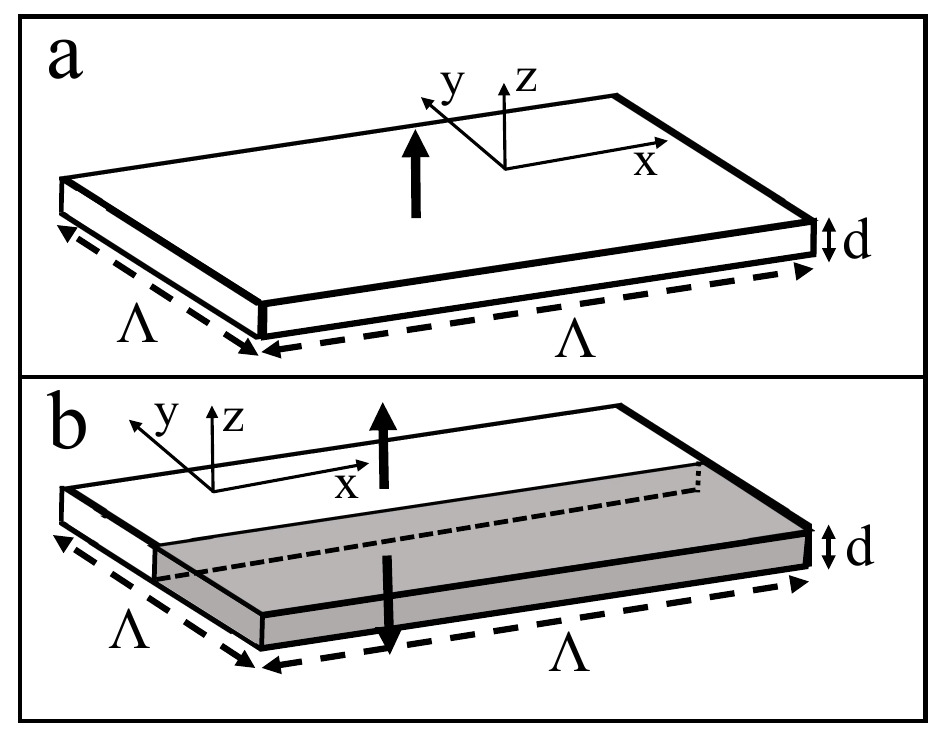}
\end{center}
\caption{The model used for the computation of the topological instability. a: the film is filled with ''up'' spins. b: a wall along $x$ separates domains with opposite perpendicular spin polarization.}
\label{Fig:Top_inst}
\end{figure}

We proceed in describing this instability on the basis of a simple model. We assume a thin film as being a slab with area $\Lambda^2$ and thickness $d\!<\!<\!\Lambda$. The slab hosts a continuous magnetization distribution.
The use of a continuous model requires establishing a link to ``real'' samples, which host magnetic moments on a lattice. We establish the link by assuming that the magnetization is measured in units of $M_0\doteq\frac{g\cdot \mu_B\cdot
S}{a^3}$, thereby introducing the atomic scale quantities $g$ (the $g$-factor),
$S$ (the spin at a lattice site, in units of $\hbar$) and $a$ (the lattice
constant for a simple cubic lattice). By these assumptions, the thickness of the slab corresponding to ``1 Monolayer'' (1ML) is $a$. Finally, we assume that the magnetization is independent of the coordinate $z$ perpendicular to the film plane. This assumption corresponds to a 2D spatial spin degree of freedom: the spin distribution is rigid along $z$ and can only assume some profile along the $xy$-plane. This requirement is only fulfilled if $d$ is sufficiently small so that the spins along $z$ can be kept locked in by the exchange interaction, i.e. in the ''ultrathin limit''.\\
Our argument requires comparing the gain in dipolar energy to the loss of exchange energy involved in creating domains with opposite magnetization. We deal with two spin states. In a state of  spontaneous perpendicular magnetization, all spins in the slab point along one of the two $z$-directions perpendicular to the slab (Fig.\ref{Fig:Top_inst}a), e.g. the $+z$-direction.
In Fig.\ref{Fig:Top_inst}a, this state is rendered with a white color and the spin vector is given by a black arrow. A possible, elementary state of vanishing spontaneous magnetization is shown in Fig.\ref{Fig:Top_inst}b. One half of the slab is still filled with spins pointing upwards ''$\uparrow$'' but the other half (gray in Fig.\ref{Fig:Top_inst}b) contains spins pointing downwards ''$\downarrow$'' (indicated as state $\uparrow\downarrow$ henceforth). A domain wall of width $w$ shall separate the two domains. Within the domain wall, the spin shall change linearly from $+1$ on one side to $-1$ on the other. These assumptions, of course, are made to make the model computable analytically.
\paragraph{The magnetostatic energy.} For computing the magnetostatic energy in this situation, we use the general formula for ultrathin films
\begin{eqnarray}
E_M[n_z]&=& \Omega\cdot \frac{d}{a}\cdot \frac{1}{a^2}\cdot \int d^2\rho \cdot n^2_z(\vec \rho)\nonumber\\
&-&\frac{\Omega}{4\pi}\cdot \frac{1}{a^3}\int d^2\rho \int d^2\rho' n_z(\vec \rho)\cdot n_z(\vec \rho')\cdot \big[\frac{\partial}{\partial x}\frac{\partial}{\partial x'}+\frac{\partial}{\partial y}\frac{\partial}{\partial y'}\big]\nonumber\\
&&\int_0^d dz\int_0^d dz'\frac{1}{\sqrt{(\vec \rho-\vec \rho')^2 + (z-z')^2}}
\end{eqnarray}
and redirect the in-plane derivatives. For simplicity, we think of the borders as being pushed to infinity. Accordingly, the spin distribution does not have a gradient and the magnetostatic energy of the state in Fig.\ref{Fig:Top_inst}a amounts to the local component
\begin{equation}
\Omega\cdot \frac{d}{a}\cdot \frac{\Lambda^2}{a^2}
\end{equation}
The situation is different for the state in Fig.\ref{Fig:Top_inst}b. Within the
wall $n_z=-\frac{2y}{w}$, so that $n_z^2$ falls below $1$ there; the local
contribution amounts to
\begin{eqnarray}
\Omega\cdot \frac{d}{a}\cdot \frac{\Lambda^2}{a^2}-\Omega\cdot \frac{d}{a}\cdot \frac{\Lambda}{a}\cdot \frac{1}{a}\cdot\int_{-\frac{w}{2}}^{+\frac{w}{2}}dy\cdot \Big[1-\big(\frac{2y}{w}\big)^2\Big]\! =\!
\Omega\cdot \frac{d}{a}\cdot \frac{\Lambda^2}{a^2}- \frac{2}{3}\cdot\Omega\cdot \frac{d}{a}\cdot\frac{\Lambda}{a}\cdot \frac{w}{a}
\end{eqnarray}
The long range component also contributes, owing to the presence of the wall.
The transition from a state of spin ''1'' to a state of spin ''-1'' introduces an effective ''charge''
\begin{equation}
-\frac{\partial n_z}{\partial y} = \frac{2}{w}
\end{equation}
localized along a thin slab with thickness $d$, width $w$ and length $\Lambda$ at about $y=0$. The non-local component of the dipolar energy for the ''up-down'' state writes, accordingly,
\begin{eqnarray}
-\frac{\Omega}{4\pi}\cdot \frac{1}{a^3}\cdot \frac{4}{w^2}\cdot \int_{0}^{\Lambda}dx\int_{0}^{\Lambda}dx' \int_0^w dy \int_0^w dy'\cdot \int_0^d dz\int_0^d dz'\frac{1}{\sqrt{(\vec \rho-\vec \rho')^2 + (z-z')^2}}
\end{eqnarray}
As we have introduced a wall with finite width, we can compute the integral over $z$ and $z'$ by setting $z=z'=0$ without needing to worry about integrating over a singularity. The parameter $d$ is, accordingly, the smallest length in the problem. This is our point of view when dealing with ultrathin films. The integral over $z$ and $z'$ leads to
\begin{equation}
d^2\cdot \frac{1}{\sqrt{(\vec \rho-\vec \rho')^2}}
\end{equation}
and the non-local component writes
\begin{equation}
-\frac{\Omega}{4\pi}\cdot \frac{1}{a^3}\cdot d^2\cdot \frac{4}{w^2}\cdot \int_{0}^{\Lambda}dx_1\int_{0}^{\Lambda}dx_2 \int_0^w dy_1 \int_0^w dy_2\frac{1}{\sqrt{(x_1-x_2)^2 + (y_1-y_2)^2}}
\end{equation}
We use the variables
\begin{equation}
\vec \rho'_1 = \frac{\vec \rho_1}{\Lambda}\quad \vec \rho'_2 = \frac{\vec \rho_2}{\Lambda}
\end{equation}
and the parameter $\omega\doteq \frac{w}{\Lambda}$
and transform the integral to
\begin{equation}
-\frac{\Omega}{4\pi}\cdot \frac{\Lambda^3}{a^3}\cdot d^2\cdot \frac{4}{w^2}\cdot \int_{0}^{1}dx'_1\int_{0}^{1}dx'_2 \int_0^\omega dy'_1 \int_0^\omega dy'_2\frac{1}{\sqrt{(x'_1-x'_2)^2 + (y'_1-y'_2)^2}}
\end{equation}
The remaining integrals are elementary ones and the result of the exact integration writes
\begin{eqnarray}
2\cdot\!
\left\{\!\frac{\omega^2}{2}
\ln\left( \frac{\sqrt{1+\omega^2}+1}{\sqrt{1+\omega^2}-1}\right)
   \!+\! \omega\, \ln\left( \sqrt{1+\omega^2}+ \omega\right)
\!-\!\frac{1}{3}\left[
\left(1+\omega^2 \right)^{\frac{3}{2}}\!-\!\omega^3\!-1\!\right]\!
 \right\}
\end{eqnarray}
We are interested in the situation where $w$ is also much smaller than $\Lambda$ while being larger than $d$. In this situation, the last expression simplifies to
 \begin{equation}
2\cdot\!
\left\{(\frac{w}{\Lambda})^2
\ln\left(\frac{1}{\frac{w}{\Lambda}}\right)  +(\frac{w}{\Lambda})^2  \left(\ln 2 +\frac{1}{2} \right) +
\frac{(\frac{w}{\Lambda})^3}{3} +
\mathcal{O}\left((\frac{w}{\Lambda})^4\right) \right\}
\end{equation}
The leading term is the logarithmic one:
\begin{equation}
2\cdot\!
\Big[(\frac{w}{\Lambda})^2
\ln\big(\frac{1}{\frac{w}{\Lambda}}\big)\Big]
\end{equation}
Using the leading term, the non-local component writes
\begin{equation}
-\frac{2}{\pi}\cdot (\Omega\cdot \frac{d}{a}) \cdot \frac{\Lambda\cdot d}{a^2}\cdot \ln\frac{\Lambda}{w}+ {\cal O}(\frac{d^2\cdot \Lambda}{a^3})
\end{equation}
The total magnetostatic energy of the ''up-down'' state writes, finally,
\begin{equation}
\Omega\cdot \frac{d}{a}\cdot \frac{\Lambda^2}{a^2}- \frac{2}{3}\cdot \Omega\cdot\frac{d}{a}\cdot\frac{\Lambda\cdot w}{a^2}- \frac{2}{\pi}\cdot (\Omega\cdot \frac{d}{a}) \cdot \frac{\Lambda\cdot d}{a^2}\cdot \ln\frac{\Lambda}{w}\pm {\cal O}(\frac{d^2\cdot \Lambda}{a^3})
\end{equation}
Subtracting from this energy the energy of the uniform state ''up-up'' cancels the leading term $\propto \Lambda^2$. The state ''up-down'' produces, with respect to the state ''up-up'', a magnetostatic energy \textbf{gain} amounting to
\begin{equation}
\frac{2}{3}\cdot \Omega\cdot\frac{d}{a}\cdot\frac{\Lambda\cdot w}{a^2}+ \frac{2}{\pi}\cdot (\Omega\cdot \frac{d}{a}) \cdot \frac{\Lambda\cdot d}{a^2}\cdot \ln\frac{\Lambda}{w}\pm {\cal O}(\frac{d^2\cdot \Lambda}{a^3})
\end{equation}
Owing to the logarithmic divergence of the non-local term, we will neglect the local term for further considerations, so that the energy gain  writes, approximately,
\begin{equation}
\frac{2}{\pi}\cdot (\Omega\cdot \frac{d}{a}) \cdot \frac{\Lambda\cdot d}{a^2}\cdot \ln\frac{\Lambda}{w}
\label{Eq:gain}
\end{equation}
\paragraph{The energy of the domain wall.}
This problem was solved originally by Landau and Lifshitz in 1935\cite{Landau_18}. We use their method but introduce explicitly a magnetostatic component originating from the dipolar interaction. We also adapt the results to our slab model of ultrathin films. We will find that the formation of a domain wall in the $\uparrow\downarrow$-state increases the
total energy of the ferromagnetic slab and therefore promotes the state of
spontaneous magnetization. For an explicit computation of the wall energy, we assume that the wall runs parallel to the $x$-direction. Within the wall, the magnetic moments rotate away from the $z$-direction. Let the rotation be characterized by an angle $\theta$, which increases from $0$ to $\pi$ when moving along $y$ within the wall. The spin rotation costs, in the first place, exchange energy. The exchange functional of Chapter 6, specialized to a distribution that varies along $y$ only, reads
\begin{equation}
A\cdot d \cdot \int d^2\rho \Big[\Big(\frac{\partial n_z}{\partial \rho_y}\Big)^2 + \Big(\frac{\partial n_x}{\partial \rho_y}\Big)^2\Big]
\end{equation}
We write
\begin{equation}
\vec n = \Big(\sin\theta(y),0,\cos\theta(y)\Big)
\end{equation}
and obtain the wall exchange energy functional as
\begin{equation}
\Lambda\cdot A\cdot d\int_{-\infty}^{\infty} \Big(\frac{\partial \theta(y)}{\partial y}\Big)^2 \cdot dy
\end{equation}
Next: the misalignment is associated with an increase of the single ion magnetic anisotropy energy that favors the perpendicular magnetization, introduced conceptually by N\'eel\cite{Neel} and computed for the first time from first principles by Gay and Richter\cite{Gay} for the monolayer of Fe. This anisotropy is proportional to $\cos^2\theta(y)$. We call the proportionality
constant $-\lambda$, the negative sign indicating that this anisotropy favors
the state of perpendicular magnetization. This anisotropy is located mainly
at the two surfaces bounding the thin film\cite{GR}. In the inside of the film, the value of the single ion uniaxial anisotropy is much weaker. However, by our assumption of a rigid profile along $z$, the turning of a spin in the inside of the film is associated with the rotation of a spin at the surface of the film. The N\'eel single ion anisotropy term can be accounted for by a surface free energy functional that writes
\begin{equation}
-\frac{\Lambda}{a}\cdot \lambda\cdot S^2\cdot \frac{1}{a}\int dy\cdot \cos^2\theta(y)
\end{equation}
If one ignores the non-local term, for simplicity, the magnetostatic energy contributes a leading functional
\begin{equation}
+\frac{\Lambda}{a}\cdot \Omega\cdot \frac{d}{a}\cdot \frac{1}{a}\int dy\cdot \cos^2\theta(y)
\end{equation}
The equilibrium spin distribution along the domain wall and its equilibrium energy are obtained by minimizing the total wall energy functional
{\small \begin{equation}
E_w[\theta(y)]=\frac{\Lambda}{a}\cdot\Big[A\cdot d\cdot a\cdot \int_{-\infty}^{\infty} \left(\frac{\partial \theta(y)}{\partial y}\right)^2 \cdot dy-(\lambda\cdot S^2-\Omega\cdot \frac{d}{a})\frac{1}{a}\int dy\cdot \cos^2\theta(y)\Big]
\end{equation}}
The corresponding Euler-Lagrange equation reads
\begin{equation}
2\cdot A\cdot d\cdot a \frac{d^2 \theta }{d y^2}
-2\frac{\lambda\cdot S^2-\Omega\frac{d}{a}}{a} \sin( \theta(y))\cos( \theta(y))=0
\end{equation}
The solution to the boundary conditions $\lim_{y\rightarrow -\infty} \theta(y) = \pi$ and $\lim_{y\rightarrow +\infty} \theta(y) = 0$
reads
\begin{equation}
\cos(\theta(y)) = \tanh\left(\frac{y}{w}\right)
\end{equation}
with the width $w$ of the wall
\begin{equation}
w=a\cdot \sqrt{\frac{A\cdot a\cdot \frac{d}{a}}{\lambda\cdot S^2-\Omega\frac{d}{a}}}
\label{Eq:w}
\end{equation}
The total energy of the wall in the slab amounts to
\begin{equation}
\frac{\Lambda\cdot d}{a^2}\cdot 4\cdot \sqrt{A\cdot a\cdot \Big(\frac{\lambda\cdot S^2-\Omega\frac{d}{a}}{\frac{d}{a}}\Big)}
\end{equation}
This last relation consists of the total number of unit cells in the wall surface multiplied by the energy per unit cell
\begin{equation}
\epsilon_w \doteq 4\cdot \sqrt{A\cdot a\cdot \Big(\frac{\lambda\cdot S^2-\Omega\frac{d}{a}}{\frac{d}{a}}\Big)}
\label{Eq:epsw}
\end{equation}
Regarding this equation: suppose that we fill the slab with a uniform magnetization distribution $\vec n= (\sin\theta,0,\cos\theta)$ ($\theta$ being the angle with respect to the $z$-direction). The total anisotropy energy per unit surface is given by
$(-\lambda\cdot S^2+\Omega\cdot \frac{d}{a})\cdot\cos^2 \theta$. When $\lambda\cdot S^2\!>\!\Omega\!\cdot\!\frac{d}{a}$, the state of perpendicular magnetization is the preferred one. In this state, the argument of the square roots in the equation for the wall energy is positive: $w$ and the wall energy have finite values. However, as $d\!\rightarrow\! d_R\doteq \frac{\lambda\cdot S^2}{\Omega}\cdot a$, $w$ tends to infinity and the wall energy to zero. Notice that spin wave excitations produce a slightly different renormalization of  $\lambda$ and $\Omega$ as a function of temperature and thickness \cite{Politi,Shi} so that $d_R$ is itself a function of the temperature. Previous works\cite{Shi,Politi,Qiu} have established
that $d_R(T)$ defines a line of phase transitions at which the perpendicular magnetization turns into the plane of the slab.
\paragraph{Topological instability, absence of spontaneous magnetization and crossover length.}
We are now able to draw conclusions about the absence of spontaneous
magnetization and the crossover length. Comparing the domain wall energy cost
to the magnetostatic energy gain, we recognize that the logarithmic term always
favors the building of domains for sufficiently large $\Lambda$. Accordingly,
Griffiths' theorem about the absence of spontaneous magnetization in the
thermodynamic limit holds true in the slab geometry with perpendicular
magnetization. One interesting outcome of our argument is the estimate of the
cross-over length $\Lambda_c$ at which a slab will transit from a  monodomain
state to a multi-domain state, resulting from equating the magnetostatic energy
gain (\ref{Eq:gain}) to the wall energy:
\begin{equation}
\Lambda_c\approx w\cdot e^{\frac{2\pi\cdot \sqrt{\Big(\frac{\lambda\cdot S^2-\Omega\!\cdot\! \frac{d}{a}}{\frac{d}{a}}\Big)\cdot A\cdot a}}{\Omega\!\cdot\!\frac{d}{a}}}= w\cdot e^{\frac{\pi}{2}\cdot \frac{\epsilon_w}{\Omega\!\cdot\!\frac{d}{a}}}
\label{Eq:Lambdac}
\end{equation}
We have some comments regarding this result that we formulate with the help of a Table containing some typical values for the parameters used in this Section.
\begin{table}[H]
\begin{center}
\scalebox{0.8}{
\begin{tabular}{|c|c|c|c|c|c|c|}
\hline
Perp.&$J\!\cdot\!S^2$&$^2$$\lambda$  &$^3$$\Omega$  &$w_\perp$&$L_c$&$d_R$\\
&[meV]&[meV]&[meV]&[nm]&[m]&[ML]\\
\hline
$^4$&$^1$46\cite{Small}&0.38\cite{Gay}&0.3\cite{Gay}&&&1.5\cite{Gay}\\
\hline
$^5$&$^1$36\cite{Small}&&&&&\\
\hline
$^6$&&&$\underline{0.28}$&\underline{5}&\underline{$\approx 10^5$} &\underline{1.26}\\
\hline
$^7$&&$0.7\!-\!1.1$\cite{Shi}&&&&5-8\cite{Shi}\\
\hline
$^8$&&&&&&6\cite{Qiu}\\
\hline
$^9$&&&&&&3\cite{Pappas}\\
\hline
$^{10}$&&&&$\approx 40$\cite{Rolf}&&10\cite{Rolf_2024}\\
\hline
$^{11}$&&$\approx\!1\!$\cite{FMR}&&&&\\
\hline
\end{tabular}}
\end{center}
\caption{
{\footnotesize The Table summarizes typical experimental and theoretical values (perpendicular magnetization).\\
$^1$: Theoretical values for bulk bcc Fe. Realistic computations find values for the exchange interaction between atoms which are further away than nearest neighbors. The value we use here is obtained by simplifying Eq. 4.10 of Ref.\cite{Small} for the spin wave stiffness constant to $D\approx S\cdot J\cdot a^2$. An effective value for $J$ is then obtained using the values of Table 1 in Ref.\cite{Small}. $S\approx 1.1$\cite{Small}.\\
$^2$: Per unit surface cell.\\
$^3$: Per bulk unit cell. Fe atoms occupying a bcc lattice, i.e. 2 atoms in the unit cell, each carrying a magnetic moment of $2.2 \mu_B$ ($g\approx 2$\cite{Small}, $S\approx 1.1$\cite{Small}). $\mu_0 = 4\pi\cdot 10^{-7} \cdot \frac{T\cdot m}{A}$, $\mu_B = 9.3\cdot 10^{-24}\cdot \frac{\text{Joule}}{T}$,
$a= 2.83\cdot 10^{-10}\cdot m$, $1\cdot \text{Joule} = 6.2\cdot 10^{18} \cdot eV$.
\\
$^4$: Values obtained from the quoted calculations.\\
$^5$: The experimental value is obtained from the experimental value for $D$ quoted in Ref.\cite{Small}.\\
$^6$: Estimated values from the equations described in the present paper, using the values of line $^4$ ($d=a$).\\
$^7$: fcc Fe, $a=0.361$ nm.\\
$^8$: bcc Fe on a Ag(100) surface, room temperature.\\
$^9$: fcc Fe on Cu(100)($a=0.361$ nm), @150 K.\\
$^{10}$: Ni on Cu(100), $d$ between 2 to 7 nm.\\
$^{11}$: 3 ML Fe on Cu(100), room temperature (estimated from Table 1 in Ref.\cite{FMR})
 }}
\label{Tab:Topo}
\end{table}
\noindent In Table \ref{Tab:Topo}, we observe that, sufficiently away from $d_R$, $\Lambda_c$ assumes values that exceed by far the lateral size of common laboratory samples (mm). When moving toward $d_R$, the exponential factor will produce a $\Lambda_c$ that decreases to a few tens of micrometers (see e.g. Ref.\cite{NiculinPRB}). These are the lateral lengths over which exfoliated two-dimensional magnets are believed to be almost perfectly flat\cite{Gong,Huang,Burch,Li,Novo,Dai}. Accordingly, a sequence of exfoliated
samples with suitable thickness and with increasing lateral size $\Lambda$ should allow an insight into the yet unexplored  mechanism of the topological instability produced by the penetration of magnetic domains in two-dimensional ferromagnetic elements as a function of their size
$\Lambda$ (some preliminary results in this direction were reported in Ref.\cite{Oliver} on epitaxially grown ultrathin films).

\section{The stripe domain structure.}
\begin{figure}[H]
\begin{center}
\includegraphics[width=0.5\textwidth]{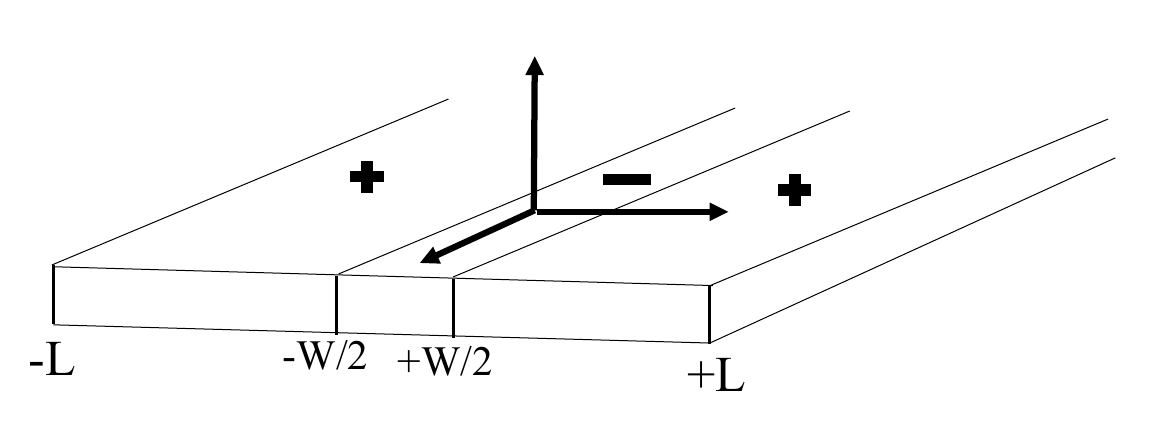}
\end{center}
\caption{Sketch of the stripe structure computed in this section. The profile runs along $x$, the stripes along $y$; $2L$ is the period, $W$ the width of the minority stripe, $d$ the thickness of the slab. In zero applied field $W=L$.}
\label{Fig:Stripes}
\end{figure}

The topological instability suggests that the ground state of perpendicularly
magnetized ultrathin films contains some kind of spin configuration and that
this spin configuration renders the spontaneous magnetization vanishing.
Experiments find indeed a stripe domain structure, see e.g. Ref.\cite{Saratz,Enc}; the theory of such structures has a long history\cite{Kaplan,Kashuba_PRB_1993,Port,Yafet,Scaling,Kron,Mobile,NiculinPRB,Saratz,Garel,Seul,Giuliani}. In this section, we show that the total energy of a one dimensional stripe domain structure, taken as a simple model of a possible ground state spin configuration, is indeed lower than that of the state of uniform spontaneous magnetization.
\paragraph{The energy of the wall system.} There are two components to the
total energy. The first component we deal with is the domain wall energy.
The set of domain walls between stripes of opposite spin
configuration provides an increase of the total energy with respect to that state of uniform magnetization. Let $L$ be the sought-for width of a single stripe\footnote{The period of the stripe structure is, accordingly, $2L$. For mathematical simplicity, we use for the lateral size of the system the parameter $2\Lambda$}. Assuming that the walls are not interacting, the energy of the wall system is found by multiplying the energy of a single wall per unit wall surface cell ($\epsilon_w$), with the number of surface unit cells within the slab domain wall ($=\frac{2\Lambda\cdot d}{a^2}$) and the number of walls
($=2\cdot \frac{2\Lambda}{2L}$)
\begin{equation}
E_w(L)=\frac{(2\Lambda)^2}{a^2}\cdot \Big\{\epsilon_w\cdot \frac{d}{L}\Big\}
\end{equation}
In this formula, we set the parameter $\frac{d}{L}$ in evidence because it will be used later to minimize the total energy of the stripe system.
\paragraph{The magnetostatic energy of the stripe system.}
The second component of the total energy is the dipolar energy. In order to compute this component we use the formula for a spin distribution $n_z(x,y)$
\begin{eqnarray}
E_M[n_z]&=& +\frac{\Omega}{4\pi}\cdot \frac{1}{a^3}\cdot \int d^2\rho \int d^2\rho' \cdot n_z(\vec \rho)\cdot n_z(\vec \rho')\cdot \nonumber\\
&\cdot &\int_0^d dz\int_0^d dz' \frac{\partial}{\partial z}\frac{\partial}{\partial z'}\frac{1}{\sqrt{(\vec \rho-\vec \rho')^2 + (z-z')^2}}
\end{eqnarray}
It will be convenient to Fourier transform the kernel
\begin{equation}
\int_0^d dz\int_0^d dz' \frac{\partial}{\partial z}\frac{\partial}{\partial z'}\frac{1}{\sqrt{(\vec \rho-\vec \rho')^2 + (z-z')^2}}
\end{equation}
Performing the integrals over $z$ and $z'$ from $0$ to $d$ one obtains
\begin{eqnarray}
\int_0^d dz\int_0^d dz' \frac{\partial}{\partial z}\frac{\partial}{\partial z'}\frac{1}{\sqrt{(\vec \rho-\vec \rho')^2 + (z-z')^2}} &=&\nonumber\\
2\cdot \Big[
\frac{1}{\sqrt{(\vec \rho-\vec \rho')^2}}-\frac{1}{\sqrt{(\vec \rho-\vec \rho')^2 + d^2}}\Big]
\end{eqnarray}
We use now the identities
\begin{eqnarray}
\frac{1}{\sqrt{(\vec \rho-\vec \rho')^2}}&=& \frac{1}{2\pi}\int d^2q
\cdot \frac{e^{i\cdot \vec q\cdot (\vec \rho-\vec \rho')}}{q}\nonumber\\
\frac{1}{\sqrt{(\vec \rho-\vec \rho')^2+d^2}}&=& \frac{1}{2\pi}\int d^2q
\cdot \frac{e^{i\cdot \vec q\cdot (\vec \rho-\vec \rho')}}{q}\cdot e^{-q\cdot d}
\end{eqnarray}
to obtain an equivalent general formula for the magnetostatic energy of a spin distribution $n_z(x,y)$:
\begin{eqnarray}
E_M[n_z]&=& +\frac{1}{2\pi}\cdot \frac{\Omega}{2\pi}\cdot \frac{1}{a^3}\cdot \int d^2\rho \int d^2\rho' \cdot n_z(\vec \rho)\cdot n_z(\vec \rho')\cdot \nonumber\\
&\cdot & \int d^2q
\cdot e^{i\cdot \vec q\cdot (\vec \rho-\vec \rho')}\cdot \frac{1-e^{-q\cdot d}}{q}
\label{Eq:EMq}
\end{eqnarray}
We apply this formula to a spin distribution that is uniform along $y$. Its profile along $x$ is sketched in Fig.\ref{Fig:Stripes}, together with the main parameters specifying its geometry. For computing the dipolar energy, we use a model where the stripes with opposite spin polarization are  separated by infinitely sharp domain walls\footnote{Notice that the expression for the energy of a domain wall $\epsilon_w$ with finite width does not apply to the sharp walls used in this abstract model.  However, we will continue using the
symbol $\epsilon_w$.} This is an abstract situation which is only realized in the context of the Ising model but the results obtained with it are paramount for a more realistic profile where $n_z(x)$ changes continuously between adjacent stripes. In view of the fact that we will also include a magnetic field applied along the $z$-axis, we will allow for the stripes with minority spins to have a smaller width than the stripes with majority spins. This leads us to a profile defined as follows:
\begin{eqnarray}
n_z(x)&=& 1\quad [-L,-\frac{W}{2}]\nonumber\\
n_z(x)&=& -1\quad [-\frac{W}{2},+\frac{W}{2}]\nonumber\\
n_z(x)&=& 1\quad [+\frac{W}{2},+L]
\end{eqnarray}
The details of the computation of the total magnetostatic energy can be found in Appendix \ref{App:Stripes}. The total magnetostatic energy of the slab containing the rectangular stripe profile is
\begin{equation}
E_M[L,W]\approx \frac{(2\Lambda)^2}{a^2}\cdot \Big\{\Omega\cdot \frac{d}{a} + \frac{2}{\pi}\cdot \Omega\cdot \frac{d}{a}\cdot \frac{d}{L}\Big[\ln \left(\frac{\pi d/L}{2\sin (\pi W/2L)}\right)-\frac{3}{2}\Big]\Big\}
\label{Eq:EMstripe}
\end{equation}
\paragraph{The Zeeman energy.}
The energy of the magnetic field applied uniformly along the $-z$ direction favors the majority spin stripes over the minority ones ($W<L$). Using our terminology, this energy component for the slab amounts to
\begin{equation}
E_B[L,W]= \frac{(2\Lambda)^2}{a^2}\cdot \Big\{-M_0\cdot B\cdot (1-\frac{W}{L})\cdot d\cdot a^2\Big\}
\end{equation}
RESULTS.\\
I. We compute first the equilibrium stripe width $L_0$ in zero applied magnetic field. This means finding the minimum of the surface energy density
\begin{equation}
\frac{E_w(L) + E_M(W=L)}{\frac{(2\Lambda)^2}{a^2}}\doteq \varepsilon_w(L) + \varepsilon_M(L)
\end{equation}
The resulting algebraic equation for $x\doteq\frac{d}{L}$ writes
\begin{equation}
\frac{d}{dx}\Big[\epsilon_w\cdot x+ \Omega\cdot \frac{d}{a}+ \frac{2}{\pi}\cdot \Omega\cdot \frac{d}{a}\cdot x\cdot \big[\ln(\frac{\pi}{2}\cdot x) -\frac{3}{2}\big]\Big]\stackrel{!}{=}0
\end{equation}
The solution $x_0$ of this equation (written with the original parameters) writes
\begin{equation}
L_0= \frac{\pi}{2\cdot \sqrt{e}}\cdot d\cdot e^{\frac{\pi}{2}\cdot \frac{\epsilon_w}{\Omega\cdot\frac{d}{a}}}
\label{Eq:L0}
\end{equation}
This result is in agreement with the numerical work by Kaplan and Gehring\cite{Kaplan}.\\
II. Next we compute a property of the system that appears in small applied magnetic fields. In this situation, we can set $L=L_0$ and $\frac{W}{L_0}= 1-\delta$ in
$\varepsilon_M[L,W]$ and $\varepsilon_B(L,W)$ and minimize
\begin{equation}
\varepsilon_w(L)+\varepsilon_M(L_0,\delta) +\varepsilon_B(\delta)\doteq\varepsilon(L_0,\delta)
\end{equation}
with respect to the small parameter $\delta$. As $\delta$ is small, the $\ln-$ function in $\varepsilon_M[L_0,\delta]$ can be developed as a Taylor series of the small parameter $\delta$ -- with $\sin\frac{\pi}{2}(1-\delta)=\cos\frac{\pi\delta}{2}\approx 1-\frac{\pi^2\delta^2}{8}$ -- resulting in
\begin{equation}
\varepsilon(L_0,\delta)\approx \epsilon_w\cdot \frac{d}{L_0}- M_0\cdot B\cdot d\cdot a^2 \cdot \delta + \Omega \cdot \frac{d}{a} + \frac{2}{\pi}\cdot \Omega \cdot \frac{d}{a}\cdot \frac{d}{L_0}\cdot \Big[\ln(\frac{\pi}{2}\cdot \frac{d}{L_0}) + \pi^2\cdot \frac{\delta^2}{8} -\frac{3}{2}\Big]
\end{equation}
Minimizing with respect to $\delta$ we find
\begin{equation}
\mu_0\cdot M_0\cdot\delta = \frac{4}{\pi}\cdot \frac{L_0}{d}\cdot B
\end{equation}
This equation expresses a linearity of the magnetization $\mu_0\cdot M_0\cdot \delta$  with respect to the magnetic field in small applied magnetic fields. The initial slope is just $\frac{4}{\pi}\cdot \frac{L_0}{d}$. This relationship was demonstrated experimentally in Ref.\cite{Saratz}.\\
III. We solve now the coupled algebraic equations arising from the minimization of the Landau functional $\varepsilon(L,W)$ with respect to the variables  $x=\frac{d}{L}$ and $y\doteq\frac{W}{d}$:
\begin{equation}
\varepsilon[x,y]= \epsilon_w\cdot x -M_0\cdot B\cdot d\cdot a^2\cdot (1-x\cdot y) + \Omega\frac{d}{a} + \frac{2}{\pi}\cdot \Omega\frac{d}{a}\cdot x\cdot\Big[\ln\big(\frac{\pi}{2}\cdot \frac{x}{\sin\frac{\pi}{2}(x\cdot y)}\big) -\frac{3}{2}\Big]
\end{equation}
The Euler equations to this variational principle write
\begin{eqnarray}
\frac{\partial}{\partial x}(..)&=&\epsilon_w + M_0\cdot B\cdot d\cdot a^2\cdot y + \frac{2}{\pi}\cdot \Omega\frac{d}{a}\cdot \Big[\ln\big(\frac{\pi}{2}\cdot \frac{x}{\sin\frac{\pi}{2}(x\cdot y)}\big) -\frac{3}{2}\Big]\nonumber\\
&+& \frac{2}{\pi}\cdot \Omega\frac{d}{a}\cdot x\cdot \Big[\frac{1}{x}-
\frac{\pi}{2}\cdot y\cdot \frac{\cos\frac{\pi}{2}(x\cdot y)}{\sin\frac{\pi}{2}(x\cdot y)}\Big]=0\nonumber\\
\frac{\partial}{\partial y}(..)&=&M_0\cdot B\cdot d\cdot a^2\cdot x-\frac{2}{\pi}\cdot \Omega\cdot \frac{d}{a}\cdot x\cdot\frac{\pi}{2}\cdot x\cdot\frac{\cos\frac{\pi}{2}(x\cdot y)}{\sin\frac{\pi}{2}(x\cdot y)}=0
\end{eqnarray}
Solving the $y$-Euler equation produces the relation
\begin{equation}
\Omega\cdot \frac{d}{a}\cdot x\cdot \frac{\cos\frac{\pi}{2}(x\cdot y)}{\sin\frac{\pi}{2}(x\cdot y)}= M_0\cdot B\cdot d\cdot a^2
\label{Eq:yEuler}
\end{equation}
Inserting this relation into the $x$-Euler equation produces a remarkable cancellation of the terms containing the magnetic field and leads to the equation
\begin{equation}
\ln\big(\frac{\pi}{2}\cdot \frac{x}{\sin\frac{\pi}{2}(x\cdot y)}\big) = \frac{-\epsilon_w}{\frac{2}{\pi}\cdot \Omega\cdot \frac{d}{a}} + \frac{1}{2}
\end{equation} or
\begin{equation}
\frac{x}{\sin\frac{\pi}{2}(x\cdot y)} = x_0
\label{Eq:x0}
\end{equation}
$x_0$ being the value of the parameter $x$ at $B=0$. Reusing this result in the $y$-Euler equation allows the variable $y$ to be eliminated by virtue of
\begin{equation}
\underbrace{\cos\frac{\pi}{2}(x\cdot y)}_{\sqrt{1-(\frac{x}{x_0})^2}}\cdot \Omega\cdot \frac{d}{a}\cdot x_0 = M_0\cdot B\cdot d\cdot a^2
\end{equation}
This equation can be solved with respect to $x$. Its solution can be written as (returning to the well-known parameters)
\begin{equation}
\frac{L}{d}= \frac{L_0}{d}\cdot \frac{1}{\sqrt{1-\frac{B^2}{B_t^2}}}
\end{equation}
with the parameter
\begin{equation}
B_t\doteq \frac{1}{2}\cdot \mu_0\cdot M_0\cdot \frac{d}{L_0}
\end{equation}
defining a critical field $B_t$ at which $L$ diverges to infinity and the film becomes (almost) uniformly magnetized (monodomain state)\footnote{The film is almost monodomain as the relation
\begin{equation}
W = \frac{2}{\pi}\cdot L\cdot \arcsin(\frac{L_0}{L})
\end{equation}
implies that $W\rightarrow \frac{2}{\pi}\cdot L_0$ as $L\rightarrow \infty$, so that some minority spin domains persist even in the almost monodomain state.}

\setcounter{section}{0} 
\renewcommand{\thesection}{\thechapter.\Alph{section}} 
\setcounter{equation}{0}
\renewcommand{\theequation}{\thesection\arabic{equation}}
\section*{Appendices}
\section{Magnetostatic energy of the stripe system.}
\label{App:Stripes}
We provide some details of the computation of the total dipolar energy of the stripe system. We start from the formula (\ref{Eq:EMq})
\begin{eqnarray}
E_M[n_z]&=& +\frac{1}{2\pi}\cdot \frac{\Omega}{2\pi}\cdot \frac{1}{a^3}\cdot \int d^2\rho \int d^2\rho' \cdot n_z(\vec \rho)\cdot n_z(\vec \rho')\cdot \nonumber\\
&\cdot & \int d^2q
\cdot e^{i\cdot \vec q\cdot (\vec \rho-\vec \rho')}\cdot \frac{1-e^{-q\cdot d}}{q}
\end{eqnarray}
and insert a field $n_z(x,y)$ that is assumed to be uniform along $y$ and to have a profile only along the $x$-direction. The uniformity along $y$ allows the integration in the variables $y$ and $y'$ to be performed with the use of the formula
\begin{equation}
\int dy\int dy'e^{iq_y\cdot( y-y')}=2\pi\cdot 2\Lambda\cdot \delta(q_y)
\end{equation}
The expression of the dipolar energy simplifies accordingly. Let us introduce, for the simplicity of writing, the function
\begin{equation}
f(q)\doteq\frac{1-e^{-q\cdot d}}{q\cdot d}
\end{equation}
and write
\begin{eqnarray}
E_M[n_z(x)]&=& \frac{1}{2\pi}\cdot \frac{\Omega}{2\pi}\cdot \frac{1}{a^3}\cdot d\cdot 2\pi\cdot 2\Lambda \cdot \int dx\cdot n_z(x) \int dx'\cdot n_z(x')\nonumber\\
&\cdot &\int dq_x\int dq_y \delta(q_y)\cdot e^{iq_x\cdot( x-x')}\cdot f(q)
\nonumber \\
&=&
\frac{1}{2\pi}\cdot \frac{\Omega}{2\pi}\cdot \frac{1}{a^3}\cdot d\cdot 2\pi\cdot 2\Lambda \cdot \int dx\cdot  n_z(x) \int dx'\cdot  n_z(x')\nonumber\\
&\cdot & \int dq_x\cdot e^{iq_x\cdot( x-x')}\cdot f(\vert q_x\vert)\nonumber\\
&=& \frac{1}{2\pi}\cdot \frac{\Omega}{2\pi}\cdot \frac{1}{a^3}\cdot d\cdot 2\pi\cdot 2\Lambda \cdot 2\cdot \int dx \cdot n_z(x) \int dx'\cdot  n_z(x')\nonumber\\
&\cdot &\int_0^{\infty} dq\cdot \cos\big(q\cdot( x-x')\big)\cdot f(q)
\end{eqnarray}
This expression reduces to integrating only over the $n_z(x)$-profile, which must be specified for further computational steps. The profile we aim at considering shall be quite general, but have a $2L$-periodicity. By the choice of the origin of the coordinate system, the profile can be chosen to have a definite parity with respect to change of sign. We chose the parity to be positive and, accordingly, the profile can be expressed by a $\cos$-Fourier series:
\begin{eqnarray}
n_z(x) &=& \sum_{n=0}^\infty a_n \cos k_n\cdot x\nonumber\\
k_n &=& \frac{n\cdot \pi}{L}\nonumber\\
a_0 &=& \frac{1}{2L}\int_{-L}^L n_z(x)\cdot dx\nonumber\\
a_n&=& \frac{2}{2L}\int_{-L}^L n_z(x)\cdot \cos(k_n\cdot x)\cdot dx
\end{eqnarray}
As the profile is even with respect to change of sign, the sum can be
restricted to $k_n\geq 0$. The multiplicative factor of '2' appearing in the
Fourier coefficients for $k_n\not= 0$ accounts for the components with negative $k_n$ values neglected when the sum was restricted to positive $k_n$ values. The factor of ''2'' is, accordingly,  missing in the $a_0$-coefficient.
To adapt to the even profile we use the trigonometric identity
\begin{equation}
\cos\big(q\cdot( x-x')\big)= \cos(q\cdot x)\cdot \cos(q\cdot x')-\sin(q\cdot x)\cdot \sin(q\cdot x')
\end{equation}
The integrals over $x$ and $x'$ involving the $\sin$-terms vanish because of symmetry so that
\begin{eqnarray}
E_M[n_z(x)]&=& \frac{1}{2\pi}\cdot \frac{\Omega}{2\pi}\cdot \frac{1}{a^3}\cdot d\cdot 2\pi\cdot 2\Lambda \cdot 2\nonumber\\
&\!\cdot \! &\sum_{n,m}a_n\cdot a_m\nonumber\\
&\!\cdot\! &\!\int dx\!\int dx'\!\cdot \! \cos (k_n\cdot x)\cdot \cos (k_m\cdot x')\int dq\!\cdot \!\cos(q\cdot x)\!\cdot \!\cos(q\cdot x')\cdot \!f(q)
\end{eqnarray}
Using
\begin{eqnarray}
&&\int dx \cos (k_n\cdot x)\cdot \cos (q\cdot x) = \frac{1}{2}\int dx \big(\cos\big((k_n-q)\cdot x\big) + \cos\big((k_n+q)\cdot x\big) \big)\nonumber\\
&=& \pi\cdot \big(\delta(k_n-q) + \delta(k_n+q)\big)
\end{eqnarray}
we eliminate the integration over $x$ to obtain
\begin{eqnarray}
E_M[n_z]&=&\frac{1}{2\pi}\cdot \frac{\Omega}{2\pi}\cdot \frac{1}{a^3}\cdot d\cdot 2\pi\cdot 2\Lambda \cdot 2\cdot \pi\cdot \sum_{n,m}a_n\cdot a_m\nonumber\\
&\cdot & \int dx' \cos (k_m\cdot x')\cdot \int dq \cos(q\cdot x')\cdot \big(\delta(k_n-q) + \delta(k_n+q)\big)\cdot f(q)
\end{eqnarray}
When evaluating the $q$-integration, we must distinguish between $n=0$ and $n\not= 0$. In the case of $n=0$ both delta functions contribute and we obtain
\begin{equation}
\int_0^{\infty} dq \cos(q\cdot x')\cdot \big(\delta(k_n-q) + \delta(k_n+q)\big)\cdot f(q)= \frac{1}{2}\cdot 2\cdot f(0)=f(0)
\end{equation}
The contribution of the $n=0$ components amounts to
\begin{eqnarray}
E_M[n_z]&=&\frac{1}{2\pi}\cdot \frac{\Omega}{2\pi}\cdot \frac{1}{a^3}\cdot d\cdot 2\pi\cdot 2\Lambda \cdot 2\cdot \pi\nonumber\\
&\cdot & \sum_{m}a_0\cdot a_m\nonumber\\
&\cdot & \underbrace{\int dx' \cos k_m\cdot x'}_{2\Lambda\cdot \delta_{k_m,0}}\cdot f(0)\nonumber\\
&=& \frac{1}{2\pi}\cdot \frac{\Omega}{2\pi}\cdot \frac{1}{a^3}\cdot d\cdot 2\pi\cdot 2\Lambda \cdot 2\cdot \pi\cdot 2\Lambda\cdot a_0^2\cdot f(0)\nonumber\\
&=& (2\Lambda)^2\cdot d\cdot \frac{1}{a^3}\cdot \Omega\cdot a_0^2\cdot f(0)
\end{eqnarray}
For $k_n>0$, when integrating over a positive $q$ only $\delta(k_n-q)$ contributes non-vanishingly. Accordingly, the $n\not=0$ components write
\begin{eqnarray}
&&\frac{1}{2\pi}\cdot \frac{\Omega}{2\pi}\cdot \frac{1}{a^3}\cdot d\cdot 2\pi\cdot 2\Lambda \cdot 2\cdot \pi\cdot \sum_{n,m}a_n\cdot a_m\nonumber\\
&\cdot & \underbrace{\int dx' \cos k_m\cdot x'\cdot \cos(k_n\cdot x')}_{\frac{2\Lambda}{2}\cdot \big(\delta_{k_n,k_m}+\delta_{k_n,-k_m}\big)}\cdot f(q)\nonumber\\
&=& (2\Lambda)^2\cdot d\cdot \frac{1}{2}\Omega\cdot \frac{1}{a^3}\cdot \sum_{n=1}^\infty a_n^2\cdot f(k_n)
\end{eqnarray}
Summarizing, the dipolar energy functional over a general profile along $x$ with $2L$-periodicity writes
\begin{equation}
E_M[\{a_n\}]= (2\Lambda)^2\cdot d\cdot \frac{1}{a^3}\cdot \frac{1}{2}\cdot \Omega\cdot \Big(2\cdot a_0^2\cdot f(0) + \sum_{n=1}^\infty a_n^2\cdot f(k_n)\Big)
\label{Eq:EMgeneral}
\end{equation}
We evaluate the dipolar energy for a special profile consisting of stripes with opposite spin polarization, separated by infinitely sharp domain walls. This is an abstract situation which is only realized in the context of the Ising model but the results obtained with it are paramount for a more realistic profile where $n_z(x)$ changes continuously between adjacent stripes. In view of the fact that we will also include a magnetic field applied along the $z$-axis, we will allow for the stripes with minority spins to have a smaller width than the stripes with majority spins. This leads us to a profile defined as follows:
\begin{eqnarray}
n_z(x)&=& 1\quad [-L,-\frac{W}{2}]\nonumber\\
n_z(x)&=& -1\quad [-\frac{W}{2},+\frac{W}{2}]\nonumber\\
n_z(x)&=& 1\quad [+\frac{W}{2},+L]
\end{eqnarray}
The Fourier coefficients read
\begin{eqnarray}
a_0 &=& \frac{L-W}{L}\nonumber\\
a_n&=& -\frac{4}{n\cdot\pi}\cdot\sin(\frac{n\pi W}{2L})
\label{Eq:ansharp}
\end{eqnarray}
and the functional for the magnetostatic energy writes
\begin{eqnarray}
E_M[\{a_n\}]&=& (2\Lambda)^2\cdot d\cdot \frac{1}{a^3}\cdot \frac{1}{2}\cdot \Omega\cdot \nonumber\\
&\cdot & \Big(2\cdot \big(\frac{L-W}{L}\big)^2 + \frac{8\cdot L}{d\cdot \pi^3}\sum_{n=1}^\infty \frac{1}{n^3}\cdot \big(1-\cos(\frac{n\cdot \pi\cdot W}{L})\big)\cdot \big(1-e^{-\frac{n\cdot \pi \cdot d}{L}}\big)\Big)
\end{eqnarray}
We evaluate
\begin{equation}
{\cal S}\doteq\sum_{n=1}^\infty \frac{1}{n^3}\cdot \big(1-\cos(\frac{n\cdot \pi\cdot W}{L})\big)\cdot \big(1-e^{-\frac{n\cdot \pi \cdot d}{L}}\big)
\end{equation}
in the limit of small $\frac{d}{L}$ with the help of AI\footnote{The author used Gemini 2.0 to perform the various sums and the analytical expansions of the trilogarithm function and related series summations. Following this assistance, the author reviewed, verified, and edited the resulting mathematical derivations for accuracy and takes full responsibility for the final content.}.\\
Step 1. The expression is expanded as a linear combination of simpler series:\\
\begin{eqnarray}
{\cal S}=\sum _{n=1}^{\infty }\frac{1}{n^{3}}
-\sum _{n=1}^{\infty }\frac{\cos (\frac{n\pi W}{L})}{n^{3}}
-\sum _{n=1}^{\infty }\frac{e^{-\frac{n\pi d}{L}}}{n^{3}}
+\sum _{n=1}^{\infty }\frac{\cos (\frac{n\pi W}{L})e^{-\frac{n\pi d}{L}}}{n^{3}}
\end{eqnarray}
Step 2: Each component can be expressed using standard functions:
\begin{equation}
\sum _{n=1}^{\infty }\frac{1}{n^{3}}=\zeta (3)
\end{equation}
\begin{equation}
\sum _{n=1}^{\infty }\frac{\cos (nx)}{n^{3}}=\text{Re}(\text{Li}_{3}(e^{ix}))
\end{equation}
(where \(x=\frac{\pi W}{L}\))
\begin{equation}
\sum _{n=1}^{\infty }\frac{e^{-ny}}{n^{3}}=\text{Li}_{3}(e^{-y})
\end{equation}
(where \(y=\frac{\pi d}{L}\))
\begin{equation}
\sum _{n=1}^{\infty }\frac{\cos (nx)e^{-ny}}{n^{3}}=\text{Re}(\text{Li}_{3}(e^{i(x+iy)}))=\text{Re}(\text{Li}_{3}(e^{i\frac{\pi W}{L}-\frac{\pi d}{L}}))
\end{equation}
Step 3: The total sum \({\cal S}\) is given by:\\
\begin{equation}
{\cal S}=\zeta (3)-\text{Re}(\text{Li}_{3}(e^{i\frac{\pi W}{L}}))-\text{Li}_{3}(e^{-\frac{\pi d}{L}})+\text{Re}(\text{Li}_{3}(e^{i\frac{\pi W}{L}-\frac{\pi d}{L}}))\end{equation}
This formula is valid if \(d/L>0\). If \(W/L\) is an even integer, the last two terms may need careful evaluation based on series convergence conditions.\\
To expand the sum \({\cal S}\) for small \(\delta =\pi d/L\) without term-by-term expansion of the exponential (which leads to divergent sums), we use the asymptotic expansion of the trilogarithm near its singularity at \(z=1\). The general expansion for \(\text{Li}_{k}(e^{-\delta })\) as \(\delta \rightarrow 0\) is given by:
\begin{equation}
\text{Li}_{k}(e^{-\delta })=\frac{(-\delta )^{k-1}}{(k-1)!}\left(H_{k-1}-\ln \delta \right)+\sum _{m=0,m\ne k-1}^{\infty }\frac{\zeta (k-m)}{m!}(-\delta )^{m}
\end{equation}
where \(H_{k-1}\) is a harmonic number (\(\text{for\ }k=3,H_{2}=3/2\))\\
Step 1: Expand the purely real term, using the formula for \(k=3\):
\begin{equation}
\text{Li}_{3}(e^{-\delta })\approx \zeta (3)-\zeta (2)\delta +\frac{\delta ^{2}}{2}\left(\frac{3}{2}-\ln \delta \right)-\frac{\zeta (0)}{6}\delta ^{3}+\dots
\end{equation}
Step 2: Expand the complex term. Let \(\theta =\frac{\pi W}{L}\). We examine \begin{equation}
\text{Re}[\text{Li}_{3}(e^{i\theta -\delta })-\text{Li}_{3}(e^{i\theta })]
\end{equation}
Since \(e^{i\theta }\ne 1\) (assuming \(W/L\) is not an even integer), the function \(\text{Li}_{3}(z)\) is analytic at \(z=e^{i\theta }\). We can therefore use a standard Taylor expansion in terms of \(\delta \):
\begin{equation}
\text{Li}_{3}(e^{i\theta -\delta })\approx \text{Li}_{3}(e^{i\theta })+\frac{d}{dz}\text{Li}_{3}(z)\Big|_{z=e^{i\theta }}(e^{i\theta -\delta }-e^{i\theta })+\dots
\end{equation}
Using the property \(\frac{d}{dz}\text{Li}_{n}(z)=\frac{1}{z}\text{Li}_{n-1}(z)\):
\begin{equation}
\text{Li}_{3}(e^{i\theta }e^{-\delta })\approx \text{Li}_{3}(e^{i\theta })-\delta \text{Li}_{2}(e^{i\theta })+\frac{\delta ^{2}}{2}\left(\text{Li}_{1}(e^{i\theta })\right)+\dots
\end{equation}
Taking the real part:
\begin{equation}
\text{Re}(\text{Li}_{3}(e^{i\theta -\delta }))\approx \text{Re}(\text{Li}_{3}(e^{i\theta }))-\delta \text{Re}(\text{Li}_{2}(e^{i\theta }))+\frac{\delta ^{2}}{2}\text{Re}(\text{Li}_{1}(e^{i\theta }))
\end{equation}
where:
\begin{equation}
\text{Re}(\text{Li}_{2}(e^{i\theta }))=\sum \frac{\cos (n\theta )}{n^{2}}=\frac{\theta ^{2}}{4}-\frac{\pi \theta }{2}+\frac{\pi ^{2}}{6}
\end{equation}
\begin{equation}
\text{Re}(\text{Li}_{1}(e^{i\theta }))=-\ln (2\sin (\theta /2))
\end{equation}
Step 3: Combine into the final result. Substituting these back into
\begin{equation}
{\cal S}=[\zeta (3)-\text{Li}_{3}(e^{-\delta })]-[\text{Re}(\text{Li}_{3}(e^{i\theta }))-\text{Re}(\text{Li}_{3}(e^{i\theta -\delta }))]\end{equation}
\begin{equation}
{\cal S}\approx \left[\zeta (2)\delta -\frac{\delta ^{2}}{2}\left(\frac{3}{2}-\ln \delta \right)\right]-\left[\delta \text{Re}(\text{Li}_{2}(e^{i\theta }))-\frac{\delta ^{2}}{2}\text{Re}(\text{Li}_{1}(e^{i\theta }))\right]
\end{equation}
\begin{equation}
{\cal S}\approx \delta \left(\frac{\pi ^{2}}{6}-\left(\frac{\theta ^{2}}{4}-\frac{\pi \theta }{2}+\frac{\pi ^{2}}{6}\right)\right)+\frac{\delta ^{2}}{2}\left(\ln \delta -\frac{3}{2}-\ln (2\sin (\theta /2))\right)
\end{equation}
Final Expansion for Small \(d/L\):
\begin{eqnarray}
{\cal S}&\approx &\frac{\pi d}{L}\left(\frac{\pi ^{2}W}{2L}-\frac{\pi ^{2}W^{2}}{4L^{2}}\right)\nonumber\\
&+&\frac{1}{2}\left(\frac{\pi d}{L}\right)^{2}\left[\ln \left(\frac{\pi d/L}{2\sin (\pi W/2L)}\right)-\frac{3}{2}\right]\nonumber\\
&+&O\left(\left(\frac{d}{L}\right)^{3}\right)
\end{eqnarray}
When considering ${\cal S}$ for finding the total dipolar energy, a remarkable cancellation of terms means that adding $2\cdot \big(\frac{L-W}{L}\big)^2$ to
\begin{equation}
\frac{8\cdot L}{d\cdot \pi^3}\cdot \frac{\pi d}{L}\left(\frac{\pi ^{2}W}{2L}-\frac{\pi ^{2}W^{2}}{4L^{2}}\right)
\end{equation}
gives just ''2'', so that the total dipolar energy for the special profile we have imposed writes, in the variables $L,W$
\begin{eqnarray}
E_M[L,W]\approx \frac{(2\Lambda)^2}{a^2}\cdot\Big\{\Omega\cdot \frac{d}{a} + \frac{2}{\pi}\cdot \Omega\cdot \frac{d}{a}\cdot \frac{d}{L}\Big[\ln \left(\frac{\pi d/L}{2\sin (\pi W/2L)}\right)-\frac{3}{2}\Big]\Big\}
\label{Eq:EMsharp}
\end{eqnarray}

\section{Magnetostatic energy of a stripe system with walls of finite width.}
\label{App:Ramp}
\setcounter{equation}{0}
The profile (\ref{Eq:ansharp}) switches abruptly from one domain to the next.
A real wall has a finite width, and it is natural to ask what the total
magnetostatic energy (\ref{Eq:EMsharp}) becomes when the switch is spread over
a wall of width $t$.\footnote{This Appendix was worked out with the assistance
of Claude (Opus 5, Anthropic), August 2026, which proposed the convolution
argument, evaluated the two series and checked the closed forms numerically;
the script that performs the check is deposited with this document
(\texttt{anc/verify\_11B\_claude.pl}). Following this assistance, the author
reviewed, verified and edited the resulting derivation for accuracy and takes
full responsibility for the final content.} We treat here the case
\begin{equation}
d\ll t\ll L
\end{equation}
i.e. a wall that is wide compared with the thickness of the slab but narrow
compared with the width of a stripe, and we let $n_z(x)$ change linearly across
it, as in section \ref{Chap:Topo}.1. The two ramps must not overlap, which
requires $t<W$ and $t<2L-W$.\\
The computation is spared entirely if one notices that the smoothed profile is
the convolution of the abrupt one with a normalized box of width $t$:
\begin{equation}
n_z^{(t)}(x)= \frac{1}{t}\int_{-\frac{t}{2}}^{+\frac{t}{2}}n_z(x-u)\cdot du
\end{equation}
Away from a wall the running average of a constant is that constant; across a
jump it interpolates linearly over exactly the width $t$. As a convolution in
real space is a product in Fourier space, the coefficients (\ref{Eq:ansharp})
are simply multiplied by the form factor of the box. With
\begin{equation}
\tau\doteq \frac{\pi\cdot t}{2L}
\end{equation}
one has
\begin{eqnarray}
a_0^{(t)} &=& \frac{L-W}{L}\nonumber\\
a_n^{(t)} &=& -\frac{4}{n\cdot \pi}\cdot\sin\Big(\frac{n\pi W}{2L}\Big)\cdot \frac{\sin (n\tau)}{n\tau}
\end{eqnarray}
The coefficient $a_0$ is untouched, as it must be: the ramp is antisymmetric
about the centre of the wall and therefore moves no net magnetization. The new
factor $\frac{\sin n\tau}{n\tau}$ cuts the series off at $n\approx
\frac{1}{\tau}$, and because $t\gg d$ this happens \textbf{before} the cut-off
at $n\approx\frac{L}{\pi d}$ carried by $f(k_n)$. This is the whole physical
content of what follows: the short distance cut-off of the dipolar sum is no
longer the thickness of the slab but the width of the wall.\\
We insert the new coefficients into (\ref{Eq:EMgeneral}). Since $d\ll t$, the
argument $n\delta$ of $f$, with $\delta=\frac{\pi d}{L}$ as in Appendix
\ref{App:Stripes}, is small over the whole range of $n$ in which the summand
survives, so that $f(k_n)=\frac{1-e^{-n\delta}}{n\delta}\approx
1-\frac{n\delta}{2}$ may be used. With $\theta=\frac{\pi W}{L}$,
\begin{equation}
\sum_{n=1}^{\infty}\big(a_n^{(t)}\big)^2\cdot f(k_n)= \frac{8}{\pi^2\cdot \tau^2}\cdot\Big[P-\frac{\delta}{2}\cdot Q\Big]
\end{equation}
with
\begin{equation}
P\doteq \sum_{n=1}^{\infty}\frac{\big(1-\cos n\theta\big)\cdot \sin^2 (n\tau)}{n^{4}}\quad\quad
Q\doteq \sum_{n=1}^{\infty}\frac{\big(1-\cos n\theta\big)\cdot \sin^2 (n\tau)}{n^{3}}
\end{equation}
$P$ is elementary. Writing $\sin^2 n\tau = \frac{1-\cos 2n\tau}{2}$ turns it
into second differences of $\sum_n \frac{\cos n\phi}{n^4}$, which is a
polynomial of the fourth degree in $\phi$ on $[0,2\pi]$, so that the expansion
in $\tau$ terminates and
\begin{equation}
\frac{P}{\tau^2}= \frac{\pi\cdot \theta}{2}-\frac{\theta^2}{4}-\frac{\pi\cdot \tau}{3}
\end{equation}
holds \textbf{exactly}. The first two terms are the value of the abrupt
profile, the ones that cancelled against $2\cdot a_0^2$ in Appendix
\ref{App:Stripes}; the third one is new. The same manipulation applied to $Q$
produces second differences of $\text{Re}(\text{Li}_{3}(e^{i\phi}))$, which is
not a polynomial: it is precisely its logarithmic singularity at $\phi=0$,
\begin{equation}
\text{Li}_{3}(e^{-\delta })-\zeta(3)\approx -\frac{\delta^2}{2}\Big(\frac{3}{2}-\ln \delta\Big)
\end{equation}
together with $\frac{d^2}{d\phi^2}\text{Re}(\text{Li}_{3}(e^{i\phi}))=\ln
\big(2\sin\frac{\phi}{2}\big)$, that gives
\begin{equation}
\frac{Q}{\tau^2}= -\Big[\ln \frac{2\tau}{2\sin\frac{\theta}{2}}-\frac{3}{2}\Big]
\end{equation}
Collecting the two contributions and adding $2\cdot a_0^2$, the same
cancellation as in Appendix \ref{App:Stripes} removes everything of order
$\big(\frac{W}{L}\big)^{0,1,2}$ and leaves
\begin{equation}
2\cdot a_0^2+\sum_{n=1}^{\infty}\big(a_n^{(t)}\big)^2 f(k_n)= 2-\frac{4}{3}\cdot\frac{t}{L}+\frac{4}{\pi}\cdot\frac{d}{L}\cdot\Big[\ln \left(\frac{\pi t/L}{2\sin (\pi W/2L)}\right)-\frac{3}{2}\Big]
\end{equation}
The middle term of this expression, $-\frac{4}{3}\cdot\frac{t}{L}$, is the loss
of shape anisotropy energy of the spins that lie inside the wall and no longer
point along $z$. Per unit cell of wall surface it amounts to
$-\frac{2}{3}\cdot\Omega\cdot\frac{t}{a}$, which is exactly the magnetostatic
term $+\Omega\cdot\frac{d}{a}\cdot\cos^2\theta(y)$ of section
\ref{Chap:Topo}.1 evaluated on a linear profile. It is therefore already
contained in the wall energy $\epsilon_w$, through the combination
$\lambda\cdot S^2-\Omega\cdot\frac{d}{a}$ of (\ref{Eq:epsw}), and must not be
counted a second time. Leaving it there, what is to be combined with
$\epsilon_w\cdot\frac{d}{L}$ in section \ref{Chap:Topo}.2 is the remainder
\begin{equation}
E_M[L,W,t]\approx \frac{(2\Lambda)^2}{a^2}\cdot\Big\{\Omega\cdot \frac{d}{a} + \frac{2}{\pi}\cdot \Omega\cdot \frac{d}{a}\cdot \frac{d}{L}\Big[\ln \left(\frac{\pi t/L}{2\sin (\pi W/2L)}\right)-\frac{3}{2}\Big]\Big\}
\label{Eq:EMramp}
\end{equation}
COMMENTS.\\
1. Compared with (\ref{Eq:EMsharp}) there is exactly one change: the thickness
$d$ inside the logarithm is replaced by the width $t$ of the wall. The short
distance cut-off of the dipolar sum is the larger of the two lengths, and
setting $t=d$ returns the two expressions to one another, as it must.\\
2. The only change to the results of section
\ref{Chap:Topo}.2 is therefore the argument of the logarithm. Repeating the minimization
that produced (\ref{Eq:L0}), at $W=L$ and with $\frac{t}{L}=\frac{t}{d}\cdot x$,
\begin{equation}
\ln \Big(\frac{\pi\cdot t\cdot x_0}{2\cdot d}\Big)= \frac{1}{2}-\frac{\pi}{2}\cdot\frac{\epsilon_w}{\Omega\cdot\frac{d}{a}}
\end{equation}
i.e.
\begin{equation}
L_0= \frac{\pi}{2\cdot \sqrt{e}}\cdot t\cdot e^{\frac{\pi}{2}\cdot \frac{\epsilon_w}{\Omega\cdot\frac{d}{a}}}
\end{equation}
The equilibrium stripe width is the one of (\ref{Eq:L0}) multiplied by
$\frac{t}{d}$. Everything downstream is unchanged in form: the logarithm enters
the minimization with respect to $\delta$ and the two Euler equations of
RESULT III only through its derivative, so that the initial slope
$\frac{4}{\pi}\cdot\frac{L_0}{d}$, the relation $\frac{x}{\sin\frac{\pi}{2}(x
\cdot y)}=x_0$, the law $\frac{L}{d}=\frac{L_0}{d}\cdot\big(1-\frac{B^2}{B_t^2}
\big)^{-\frac{1}{2}}$ and $B_t=\frac{1}{2}\cdot\mu_0\cdot M_0\cdot
\frac{d}{L_0}$ all survive verbatim, with the new $L_0$.\\
3. This also removes an asymmetry between the two sections of the Chapter. The
crossover length (\ref{Eq:Lambdac}) of section \ref{Chap:Topo}.1 carries the
wall width $w$ as its prefactor, whereas $L_0$ carried the thickness $d$; the
two expressions have the same structure once the profile of section
\ref{Chap:Topo}.2 is given a wall of finite width as well.

\backmatter

\end{document}